\documentclass[11pt,a4paper]{article}

\usepackage{graphicx}
\usepackage{float}
\usepackage{afterpage}
\usepackage{epsfig,cite}
\usepackage{amssymb}
\usepackage{amsmath}
\usepackage{dsfont}
\usepackage{multirow}
\usepackage{url,hyperref}

\usepackage{url}
\usepackage{hyperref}

\newcommand{\be}{\begin{equation}}
\newcommand{\ee}{\end{equation}}
\newcommand{\bi}{\begin{itemize}}
\newcommand{\ei}{\end{itemize}}
\newcommand{\ben}{\begin{enumerate}}
\newcommand{\een}{\end{enumerate}}
\newcommand{\la}{\left\langle}
\newcommand{\ra}{\right\rangle}
\newcommand{\lc}{\left[}
\newcommand{\rc}{\right]}
\newcommand{\lp}{\left(}
\newcommand{\rp}{\right)}

\newcommand{\gsim}{\gtrsim}
\newcommand{\lsim}{\lesssim}

\newcommand{\draft}[1]{}

\def\beq{\begin{equation}}  
\def\eeq{\end{equation}}

\usepackage{xcolor}

\begin{document}
\sloppy
\begin{figure}[h]
  \includegraphics[width=0.32\textwidth]{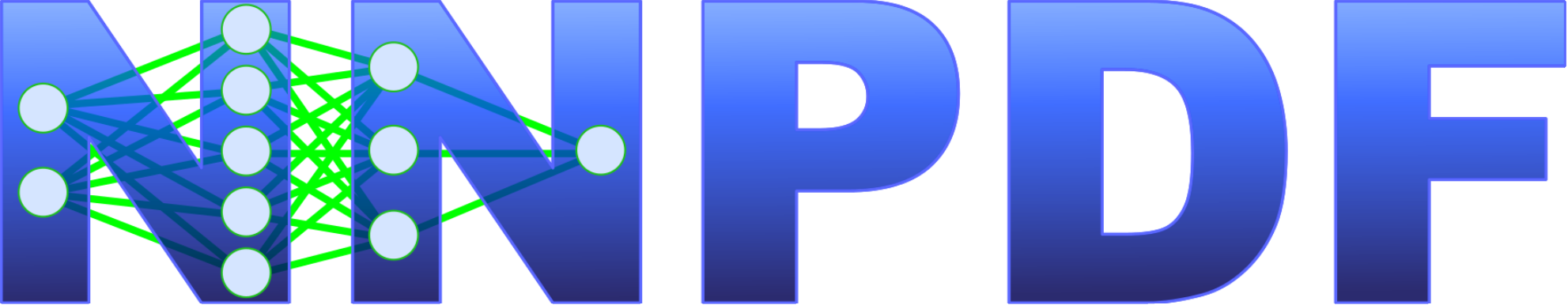}
\end{figure}
\vspace{-2.0cm}
\begin{flushright}
Edinburgh 2016/06\\
CERN-TH-2016-087\\
OUTP-16-03P \\
TIF-UNIMI-2016-3
\end{flushright}
\vspace{.3cm}

\begin{center}
{\Large \bf A Determination of the Charm Content of the Proton}
\vspace{.7cm}

{\bf  The NNPDF Collaboration:}\\

Richard~D.~Ball$^{1}$, Valerio~Bertone$^{2}$, Marco~Bonvini$^{2}$,
Stefano~Carrazza$^{3}$, Stefano~Forte$^4$, Alberto~Guffanti$^{5}$,
Nathan~P.~Hartland$^{2}$, Juan~Rojo$^{2}$ and Luca Rottoli$^{2}$

\vspace{.3cm}
{\it ~$^1$ The Higgs Centre for Theoretical Physics, University of Edinburgh,\\
  JCMB, KB, Mayfield Rd, Edinburgh EH9 3JZ, Scotland\\
  ~$^2$ Rudolf Peierls Centre for Theoretical Physics, 1 Keble Road,\\ University of Oxford, OX1 3NP Oxford, United Kingdom\\
  ~$^3$ Theory Division, CERN, Switzerland\\
~$^4$ Dipartimento di Fisica, Universit\`a di Milano and
INFN, Sezione di Milano,\\ Via Celoria 16, I-20133 Milano, Italy\\
~$^5$ Dipartimento di Fisica, Universit\`a
di Torino and INFN, Sezione di Torino,\\
Via P. Giuria 1, I-10125, Turin, Italy\\}

\vspace{.5cm}

{\bf \large Abstract}

\end{center}

We present an unbiased determination of the charm content of the proton,
in which the charm parton distribution function (PDF)
is parametrized on the same footing as the light quarks and the gluon  
in a global PDF analysis.
This determination relies on the NLO
calculation of deep-inelastic structure functions in the
FONLL scheme, generalized to account for
massive charm-initiated contributions.
When the EMC charm structure function dataset is included,
it is well described by the fit, and PDF uncertainties in the fitted charm 
PDF are significantly reduced. We then find that the fitted charm PDF vanishes 
within uncertainties at a scale $Q\sim1.6$~GeV for all  $x\lsim0.1$, 
independent of the value of $m_c$ used in the coefficient functions.
We also find some evidence that the charm PDF at
large $x\gsim0.1$ and low scales does not vanish, but rather has an
``intrinsic'' component, very weakly scale dependent and almost independent
of the value of $m_c$, carrying less than $1\%$ of the total momentum of
the proton. 
The uncertainties in all other PDFs are only slightly increased 
by the inclusion of fitted charm,
while the dependence of these PDFs on $m_c$ is  reduced.
The increased stability with respect to $m_c$ persists at
high scales and is the main implication of our results for LHC
phenomenology. Our results show that if the EMC data are correct, then
the usual approach in which charm is perturbatively generated leads to
biased results for the charm PDF, though at small $x$ this bias could
be reabsorbed if the uncertainty due to the charm mass and missing
higher orders were included.
We show that LHC data  for processes
such as  
high $p_T$ and large rapidity charm pair production and $Z+c$ production,
have the potential to confirm or disprove the implications of the EMC data.

\clearpage

\tableofcontents

\clearpage

\section{Introduction}
\label{sec:introduction}

Current general-purpose global PDF sets~\cite{Ball:2014uwa,Dulat:2015mca,Harland-Lang:2014zoa,
  Alekhin:2013nda,Abramowicz:2015mha,Accardi:2016qay,Jimenez-Delgado:2014twa}
assume that the charm
PDF is perturbatively generated through pair production 
from gluons and light quarks.
This assumption could be a limitation, and possibly a source of bias,
for at least three different reasons. First, the
charm PDF might have a non-vanishing ``intrinsic'' component 
of non-perturbative
origin, such that it does not vanish at any scale within the perturbative
region (see~\cite{Brodsky:2015fna} for a recent review).
Second, even
if the charm PDF is purely perturbative in origin and thus  vanishes below 
the physical threshold for its production, it is unclear what the value of this
physical threshold is, as it is  related to the charm pole mass, which 
in itself is not known very precisely. Finally, even if charm is
entirely perturbative, and we knew accurately its production threshold,
in practice massive charm production cross-sections are only
known at low perturbative order (at most NLO) and it is unclear whether
this leads to sufficiently accurate predictions.

All these difficulties are solved if the charm quark PDF is
parametrized and determined along with light quark and gluon
PDFs. Whether or not the PDF vanishes, and, if it does, at which scale,
will then be answered by the fit. From this point of view, the
distinction between the perturbatively generated component, and a
possible intrinsic component (claimed to be
power suppressed~\cite{Collins:1998rz,Brodsky:2015fna}  
before mixing with other PDFs
upon perturbative evolution) becomes irrelevant. This is quite
advantageous because the ensuing PDF set 
automatically incorporates in the standard PDF
uncertainty the theoretical uncertainty related to the size of the
perturbative charm component due to uncertainty in
the value of the charm mass. Also,
the possible intrinsic component, though concentrated at large
$x$ at a suitably chosen starting scale, 
will affect non-trivially PDFs at lower $x$ at
higher scale due to mixing through perturbative evolution.

The aim of this paper is to perform a first 
determination of the charm PDF of the proton in which no assumption is
made about its origin and shape, 
and charm is treated on the same  footing as the other
fitted PDFs. This
will be done using the NNPDF methodology:
we will present  a variant
of the NNPDF3.0~\cite{Ball:2014uwa} PDF determination, in which
the
charm PDF is parametrized in the same way as  the light quark and
gluon PDFs, i.e.\ with an independent
neural network with 37 free parameters. In the present analysis, we will
assume the charm and anti-charm PDFs to be equal, since there is
currently not data which can constrain their difference.

The possibility of introducing 
a non-perturbative ``intrinsic'' charm PDF has been discussed 
several times in the past, see e.g.\ Refs.~\cite
{Harris:1995jx,Pumplin:2007wg,Martin:2009iq,Dulat:2013hea,Jimenez-Delgado:2014zga,Hou:2015emq}.
In all of these earlier studies, only charm PDFs with a restrictive
parametrization based on model assumptions are considered. Moreover, 
in the CT family of PDF 
determinations~\cite{Pumplin:2007wg,Dulat:2013hea,Hou:2015emq},
intrinsic charm is introduced as a non-vanishing boundary condition to 
PDF evolution, but the massive corrections to the charm-initiated
contributions~\cite{Hoffmann:1983ah,Kretzer:1998ju} are
not included. While this would be consistent if all charm were 
generated perturbatively, as in the standard 
FONLL~\cite{Cacciari:1998it,Forte:2010ta} or S-ACOT~\cite{Kramer:2000hn}
schemes, when there is a non-perturbative charm PDF it is  justified only
 if this non-perturbative component is uniformly power-suppressed (of order
$\Lambda^2/m_c^2$, as in Ref.~\cite{Vogt:1995tf}) over the full range of $x$. 

Here,
however, as explained above, we wish to be able to parametrize the
charm PDF at any scale, without committing ourselves to any specific
hypothesis on its shape, and without having to
separate the perturbative and nonperturbative components. 
A formalism which includes the mass 
corrections~\cite{Hoffmann:1983ah,Kretzer:1998ju} 
by extending the FONLL~\cite{Cacciari:1998it} GM-VFN scheme for deep-inelastic
scattering of Ref.~\cite{Forte:2010ta} was implemented at
NLO~\cite{Ball:2015tna}, and consistently worked out to all orders 
in~\cite{Ball:2015dpa}. It is this
implementation that will be used in this paper.

In the present PDF fit
we use essentially the same data as in the NNPDF3.0 PDF determination, 
including as before the HERA  charm production
cross-section
combination~\cite{Abramowicz:1900rp}, but extended to also include the
EMC charm structure function data of Ref.~\cite{Aubert:1982tt}, 
which is the only existing measurement of the charm structure function at 
large $x$ . We  also replace all the HERA inclusive structure function data
with the final combined dataset~\cite{Abramowicz:2015mha}.

The outline of the paper is the following.
First, in Sect.~\ref{sec:fitsettings} we present the settings
of the analysis:
the dataset we use, the NLO implementation of the theory of
Refs.~\cite{Ball:2015tna,Ball:2015dpa} for the inclusion of a fitted
charm PDF, and the fit settings which have been used in the PDF fits.
In Sect.~\ref{sec:results} we present the fit results: we compare 
PDF determinations with and without fitted charm; we
discuss the stability of our results with respect to variations of the
charm mass; and we discuss the features of our best-fit charm PDF,
specifically in terms of the momentum fraction carried by charm, and
in comparison to existing models.
In Sect.~\ref{sec:pheno} we discuss the implications of our
results for LHC phenomenology, both for processes which are
particularly sensitive to the charm PDF and thus might be used for its
determination (such as $Z+c$ and charm pair production), and for LHC
standard candles (such as $W$, $Z$ and Higgs production).
Finally, in Sect.~\ref{sec:delivery} we discuss the delivery of
our results and outline future developments.

\section{Settings}
\label{sec:fitsettings}

The PDF determination presented in this paper,
which we will denote by NNPDF3IC, is based on settings which are similar to
those used for the latest  NNPDF3.0 global analysis~\cite{Ball:2014uwa},
but  with a number of differences, mostly related to the
inclusion of a fitted charm PDF. These involve the experimental data, the theory
calculations, and the
fit settings, which we now discuss in turn.

\subsection{Experimental data}
\label{sec:expdata}

The dataset used in the present analysis is the same as used for 
NNPDF3.0, with two differences.
The first has to do with HERA data: for NNPDF3.0, the combined inclusive
HERA-I data~\cite{Aaron:2009aa} were used along with the separate
HERA-II 
datasets from the  H1 and ZEUS
collaboration~\cite{Aaron:2012qi,Collaboration:2010ry,ZEUS:2012bx,Collaboration:2010xc}.
Meanwhile, the  final HERA legacy
combination~\cite{Abramowicz:2015mha} data have become
available. These have been used here. It has been
shown~\cite{Rojo:2015nxa} that, while 
the impact of the HERA-II data on top of the HERA-I combined data is
moderate but not-negligible, the impact of the global legacy
combination in comparison to HERA-I and separate HERA-II measurements
is extremely small. Nevertheless, this replacement is performed for
general consistency. 
Similar conclusions on the impact of these data 
have been reached by the MMHT group~\cite{Harland-Lang:2016yfn}.

The second difference is that we will also include
EMC charm structure function data~\cite{Aubert:1982tt}.
Since the EMC collaboration presented this measurement in the early 80s,
some studies\cite{Harris:1995jx,Martin:2009iq} have suggested that these data 
might provide direct
evidence for non-perturbative charm in the
proton~\cite{Brodsky:1991dj,Brodsky:2015fna}.
On the other hand, some previous PDF fits with intrinsic 
charm have not
been able to provide a satisfactory description of this dataset~\cite{Jimenez-Delgado:2014zga}.
Since it is known that the EMC measurements were affected by some systematic
uncertainties which were only identified after the experiment was completed,
we will perform fits both with and without it.
We will also perform fits where the EMC charm data have been rescaled
to match the current value of the branching ratio of charm quarks into muons.

Summarizing, the dataset that we will use is the following: fixed-target
neutral-current deep inelastic scattering (DIS) 
data from NMC~\cite{Arneodo:1996kd,Arneodo:1996qe}, BCDMS~\cite{bcdms1,bcdms2},
SLAC~\cite{Whitlow:1991uw} and EMC~\cite{Aubert:1982tt};
the legacy HERA
combinations for inclusive~\cite{Abramowicz:2015mha}
and charm ~\cite{Abramowicz:1900rp} reduced cross-sections;
charged-current structure
functions from CHORUS inclusive neutrino DIS~\cite{Onengut:2005kv} and from
NuTeV dimuon production data~\cite{Goncharov:2001qe,MasonPhD};
fixed target E605~\cite{Moreno:1990sf} and
E866~\cite{Webb:2003ps,Webb:2003bj,Towell:2001nh} Drell-Yan production
data;
Tevatron collider data including
the
CDF~\cite{Aaltonen:2010zza} and D0~\cite{Abazov:2007jy} $Z$ rapidity
distributions and the CDF~\cite{Aaltonen:2008eq} one-jet inclusive 
cross-sections; LHC collider data including 
ATLAS~\cite{Aad:2011dm,Aad:2013iua,Aad:2011fp}, 
CMS~\cite{Chatrchyan:2012xt,Chatrchyan:2013mza,Chatrchyan:2013uja,CMSDY} 
and LHCb~\cite{Aaij:2012vn,Aaij:2012mda}  vector boson production, 
ATLAS~\cite{Aad:2011fc,Aad:2013lpa} and CMS~\cite{Chatrchyan:2012bja}
jets, and finally, total cross section measurements for
top quark pair production data from ATLAS and CMS
at 7 and 8 TeV~\cite{ATLAS:2012aa,ATLAS:2011xha,TheATLAScollaboration:2013dja,
  Chatrchyan:2013faa,Chatrchyan:2012bra,Chatrchyan:2012ria}. 
Data with $Q<3.5$~GeV and $W^2<12.5$~GeV$^2$ are
excluded from the fit.

A final change in comparison to Ref.~\cite{Ball:2014uwa} is that 
we now impose additional cuts on the Drell-Yan fixed-target cross-section data:
\be
\tau \le 0.08 \, , \qquad |y|/y_{\rm max} \le 0.663 \, ,
\ee
where $\tau=M^2/s$ and
$y_{\rm max}=- (1/2) \log \tau$, and $y$ is the rapidity and $M$ the invariant mass of the
dilepton pair.
These cuts are meant to ensure that an unresummed perturbative
fixed-order description is adequate; the choice of values is
motivated by studies performed in Ref.~\cite{Bonvini:2015ira} in
relation to the determination of PDFs with threshold resummation,
which turns out to have a rather larger impact on Drell-Yan production than on
deep-inelastic scattering.
These cuts
 reduce by about a factor two  the number of fixed-target Drell-Yan data points
 included here in comparison to Ref.~\cite{Ball:2014uwa}, and improve the agreement
 between theory and data.

\subsection{Theory}
\label{sec:theory}

In the presence of fitted charm,
the original FONLL expressions for deep-inelastic structure  functions
of Ref.~\cite{Forte:2010ta}
need to be modified to account for the new
massive charm-initiated contributions~\cite{Ball:2015dpa,Ball:2015tna}.
Also, while in previous NNPDF determinations pole quark masses only
have been used, here we will  consider both pole and $\overline{\rm
  MS}$ heavy quark masses.
These new features have been implemented along with a major update
in the codes used to provide the theory calculations.
Indeed, in all previous NNPDF determinations,
PDF evolution and the computation of deep-inelastic structure
functions
were performed by means of the Mellin-space {\tt FKgenerator} NNPDF internal
code~\cite{Ball:2008by,Ball:2010de}. Here (and henceforth) we will use
the public $x$-space {\tt APFEL} code~\cite{Bertone:2013vaa} for the solution of
evolution equations and the computation of DIS structure functions.
For hadronic observables, PDF evolution kernels are pre-convoluted with
{\tt APPLgrid}~\cite{Carli:2010rw}
partonic cross-sections using the {\tt APFELcomb} interface~\cite{Bertone:2016lga}.

\begin{figure}[t]
  \begin{center}
   \includegraphics[width=0.49\textwidth]{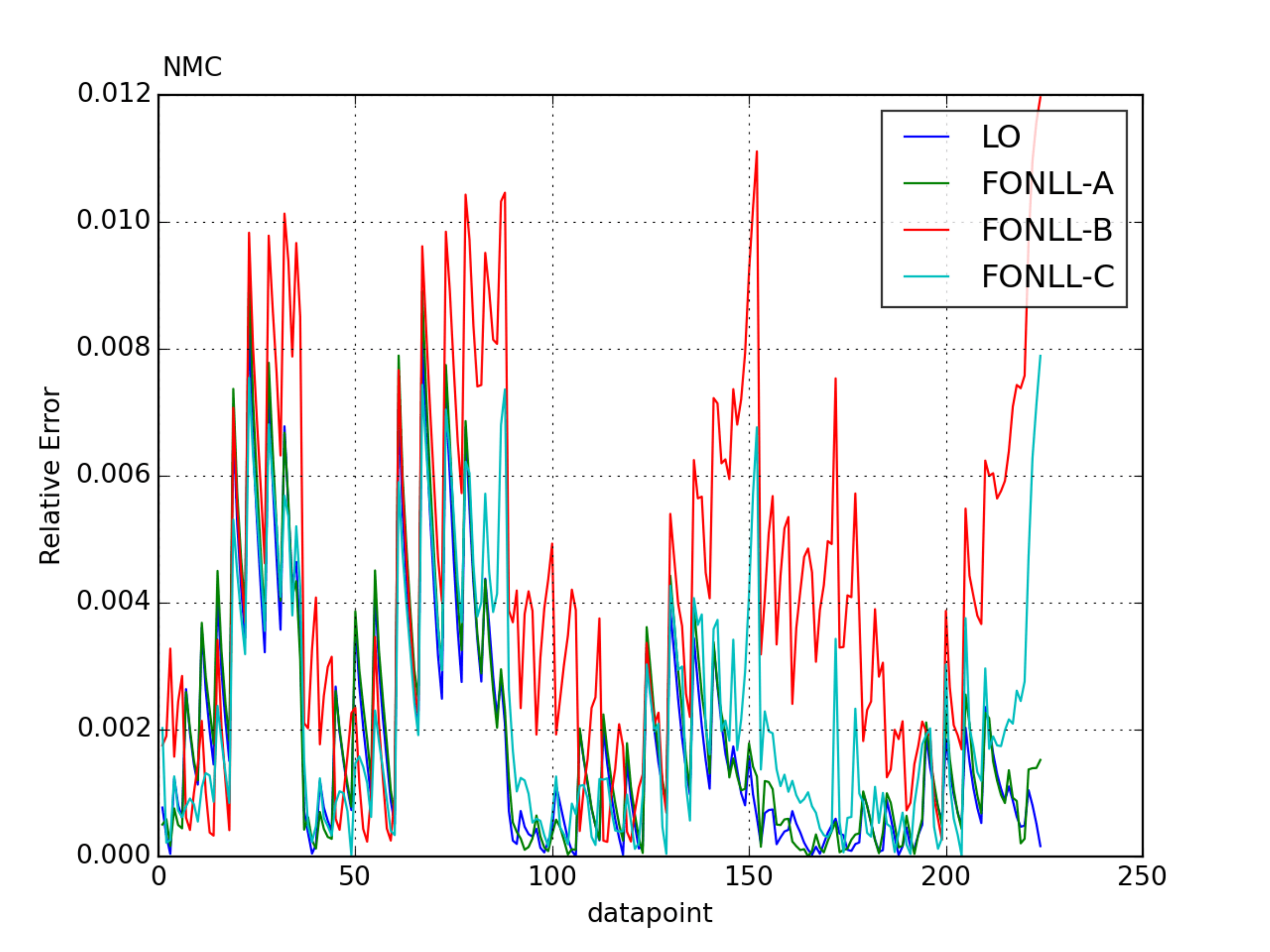}
   \includegraphics[width=0.49\textwidth]{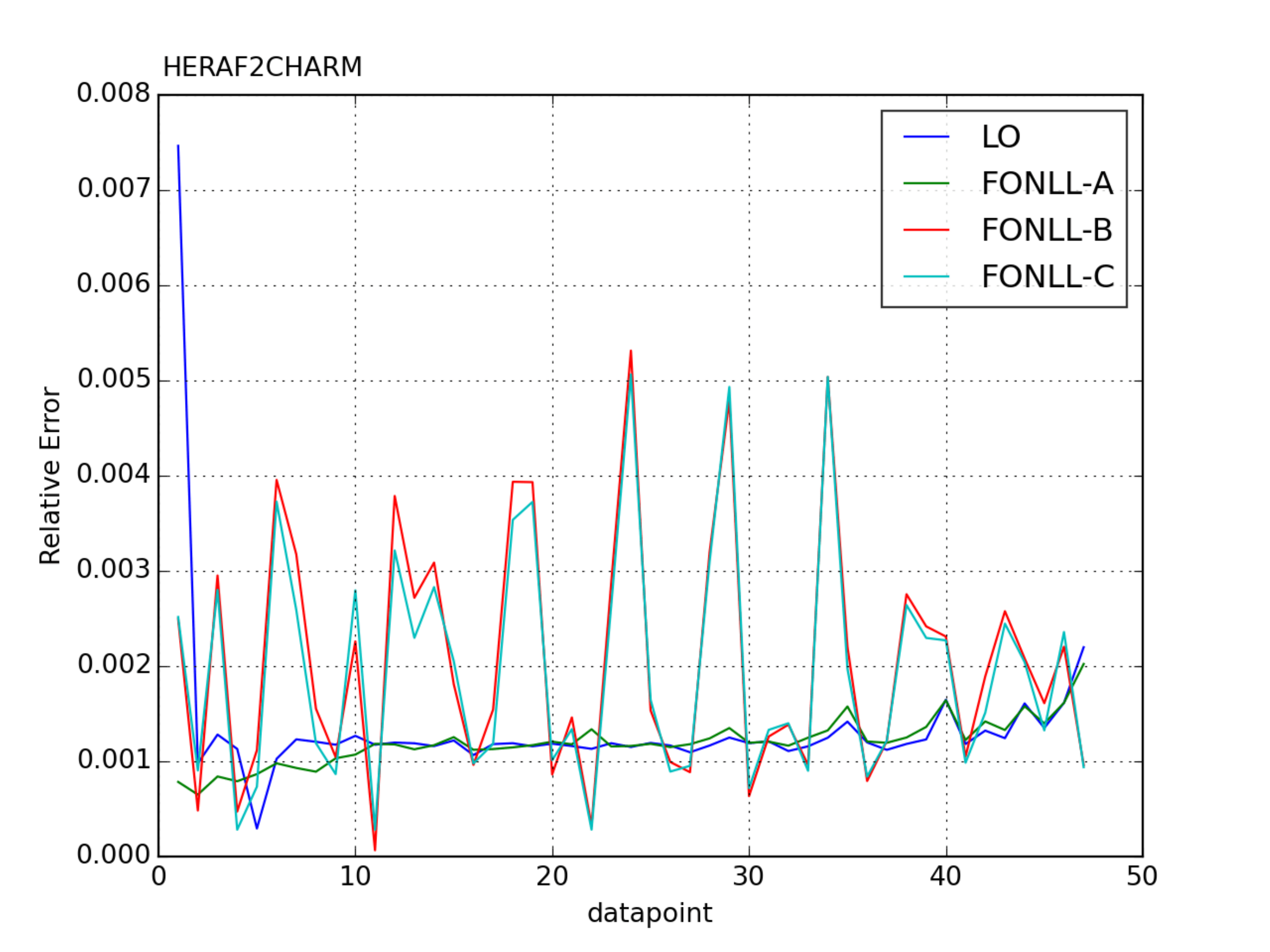}
\end{center}
\vspace{-0.8cm}
\caption{\small \label{fig:benchmarking1}
  Representative benchmark comparisons between deep-inelastic
  structure functions computed with the {\tt FKgenerator} and {\tt APFEL} programs.
  We show the relative differences between the two codes
  for $\sigma_{\rm NC}^p(x,Q^2)$ at the NMC
  data points (left) and for $\sigma_{\rm c\bar{c}}(x,Q^2)$ for the HERA
  charm  data points (right).
  In each case, we show results at  LO (massless
  calculation) and for the FONLL-A, B and C general-mass schemes.
   }
\end{figure}

The {\tt FKgenerator} and {\tt
    APFEL}  codes have been extensively benchmarked.
As an illustration, in Fig.~\ref{fig:benchmarking1}
we show  representative benchmark comparisons between deep-inelastic
  structure functions computed with the two codes.
  We plot the relative differences between the computation with either
  of these two codes
 of the inclusive neutral-current
  cross-sections  $\sigma_{\rm NC}(x,Q^2)$ at the NMC
  data points and for the charm production
  reduced cross-sections $\sigma_{\rm c\bar{c}}(x,Q^2)$ for the HERA
   data points.
  In each case we compare results obtained at LO (massless
  calculation) and using the FONLL-A, B and C general-mass schemes.
  Similar agreement is found for all other DIS experiments included in
  NNPDF3.0.

  The agreement is always better than 1\%. Differences
 can
  be traced to the interpolation  used by the {\tt FKgenerator}, as
  demonstrated by the fact that they follow roughly the same pattern
  for all theoretical computations shown, with the largest differences  
  observed for the NMC data,  in the large $x$, low $Q^2$ region where
  the interpolation is most critical.  Specifically, {\tt FKgenerator}
  uses a fixed grid in $x$ with 25 points
logarithmically spaced in $[x=10^{-5},x=10^{-1}]$ and 25 points linearly spaced
in $[x=10^{-1},x=1]$, while {\tt
    APFEL} instead optimises the distribution of the $x$-grid points
experiment by experiment. Hence we estimate that with the current {\tt
    APFEL} implementation accuracy has significantly improved to
better than 1\%. 
  %

An advantage of using {\tt APFEL} to compute DIS structure functions
is that it allows for the use of either pole
or $\overline{\rm
  MS}$  heavy quark masses~\cite{xFitter::2016pbr,Alekhin:2010sv}.
The implementation of running masses in the PDF evolution in {\tt APFEL}
has been benchmarked with the {\tt HOPPET} program~\cite{Salam:2008qg},
finding better than 0.1\% 
agreement.
In addition, the {\tt APFEL}
calculation of structure functions with running heavy quark masses in
the fixed three-flavour number scheme has been compared with the 
{\tt OpenQCDrad} code~\cite{Alekhin:2013nda}, with which it has been
found to agree at the 1\% level.

\begin{figure}[t]
  \begin{center}
   \includegraphics[width=0.49\textwidth]{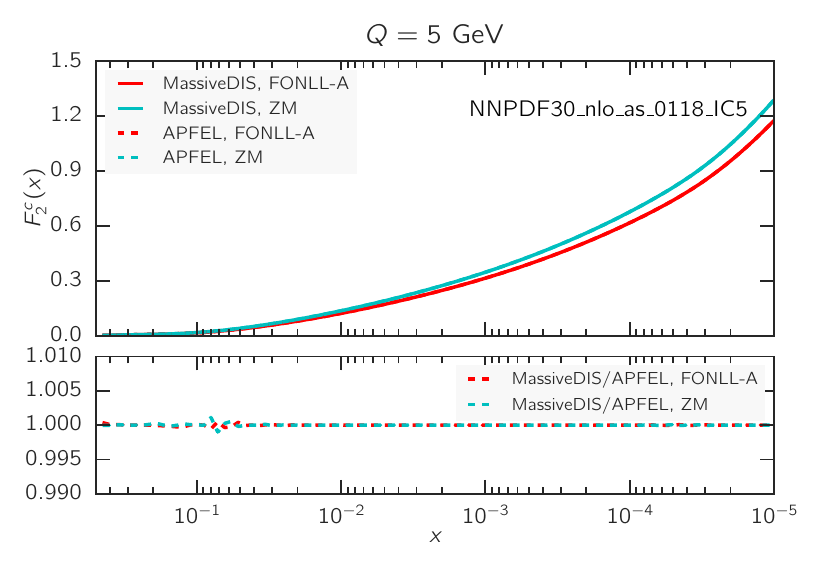}
   \includegraphics[width=0.49\textwidth]{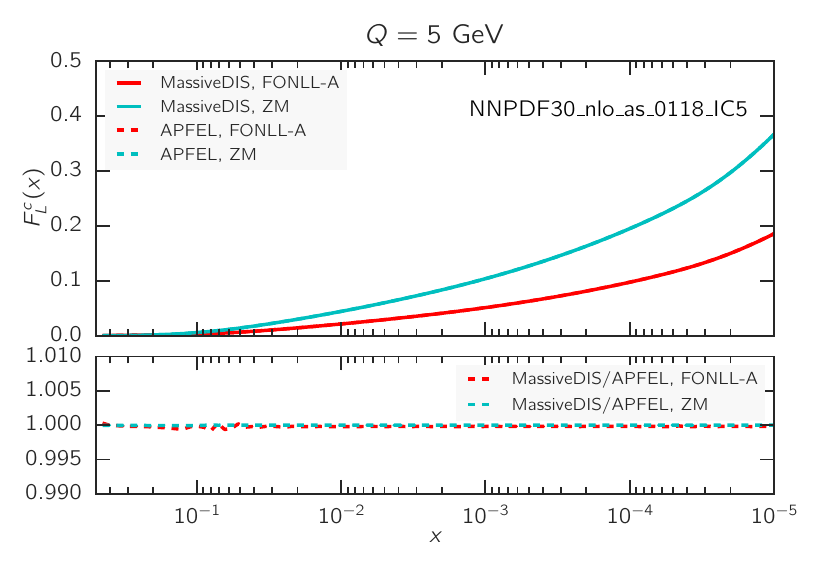}
\end{center}
\vspace{-0.8cm}
\caption{\small \label{fig:massiveDISbench}
  Benchmarking of the implementation in the
{\tt APFEL} and
  {\tt MassiveDISsFunction} codes of
deep-inelastic structure functions in the FONLL-A scheme with
intrinsic charm of Refs.~\cite{Ball:2015dpa,Ball:2015tna}.
The charm structure functions $F_2^c(x,Q^2)$ (left) and $F_L^c(x,Q^2)$
  (right) are shown as a function of $x$ for $Q=5$ GeV; the
  relative difference between the two codes is shown in the lower
  panel.
In each case we show full matched 
FONLL-A result as well as the purely massless calculation.
   }
\end{figure}

Massive charm-initiated terms for both neutral and charged current processes 
have been implemented in {\tt APFEL}  up to
$\mathcal{O}\lp \alpha_s\rp$. Target mass corrections are included throughout. 
The implementation has been validated through
benchmarking against the  public stand-alone {\tt MassiveDISsFunction}
code,\cite{ MassiveDISsFunction} which also implements the theory
calculations of 
Refs.~\cite{Ball:2015dpa,Ball:2015tna}. 
Some illustrative comparisons between  
 the charm structure functions $F_2^c(x,Q^2)$ and $F_L^c(x,Q^2)$,
computed using  {\tt APFEL} and
  {\tt MassiveDISsFunction}, are shown
in Fig.~\ref{fig:massiveDISbench}. The various inputs to the FONLL-A
scheme computation, namely the three- and four-flavour scheme results
are shown, along with the full matched result, as
a function of $x$ at the scale $Q=5$ GeV, computed using an input toy
intrinsic charm PDF, corresponding to the  {\tt NNPDF30\_nlo\_as\_0118\_IC5}
set of Ref.~\cite{Ball:2015tna}.
The two
codes turn out to agree at the 0.1\% level or better, for all
neutral-current and charged-current structure functions.

\subsection{Fit settings}
\label{sec:methodology}

We can now specify the theory settings used for the PDF fits presented
in this paper.
We will use NLO theory with $\alpha_s(M_Z)=0.118$, with a bottom mass  
of $m_b=4.18$ GeV.
We will present fits with the $\overline{\rm MS}$  charm mass set
equal to
$m_c(m_c)=1.15,\> 1.275$ and $1.40$~GeV, which corresponds
to the PDG central value and upper and lower five-sigma
variations~\cite{Beringer:1900zz}.
We will also present fits with the charm
pole mass  $m_c^{\rm pole}=1.33$, 1.47 and 1.61 GeV,
obtained from the corresponding $\overline{\rm MS}$ values using  one-loop
conversion. This conservative range of charm pole mass value allows us
to account for the large uncertainties in the one-loop conversion factor.
In addition, as a cross-check, we also perform a pole mass fit with 
$m_c^{\rm pole}=1.275$ GeV, which was the choice adopted  in NNPDF3.0.
When the charm PDF is generated perturbatively, the charm threshold is set to be the charm mass.
The input parametrization scale is $Q_0=1.1$ GeV for the fits with perturbative
charm and $Q_0=1.65$ GeV in the case of fitted charm, ensuring that the scale
where PDFs are parametrized is always above (below) the charm threshold
for the analysis with fitted (perturbative) charm in all the range of charm masses
considered.
In sum, we will consider seven charm mass values (four pole, and three
$\overline{\rm MS}$), and for each of them, we will present fits with
perturbative charm or with fitted charm.

In the NNPDF3.0 analysis,
seven independent PDF
combinations were parametrized
with artificial neural networks at the input
evolution scale $Q_0$: the gluon, the total quark
singlet $\Sigma$, the non-singlet quark triplet and octet
$T_3$ and $T_{8}$ and the quark valence combinations $V$, $V_3$
and $V_8$.
In this analysis, when we fit the charm PDF, we use the same PDF parametrization basis  supplemented
by the total charm PDF $c^+$, that is,
\be
c^+(x,Q_0)\equiv c(x,Q_0)+\bar{c}(x,Q_0) =
x^{a_{c^+}}(1-x)^{b_{c^+}}{\rm NN}_{c^+}(x) \, ,
\ee
with ${\rm NN}_{c^+}(x)$ a feed-forward neural network with the
same architecture (2-5-3-1) and number of free parameters
(37) as the other PDFs included in the fit, and
$a_{c^+}$ and $b_{c^+}$ the corresponding preprocessing exponents, whose
range is determined from an iterative procedure designed to ensure 
that the resulting PDFs are unbiased.
In addition, we assume that the  charm and anticharm PDFs are the same,
$c^-(x,Q_0)\equiv c(x,Q_0)-\bar{c}(x,Q_0)=0$. 
Since at NLO this distribution evolves multiplicatively, 
it will then vanish at all values of $Q^2$. It might be interesting 
to relax this
assumption once data able to constrain  $c^-(x,Q_0)$ become available.

The fitting methodology used in the present
fits is the same as in NNPDF3.0, with some minor
improvements.
First, we have enlarged the set of positivity constraints.
In NNPDF3.0, positivity was imposed for  
the up, down and strange structure functions,
$F_2^u$, $F_2^d$ and $F_2^s$; for the light component of the
longitudinal structure function, $F_L^l$; and for Drell-Yan rapidity
distributions with the flavour quantum numbers of $u\bar{u}$,
$d\bar{d}$, and $s\bar{s}$; and for the rapidity
distribution for Higgs production in gluon-fusion (see Section~3.2.3
of Ref.~\cite{Ball:2014uwa} for a detailed discussion). This set of
positivity observables has now been enlarged to also
include flavour non-diagonal combinations:
we now impose the positivity of the $ud$, $\bar{u}d$,
$\bar{u}\bar{d}$ and $u\bar{d}$ Drell-Yan rapidity distributions.
As in Ref.~\cite{Ball:2014uwa}, positivity is 
 imposed for all replicas at  $Q^2_{\rm pos}=5$ GeV$^2$, which ensures
 positivity for all higher scales.

Also, we have modified the way asymptotic exponents used in the
iterative determination of the preprocessing range are
computed. Specifically, we now use the definition
\begin{equation}
\alpha_{f_i}(x,Q^2)
\equiv
\frac{\partial\ln[xf_i(x,Q^2)]}{\partial\ln x}
\ \ \ \ \ \ \ \,
\beta_{f_i}(x,Q^2)
\equiv
\frac{\partial\ln[xf_i(x,Q^2)]}{\partial\ln(1-x)}
\,\mbox{,}
\label{eq:def2}
\end{equation}
suggested in Ref.~\cite{Nocera:2014uea,eff2},
which is less affected by  sub-asymptotic
terms at small and large-$x$ than the definition used in the
NNPDF3.0 analysis~\cite{Ball:2014uwa}.
This allows a more robust determination of the ranges in which the
PDF preprocessing exponents should be varied, following the iterative procedure
discussed in~\cite{Ball:2014uwa}.
This modification affects only the PDFs in the extrapolation regions where there
are little or none experimental data constraints available.
The implications of these modifications in the global
analysis will be more extensively
discussed in a forthcoming publication.

\section{Results}
\label{sec:results}

In this section we discuss the main results of this
paper, namely the NNPDF3 PDF sets with fitted charm.
After presenting  and discussing the statistical indicators of the 
fit quality, we discuss  the most significant effects of fitted charm, 
namely, its
 impact on the dependence of PDFs on  the charm mass, and its
 effect on PDF
uncertainties. We then discuss the extent to which our results are
affected by the inclusion of EMC data on the charm structure function. 
Having established the
robustness of our results, we turn to a study of the properties of
the fitted 
charm PDF: whether or not it has an intrinsic component, the size of
the momentum fraction carried by it, and how it compares to some of the 
models for intrinsic charm constructed in the past.

Here and henceforth we will refer to a fit using the FONLL-B scheme of
    Ref.~\cite{Forte:2010ta}, in which all charm is generated perturbatively, 
both at fixed order and by PDF evolution, as ``perturbative charm'', 
while ``fitted charm'' refers to fit obtained using the theory reviewed in 
Section~\ref{sec:theory}. Note that fitted charm includes a perturbative 
component, which grows above threshold until it eventually dominates: at 
high enough scales most charm is inevitably perturbative. However close 
to threshold the non-perturbative input might still be important: in 
particular  below threshold the perturbative charm vanishes by construction, 
whereas the fitted charm can still be non-zero (so-called ``intrinsic'' charm).

\begin{table}[t!]
  \centering
  \small
  \begin{tabular}{|c|c|c|}
    \hline
    \multicolumn{3}{|c|}{NNPDF3 NLO $m_c=1.47$ GeV (pole mass) } \\[0.1cm]
    \hline
    &  fitted charm  &
    perturbative charm  \\[0.1cm]
    \hline
    \hline
     $\chi^2/N_{\rm dat}$ (exp)  &  1.159   &  1.176  \\
    $\la \chi^2\ra_{\rm rep}/N_{\rm dat}$  (exp) &   $ 1.40\pm 0.24$  &   $ 1.33\pm 0.12$\\
    $\chi^2/N_{\rm dat}$ ($t_0$)  & 1.220     & 1.227    \\
    $\la \chi^2\ra_{\rm rep}/N_{\rm dat}$  ($t_0$) &   $ 1.47 \pm 0.26 $  &   $ 1.38 \pm 12  $\\
    \hline
    $\la E_{\rm tr} \ra/N_{\rm dat}$  &  $2.38 \pm 0.29 $  &
    $2.32 \pm 0.16 $\\
    $\la E_{\rm val} \ra/N_{\rm dat}$  &  $2.60 \pm 0.37 $  &
    $2.48 \pm 0.16 $\\
    $\la {\rm TL} \ra$  & $\lp 3.5 \pm 0.8\rp\cdot 10^3$  &
    $\lp 2.2 \pm 0.8\rp\cdot 10^3$\\
    \hline
    $\varphi_{\chi^2}$  &  $0.49 \pm 0.02$  & $0.40\pm 0.01$   \\
    $\la \sigma^{\rm (exp)}\ra_{\rm dat}$  &  13.1\%  &  12.2\%  \\
    $\la \sigma^{\rm (fit)}\ra_{\rm dat}$  &  7.4\%  &  4.4\%  \\
    \hline
  \end{tabular}
  \caption{\small Statistical estimators of the fitted and perturbative
    charm PDFs for the central value of the charm pole mass, for both fitted charm and perturbative charm.
    For $\chi^2$ and $\la \chi^2\ra$ we provide the results using both
    the $t_0$ and ``experimental'' definition of the $\chi^2$ (see text).
    $\la E_{\rm tr}\ra$ and $\la E_{\rm val}\ra$ are computed during the fit using the $t_0$ definition.
    \label{tab:stat}
  }
\vspace{0.4cm}
  \centering
  \small
  \begin{tabular}{|c|c|c|c|}
    \hline
    \multicolumn{4}{|c|}{NNPDF3 NLO $m_c=1.47$ GeV (pole mass) } \\[0.1cm]
    \hline
    \hline
    Experiment  & $N_{\rm dat}$  & $\chi^2/{N_{\rm dat}}$   &
    $\chi^2/{N_{\rm dat}}$   \\[0.1cm]
     &   & fitted charm  &
    perturbative charm  \\[0.1cm]
    \hline
    \hline
    NMC     &  325  & 1.36    &   1.34       \\
    SLAC    &  67   & 1.21    &   1.32       \\
    BCDMS    & 581   & 1.28    &  1.29        \\
    CHORUS   &  832  & 1.07    &  1.11        \\
    NuTeV   &  76    & 0.62    &  0.62        \\
    EMC    &  16     & 1.09    &  [7.3]        \\
    \hline
    HERA inclusive &  1145  & 1.17   & 1.19         \\
    HERA charm   &  47  & 1.14    &  1.09        \\
    \hline
    DY E605  & 104  & 0.82    &  0.84        \\
    DY E866 & 85 &  1.04      &  1.13        \\
    \hline
    CDF  & 105 & 1.07      &  1.07        \\
    D0  &  28 & 0.64       &  0.61        \\
    \hline
    ATLAS &  193 &  1.44    &   1.41       \\
    CMS  &  253 & 1.10      &   1.08       \\
    LHCb  & 19  & 0.87      &   0.83       \\
    $\sigma(t\bar{t})$  & 6   & 0.96    &  0.99          \\
    \hline
    \hline
    Total   &  3866  &  1.159   &     1.176     \\
    \hline
  \end{tabular}
  \caption{\small The $\chi^2$ per data point for the experiments included
    in the present analysis, computed using the experimental covariance 
    matrix, comparing
    the results obtained with fitted charm with those
    of perturbative
    charm. 
    We also provide the total $\chi^2/N_{\rm dat}$ of the fit, as well
    as the number of data points per experiment.
     In the case of perturbative charm, we indicate the values
    of the fit without the EMC data, and show in brackets
    the $\chi^2$  of this experiment when included in the fit.
    \label{tab:chi2}
  }
  \end{table}

\subsection{Fit results}
\label{sec:resultsPDF}

In Tables~\ref{tab:stat} and \ref{tab:chi2} we collect the 
statistical estimators for our best fit with
 central value of the charm pole mass, namely
 $m_c^{\rm pole}=1.47$~GeV, both with fitted  and perturbative charm.
A detailed discussion of statistical indicators and their meaning can
be found in
Refs.~\cite{Ball:2010de,Ball:2011mu,Ball:2011eq,Ball:2014uwa}. Here we
merely recall that  $\chi^{2}$ is computed by comparing
the central (average) fit to the original experimental data;
 $\la \chi^{2} \ra_{\rm rep}$ is computed by comparing 
each PDF replica
to the data and averaging over replicas, while $\la E \ra$ is the
quantity that is actually minimized, i.e.\ it coincides with the
$\chi^{2}$ computed by comparing each  replica to the data
replica it is fitted to, with the two values given corresponding to
the training and validation data sets respectively.
The values of $\la E \ra$  are computed using the so-called
$t_0$ definition of the $\chi^2$, while for $\chi^2$ and  $\la
\chi^{2} \ra_{\rm rep}$ we show in the table values computed using 
both the $t_0$ and
the ``experimental'' definition (see
Refs.~\cite{Ball:2009qv,Ball:2012wy} for a discussion of different
$\chi^2$ definitions); they are seen to be quite close anyway.

Moreover, $\langle {\rm TL}\rangle$ is
the training length, expressed in number of cycles (generations) of the
genetic algorithm used for minimization. 
$\varphi_{\chi^2}$~\cite{Ball:2014uwa} 
is the average over all data of uncertainties and correlations
normalized to the corresponding experimental quantities (i.e., roughly
speaking, $\varphi_{\chi^2}=0.5$ means that the PDF uncertainty is
half the uncertainty in the original data), while  $\la \sigma^{\rm
  (exp)}\ra_{\rm dat}$ is the average percentage experimental
uncertainty, and 
    $\la \sigma^{\rm (fit)}\ra_{\rm dat}$ is the average percentage
PDF uncertainty at data points.

In Table~\ref{tab:chi2} we provide a breakdown of the
 $\chi^2$ per data point for all experiments (the value
computed with the ``experimental'' definition only).
In the case of perturbative charm, the $\chi^2$ values listed
correspond to a 
    fit without  EMC data, with the $\chi^2$  for this 
    experiment if it were included in the fit given in square parenthesis.
Note that the total $\chi^2$ values in this table are significantly lower 
than those reported in our previous global fit NNPDF3.0 \cite{Ball:2014uwa}: 
this is mainly due the much lower $\chi^2$ value for HERA data, which
in turn results from using the full combined HERA dataset instead
of separate HERA-II H1 and ZEUS data.

\begin{figure}[h!]
  \begin{center}
   \includegraphics[width=0.45\textwidth]{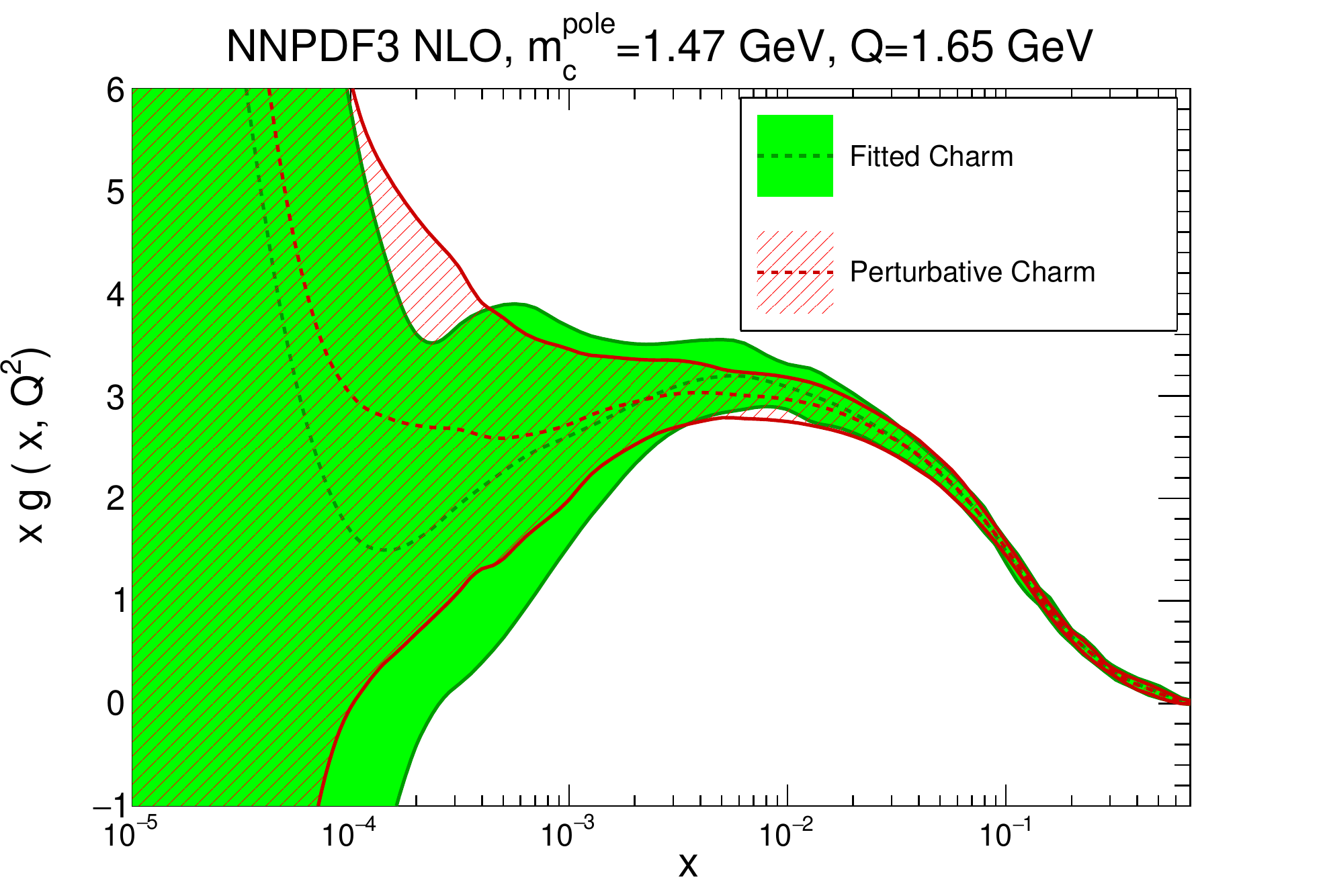}
  \includegraphics[width=0.45\textwidth]{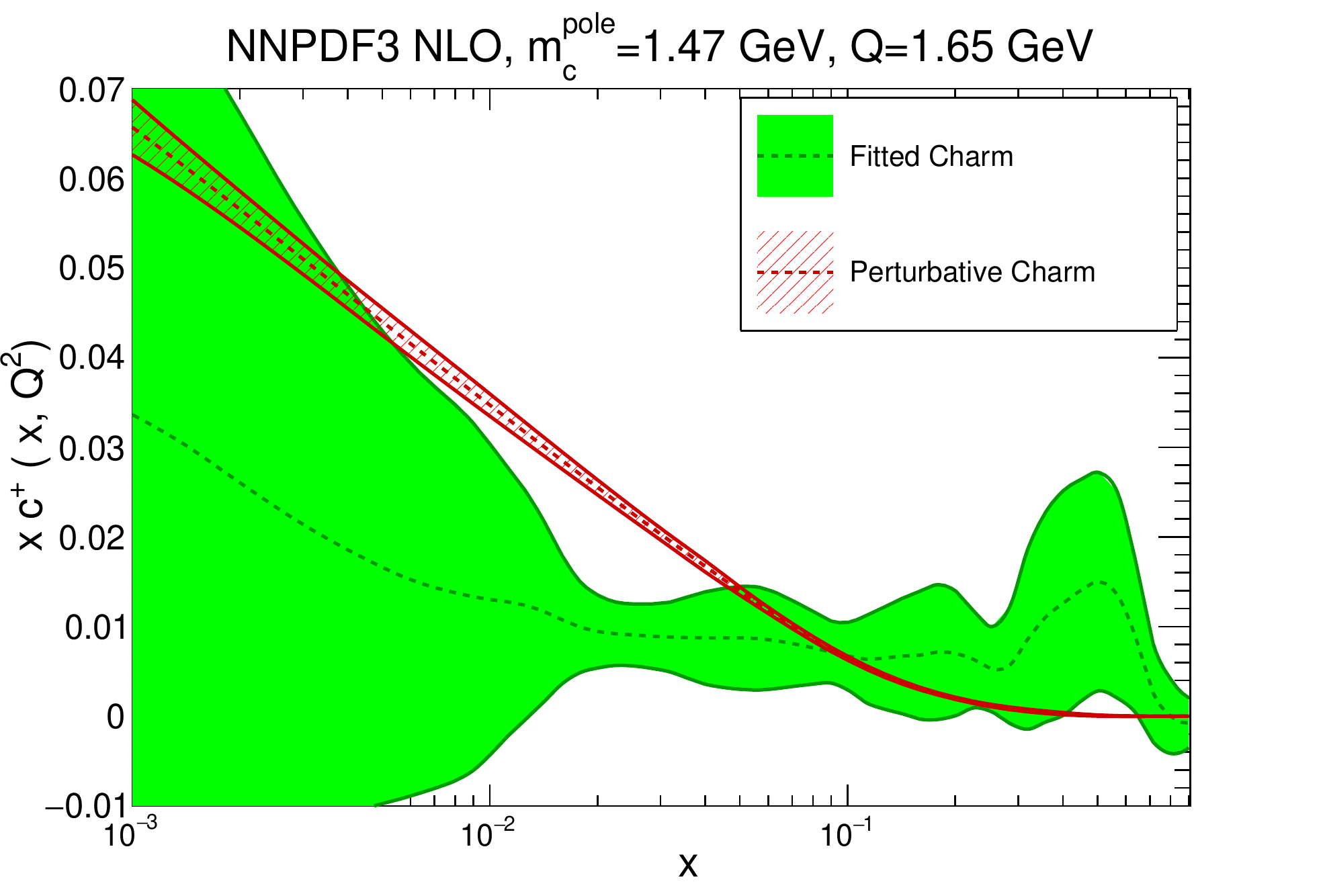}
  \includegraphics[width=0.45\textwidth]{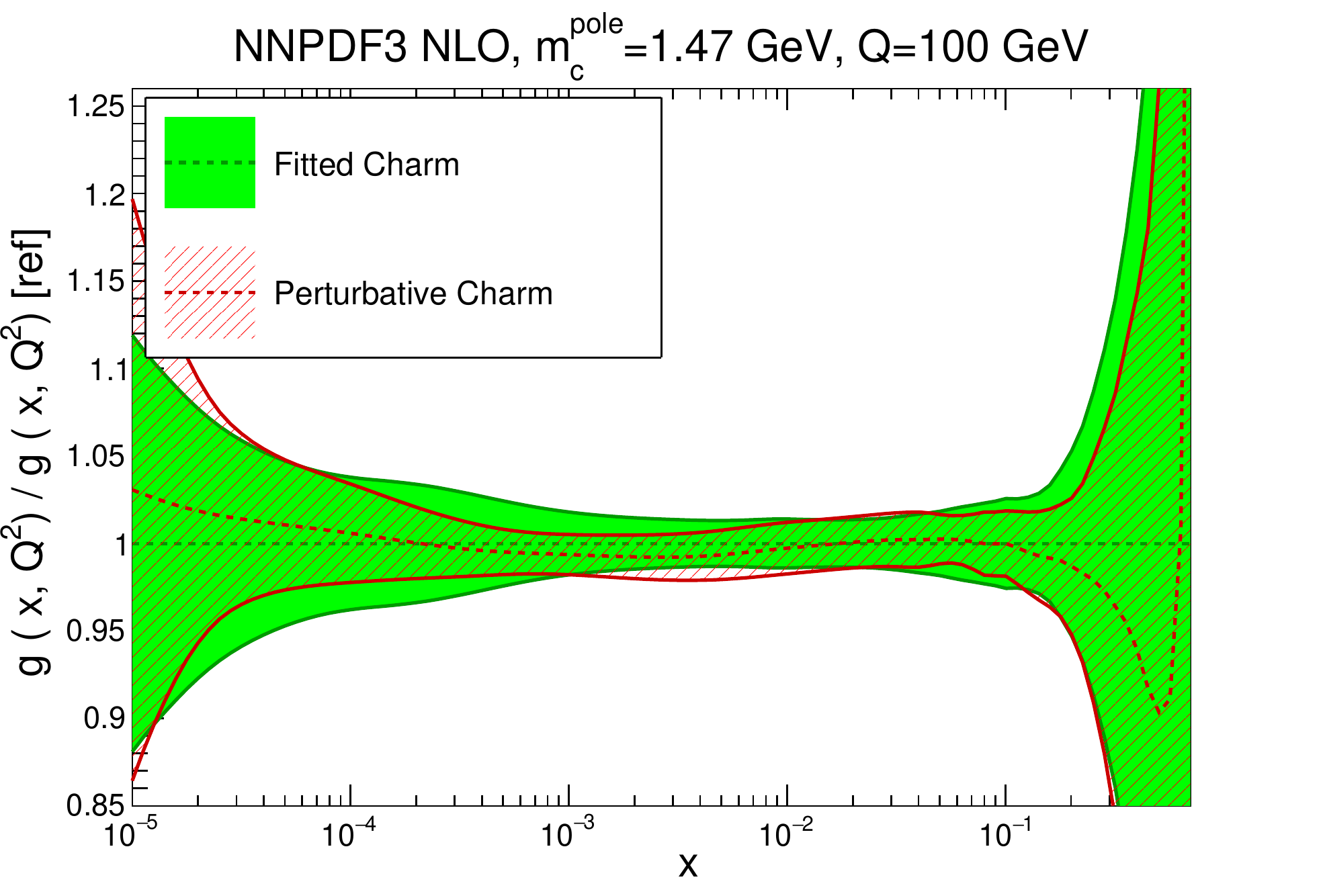}
  \includegraphics[width=0.45\textwidth]{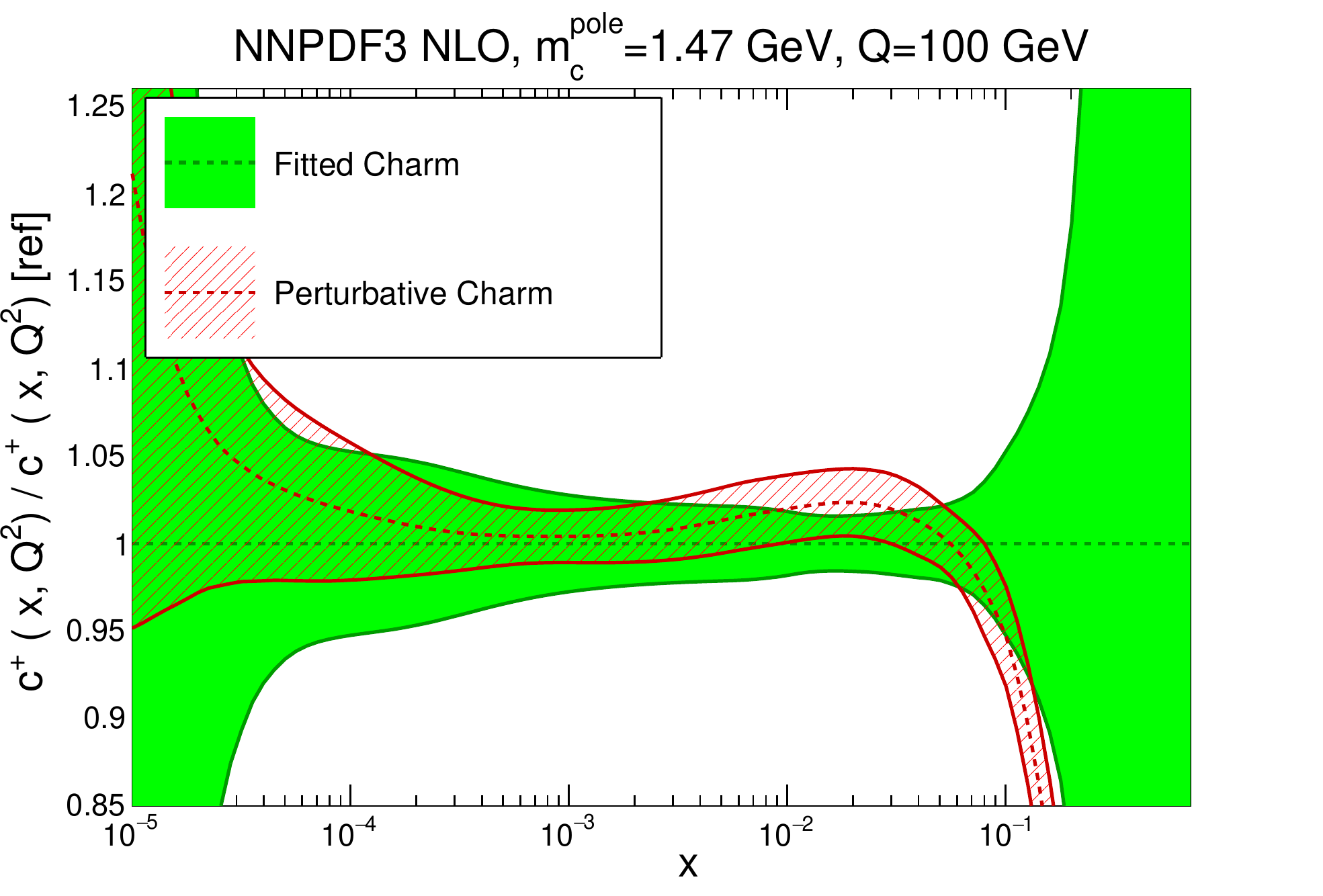}
\end{center}
\vspace{-0.5cm}
\caption{\small \label{fig:pdfbase}
  Comparison of the NNPDF3 NLO PDFs with fitted and perturbative charm,
  for a charm pole mass of $m^{\rm pole}_c=1.47$ GeV.
  We show the gluon
  (left plots) and the charm quark (right plot), at a low scale
  $Q=1.65$ GeV (upper plots) and at a high scale, $Q=100$ GeV (lower plots).
  In the latter case, results are shown normalized to the central value
  of the fitted charm PDFs.
   }
\vspace{-0.4cm}
  \begin{center}
   \includegraphics[width=0.45\textwidth]{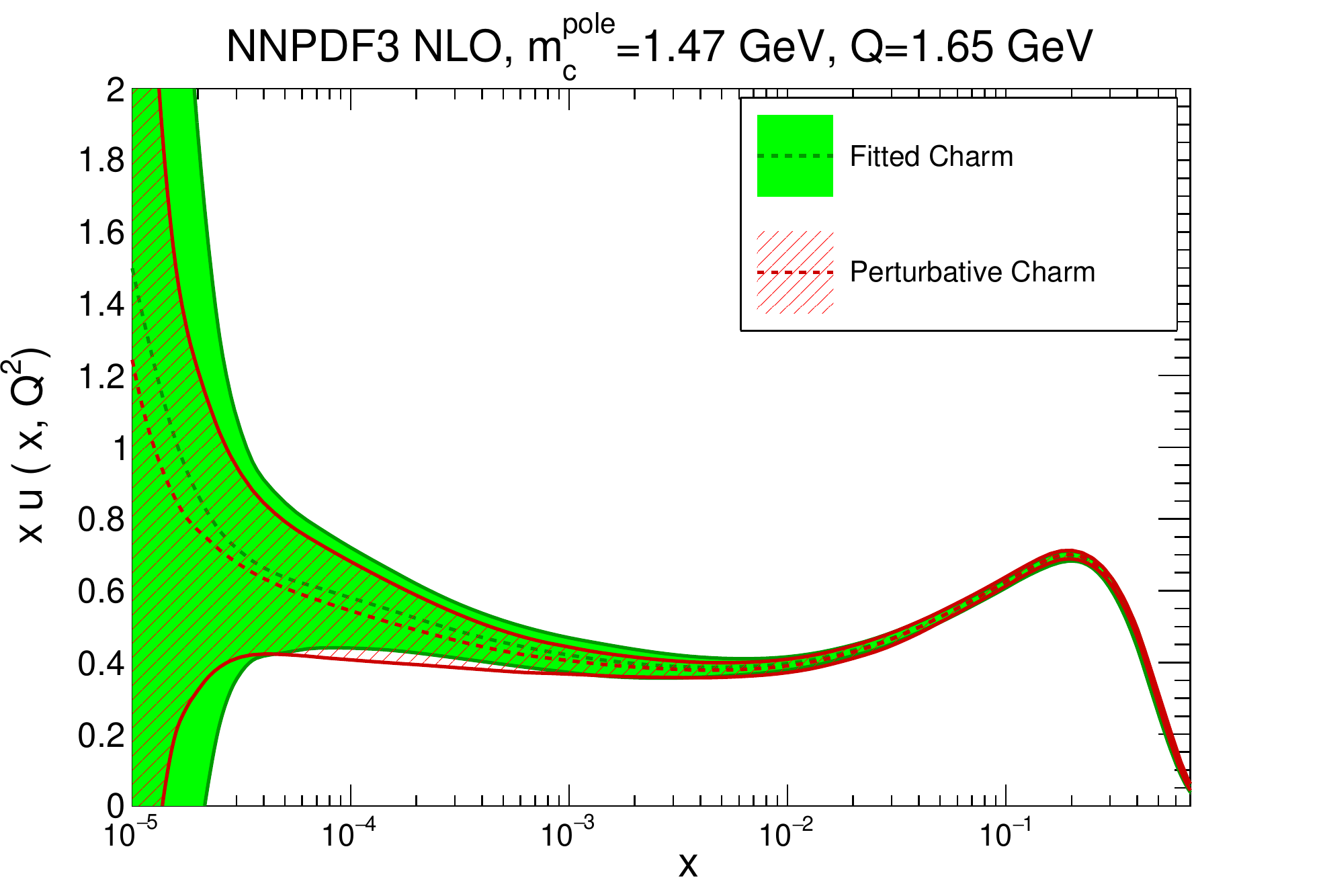}
  \includegraphics[width=0.45\textwidth]{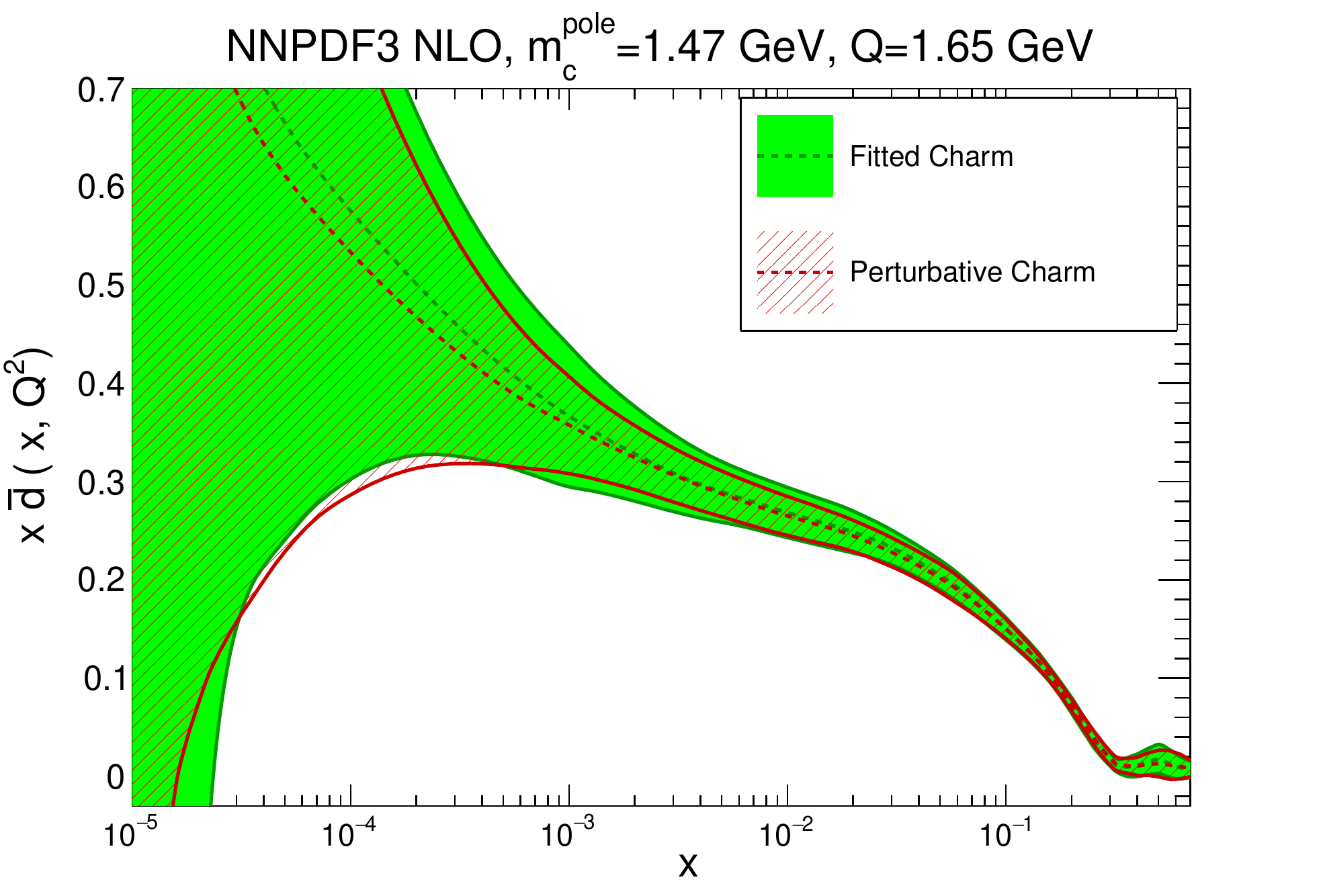}
  \includegraphics[width=0.45\textwidth]{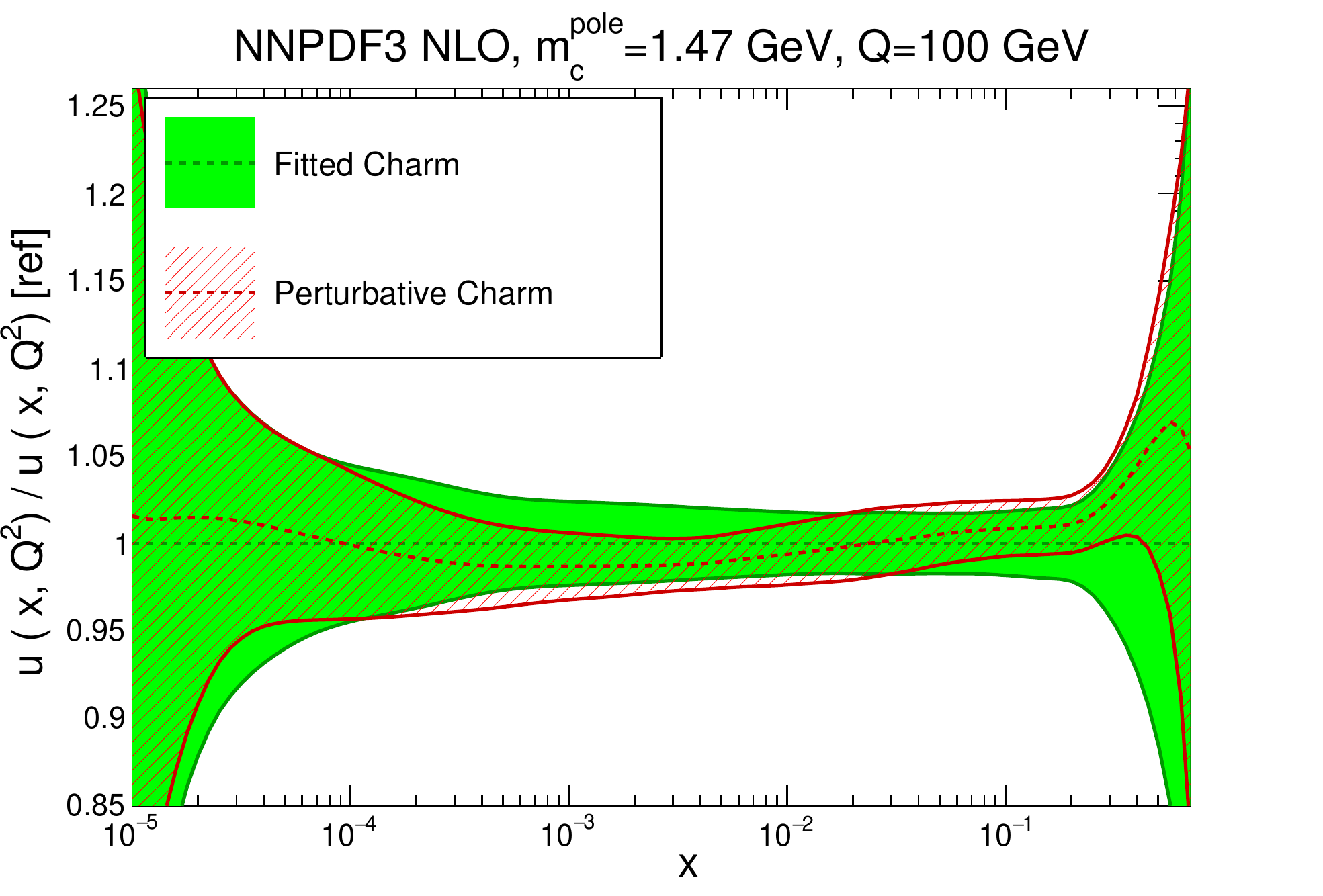}
  \includegraphics[width=0.45\textwidth]{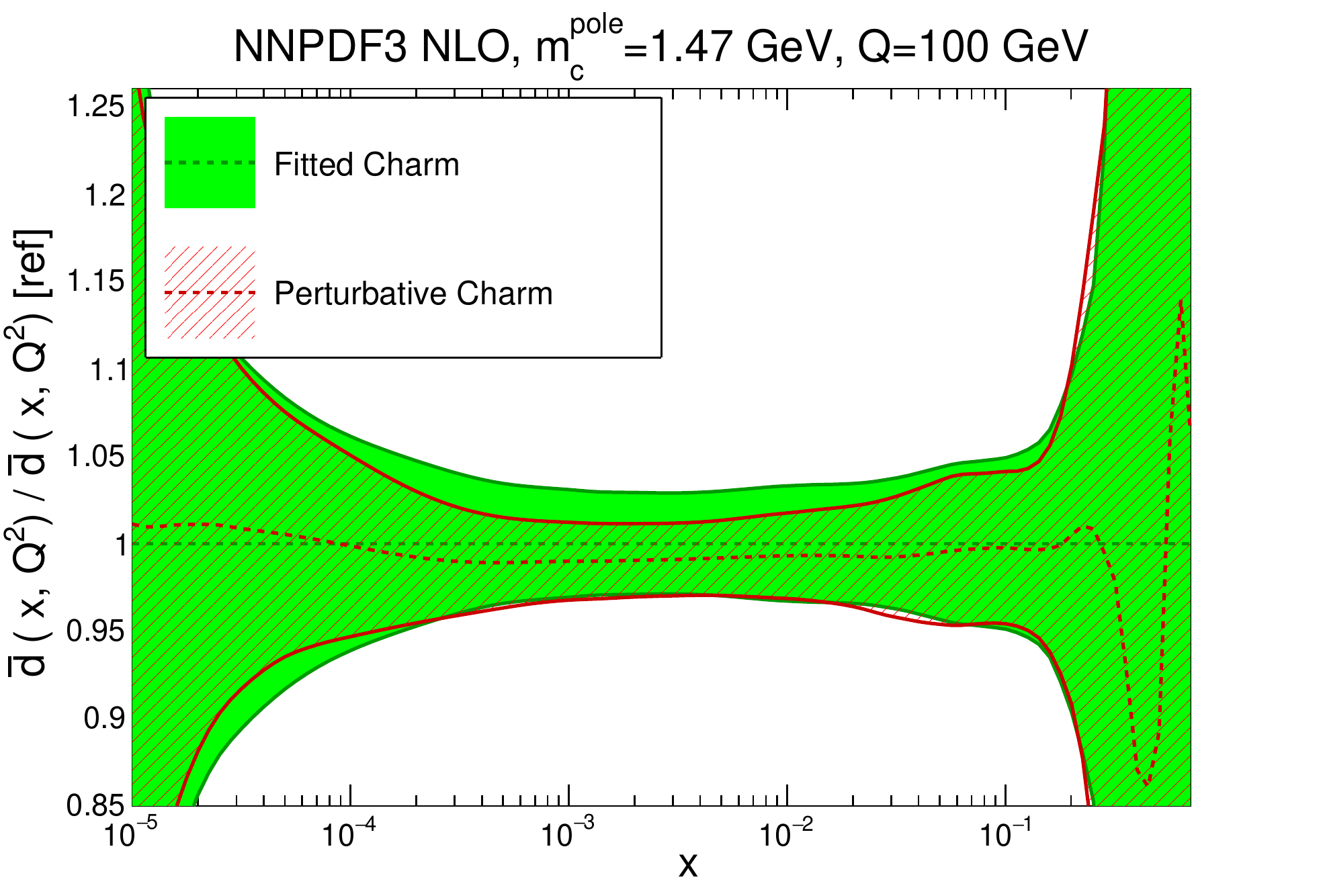}
\end{center}
\vspace{-0.8cm}
\caption{\small \label{fig:pdfbase1}
Same as Fig.~\ref{fig:pdfbase}, but now showing the up (left) and
anti-down (right) PDFs.
   }
\end{figure}

It is clear from these comparisons that fitting charm has a moderate
impact on the global fit: the fit is somewhat longer (by less than two
sigma), and uncertainties on predictions are a little larger. 
However the overall quality of the 
global fit is
somewhat improved: at the level of individual experiments, in most
cases the fit quality is similar, with the improvements in the 
case of fitted charm more marked for the HERA inclusive, SLAC, CHORUS 
and E866 data. The $\chi^2/N_{\rm dat}$ of the HERA charm combination 
is essentially the same in the fitted and perturbative charm cases, and
the fit quality to the LHC experiments is mostly unaffected, as expected
since the  measurements included have very limited direct
sensitivity to the charm PDF.

On the other hand,
the EMC charm structure function data cannot be fitted in a
satisfactory way with perturbative charm: the best we can do without 
fitted charm is $\chi^2/N_{\rm dat}=7.3$, corresponding to an increase in 
the total $\chi^2$ of over $100$ units.
However, the $\chi^2$ to these data improves dramatically when charm is fitted,
and an excellent description with $\chi^2/N_{\rm dat}=1.09$ is
achieved.
It is interesting to note that some previous PDF determinations with
intrinsic charm had difficulties in providing a satisfactory
description of the EMC charm structure function data 
(see e.g. Ref.~\cite{Jimenez-Delgado:2014zga}).
In the following, the EMC charm data will be excluded
from the default  fits  with perturbative charm, though we will come
back to the issue of including these data when charm is purely
perturbative when discussing charm mass dependence in
Sect.~\ref{sec:hqmass}, and when specifically analyzing the impact of
these data in Sect.~\ref{sec:emc}.

In Figs.~\ref{fig:pdfbase} and \ref{fig:pdfbase1} we compare several PDFs
with fitted and perturbative charm, both
at a low scale,
  $Q=1.65$ GeV (just above the scale at which charm is generated in the purely 
perturbative fit),  and at a high scale, $Q=100$ GeV. It is clear that light
quarks and especially the gluon are moderately affected by the
inclusion of fitted charm, with a barely visible increase in the PDF
uncertainty. The charm PDF and especially its uncertainty are affected 
more substantially: we will discuss this in
detail in Sect.~\ref{sec:intrinsic}.

\subsection{Dependence on the charm quark mass and fit stability}
\label{sec:hqmass}

As discussed in the introduction, one of the motivations for
introducing a fitted charm PDF is to separate the role of the charm
mass as a physical parameter from its role in determining the
boundary condition of the charm PDF. This dual role played by the
charm mass can be disentangled by studying the dependence of the
fit results (and in particular the charm PDF) on the value of the
charm mass when charm is perturbative or fitted.
To this purpose, we compare fit results obtained when the charm mass
is varied between $m_c^{\rm pole}=1.33$ and $1.61$~GeV about our
central  $m_c^{\rm pole}=1.47$, corresponding to a five-sigma
variation in units of the
PDG uncertainty on the  $\overline{\rm MS}$ mass $m_c(m_c)$ 
using  one-loop conversion to pole.
After examining the stability of our results on the charm mass
value, we discuss their stability with respect to different
theoretical treatments. First, we  show  results for a fit with $m_c^{\rm pole}=1.275$ GeV,
produced in order to compare with a fit with $\overline{\rm MS}$ masses with the
same numerical value of $m_c$, and then, we  discuss how the fit results change if we switch from 
pole to $\overline{\rm MS}$ masses. Finally, we discuss how fit results
would change if an S-ACOT-like treatment of the heavy quark was
adopted, in which massive corrections to charm-initiated contributions
are neglected.

\begin{figure}
\begin{center}
  \includegraphics[width=0.49\textwidth]{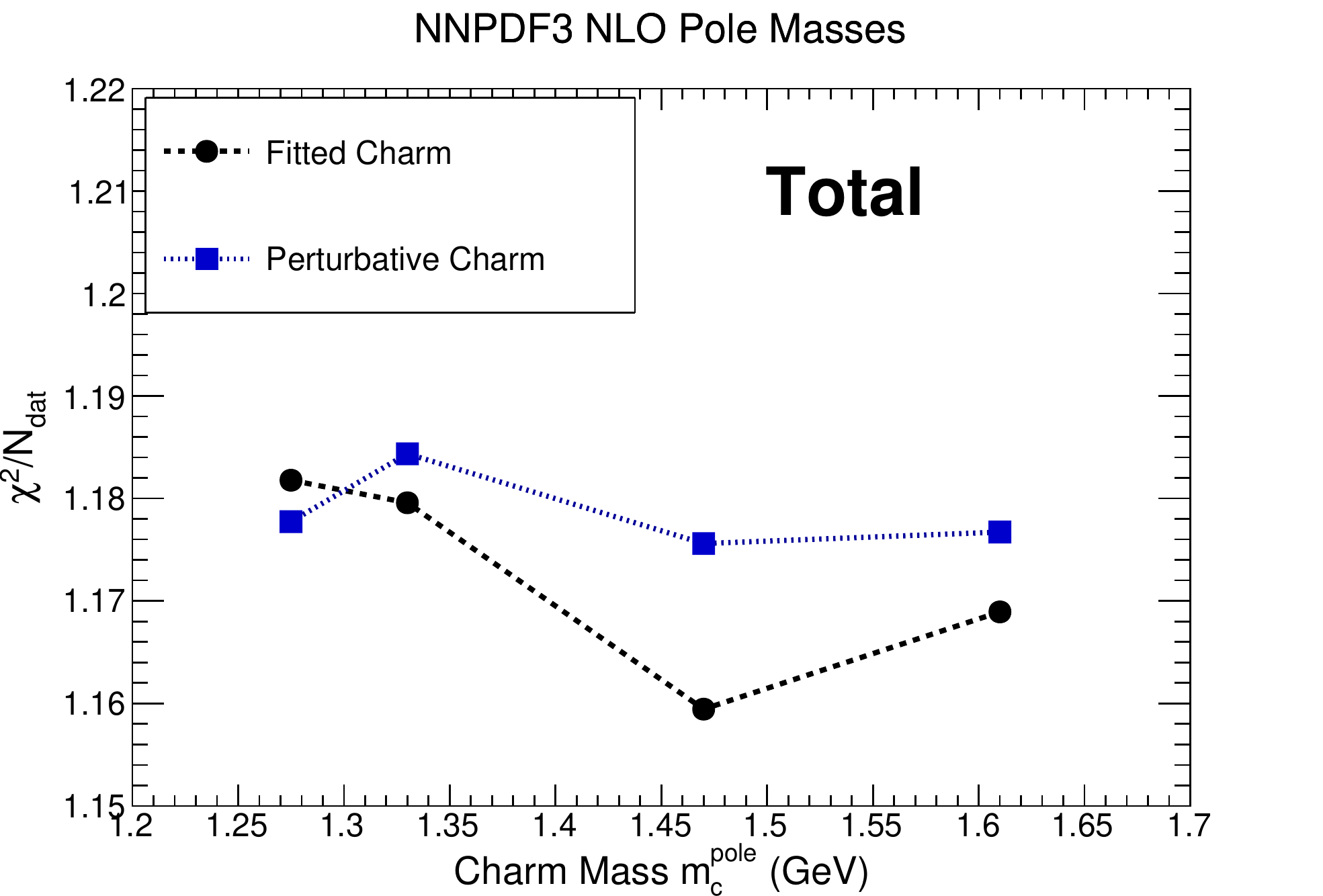}
  \includegraphics[width=0.49\textwidth]{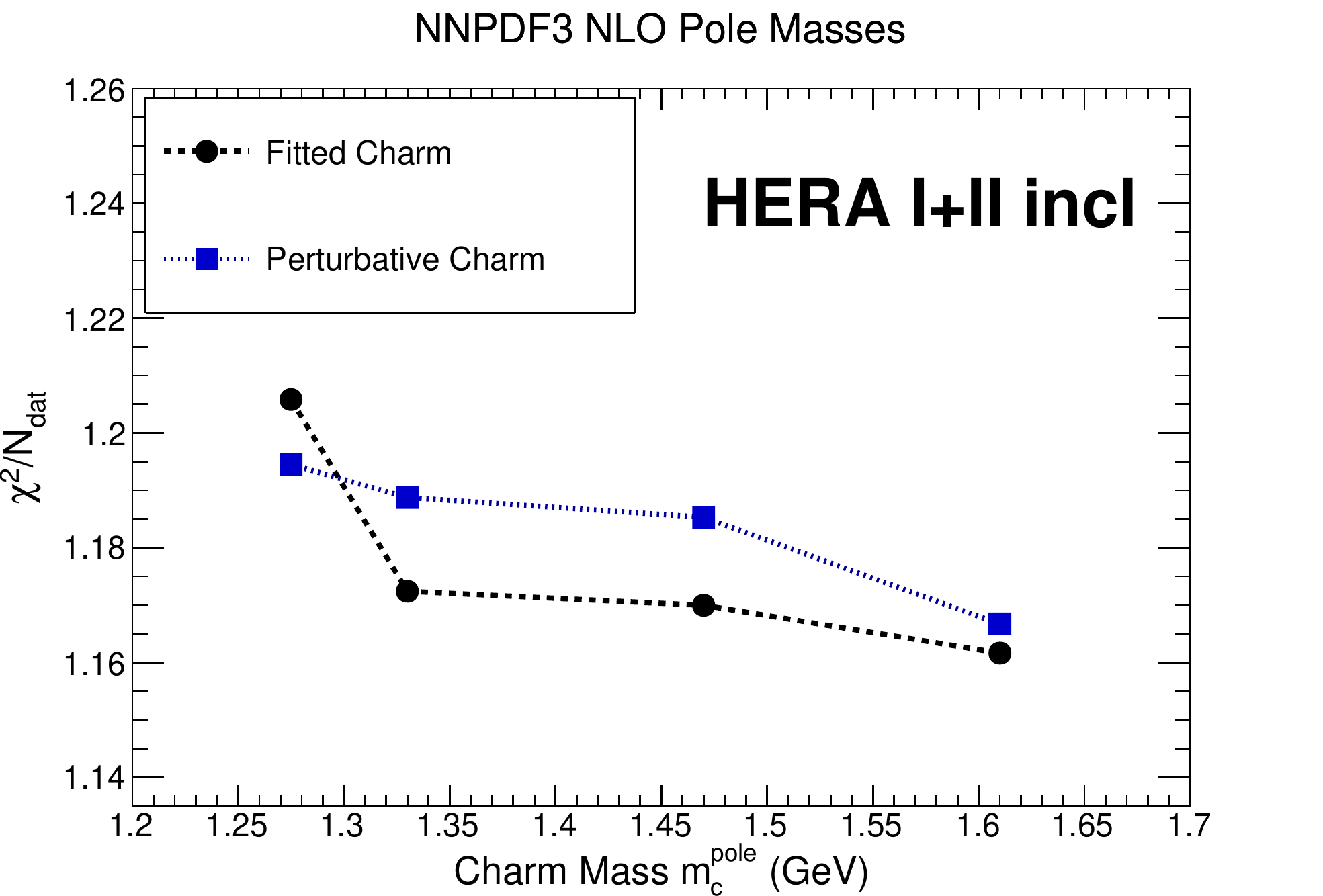}
  \includegraphics[width=0.49\textwidth]{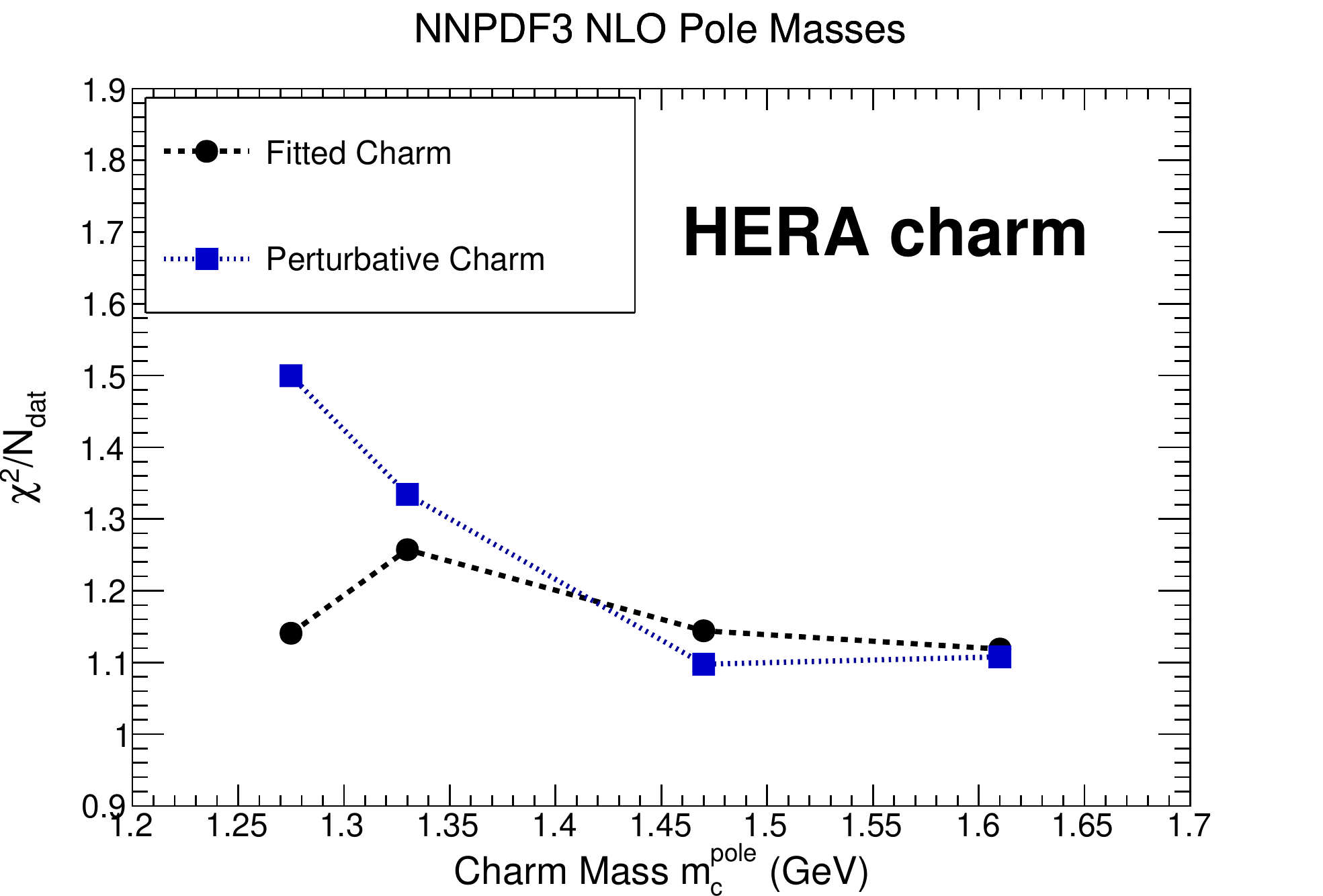}
  \includegraphics[width=0.49\textwidth]{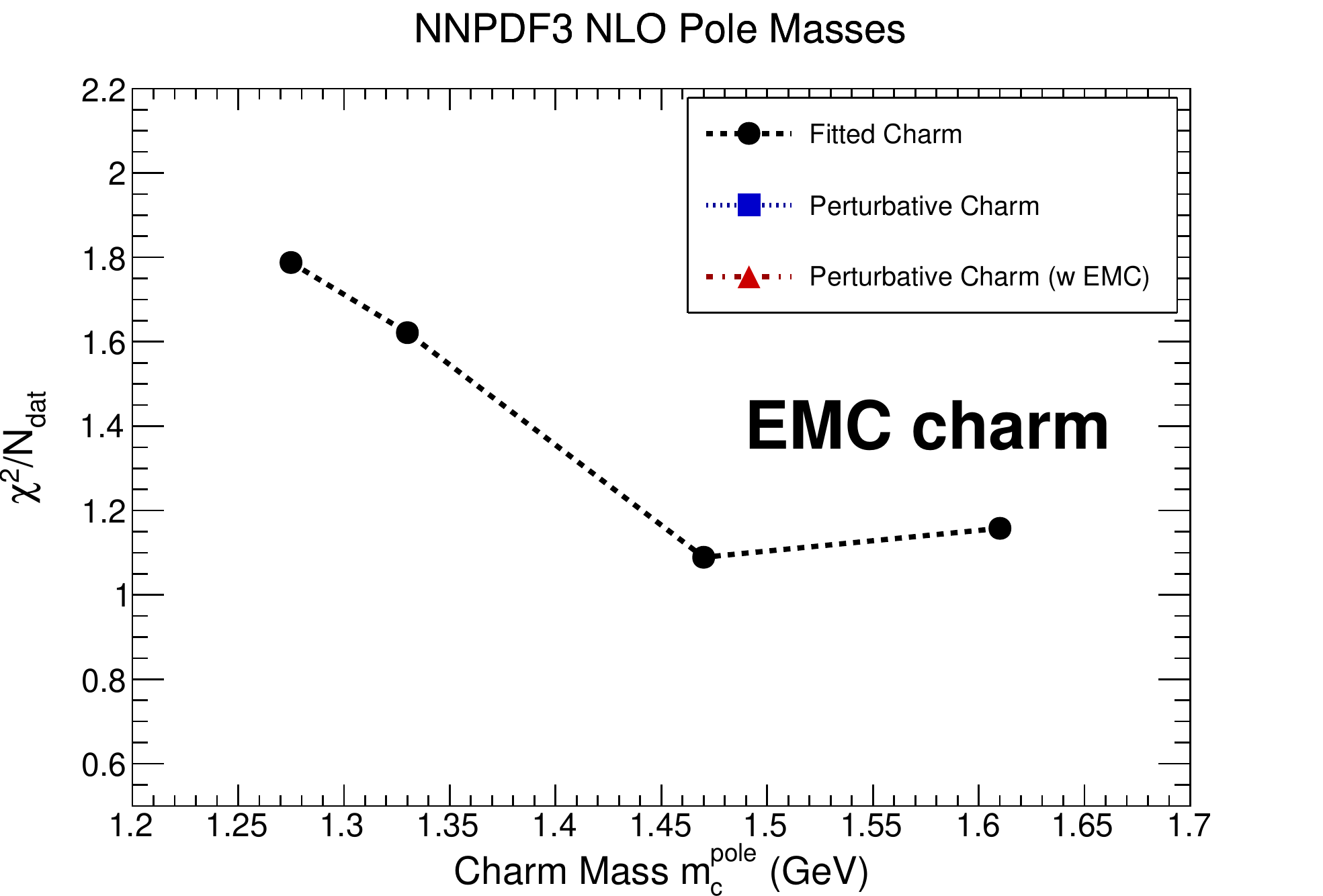}
  \includegraphics[width=0.49\textwidth]{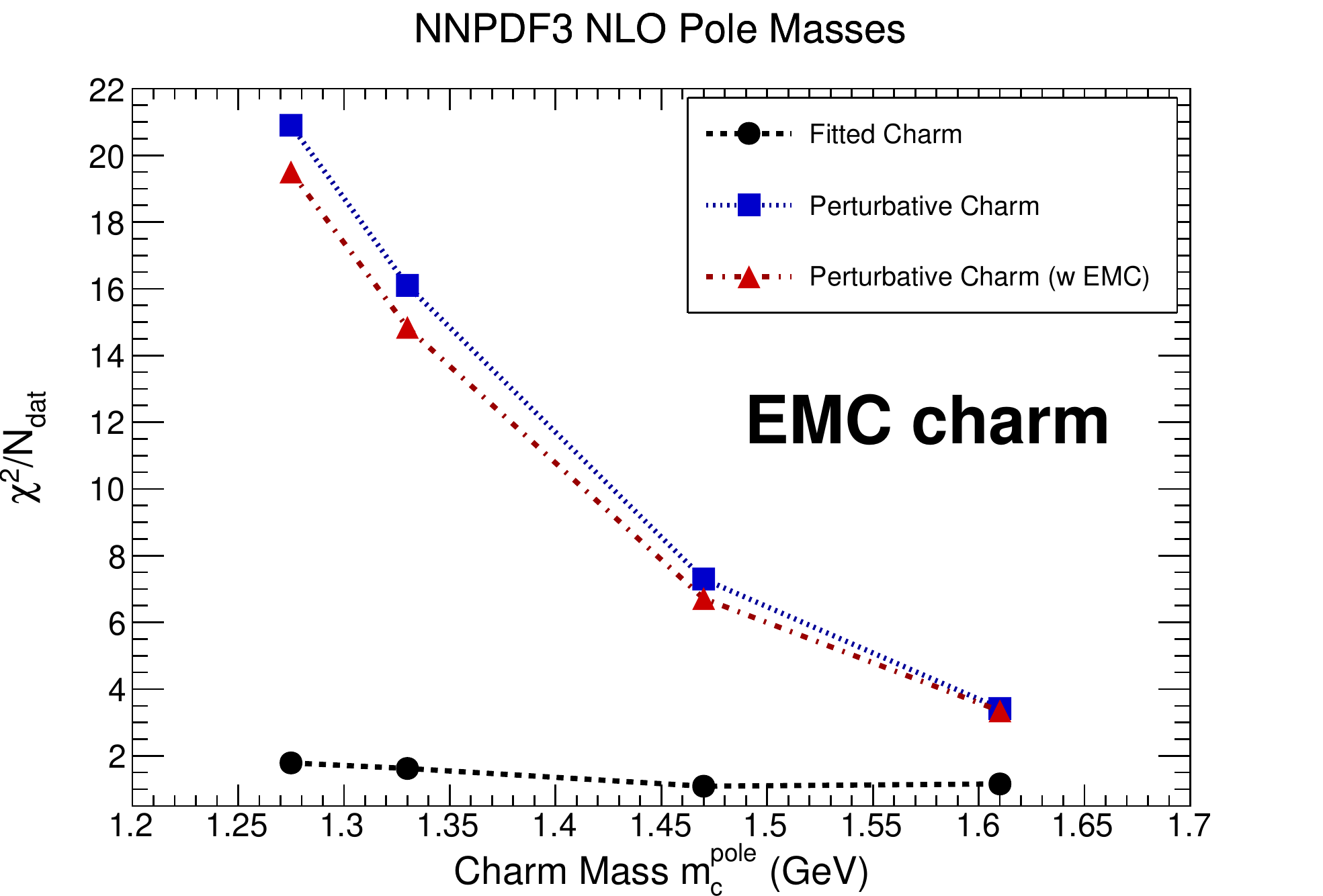}
\end{center}
\vspace{-0.3cm}
\caption{\small \label{fig:chi2profiles}
  The $\chi^2$ per data point for the total dataset (top left); 
for the HERA inclusive (top right) and charm structure function
(center left) combined datasets and for the  EMC charm data (center
right), for fits  
with perturbative and fitted charm, as a function of the
value of the charm pole mass $m_c^{\rm pole}$. In the bottom row the
$\chi^2$ for the  EMC charm data is shown again with an enlarged scale
which enables the inclusion of the values for perturbative charm; in
this plot only  
for fits with perturbative charm we show results both with and without the EMC  
data included in the fit.  In all other plots, the perturbative charm  
results are for fits without EMC data. The fitted charm fits always  
include the EMC data.  
   }
\end{figure}
\pagebreak

For a first assessment of the relative fit quality, 
in Fig.~\ref{fig:chi2profiles} we show $\chi^2/N_{\rm dat}$
as a function of the pole charm mass value, in the fits both with perturbative and
fitted charm. 
The plot has been produced  using  the experimental 
definition of the $\chi^2$.
The values shown here  correspond to the full dataset, the inclusive and charm
HERA structure function combined data, and EMC structure function data.
In the case of perturbative charm, we generally show the results of a
fit in which the EMC data are not included, except  in the plot of the
$\chi^2$ to the EMC data themselves, where we show both fits with EMC
data included and not included. It is seen that the EMC data cannot be
fitted when charm is perturbative in the sense that their poor
$\chi^2$ does not significantly improve upon their inclusion in the
fit.  We will accordingly henceforth exclude the EMC data from all fits
with perturbative charm, as their only possible effect would be to
distort fit results without any significant effect on fit quality.

 It is interesting to observe that while with fitted charm the EMC
data seem to favour a value of the charm mass around $1.5$~GeV, 
close to the current PDG
average, with perturbative charm they would favour an unphysically
large value.
These results also suggest
that a determination of the charm mass from a global fit with 
fitted charm might in principle be possible, but that this requires high 
statistics and precision
analysis techniques, such as those used in
Refs.~\cite{Lionetti:2011pw,Ball:2011us} for the determination of the
strong coupling $\alpha_s$.
%

\begin{figure}[h!]
  \begin{center}
   \includegraphics[width=0.45\textwidth]{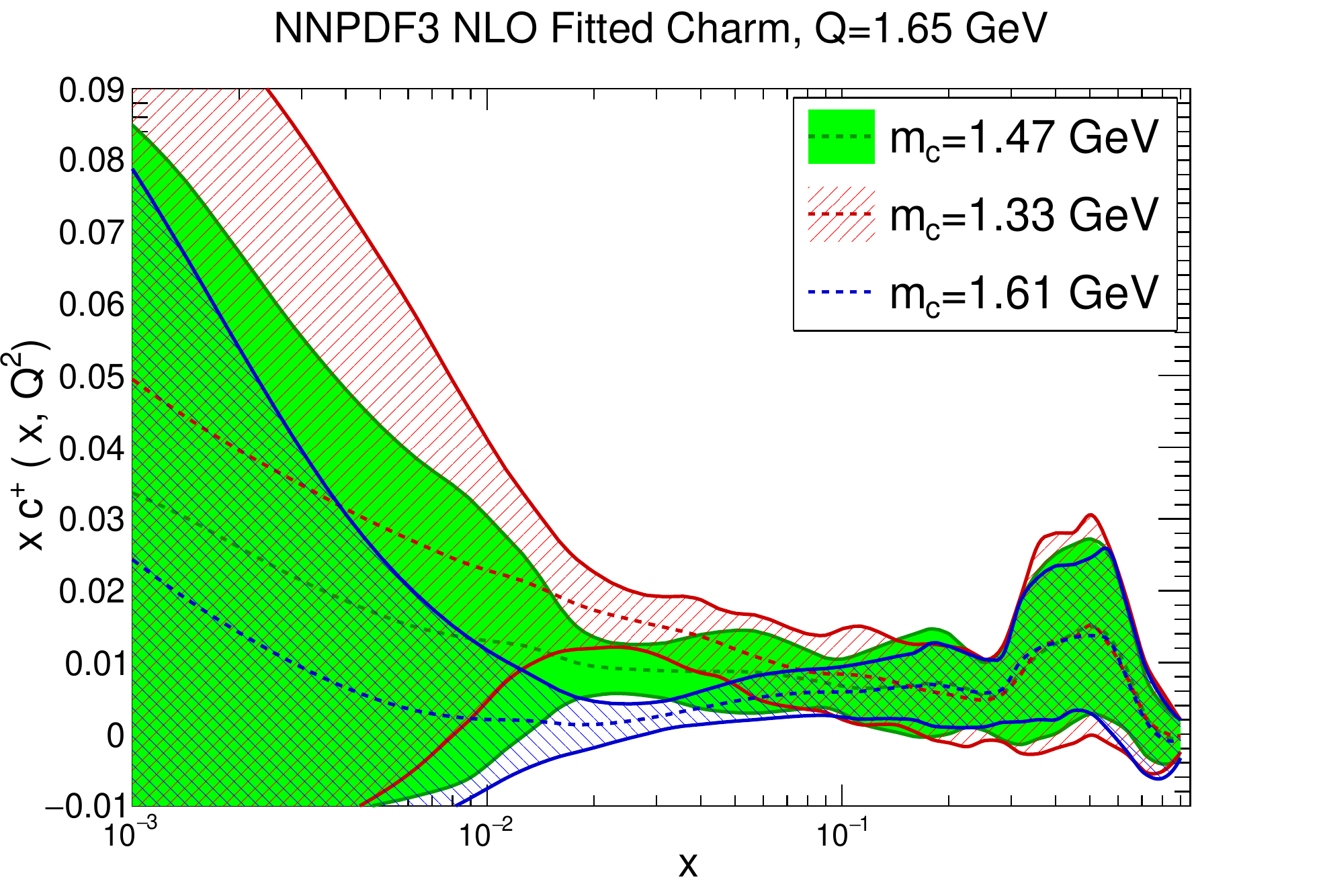}
  \includegraphics[width=0.45\textwidth]{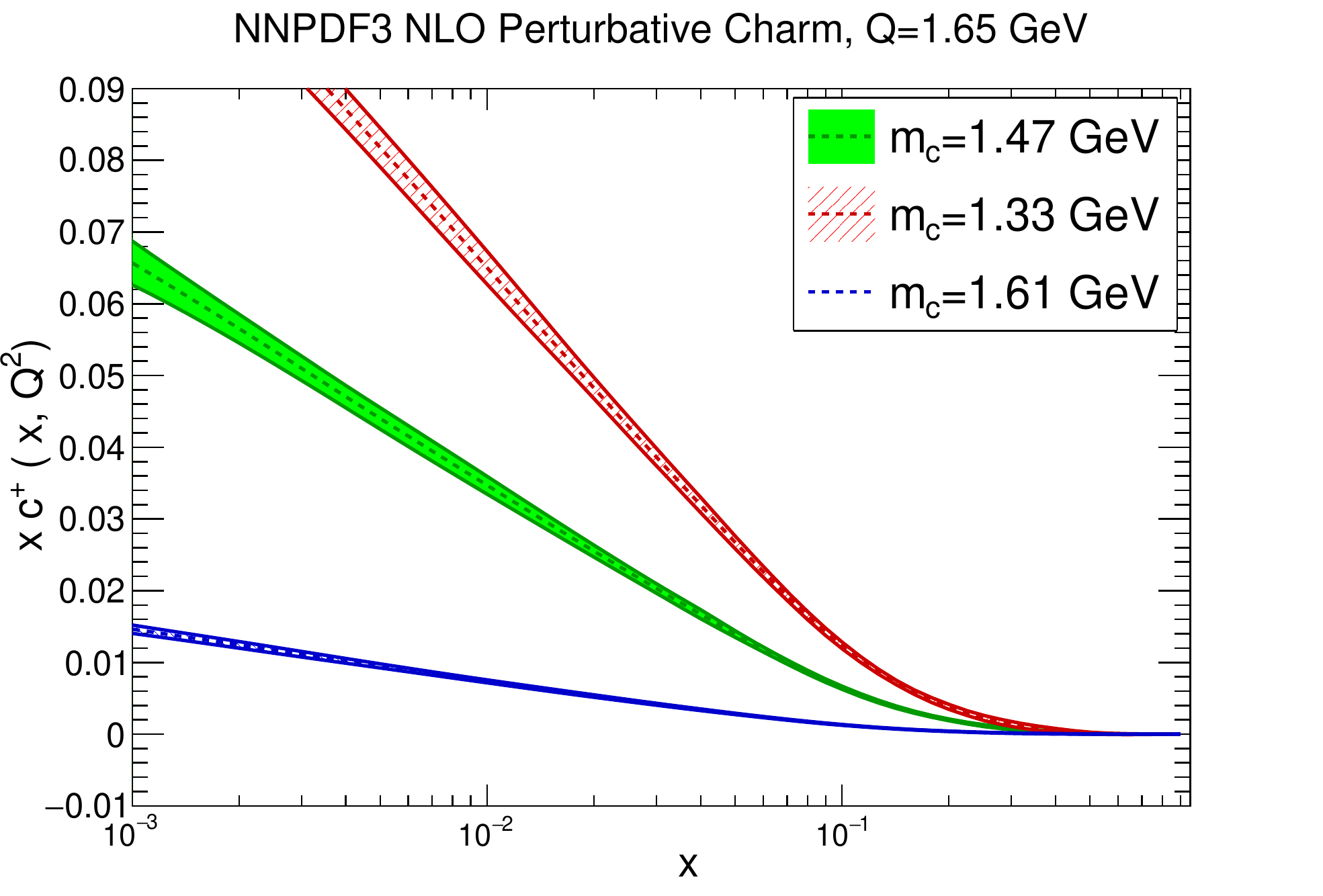}
  \includegraphics[width=0.45\textwidth]{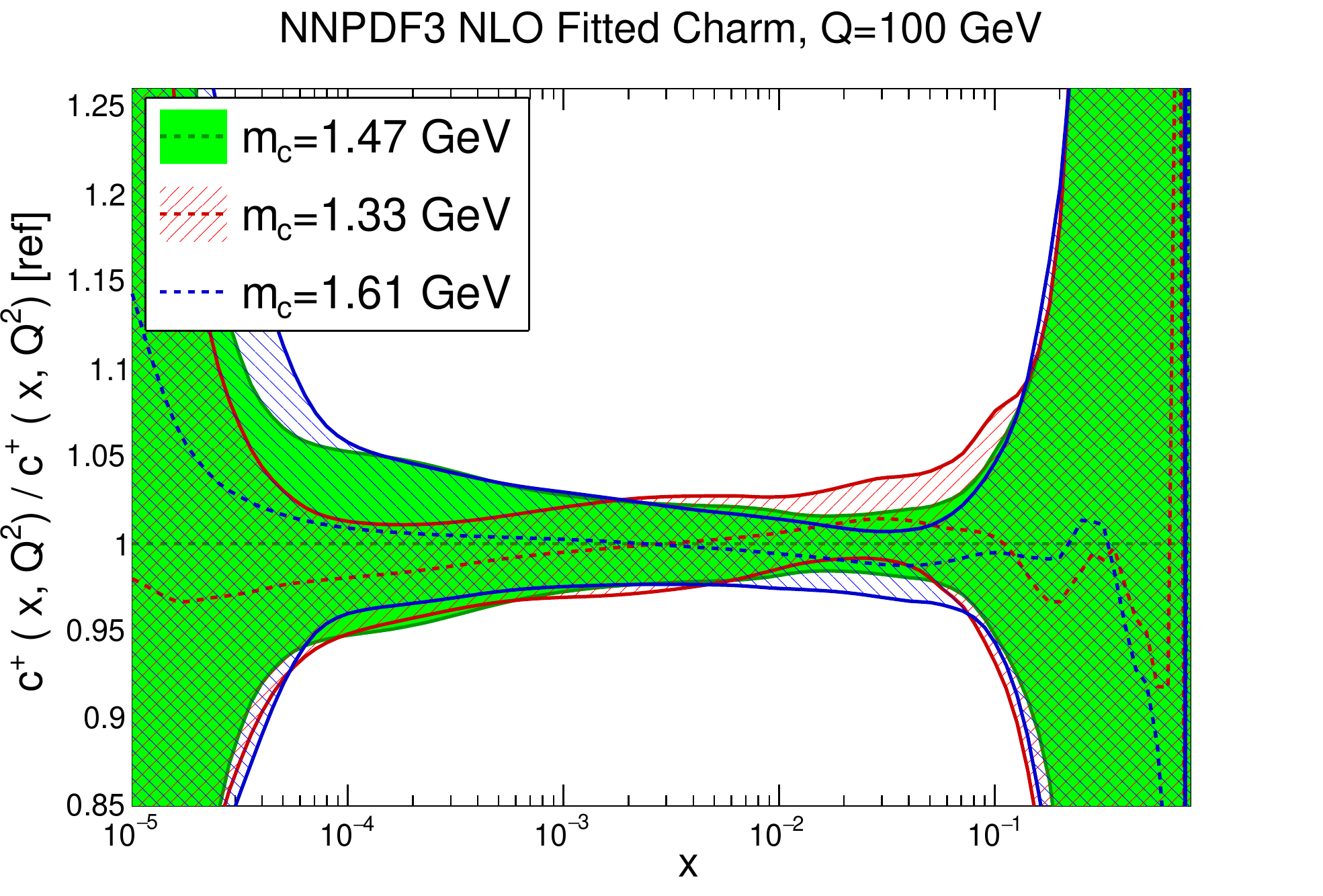}
  \includegraphics[width=0.45\textwidth]{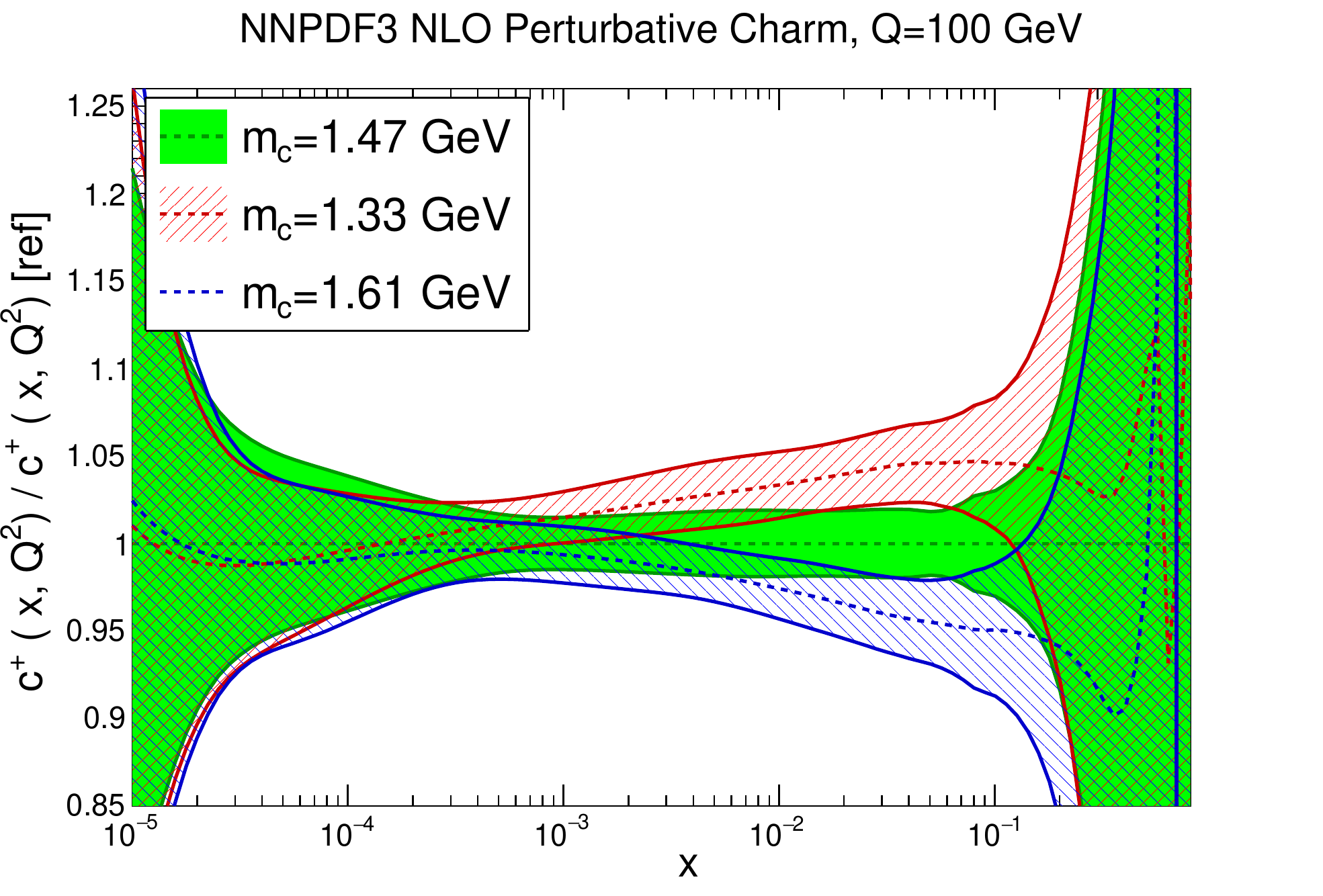}
\end{center}
\vspace{-0.8cm}
\caption{\small \label{fig:pdfmcvar}
 Dependence of the charm PDF on the value of the pole charm
  mass $m_c^{\rm pole}$: the charm PDF obtained with fitted  charm  (left) and  perturbative charm
   (right) are compared for   
  $m_c^{\rm pole}=1.33,~1.47$ and $1.61$ GeV, at a low scale $Q=1.65$ GeV
  (top) and at a high scale $Q=100$ GeV (bottom). At high scale, PDFs
  are shown as a ratio to the fit with central $m_c^{\rm pole}=1.47$~GeV.
   }
\vspace{-0.1cm}
  \begin{center}
   \includegraphics[width=0.45\textwidth]{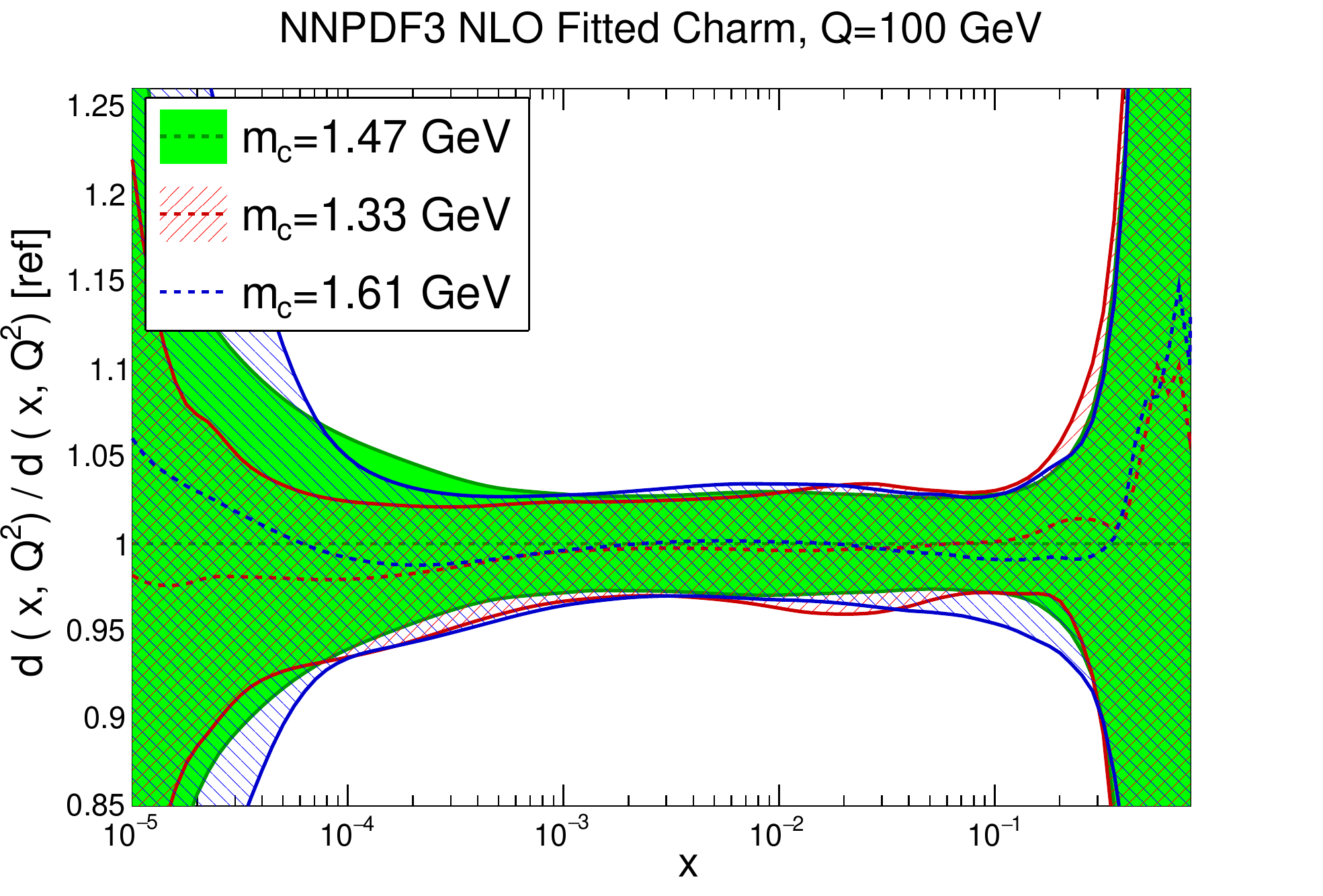}
  \includegraphics[width=0.45\textwidth]{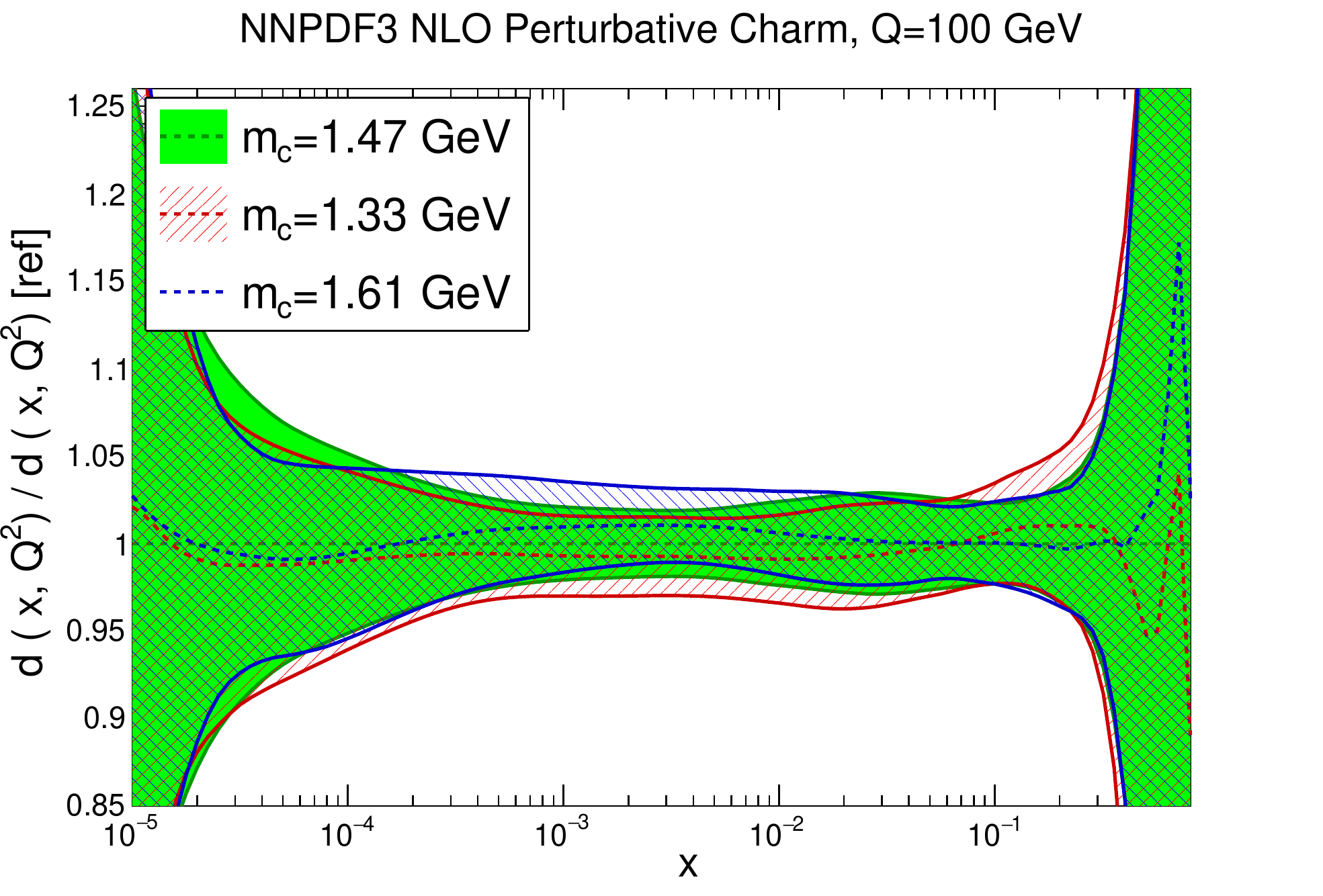}
  \includegraphics[width=0.45\textwidth]{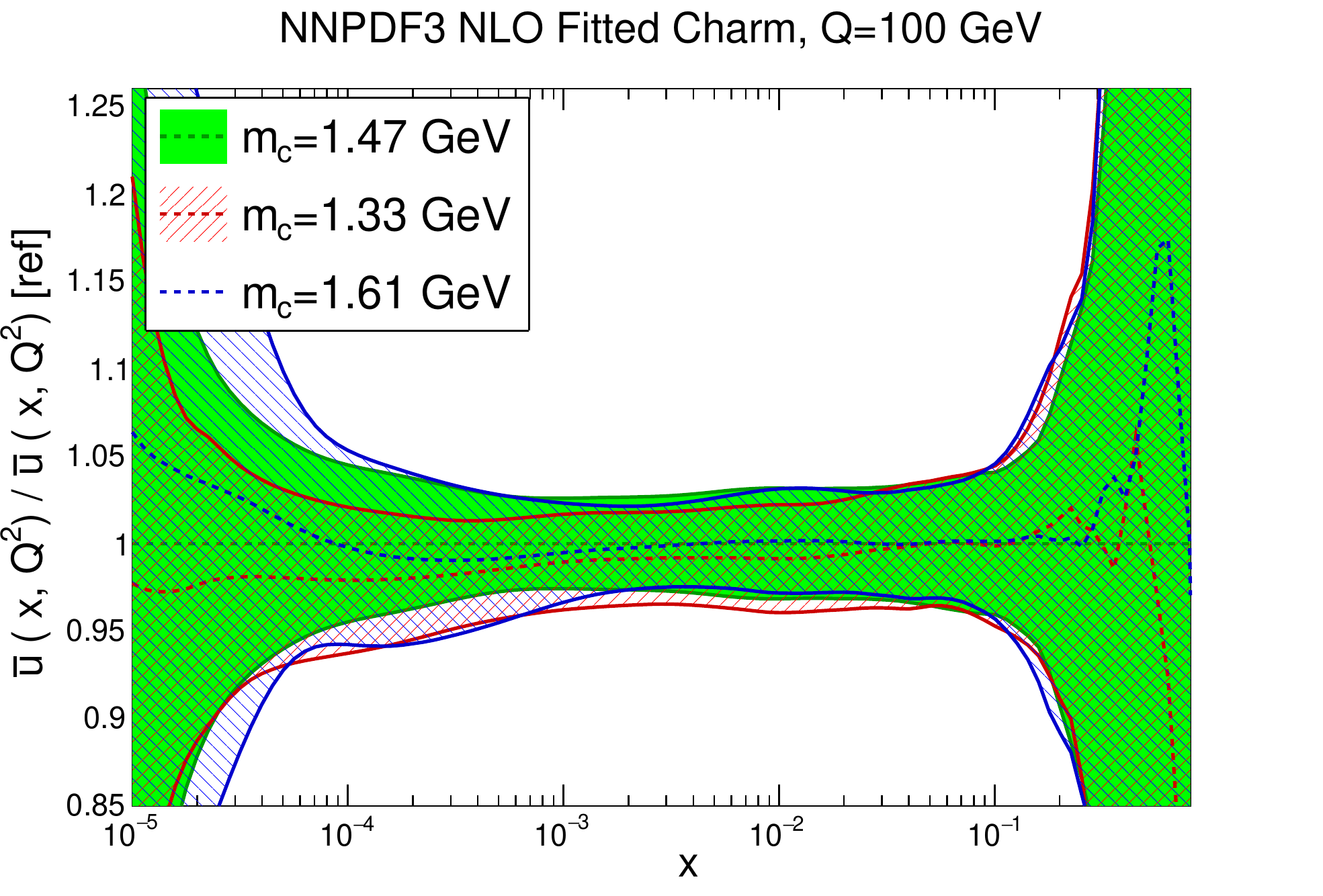}
  \includegraphics[width=0.45\textwidth]{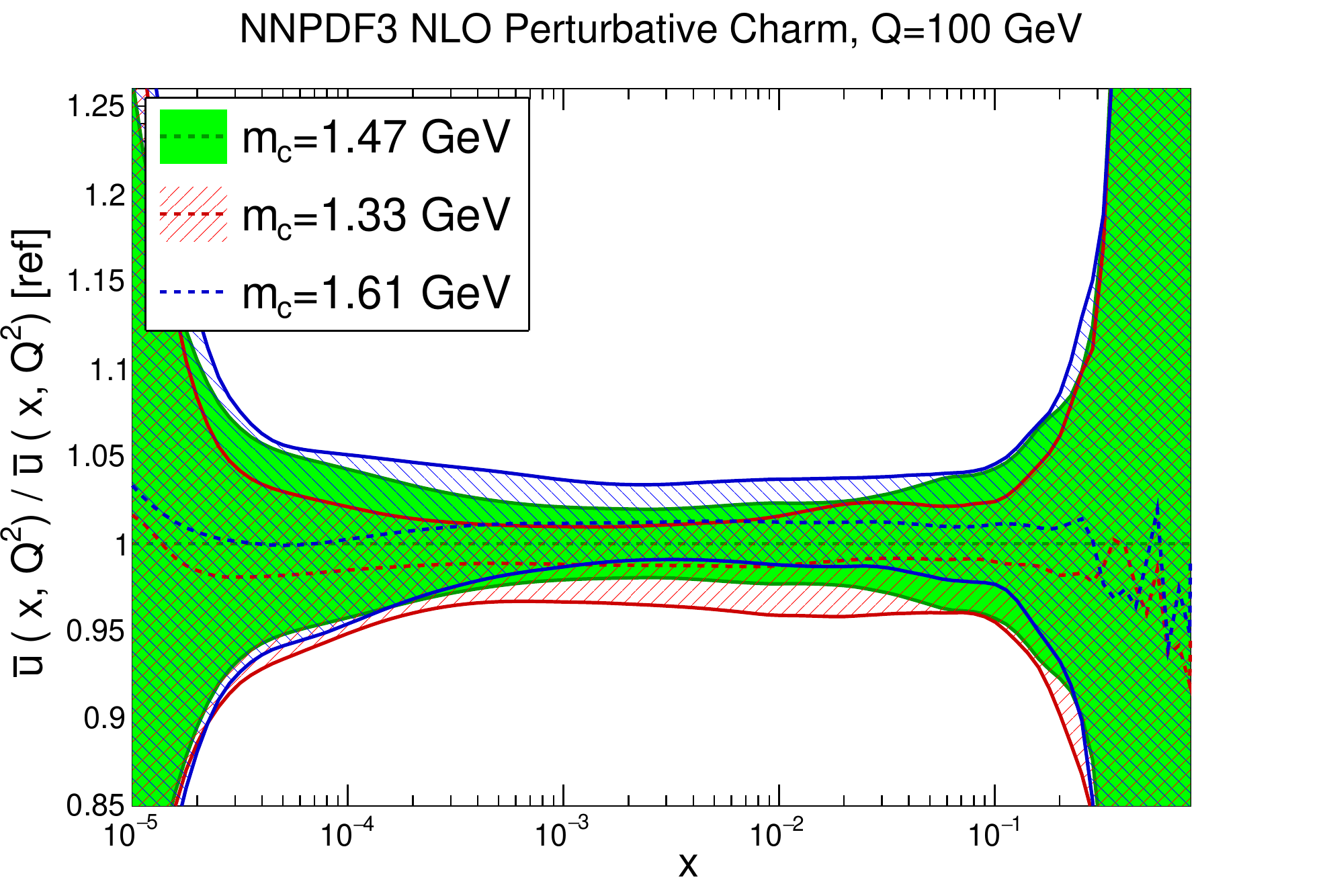}
\end{center}
\vspace{-0.8cm}
\caption{\small \label{fig:pdfmcvar2}
 Same as the bottom row of Fig.~\ref{fig:pdfmcvar}, but now for 
the down (top) and anti-up (bottom) PDFs.
   }
\end{figure}
    
We now compare the PDFs obtained with different values of the 
charm mass both with
perturbative and fitted charm: in
Fig.~\ref{fig:pdfmcvar} we show gluon and
charm, and in Fig.~\ref{fig:pdfmcvar2} up and
anti-down quarks. Results are shown at low and high scale
(respectively $Q=1.65$~GeV and $Q=100$~GeV) for charm, and at a high
scale only for the light quarks.
Of course, with perturbative charm the size of the charm PDF at any given scale
depends significantly on the value of the charm mass that sets the
evolution length: the lower the mass, the lower the starting scale,
and the larger the charm PDF at any higher scale. The percentage shift of
the PDF as
the mass is varied is of course very large close to threshold, but it
persists as a sizable effect even at high scale. Remarkably, this
dependence all but disappears when charm is fitted: both at low and
high scale the fitted charm PDF is extremely stable as the charm mass
is varied. This means that indeed once charm is fitted, its size is
mostly determined by the data, rather than by the (possibly
inaccurate) value at which we set the threshold for its production.
Interestingly, the other PDFs, and specifically the light quark PDFs,
also become generally less dependent on the value of the heavy quark masses,
even at high scale, thereby making LHC phenomenology somewhat more reliable.

\begin{figure}[t]
  \begin{center}
   \includegraphics[width=0.46\textwidth]{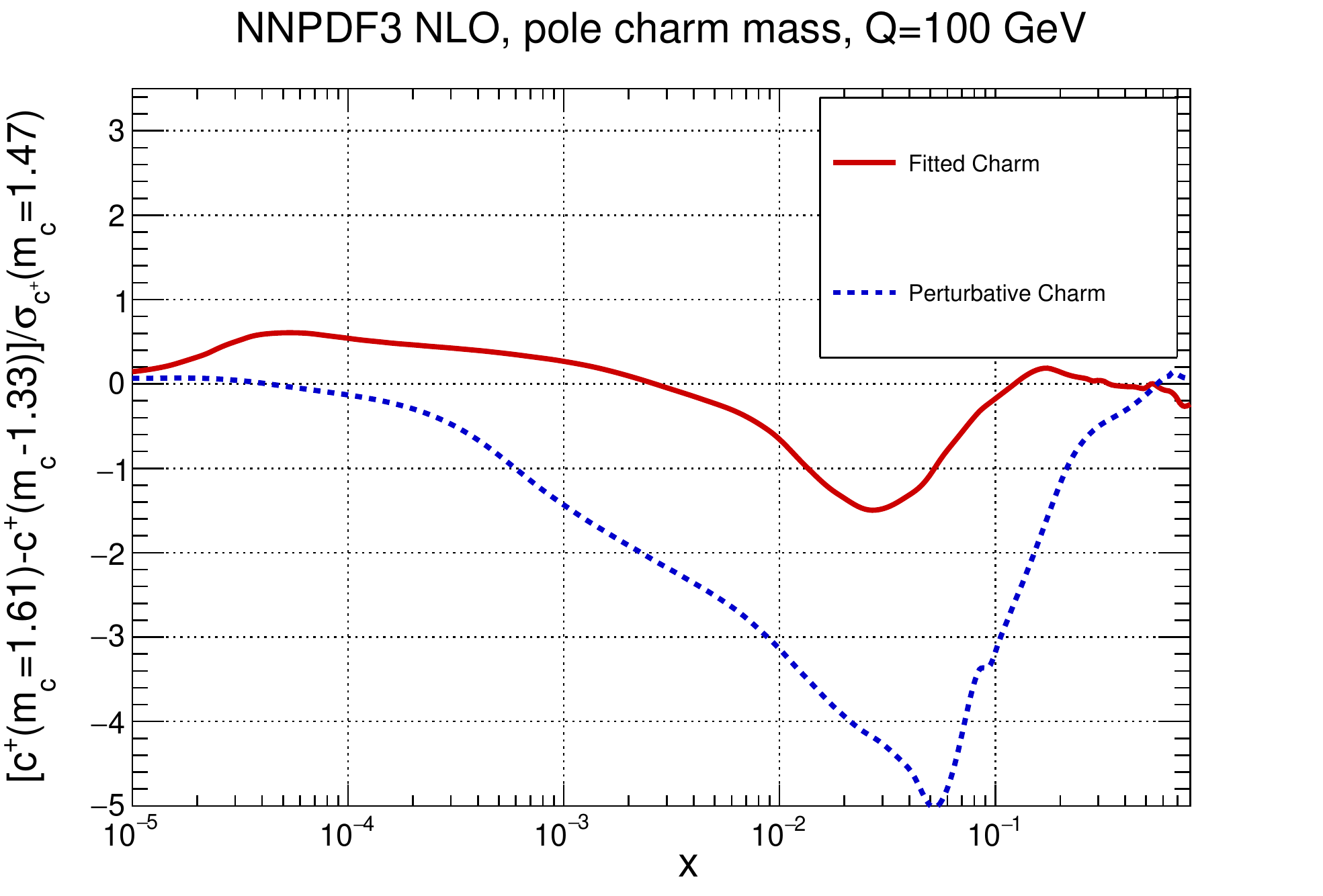}
  \includegraphics[width=0.46\textwidth]{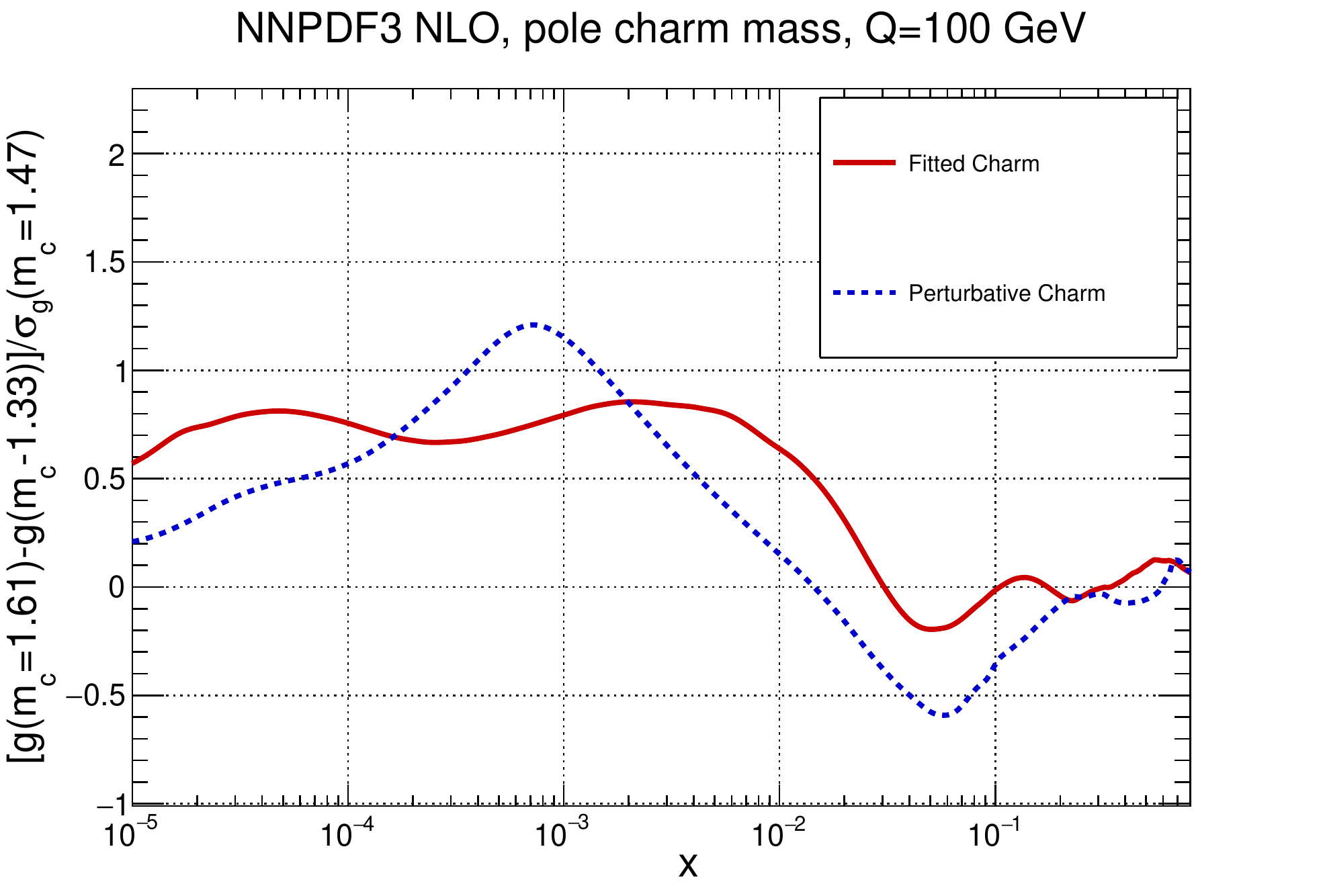}
  \includegraphics[width=0.46\textwidth]{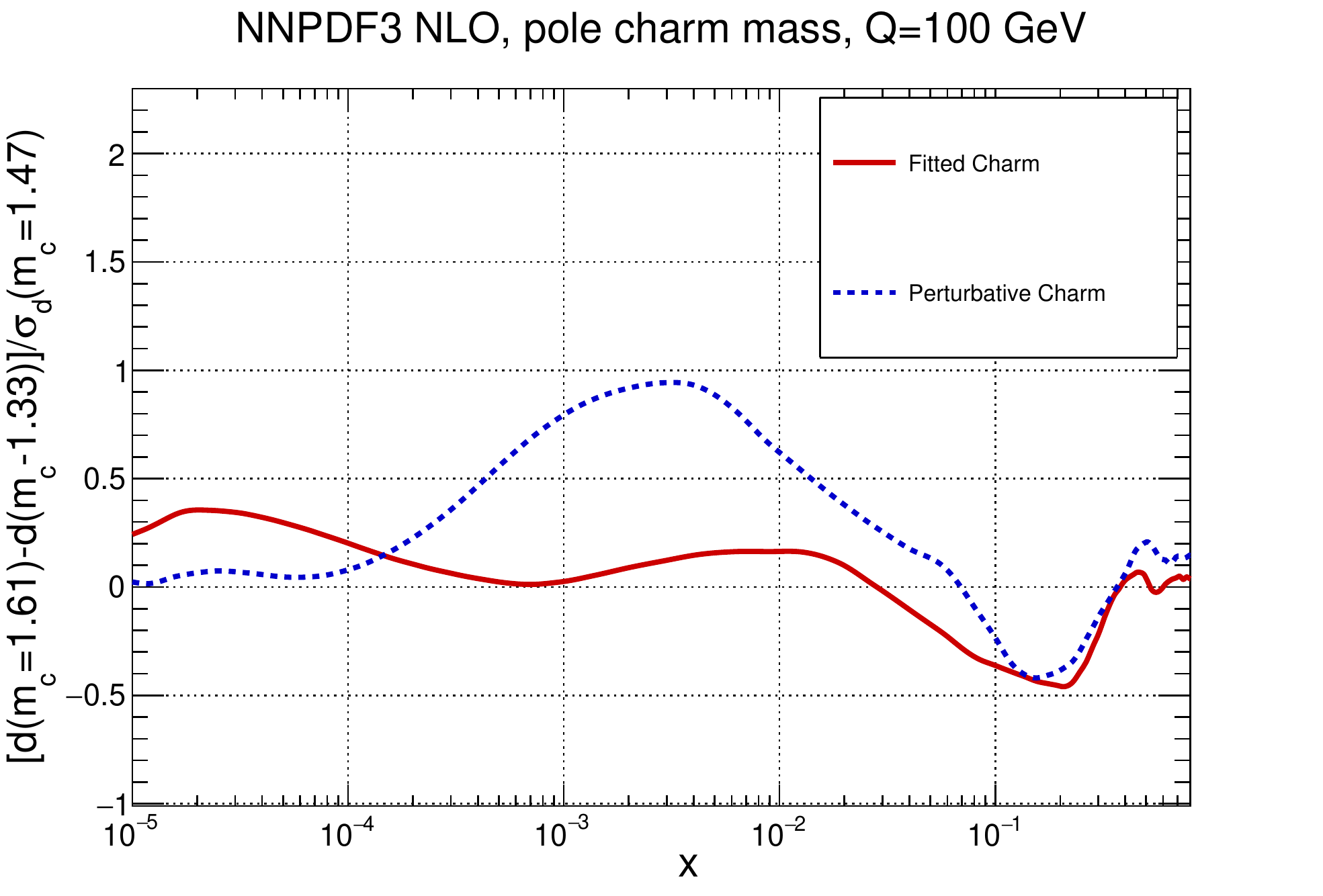}
  \includegraphics[width=0.46\textwidth]{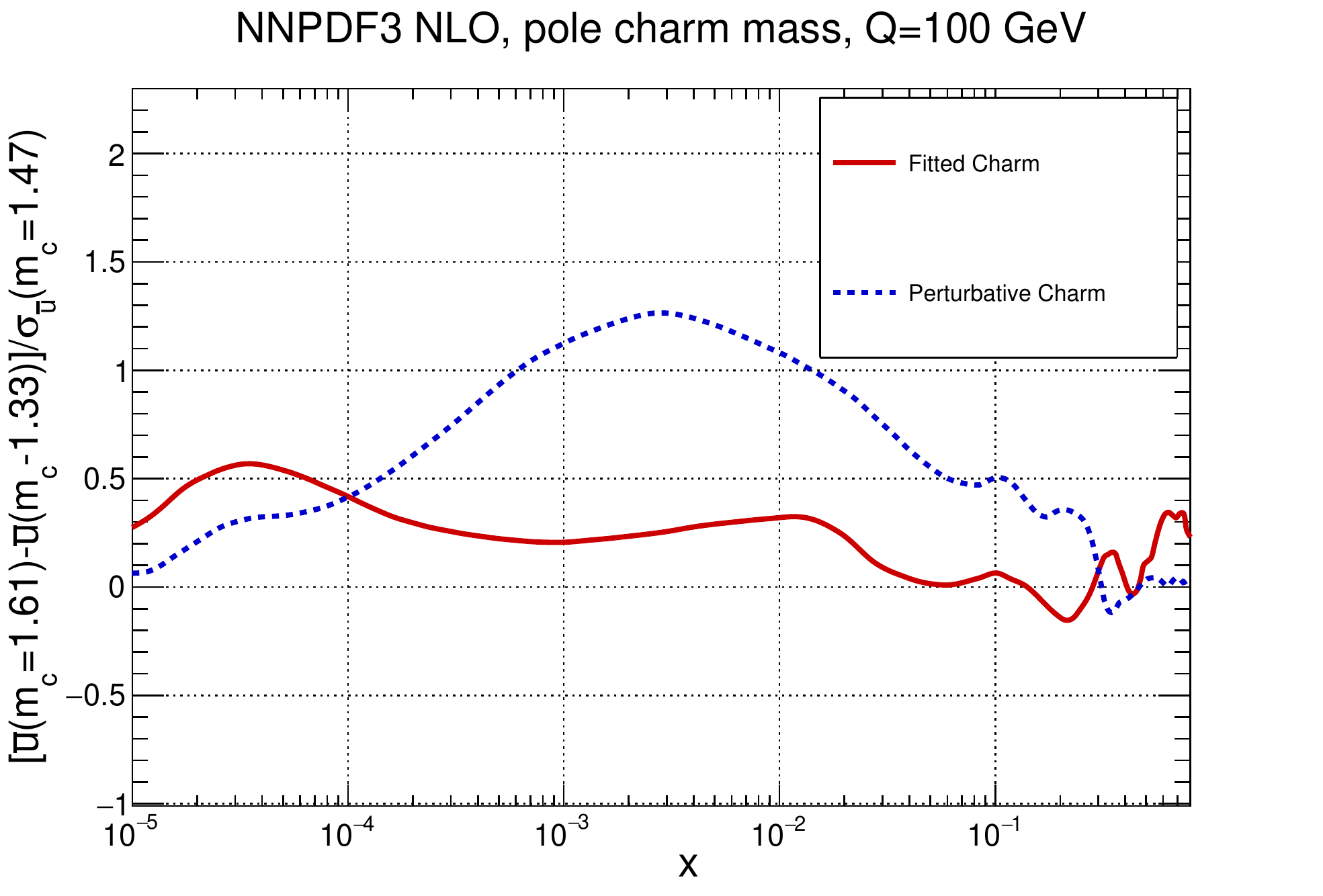}
\end{center}
\vspace{-0.8cm}
\caption{\small \label{fig:pdfmcvarpull}
  The pull Eq.~(\ref{eq:pull}) between PDFs determined with the two
  outer values of the quark mass ($m_c=1.61;\>1.33$~GeV),  in units of
  the PDF uncertainty 
  plotted as a function of $x$ at  $Q=100$ GeV. Results are shown for
  charm (top left), gluon (top right), down (bottom left) and anti-up
  (bottom right). Note the different scale on the $y$ axis in the
  different plots.
   }
\end{figure}

This improved stability upon heavy quark mass variation can be seen in
a more quantitative way by computing the pulls 
between the PDFs obtained using the two outer values
of the charm mass, defined as
\be
\label{eq:pull}
P_q(x,Q^2) \equiv \frac{q(x,Q^2)\vert_{m_{c}=1.61~{\rm GeV}} -
  q(x,Q^2)\vert_{m_{c}=1.33~{\rm GeV}}}{
\sigma_q(x,Q^2)\vert_{m_c=1.47~{\rm GeV}}} \, ,
\ee
where $q$ stands for a generic
PDF flavour, and $\sigma_q$ is the PDF uncertainty on 
the fit with the central $m_c$ value.
The  pull Eq.~(\ref{eq:pull}) evaluated at $Q=100$ GeV is plotted in
 Fig.~\ref{fig:pdfmcvarpull} as a function of $x$ for the charm,
 gluon, down and anti-up PDFs.
 It is clear that once charm is
  fitted the pull is essentially always less than one (that is, the PDF central value
  varies by less than one sigma when the mass is varied in the given range), while
  it is somewhat larger for light quarks and gluon, and much larger
  (up to five sigma) for the charm PDF if charm is purely
  perturbative. The smallest difference is seen for the gluon, for
  which the pull is less than one in both cases, and in fact slightly
  larger for fitted charm when $x\sim10^{-2}$.

\begin{figure}[t!]
  \begin{center}
    \includegraphics[width=0.46\textwidth]{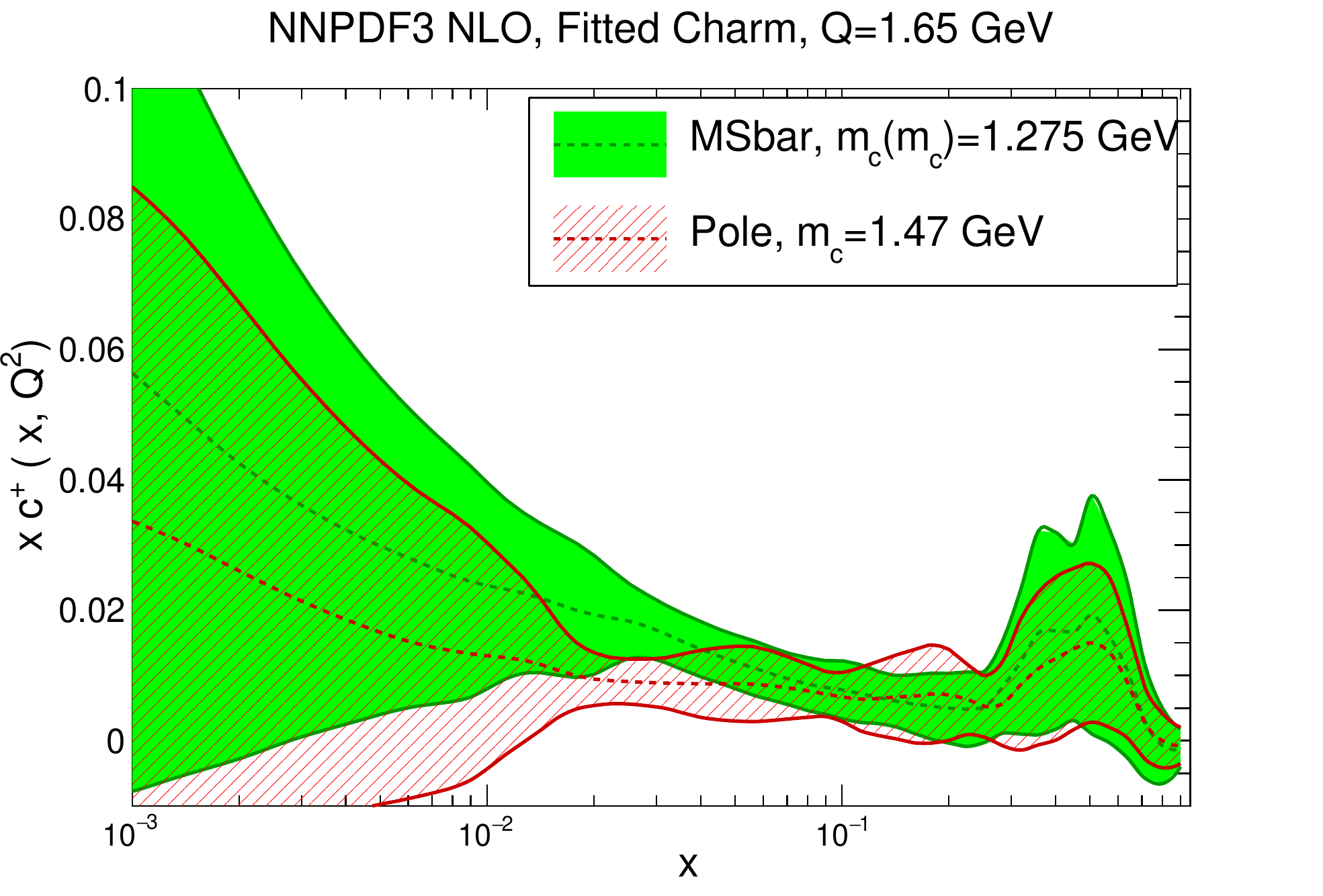}
    \includegraphics[width=0.46\textwidth]{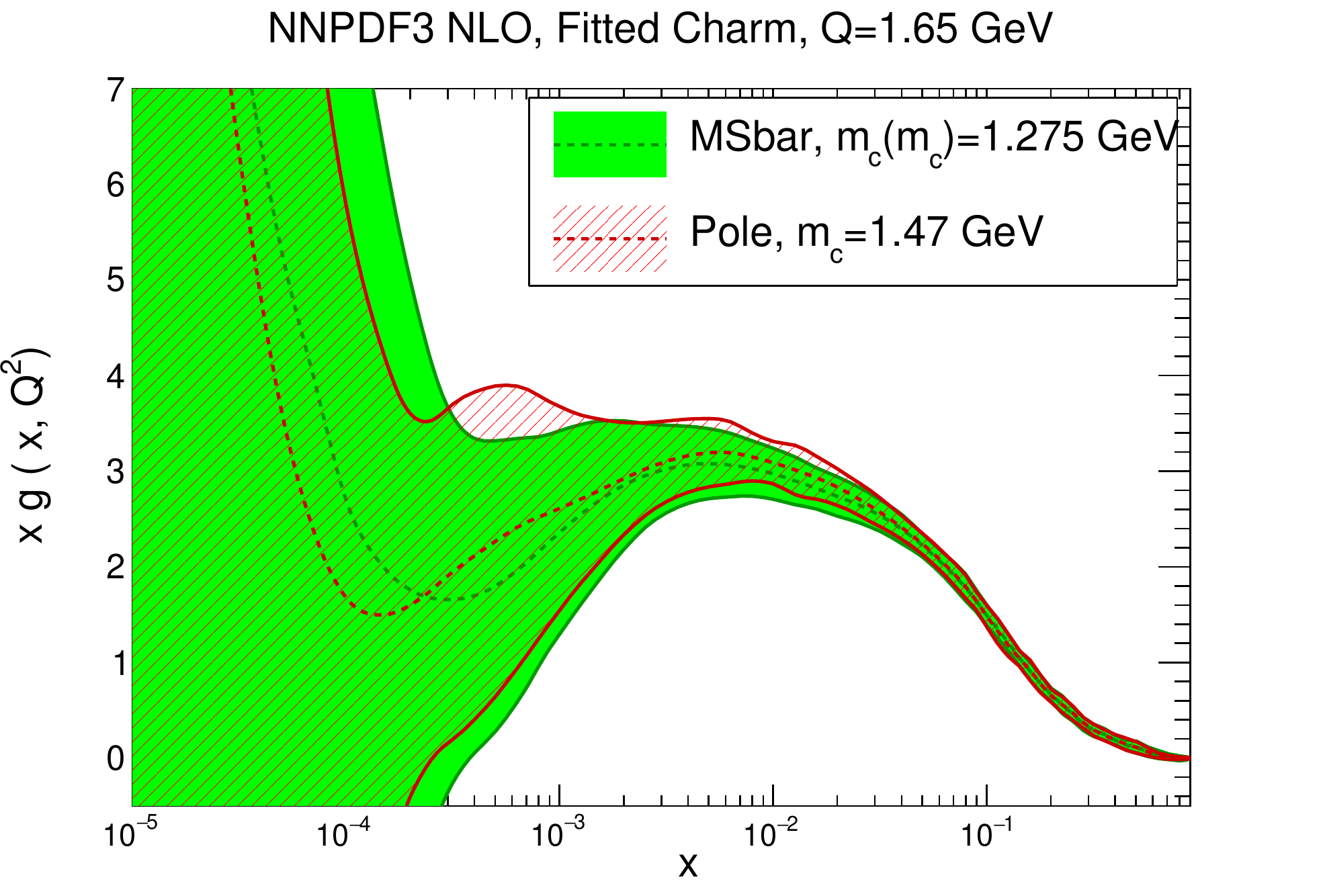}
    \includegraphics[width=0.46\textwidth]{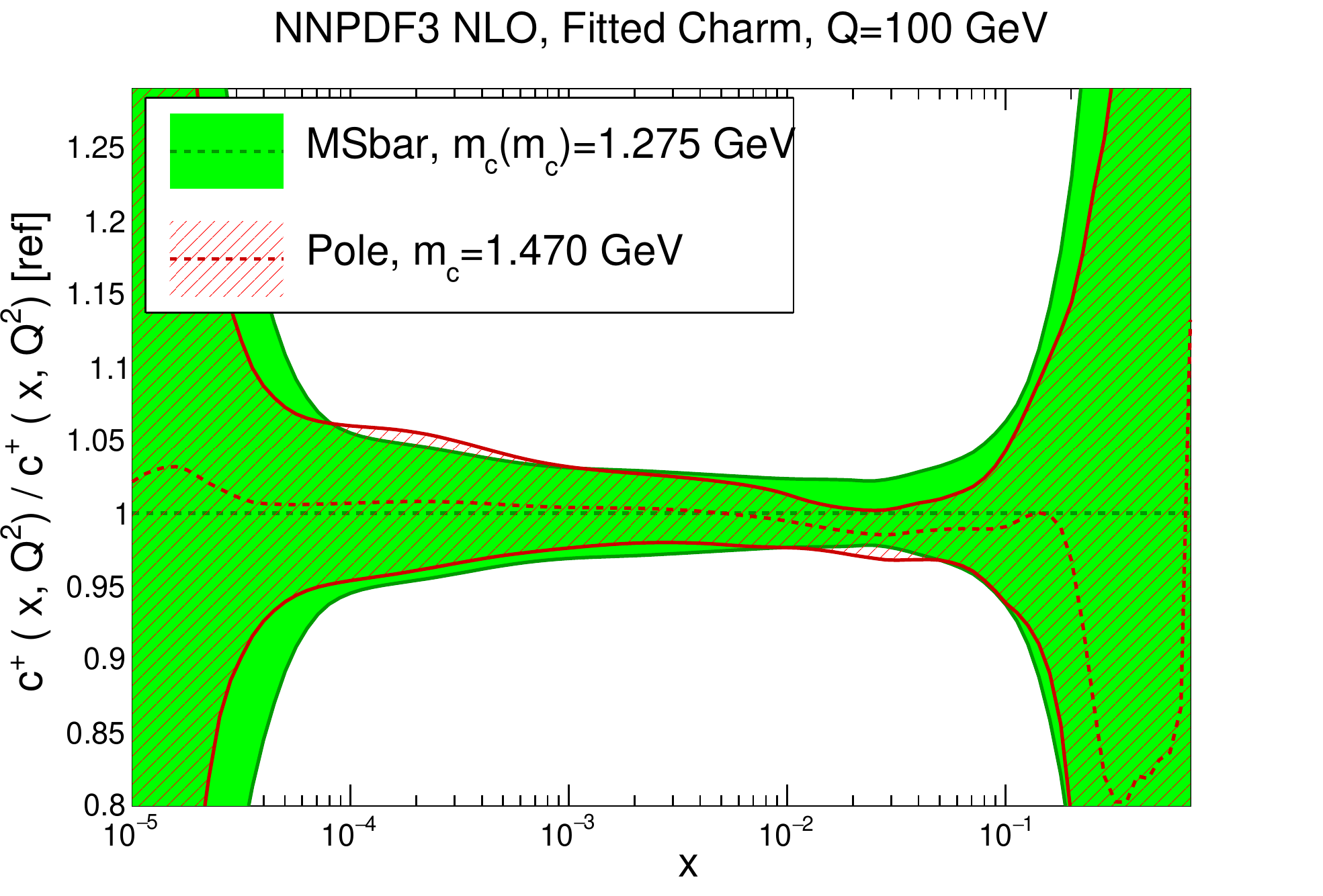}
    \includegraphics[width=0.46\textwidth]{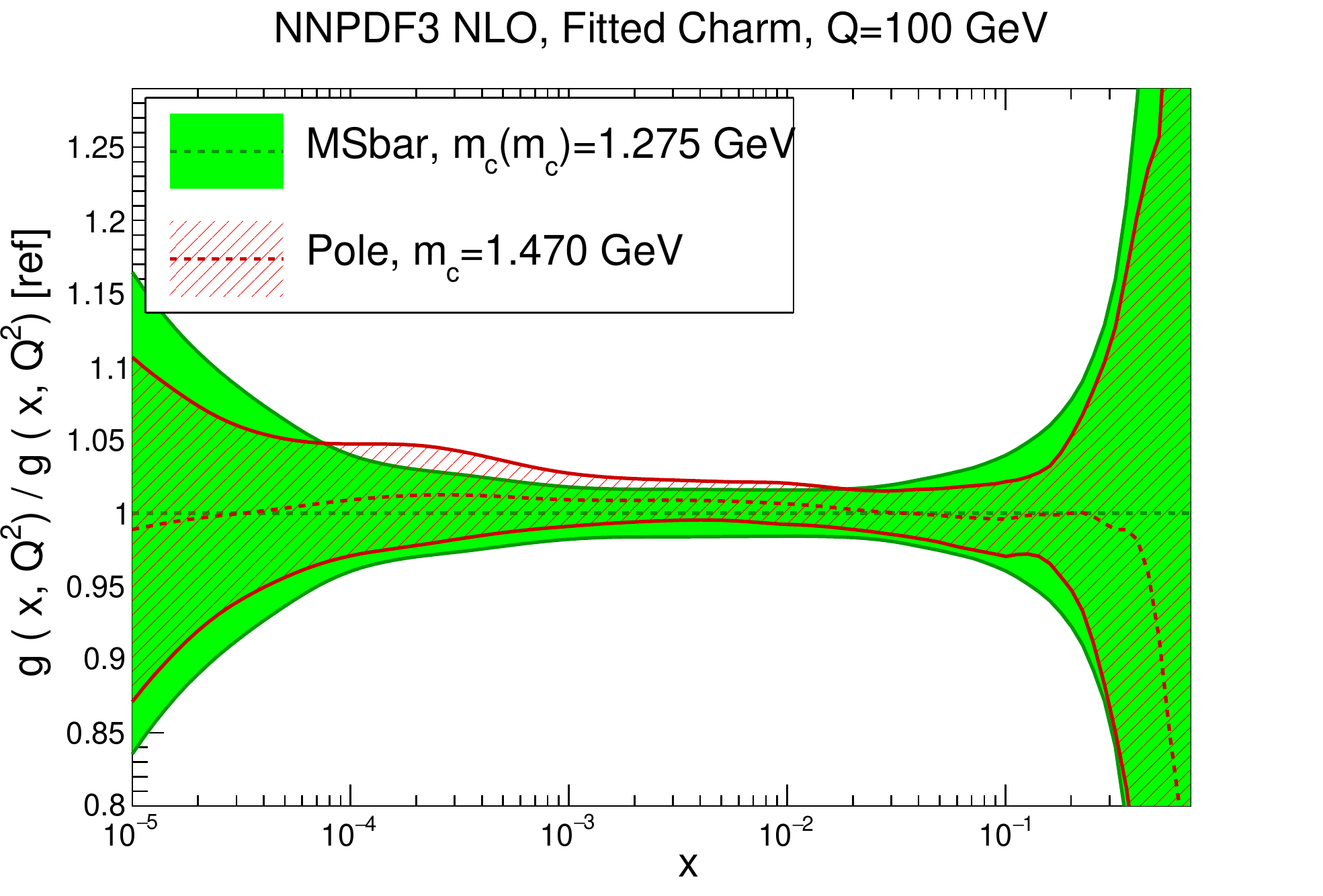}
\end{center}
\vspace{-0.8cm}
\caption{\small \label{fig:pdf-hqscheme-conv}
Comparison of PDFs determined with  $\overline{\rm{MS}}$ vs. pole
mass, for corresponding values of the mass obtained by one-loop conversion:
$m_c^{\rm pole}=1.47$~GeV and
 $m_c(m_c)=1.275$~GeV.
 The charm (left) and gluon (right) PDFs are shown, at low scale 
 $Q=1.65$ GeV (top) and  high scale $Q=100$~GeV (bottom). In the
 high-scale plots, results are shown as a ratio to the
 $\overline{\rm{MS}}$ mass result.
   }
\end{figure}

We next check the impact of switching from pole to  
$\overline{\mathrm{MS}}$ masses. 
In Fig.~\ref{fig:pdf-hqscheme-conv} we compare PDFs obtained 
using  pole mass $m_c^{\rm pole}=1.47$~GeV, or 
 $\overline{\rm{MS}}$ mass $m_c(m_c)=1.275$~GeV, the two values being
related by one-loop
  perturbative conversion. The charm and gluon PDFs are shown, at low
  and high scale. It is clear that the change in results is compatible
  with a statistical fluctuation. Similar results hold for
  other PDFs.

\begin{figure}[t!]
  \begin{center}
    \includegraphics[width=0.46\textwidth]{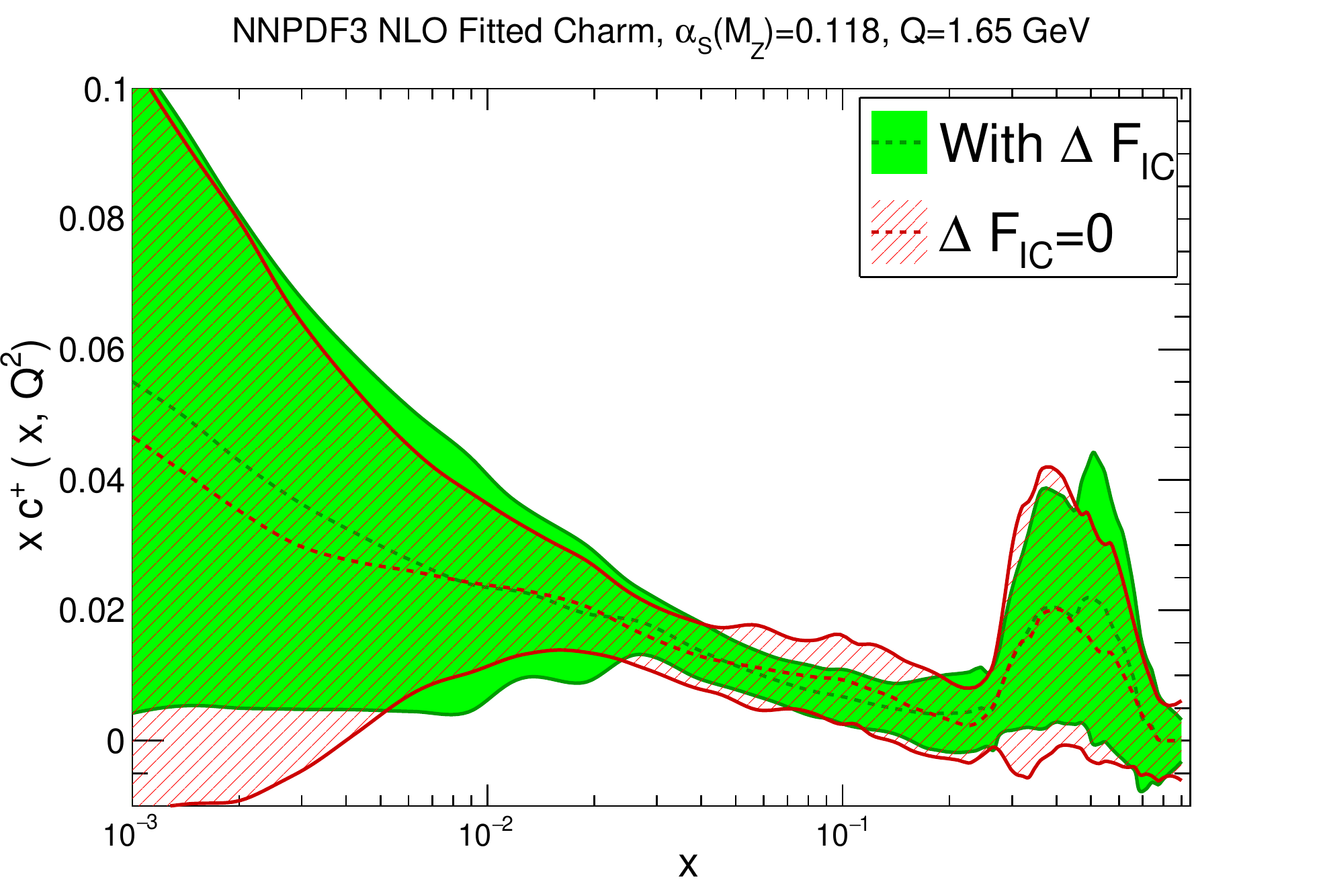}
    \includegraphics[width=0.46\textwidth]{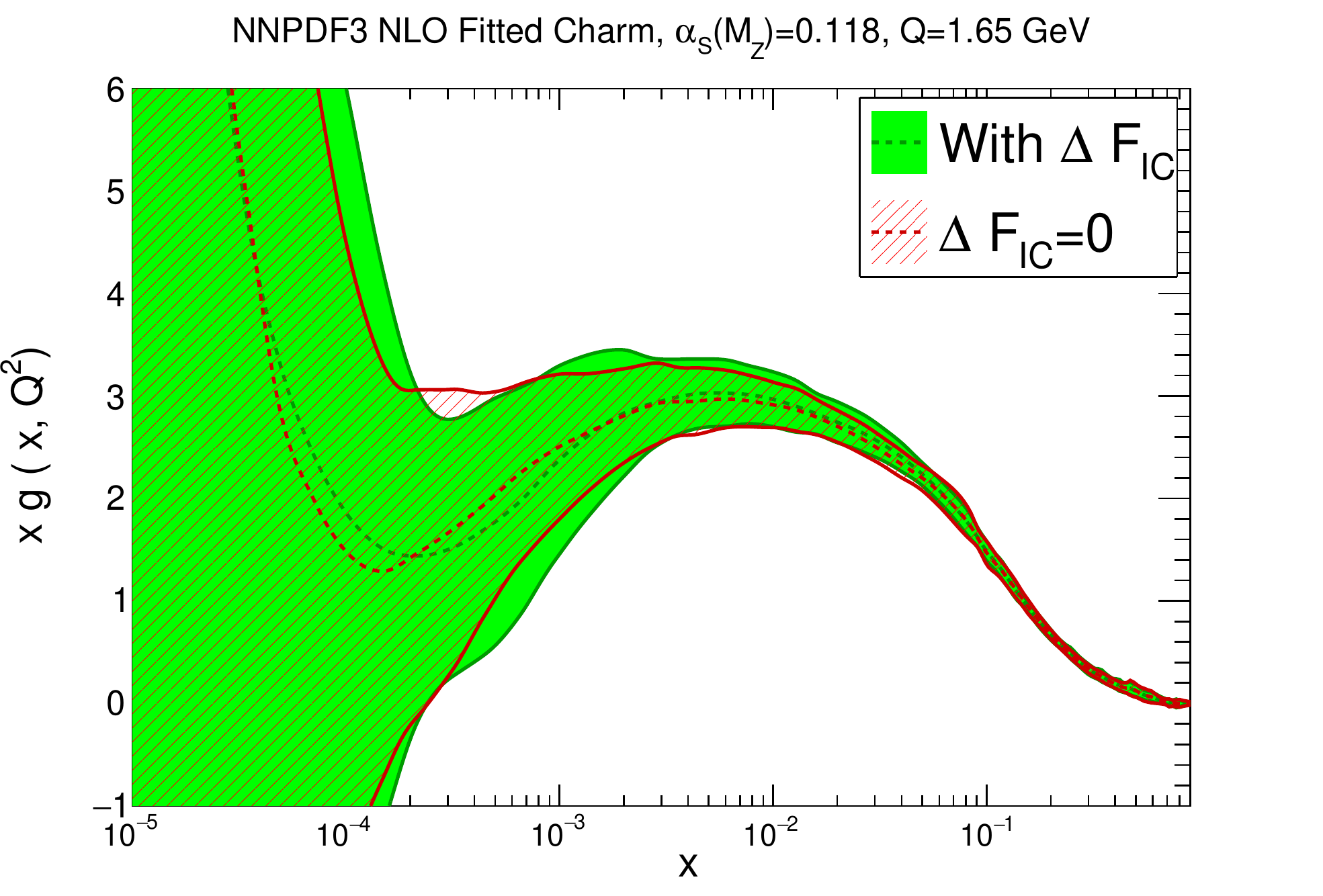}
    \includegraphics[width=0.46\textwidth]{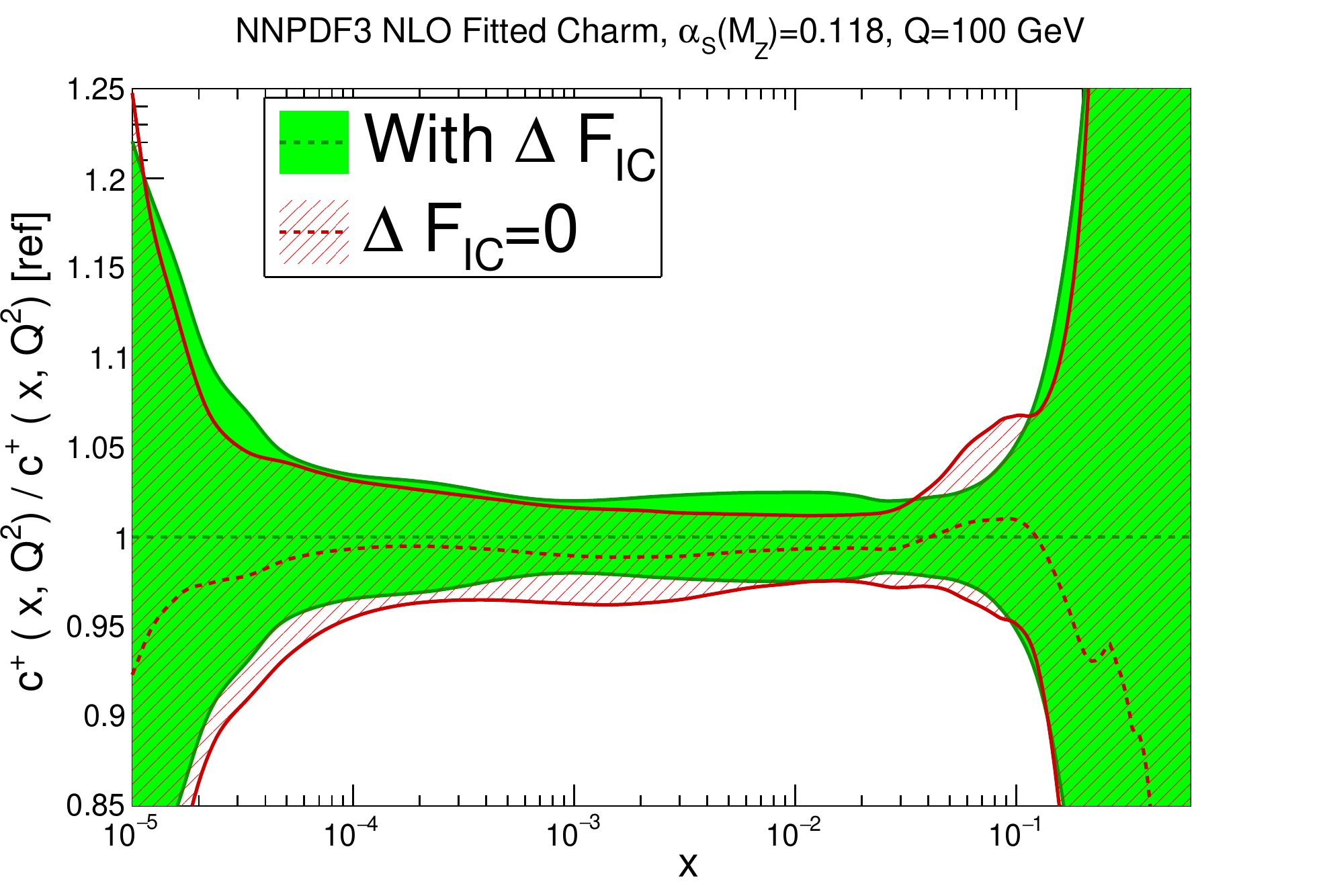}
    \includegraphics[width=0.46\textwidth]{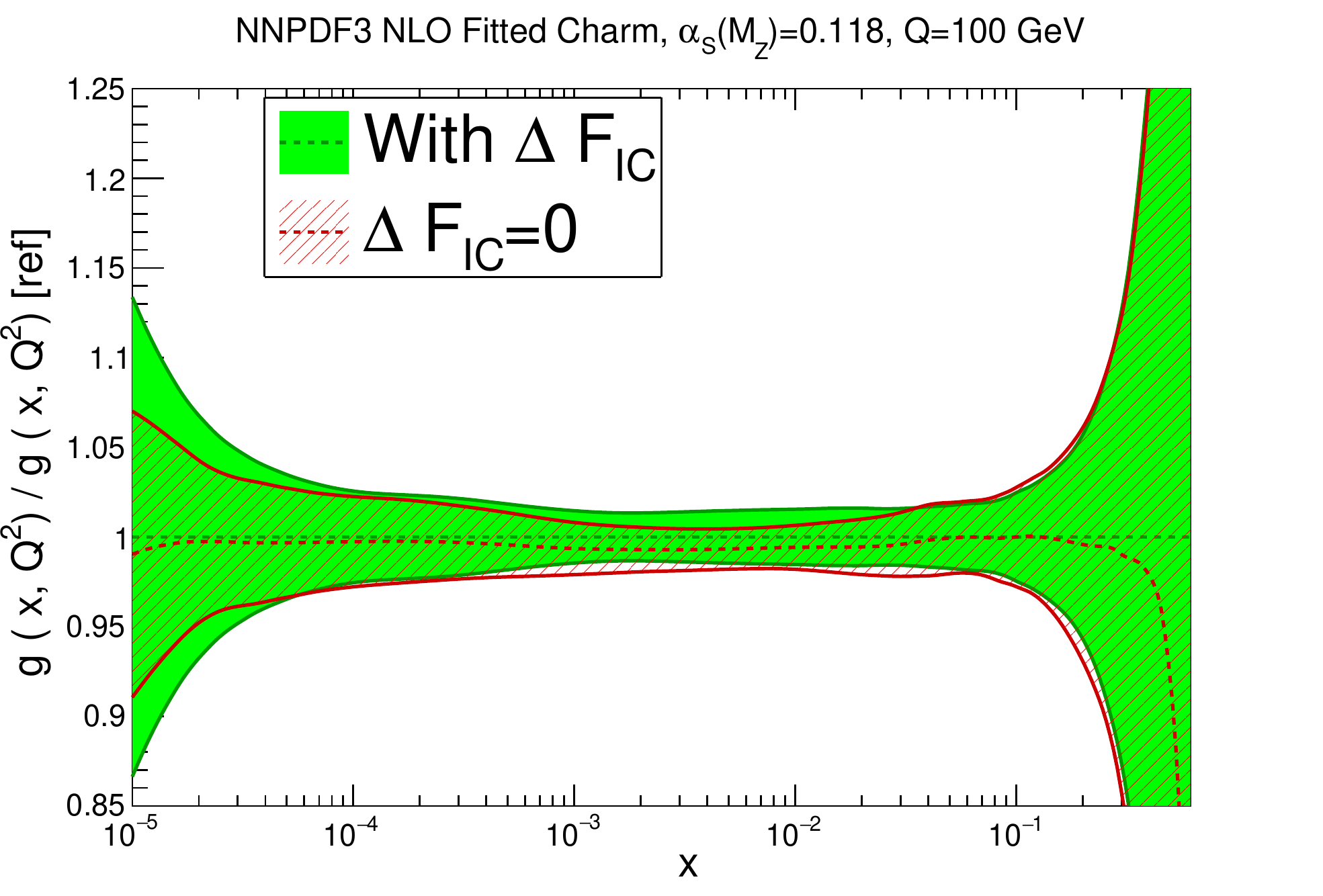}
\end{center}
\vspace{-0.8cm}
\caption{\small \label{fig:deltaficzero}
Same as Fig.~\ref{fig:pdf-hqscheme-conv}, but now comparing our
default results to the case in which massive charm-initiated
contributions are neglected (original FONLL-B of Ref.~\cite{Forte:2010ta} 
or S-ACOT, see text).
   }
\end{figure}

Finally, we study how our results would change if massive
charm-initiated contributions are neglected, i.e., if the original FONLL-B
scheme of Ref.~\cite{Forte:2010ta} is used. This corresponds to
 setting to zero the correction term $\Delta F_h$ (Eq.~(11) of
 Ref.~\cite{Ball:2015tna}), it is~\cite{Ball:2015tna,Ball:2015dpa}
 completely equivalent to the S-ACOT scheme used in intrinsic charm
 studies by the CT
 collaboration~\cite{Pumplin:2007wg,Dulat:2013hea,Hou:2015emq}, and,
 as mentioned in the introduction, it might be justified if the intrinsic
 charm contribution is power-suppressed. Results are shown in
 Fig.~\ref{fig:deltaficzero}: again, the change in results is
 compatible with a statistical fluctuation. This fact has some
 interesting implications. First, it shows that the size our best-fit
 charm is moderate, and compatible with a power-suppressed intrinsic
 charm. Also, it suggests that  the approximate  NNLO treatment of
 fitted charm proposed in Ref.~\cite{Ball:2015tna}, in which these
 terms are actually only included up to NLO (given that the massive
 charm-initiated coefficient functions are only known to this
 order~\cite{Hoffmann:1983ah,Kretzer:1998ju}), should actually be
 quite reliable. Finally, it should be noted that for the
 charm-initiated contribution the charm production threshold is set by
 $m_c$, but for the overall process, including the proton remnant,
 the threshold is set by $2m_c$, so there must be nonperturbative 
 contributions which restore momentum conservation: these would appear
 as power-suppressed corrections which should be resummed to all
 orders when $W^2\sim m_c^2$. In our case $W^2\gg m_c^2$ for all $x$,
 and the  charm-initiated contribution is seen to be sufficiently
 small that this issue should be of no  concern.

\subsection{Impact of the EMC data}
\label{sec:emc}

\begin{figure}[t!]
\begin{center}
 \includegraphics[width=0.8\textwidth]{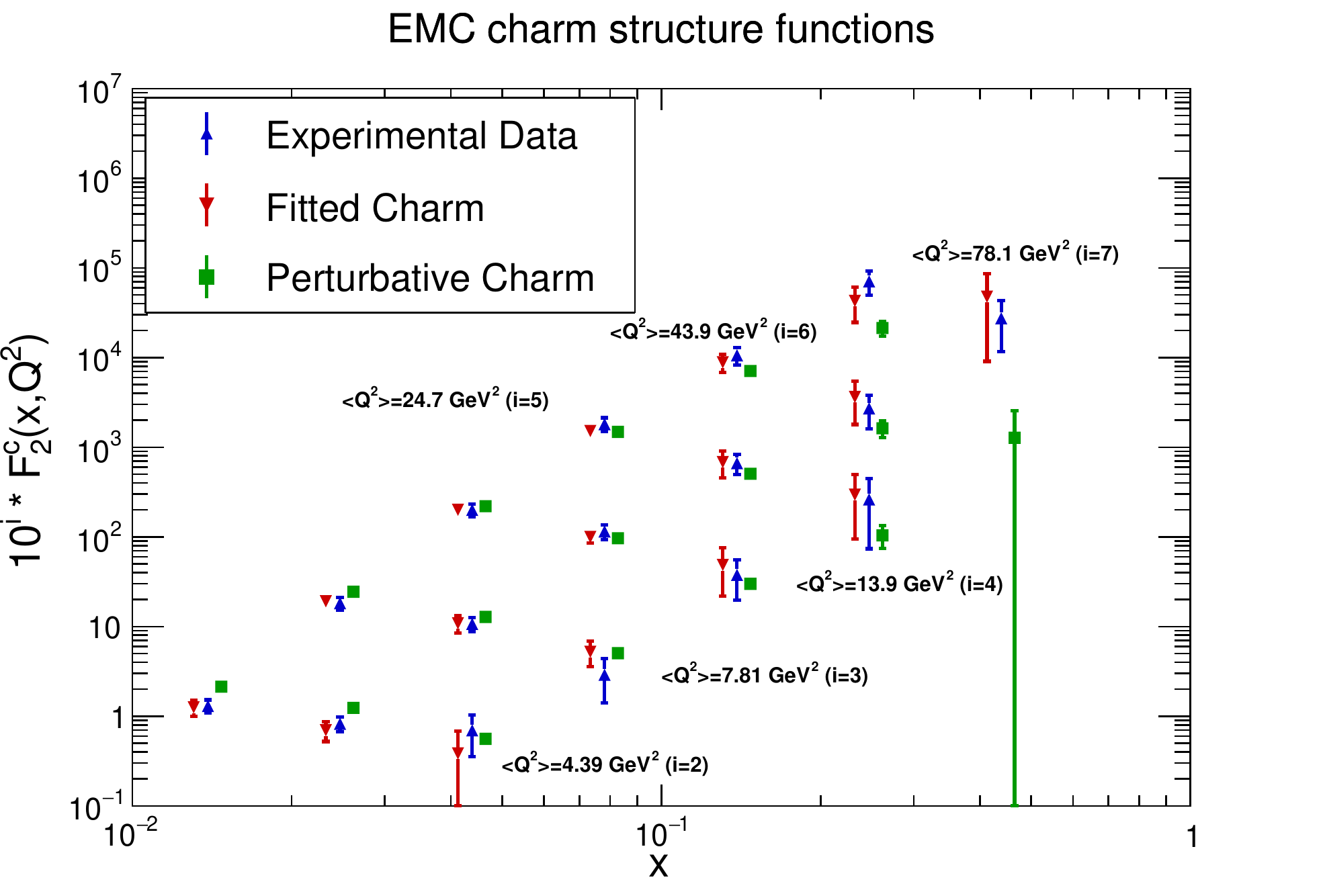} 
  \includegraphics[width=0.8\textwidth]{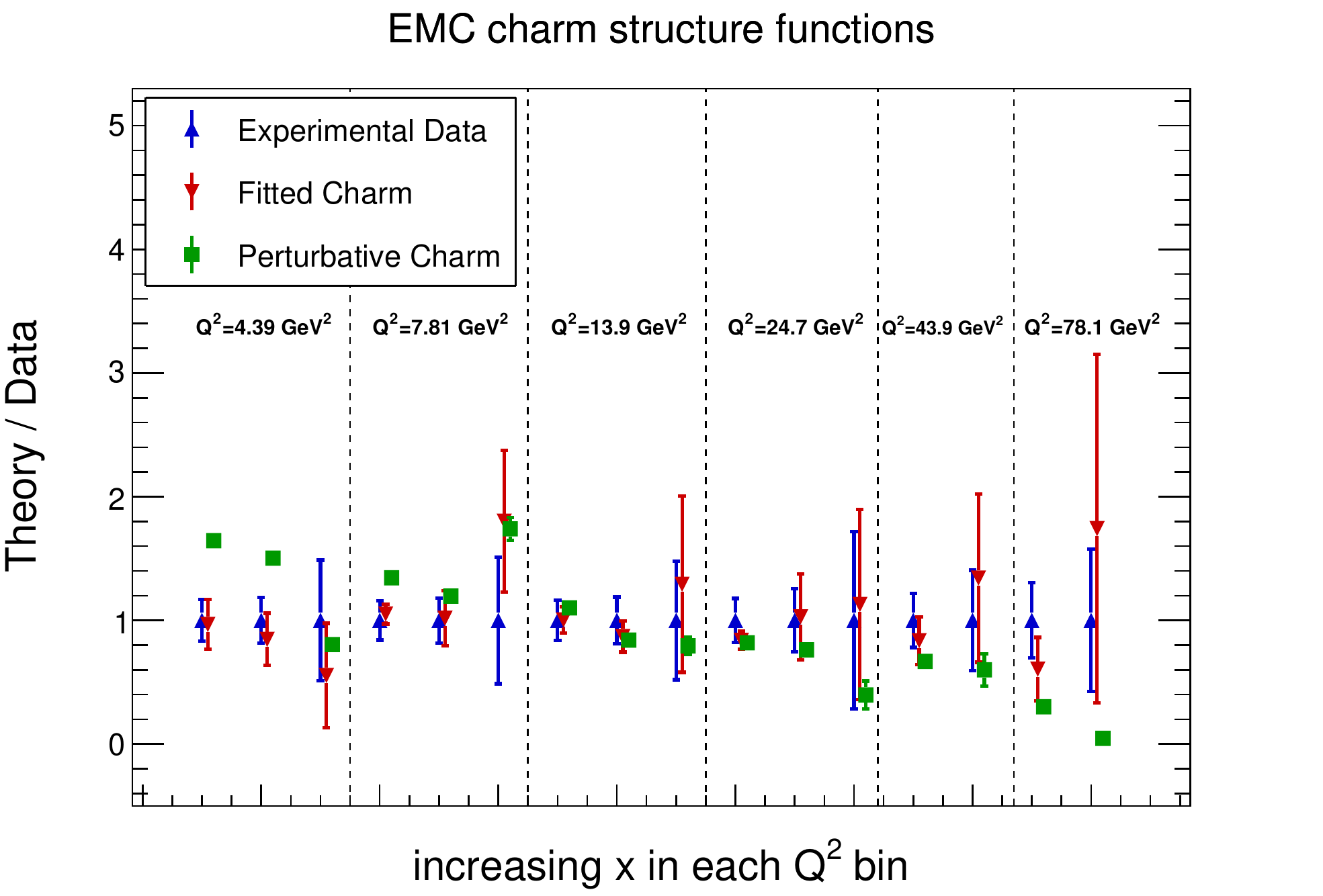}
 \end{center}
\vspace{-0.3cm}
\caption{\small \label{fig:EMCdata}
Comparison of the  best-fit theoretical result to 
the  experimental result for the EMC $F_2^c$ structure function data
  with fitted and with perturbative charm.
  The uncertainties shown are the  total PDF uncertainty in each data
  bin for theory, and the total experimental uncertainty for the data.
We show both the data vs. $x$ in $Q^2$ bins, offset to improve
readability (top) and the ratio of theory to data (bottom):
here the order in each bin is from small to large
values of $x$.
   }
\end{figure}

As already noted, it is not possible to fit the EMC
$F_2^c$ data of Ref.~\cite{Aubert:1982tt} with perturbative charm. It is
then important to assess carefully the effect of these data when we
fit charm.
The purpose of this assessment is twofold. First, we have the
phenomenological goal of assessing to which extent conclusions may be
affected if the EMC data are entirely or in part unreliable, or
perhaps have underestimated uncertainties. Second, perhaps more
interestingly, we would like to understand whether, quite
independently of the issue of their reliability, the EMC data might
provide a realistic scenario   in which not
fitting charm would lead to biased fit results.

The agreement between data and theory when charm is fitted 
is illustrated in
Fig.~\ref{fig:EMCdata}, where we compare the EMC charm structure function data
with the structure function computed using the best-fit PDFs, with
either fitted or perturbative charm.
Both the absolute structure
function (top) and the theory to data ratio (bottom) are shown.
It is interesting to observe that the discrepancy between the data and
the perturbative charm PDFs is large, and it is not confined in any
specific region of $x$ or $Q^2$, making an explanation of the
discrepancy based on a single cause 
such as resummation or higher order corrections
rather unlikely. More specifically,  
it is clear that the data at large $x$ in the highest $Q^2$ bins
cannot be reproduced by  perturbative charm, which gives a very small
contribution in this region. Interestingly, in this region one has $Q^2\gsim
25$~GeV$^2$, so a possible  higher-twist component that might
imitate the charm contribution~\cite{Blumlein:2015qcn} would be quite suppressed.
Likewise, in the small $x$ region, $x\lesssim 0.1$, perturbative charm 
overshoots the data. Here again, higher twist is expected to be small since, 
although $Q^2$ is quite low, $W^2 \gsim 50$ GeV$^2$. The fitted charm PDF 
corrects both these discrepancies rather neatly, by increasing the charm 
content at large $x$, and reducing it at small $x$, to produce a 
perfectly satisfactory fit. This leads to the perhaps surprising
conclusion that in order to fit the EMC data both a large-$x$ positive
bump (possibly of nonperturbative origin), and a small $x$ undershoot
(possibly mimicking missing higher-order corrections) are
needed.  Both the
way the large $x$ behaviour of our best-fit charm compares to existing
models, and its small $x$ component compares to what we expect from
missing higher orders will be discussed in Sect.~\ref{sec:intrinsic}
below.

\begin{figure}[t!]
  \begin{center}
   \includegraphics[width=0.45\textwidth]{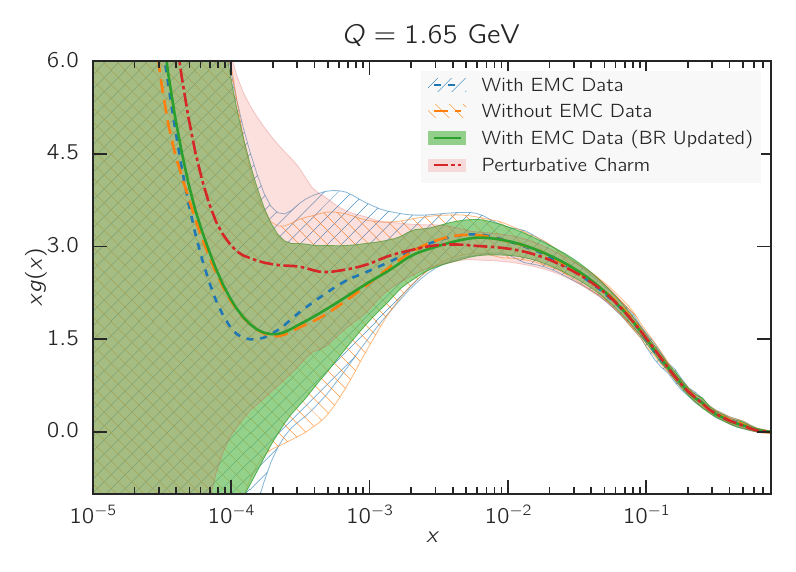}
  \includegraphics[width=0.45\textwidth]{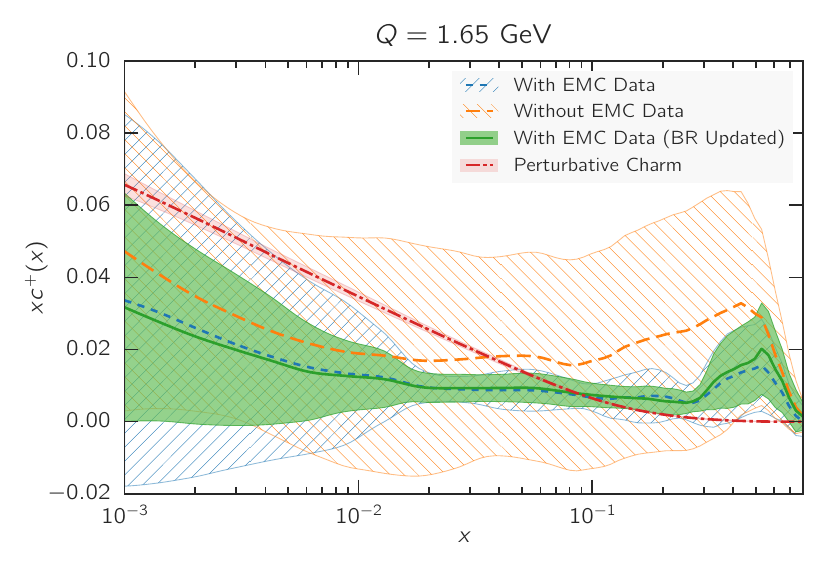}
  \includegraphics[width=0.45\textwidth]{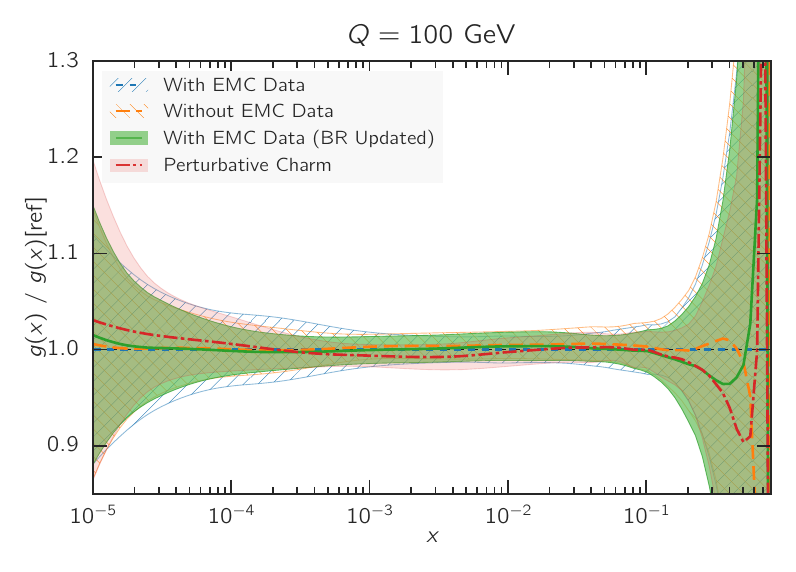}
  \includegraphics[width=0.45\textwidth]{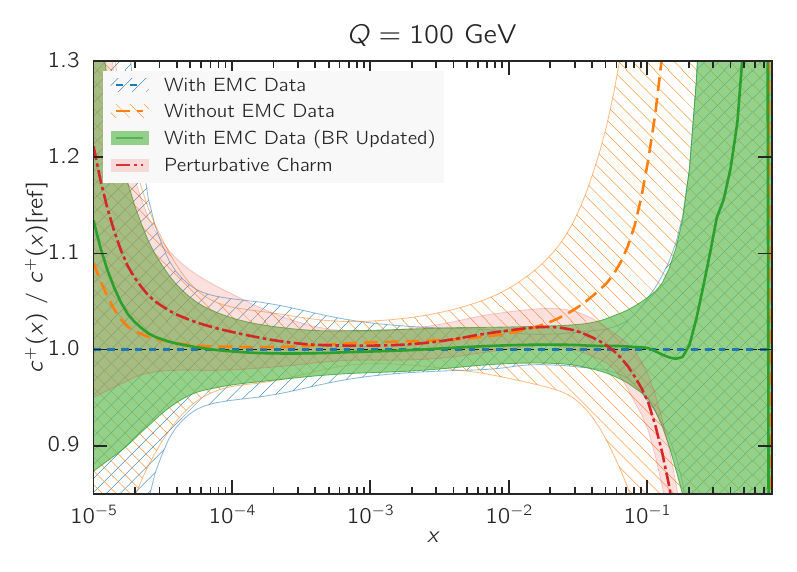}
\end{center}
\vspace{-0.3cm}
\caption{\small \label{fig:pdfemc}
  Same as Fig.~\ref{fig:pdfbase}, but now, when charm is fitted, also
  showing results obtained when EMC data are  rescaled to match updated branching fraction of $D$ mesons
  into muons (see text), or excluded altogether.
   }
\end{figure}

The impact of the EMC data on the PDFs is 
illustrated in
 Fig.~\ref{fig:pdfemc}, where we compare the charm and gluon PDFs 
with and without the EMC data included in the fit, everything else 
being unchanged, with the perturbative charm fit also being shown for
reference.
It is clear that for all $x\gsim 10^{-2}$ the uncertainty on the fitted charm
PDF is greatly increased in the absence of the EMC data. Reassuringly,
the qualitative features of the central charm PDF (to be discussed more
extensively in Sect.~\ref{sec:intrinsic} below) do not change
substantially: in particular it is still true that the central PDF at
large $x$ displays a bump, while at small $x$ it lies below the
perturbatively generated charm --- though uncertainties are now so
large that neither effect can be considered statistically
significant.  The other PDFs change very little.

We now specifically address the phenomenological issue of the
reliability of the EMC data. First of all, it should be noticed that
the published uncertainty in the EMC data is quite large to begin
with: the average uncertainty is about 27\%.
This said, various issues have been raised concerning this dataset.
Firstly, the inclusive EMC structure function data are known to be
inconsistent with  
BCDMS data (see e.g.~\cite{Feltesse:2012yaa}), but this was due to
underestimated backgrounds in drift chambers. Therefore,  this 
problem is expected to be 
 absent in the
charm structure function data which were taken with a calorimetric
target~\cite{arneodopriv}. The correction is anyway  never more than
20\%~\cite{Feltesse:2012yaa}, hence much smaller than the effect seen
in Fig.~\ref{fig:EMCdata}. In Ref.~\cite{Rottoli:2016lsg} it was
checked explicitly that if the inclusive EMC data are added to the
fit they  have essentially no impact.

\begin{figure}[t!]
  \begin{center}
   \includegraphics[width=0.45\textwidth]{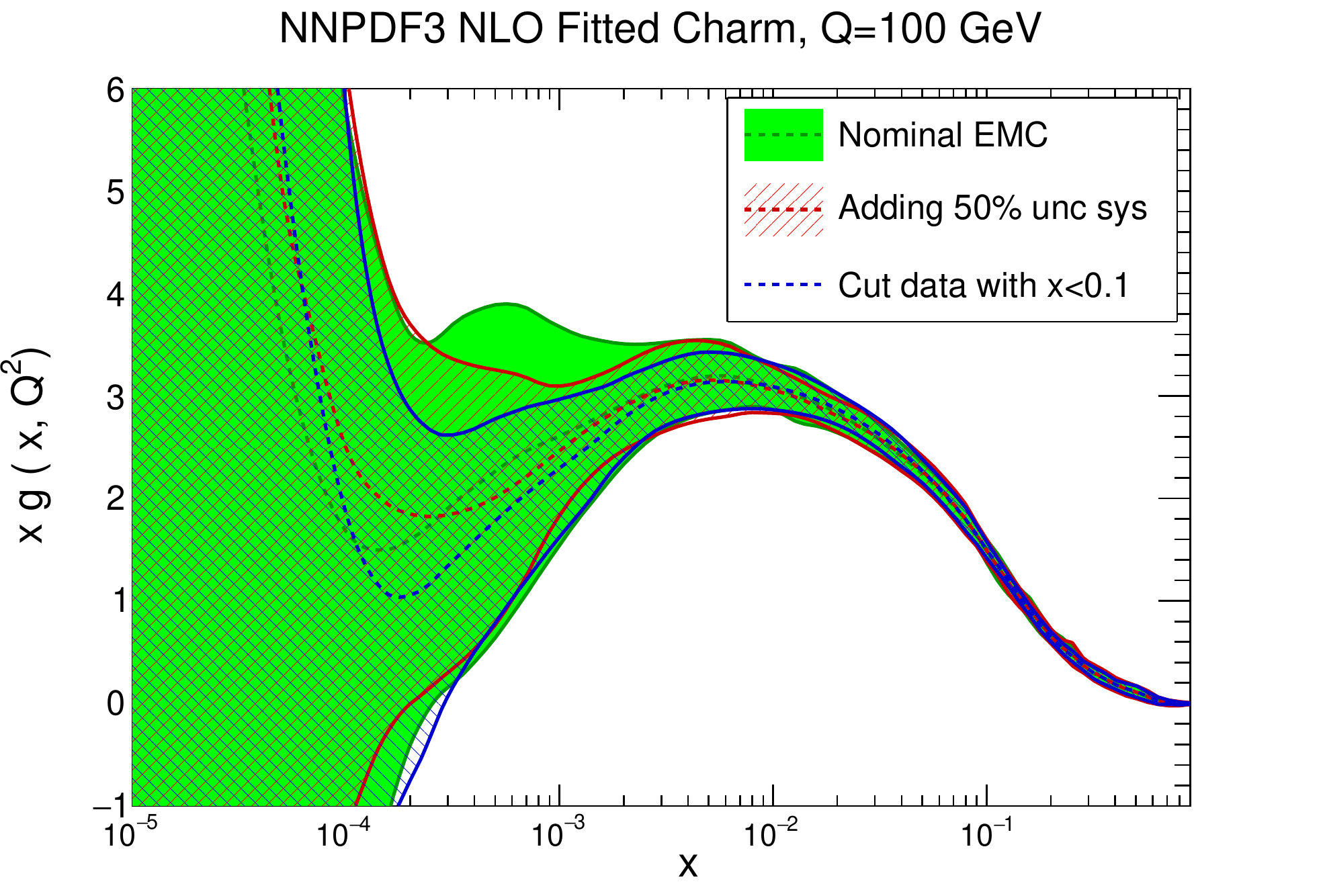}
  \includegraphics[width=0.45\textwidth]{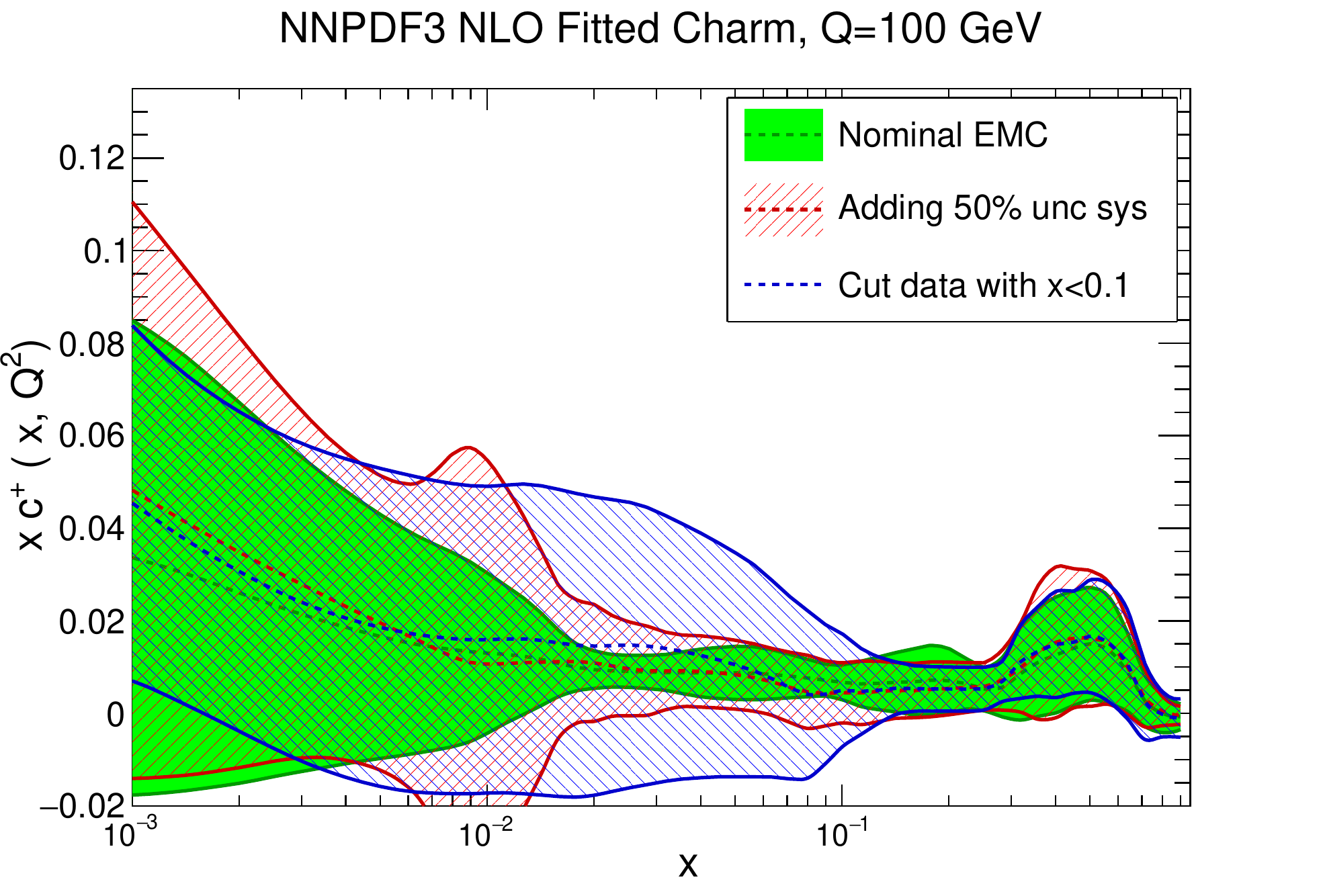}
  \includegraphics[width=0.45\textwidth]{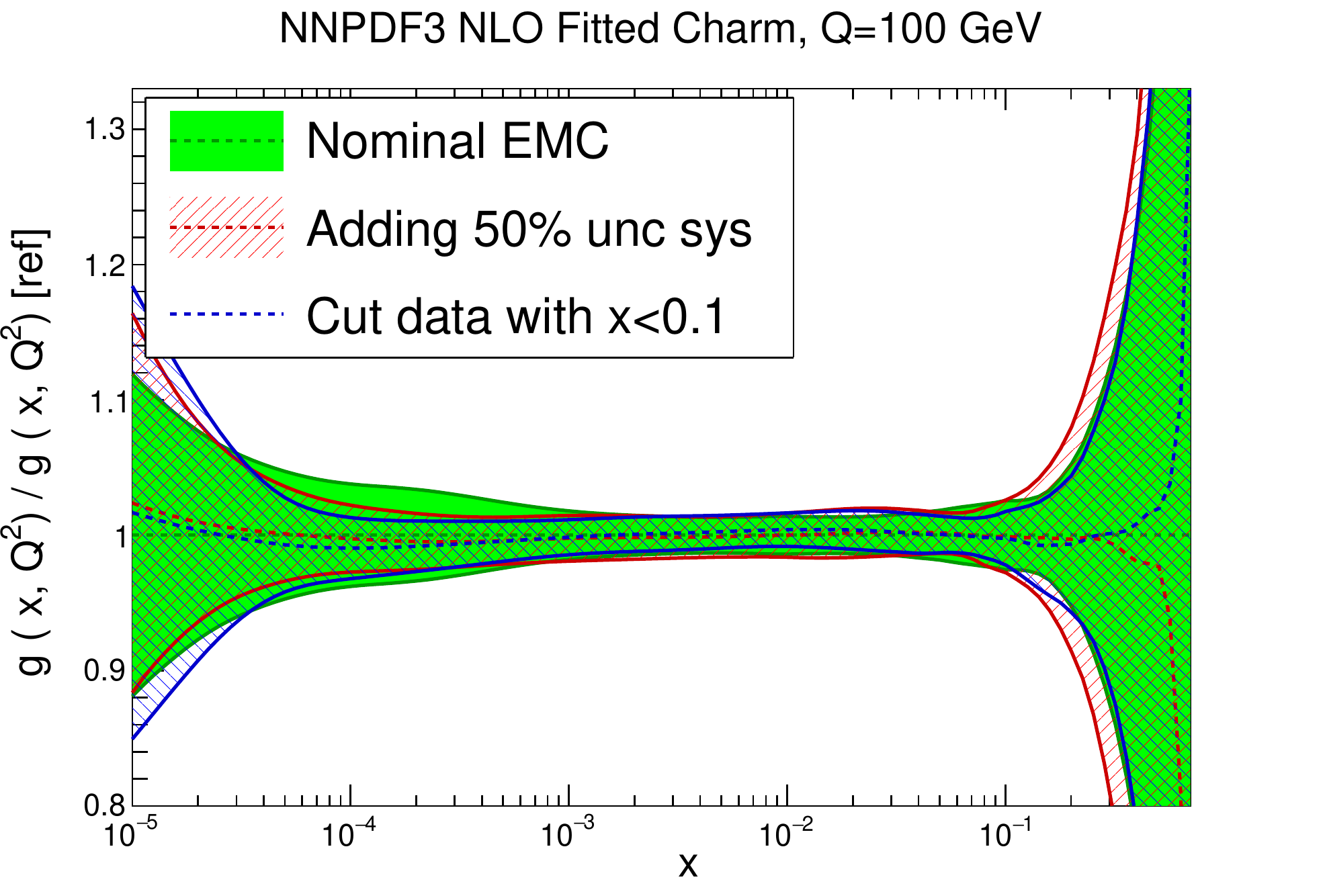}
  \includegraphics[width=0.45\textwidth]{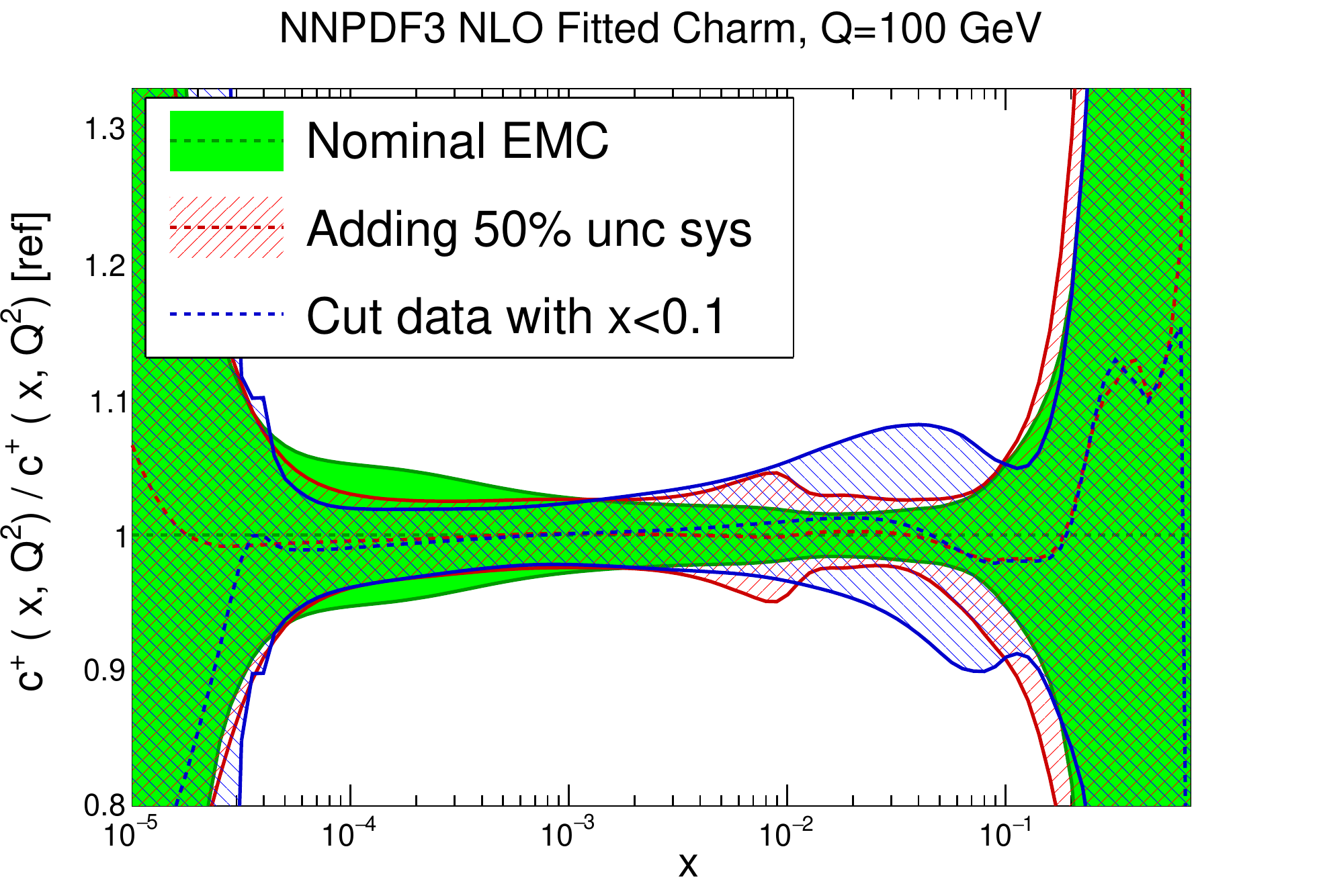}
\end{center}
\vspace{-0.3cm}
\caption{\small \label{fig:pdfemcvar}
  Same as Fig.~\ref{fig:pdfbase}, but now comparing the default results
  with fitted charm with those obtained removing all EMC data with
  $x<0.1$, or adding and extra 50\% systematics to all EMC data.
   }
\end{figure}
The 
original EMC charm structure functions were obtained assuming an
inclusive branching fraction of $D$ mesons into muons, 
${\rm BR}(D\to \mu+X)=8.2\%$,
which differs from the current PDG average~\cite{Agashe:2014kda}
and the latest direct measurements
from LHCb~\cite{Gauld:2015yia,Aaij:2013mga} of the fragmentation 
probabilities and
branching fractions of $D$ mesons, which give a value of around 10\%.
To verify the impact of using these updated branching fractions, and 
estimate also the possible impact of the other effects, 
we have rescaled the EMC data by a factor 0.82 and added an additional
uncorrelated  
15\% systematic uncertainty due to ${\rm BR}(D\to \mu+X)$.
The results are also shown in Fig.~\ref{fig:pdfemc}, where we 
see that this rescaling
has a only small impact on the charm PDF. The impact becomes
completely negligible if the systematics is taken to be
correlated~\cite{Rottoli:2016lsg}.

Since the charm data were taken 
on an iron target, nuclear corrections should be applied, as is 
the case also for the various fixed target neutrino datasets included in 
our global fit: in fact, in the smallest $x$ bins, shadowing
corrections could be as large as 10-20\% (see
e.g. Ref.~\cite{Rith:2014tma}). Furthermore, it was argued in
Ref.~\cite{Martin:2009iq} that higher twist corrections obtained by
replacing $m_c^2$ by $m_c^2\left(1+\frac{\Lambda^2}{m_c^2}\right)$
(where $\Lambda\sim200$~MeV is a binding energy scale) may have a
substantial effect on the lowest $Q^2$ (and thus smallest $x$) EMC
data. Finally, of course, the EMC data have been obtained using
analysis techniques which are  quite crude to modern standards, for
example only relying on  LO QCD computations. This latter
caveat, however, 
is in fact common to all  the oldest fixed-target deep-inelastic
scattering data which are still currently used for PDF determination,
such as SLAC~\cite{Whitlow:1991uw} and
BCDMS~\cite{Benvenuti:1989rh,Benvenuti:1989fm}, for which there is
no evidence (see in particular Table~10 of
Ref.~\cite{Ball:2014uwa}) that systematics are significantly
underestimated, though, of course, specific issues only affecting EMC
(such as the aforementioned background estimation) cannot be excluded.

In order to explore possible consequences of missing corrections (such
as nuclear or higher twist), or uncertainty underestimation, we have
performed two more fits. In the first, we have removed all EMC data
with $x<0.1$, namely the region where nuclear and higher twist
corrections are largest. In the second, we have have retained all EMC
data, but with an extra 50\%
correlated systematics. Results are shown in Fig.~\ref{fig:pdfemcvar}.
It is clear that the effect of the added systematics is minor: the
percentage increase of 
uncertainties is moderate, and the central value changes very
little. On the other hand, as one might expect, 
removing the small-$x$  EMC data leaves the
best fit charm unchanged for $x>0.1$, but for smaller $x$ it leads to
results which are similar to those  (shown in Fig.~\ref{fig:pdfemc})
when the EMC data are not included. This shows that the large $x$ EMC
data are responsible for the large $x$ bump, while the small $x$ EMC
data are responsible for the small $x$ undershoot in comparison to the
perturbative charm case.

We conclude that while we have no direct evidence that uncertainties
in the EMC data
might be underestimated, and specifically not more than
for any other old deep-inelastic scattering dataset, there are
persuasive theoretical arguments which suggest that these data might
be affected by significant nuclear or higher twist corrections,
especially at small $x$. However, we find that even a very substantial
increase of the systematic uncertainty of this data does not change
its qualitative impact, as one might perhaps expect given the very
large discrepancy between the data and predictions obtained with
purely perturbative charm at small and large $x$. On the other hand,
until more data are available phenomenological conclusions based on
this data should be taken with a grain of salt, as is always the case
when only a single dataset is responsible for a particular effect: as
seen in Fig.~\ref{fig:EMCdata}, about half a dozen points are mostly
responsible for the effect seen at small $x$ and as many at large
$x$. However, regardless of the actual reliability of these data, 
there remains an issue of principle:
if the EMC results were true, to what extent might the assumption of
perturbative 
charm bias the fit result? This question is addressed in the next subsection.

\subsection{The charm PDF and its intrinsic component}
\label{sec:intrinsic}

\begin{figure}[t!]
  \begin{center}
   \includegraphics[width=0.45\textwidth]{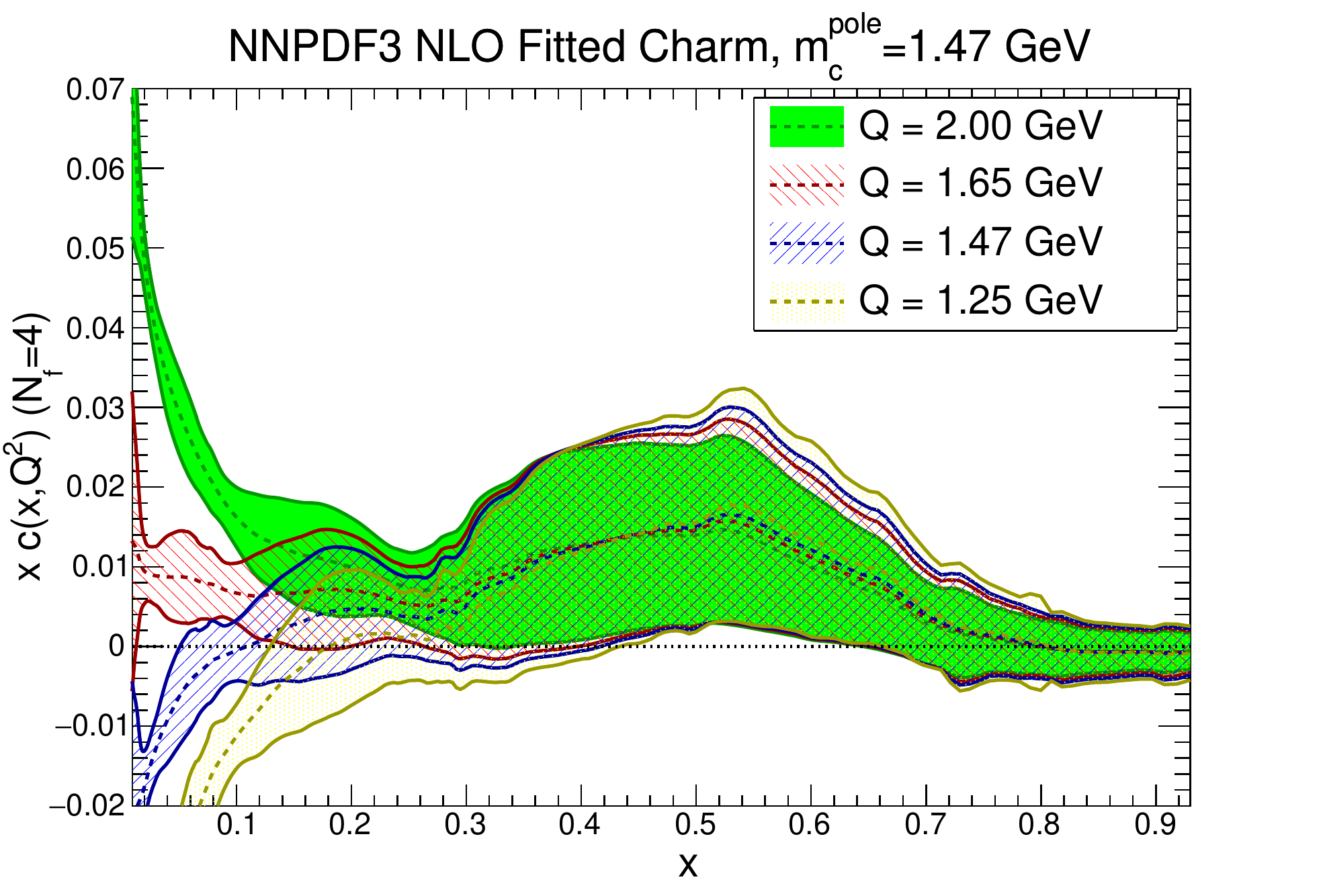}
   \includegraphics[width=0.45\textwidth]{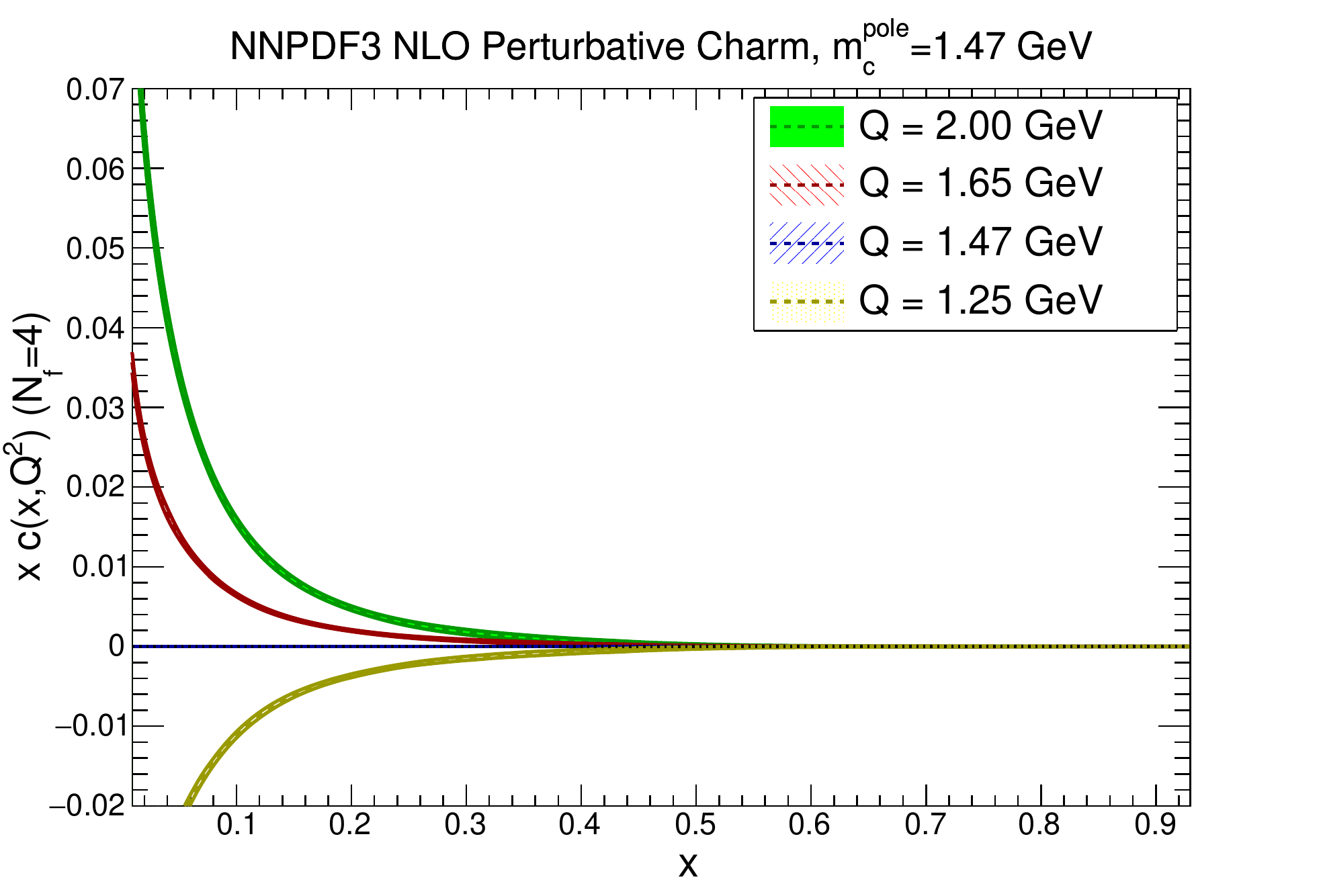}
    \includegraphics[width=0.45\textwidth]{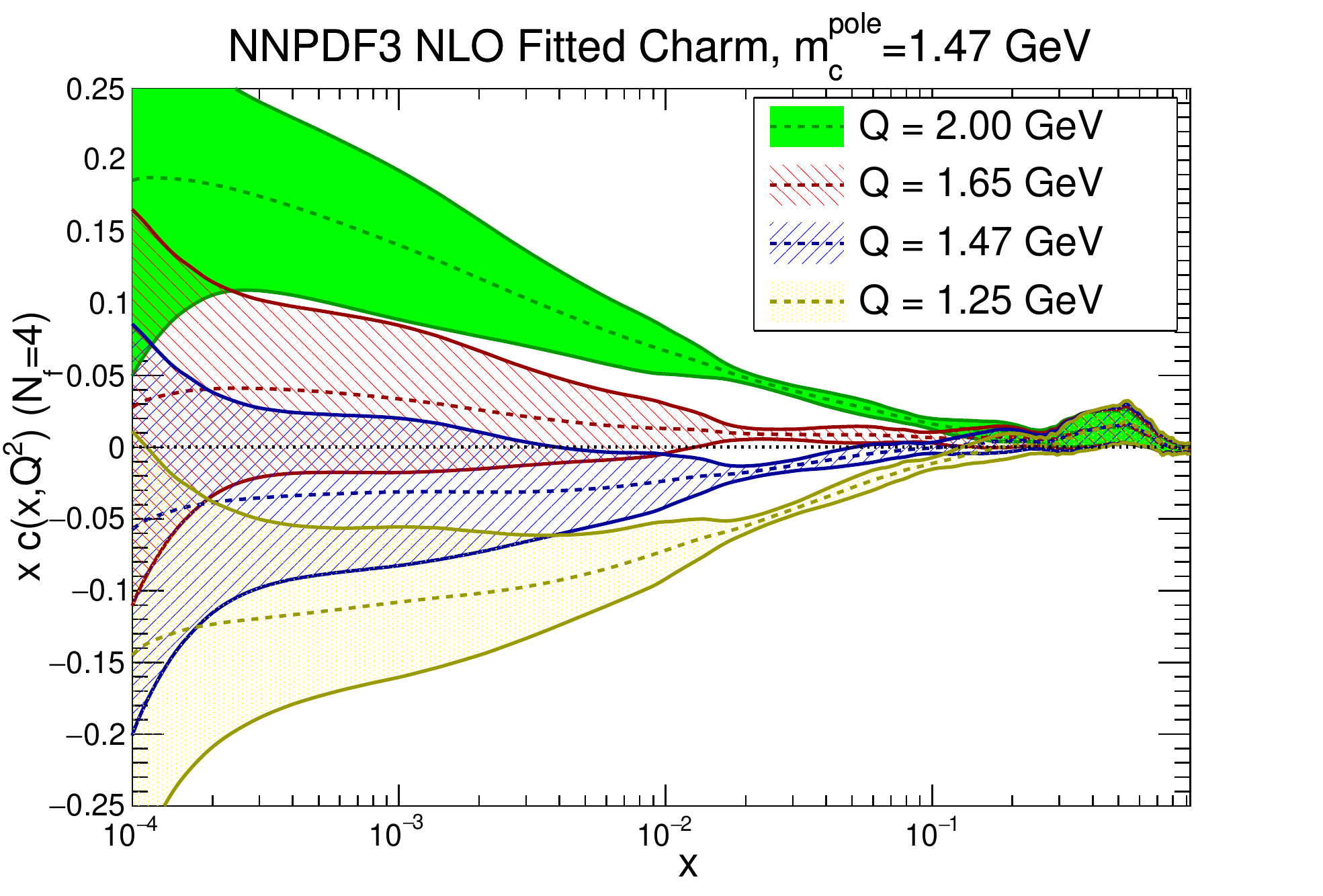}
  \includegraphics[width=0.45\textwidth]{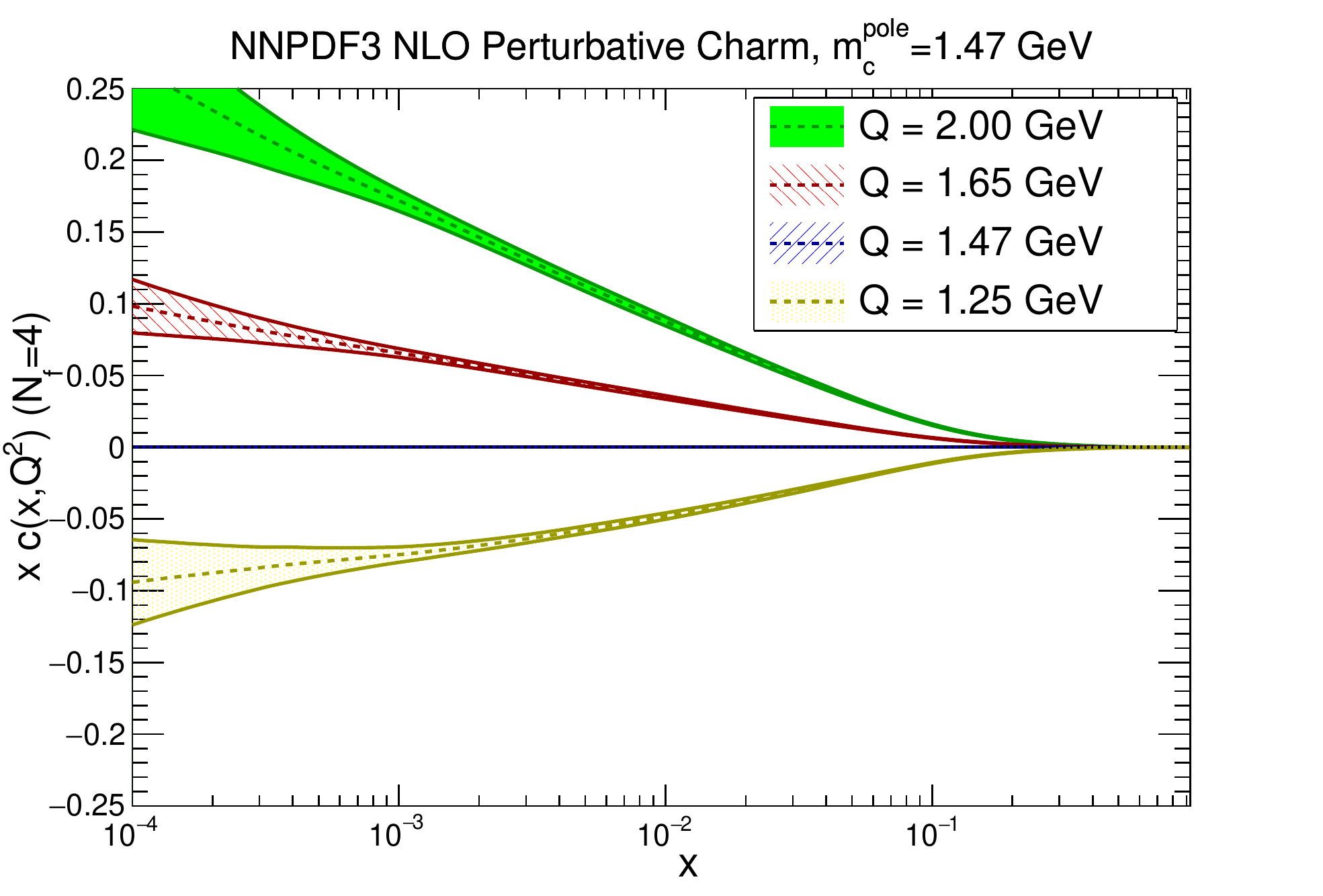}
\end{center}
\vspace{-0.5cm}
\caption{\small \label{fig:charm-Qdep}
The charm PDF (when $m_c=1.47$~GeV) plotted 
as a function $x$ on a linear (top) or
logarithmic scale (bottom) for four low scale values
$Q=1.25$, $1.47$, 1.65 and 2~GeV in the four-flavour scheme. 
Both fitted (left) and perturbative (right) charm are shown.
Note that in a matched scheme the charm PDF would become scale independent for $Q<m_c$. 
   }
\end{figure}

We now discuss the qualitative features of the best-fit charm PDF. Our
goal here is not to assess the reliability of the data on which it is
based (which was discussed in the previous subsection) but rather to
examine the implication of a scenario in which such data are assumed
to be true. Such a scenario does not appear to be forbidden or
unphysical in any sense, so it is interesting to ask whether in this secenario
 a PDF determination without fitted charm would lead to biased
results.

\begin{figure}[t!]
  \begin{center}
    \includegraphics[width=0.45\textwidth]{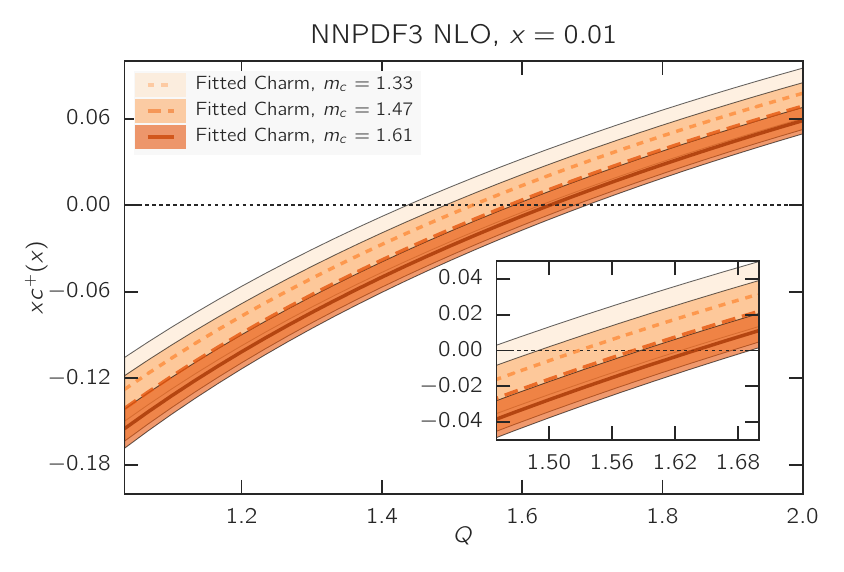}
   \includegraphics[width=0.45\textwidth]{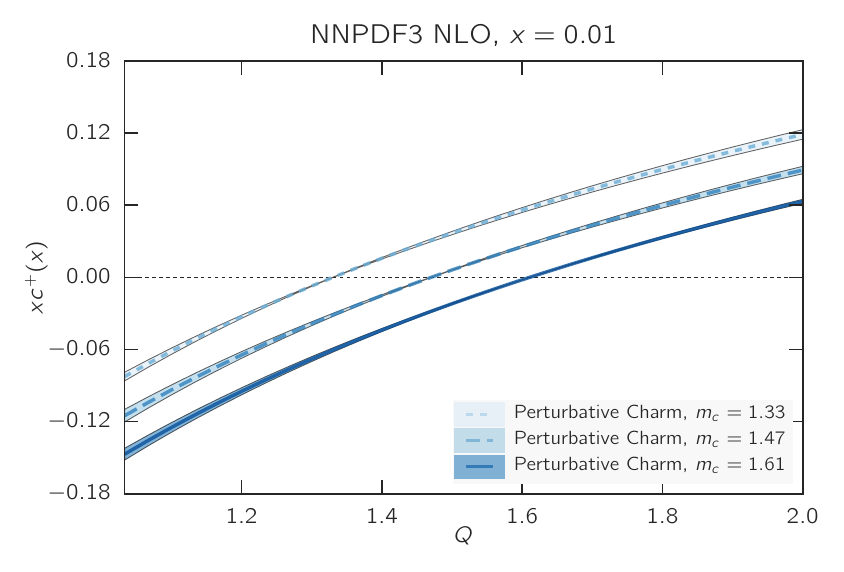}
 \end{center}
\vspace{-0.5cm}
\caption{\small \label{fig:charm-pos} The charm PDF   in the
  four-flavour scheme as a function of scale
 at $x=0.01$ for different values of the heavy quark mass with fitted
 (left) and perturbative (right) charm.
   }
\end{figure}

In order to get a first qualitative assessment, 
 in Fig.~\ref{fig:charm-Qdep} the charm PDF is plotted as a function
of $x$ for various scales close to the threshold. 
Results are shown, for illustrative purposes, in the
four-flavour scheme: in the three-flavour scheme the PDF would
become scale independent.
Both the fitted (left) and the perturbative (right) charm PDF are shown.
The plot is produced from the fitted PDFs
by backward evolution using {\tt APFEL} from the scale $Q=1.65$~GeV.
Recall that
the independence of NNPDF results on the scale at which
PDF are parametrized is a feature of the NNPDF approach which has been
repeatedly verified, see e.g. Ref.~\cite{Ball:2014uwa}.

The  plot vs.\ $x$ on a logarithmic scale, in which the small $x$
region is emphasized, shows 
 that for all $x\lsim 10^{-1}$ 
fitted charm lies below the perturbative charm. However, 
a scale $Q_0$ at which fitted charm vanishes for all $x$ in this region
does appear to exist, but it is rather higher, around $Q_0\sim1.6$~GeV.
Recalling  that the dependence of the
size of the charm PDF at small $x$ on the value of charm mass is very
considerably reduced  when charm is fitted (see
Fig.~\ref{fig:pdfmcvar}), this is a genuine feature, which
follows from the data. Of course, in the case of perturbative charm
the scale at which the PDF vanishes is instead determined  by the
value of the mass, as is  
clear from the right plots of Fig.~\ref{fig:charm-Qdep}.

The  plot vs. $x$ on a linear scale, in which  the large $x$ region
is emphasized, in turn shows
that the fitted charm PDF displays an `intrinsic' bump, peaked
at $x\sim0.5$ and very weakly scale dependent. This bump is of course 
absent when charm is generated perturbatively.

The impact of the EMC data on the features of the charm PDF shown in 
Fig.~\ref{fig:charm-Qdep} can be traced to the behaviour shown in
 Fig.~\ref{fig:EMCdata} and discussed in Sect.~\ref{sec:emc}.
Namely, at medium-$x$ and low-$Q^2$  the  EMC data undershoot the prediction
obtained using perturbative charm, while at large-$x$ and large $Q^2$
they overshoot it. This leads to a fitted charm which is significantly
larger than the perturbative one at large $x$, but somewhat smaller  
at low $x$.

\begin{figure}[t!]
  \begin{center}
  \includegraphics[width=0.45\textwidth,page=2]{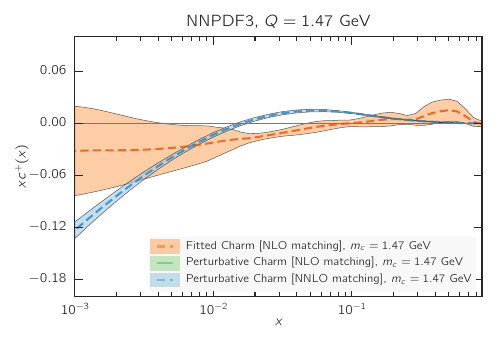}
    \includegraphics[width=0.45\textwidth,page=2]{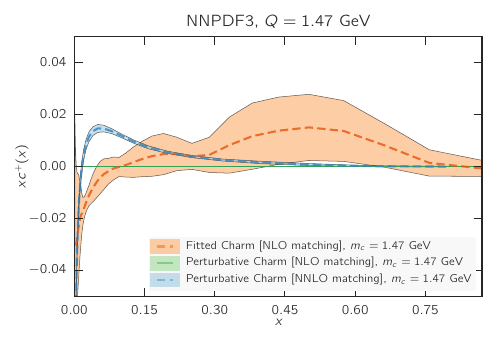}
 \end{center}
\vspace{-0.5cm}
\caption{\small \label{fig:charm-nnlo}  The charm PDF plotted vs. $x$
  on a logarithmic (left) or linear (right) scale, when
  $Q=m_c=1.47$~GeV. The fitted and perturbative NLO and NNLO (see
  text) results are compared.
   }
\end{figure}

We now discuss each of these features in turn.
To elucidate the small $x$ behaviour, 
in Fig.~\ref{fig:charm-pos} we plot
the charm PDF as a function of the scale $Q$  for fixed $x=0.01$, for
the three values of 
the charm mass that have been considered above in
Sect.~\ref{sec:hqmass}.  It is clear that, as mentioned, 
when charm is fitted (left) the scale 
at which the PDF vanishes is quite stable, while when charm is 
perturbative (right)
the PDF is very sensitive to the value of the mass since the PDF 
is constrained to vanish at $Q=m_c$.  Specifically the exact scale
at which fitted charm vanishes at $x=0.01$ turns out to be 
$Q_0=1.59$~GeV (when
$m_c=1.47$~GeV).

In order to better understand the meaning of this result, in
Fig.~\ref{fig:charm-nnlo} we compare at the scale $Q=m_c=1.47$ the
fitted charm to its perturbative counterpart determined at NLO and
NNLO. While the NLO result vanishes by construction, the NNLO result
(which will refer to as ``NNLO perturbative charm'' for short)
is obtained using NNLO matching conditions~\cite{Buza:1995ie,pdfnnlo}
from our best fit perturbative charm NLO PDF set. Within the FONLL-B accuracy
of our calculation, this NNLO charm is subleading, hence
it provides an estimate of the expected size of missing higher order
corrections on perturbatively generated charm.  

It is interesting to
observe that fitted charm for $x\lsim 0.2$ is similar in size to NNLO
perturbative charm, and it has in fact the same (negative) sign for
$x\lsim 0.02$. Of course, to the extent that fitted charm might
reabsorb missing higher order corrections, it would do so not only for
matching terms but also for missing corrections to hard matrix
elements, which are of the same order and likely of similar size. It
is nevertheless intriguing that the observed undershoot of fitted
charm when compared to perturbative charm is a feature of the NNLO
matching condition at sufficiently small $x$. 

All this suggests that
our best-fit  fitted charm at small $x$ is compatible with
perturbative behaviour with either a somewhat larger value of the
charm mass, or missing higher order corrections reabsorbed into the
initial PDF or a combination of both. This means that if uncertainties
related to missing higher orders and the charm mass value were
included in perturbative charm, then our fitted charm would be
compatible with perturbative charm, but possibly more accurate (in view
of the greater stability seen in Fig.~\ref{fig:charm-pos} of the fitted charm in
comparison to the perturbative one). If instead uncertainties related to
missing higher orders and the charm mass value are not included (as it
is now the case for most PDF sets, including  NNPDF3.0) then the charm
PDF, within the given uncertainty, is biased (assuming the EMC data
are correct).

\begin{figure}[t]
\begin{center}
\includegraphics[width=0.60\textwidth]{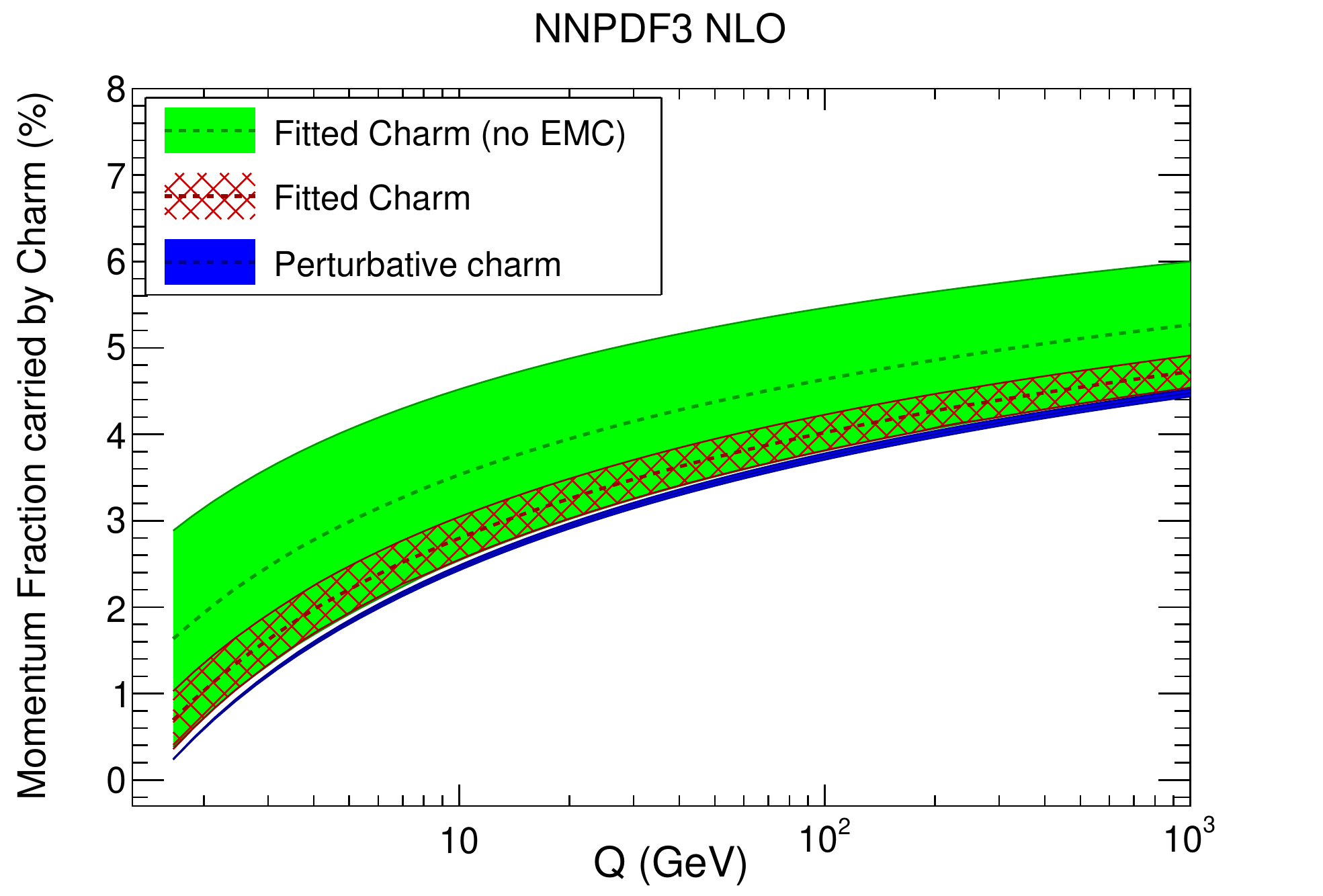}
\end{center}
\vspace{-0.3cm}
\caption{\small \label{fig:momfrac}
    The charm momentum fraction  $C(Q^2)$
    Eq.~(\ref{eq:momfrac}) as a function of scale
with perturbative and with fitted charm, with and
 without the EMC data included in the fit.
   }
\begin{center}
  \includegraphics[width=0.49\textwidth]{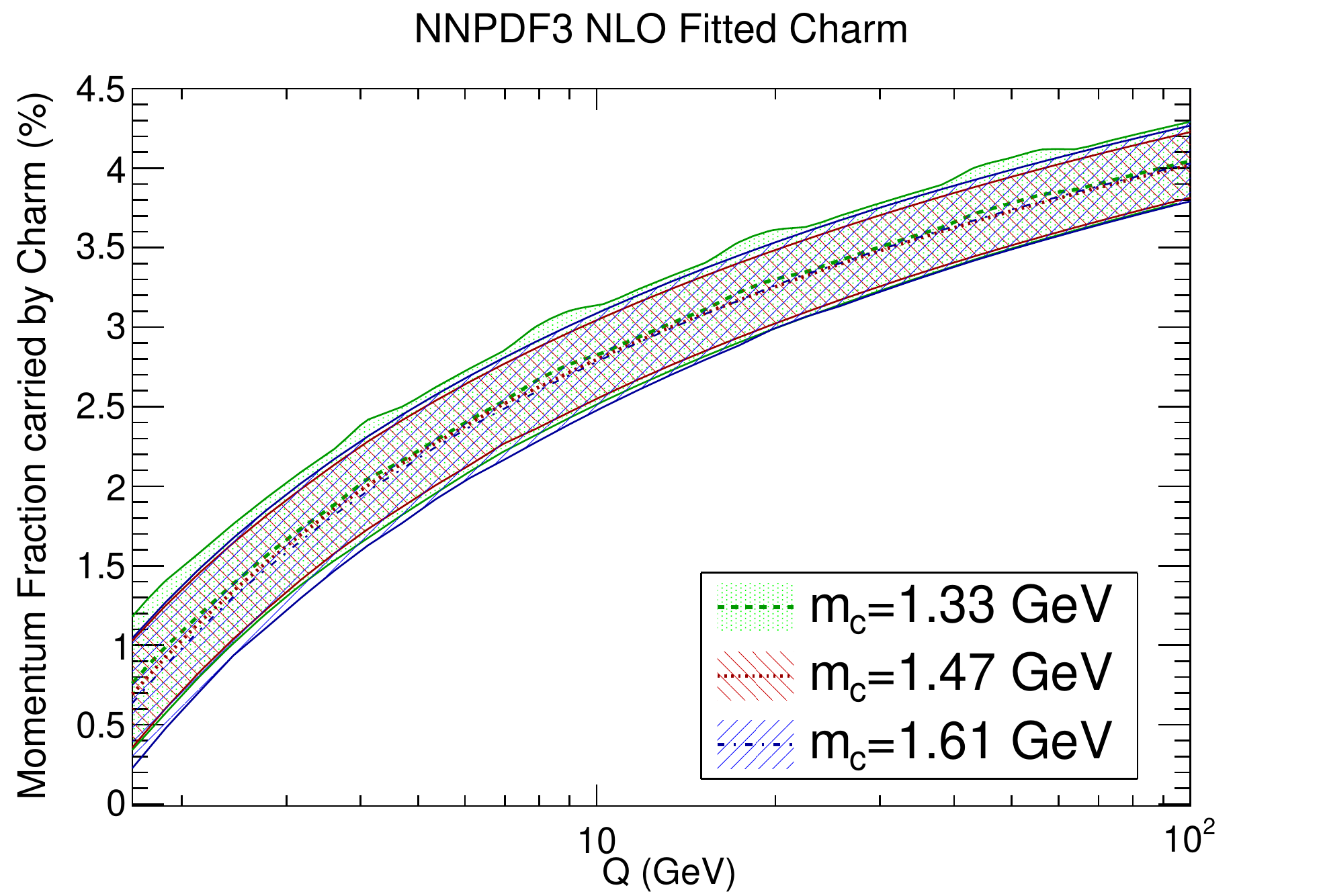}
  \includegraphics[width=0.49\textwidth]{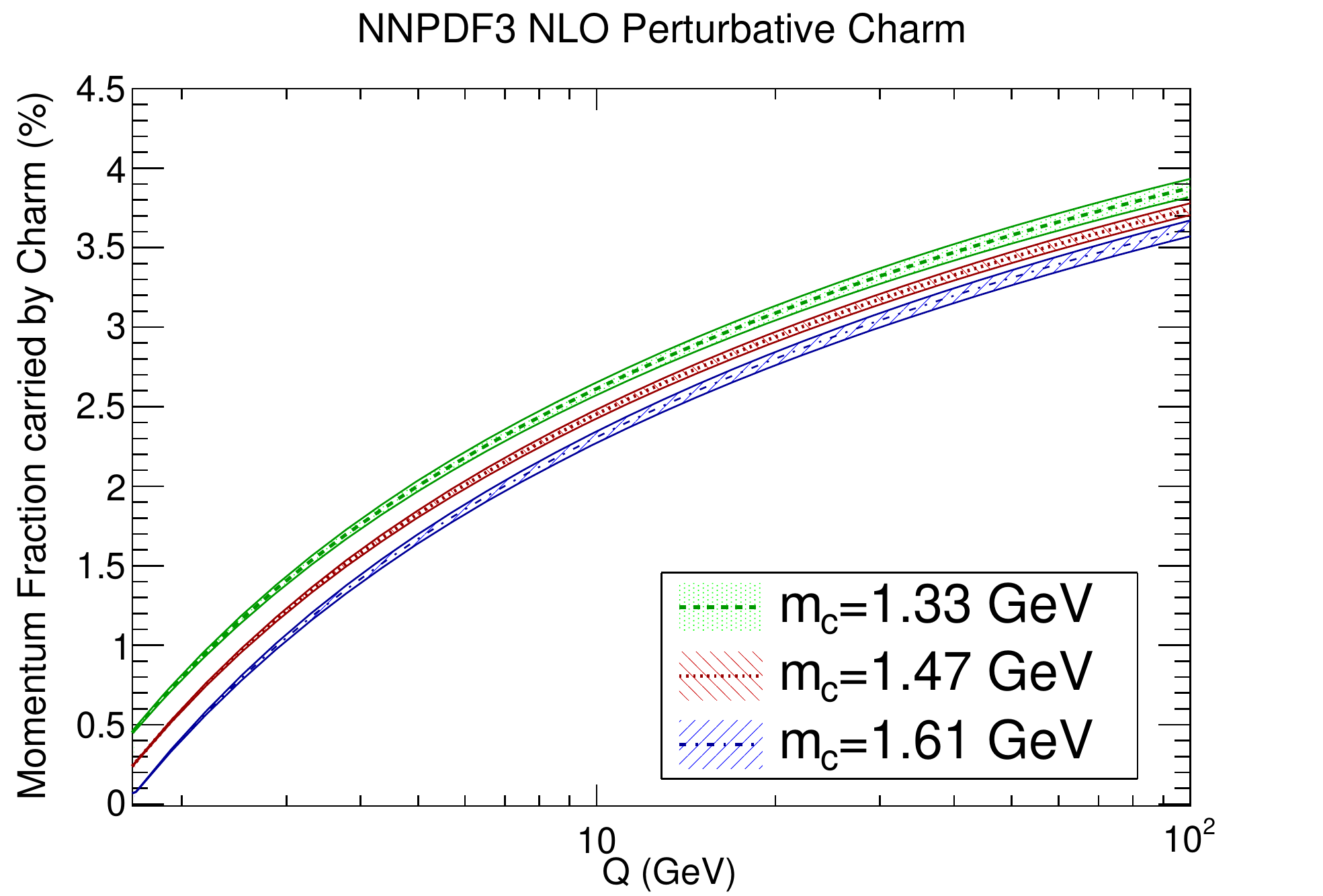}
\end{center}
\vspace{-0.3cm}
\caption{\small \label{fig:momfrac2}
Same as Fig.~\ref{fig:momfrac}, with three different values of the
pole charm mass, for fitted (left) and perturbative (right) charm.
   }
\end{figure}

We now turn to the large $x$ behaviour. The fact that our fitted charm
has an ``intrinsic'' component means that it carries a non-negligible
fraction of the proton's momentum.
In order to  quantify this, we compute
the momentum fraction carried by charm,
defined as
\be
\label{eq:momfrac}
C(Q^2) \equiv \int_0^1 dx\, x \lc c(x,Q^2)+\bar{c}(x,Q^2)\rc \, .
\ee
Of course for scales significantly above threshold, both the 
intrinsic and perturbative
components of the charm PDF will contribute.
The momentum fraction $C(Q^2)$ Eq.~(\ref{eq:momfrac}) is plotted as a
function of the scale $Q$ in  Fig.~\ref{fig:momfrac}, both for fitted and
perturbative charm. In the case of fitted charm, results are shown both
with and without the EMC data. 
In Fig.~\ref{fig:momfrac2} we then show the momentum fraction  with the three
different values of the charm mass considered in
Sect.~\ref{sec:hqmass}. 

\begin{table}[t]
  \center
  \begin{tabular}{|l|c|}
    \hline
    PDF set & $C(Q=1.65~{\rm GeV})$ \\
    \hline
    \hline
    NNPDF3 perturbative charm        & $\lp 0.239 \pm 0.003\rp $\%  \\
    NNPDF3 fitted charm           & $\lp 0.7 \pm 0.3\rp $\%  \\
    NNPDF3 fitted charm (no EMC)  &   $\lp 1.6 \pm 1.2\rp $\%  \\
    \hline
    \hline
    CT14IC BHPS1  & 1.3\%   \\
    CT14IC BHPS2  &   2.6\%  \\
    CT14IC SEA1  &    1.3\%  \\
    CT14IC SEA2  &    2.2\%  \\
    \hline
  \end{tabular}
  \caption{\small \label{tab:momsr}
    The charm momentum fraction  $C(Q^2)$ at a low scale  $Q=1.65$~GeV
with perturbative charm, and with fitted charm with and
 without the EMC data included.
The momentum fractions for several CT14IC PDF sets are also given for
comparison (see text).
  }
  \end{table}

The values of the momentum fraction at a low scale
$Q=1.65$~GeV just above the charm mass, using the central value
$m_c=1.47$~GeV are collected
in Table~\ref{tab:momsr}: they shows that both with and without the EMC data we find evidence
for intrinsic charm at about the one sigma level. The intrinsic charm contribution to the momentum  
fraction, when the EMC data are included, is then around $0.5\pm 0.3\%$ (after subtracting the
perturbative contribution  at this scale),  
entirely consistent with a power suppression of order $\Lambda^2/m_c^2$.  
Without the EMC data, the fraction increases to $1.4 \pm 1.2\%$, so the allowed range
for $C(Q)$ is reduced once the EMC data are included. 

At high scale, as shown in
Fig.~\ref{fig:momfrac}, the  momentum
fraction carried by the charm PDF is dominated by its perturbative component, and it becomes about
$5\%$ at $Q=1$~TeV. However, it
is clear from Fig.~\ref{fig:momfrac2} that the momentum fraction of
fitted charm is essentially independent of the charm mass at all
scales, and is thus determined exclusively by the data. On the other hand, with 
perturbative charm the momentum fractions
obtained for different values of the mass do not overlap at the
one-sigma level, even at high scale, and are thus instead determined 
by the assumed  value of the mass.

In order to further understand the features of our fitted intrinsic
component we compare it  to previous determinations  based on
models. To this purpose, 
we compare our fitted charm with the charm  PDFs recently given in
Refs.~\cite{Hou:2015emq,Dulat:2015mca} within the framework of the CT14
NNLO PDF determination.
 In this analysis two different models for intrinsic charm
  were considered: a BHPS scenario~\cite{Brodsky:1980pb} 
in which charm at $Q_0=1.3$~GeV has a valence-like shape
  \be
  c(x,Q_0)=Ax^2\lc6x(1+x)\ln x+ (1-x)(1+10x+x^2)\rc \, ,
  \ee
which peaks around $x\sim 0.25$, and a SEA model in which charm is 
assumed to have the same shape as the light quark sea:
  \be
  c(x,Q_0)=A\lc \bar{d}(x,Q_0)+\bar{u}(x,Q_0)\rc \, . 
  \ee
In both cases, the only free parameter of the model is the
positive-definite normalization $A$, for which two different values, corresponding to
two different momentum fractions, are considered (see  Table~\ref{tab:momsr}).

\begin{figure}[ht!]
\begin{center}
  \includegraphics[width=0.49\textwidth]{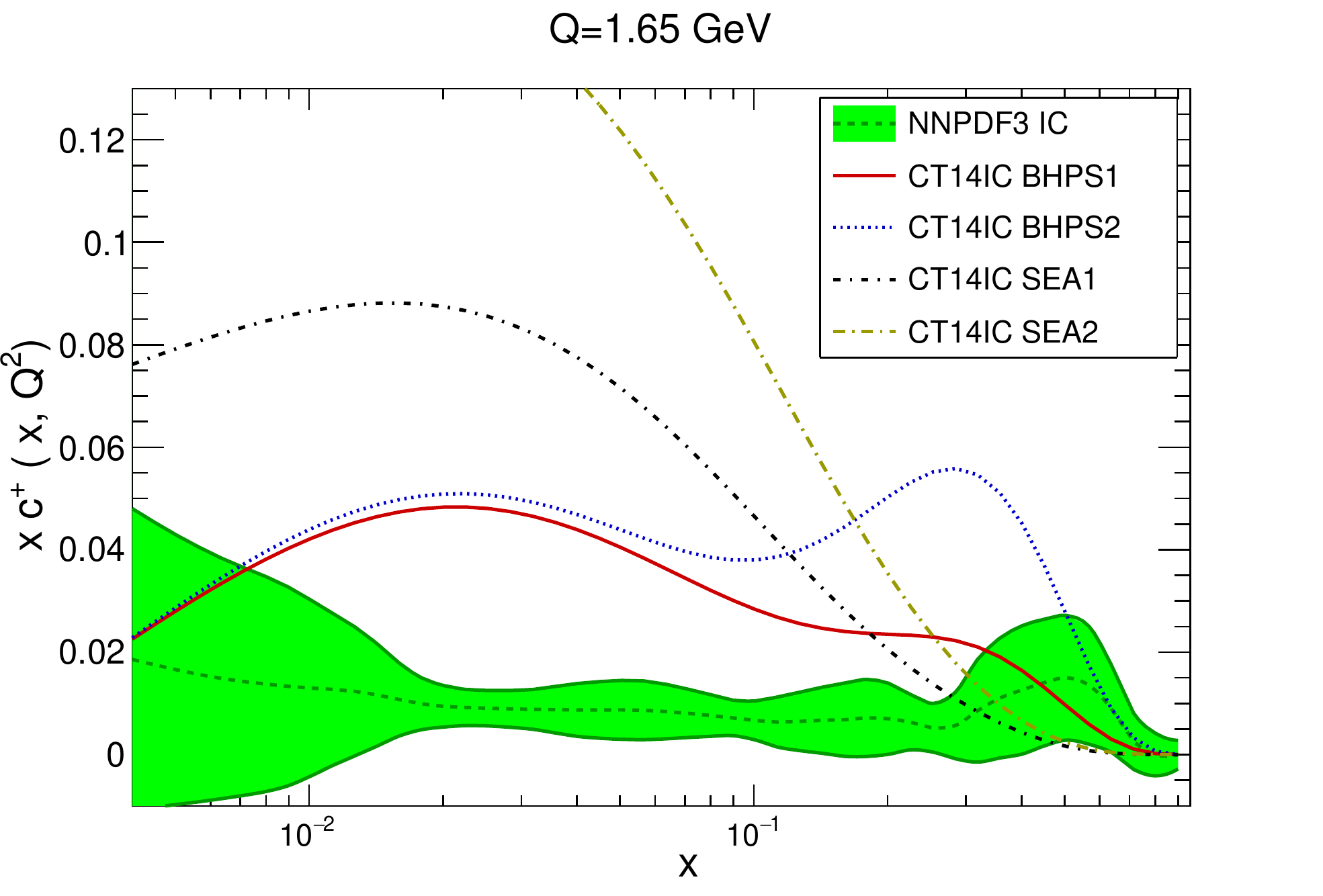}
  \includegraphics[width=0.49\textwidth]{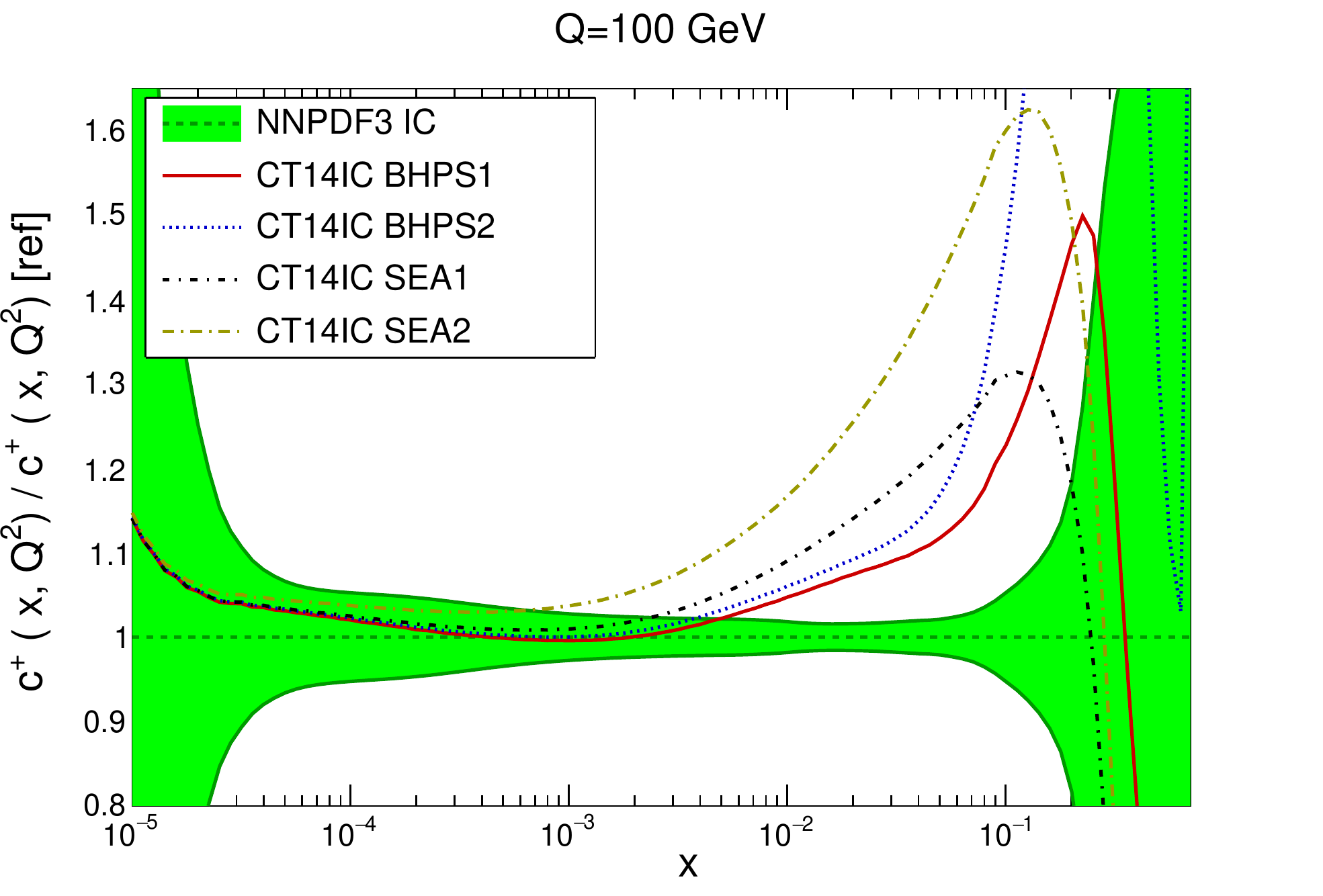}
\end{center}
\vspace{-0.3cm}
\caption{\small \label{fig:ct14}
  Comparison of the NNPDF3 fitted charm PDF
  with the different CT14IC models of~\cite{Hou:2015emq,Dulat:2015mca} 
  at a low scale $Q=1.65$ GeV (left) and at
  a high scale $Q=100$ GeV (right).
 }
\end{figure}

  In Fig.~\ref{fig:ct14} we compare
the NNPDF3 fitted charm PDF with the four CT14 IC  models
  both at a low scale $Q=1.65$ GeV and at a high scale $Q=100$ GeV.
While the fitted charm is qualitatively similar to the BHPS 
model~\cite{Brodsky:1980pb}, it is entirely different to the SEA model. 
 At small $x$ the NNPDF3 fitted charm is smaller
  than all the models, and it peaks at larger values of $x$
  than the BHPS model. At high scale, there is good agreement between our
  fitted charm and the models in the region where perturbative evolution
  dominates, $x\lsim 10^{-3}$, with more substantial differences
  at medium and large-$x$: for example, for $x\simeq 0.2$ the charm
  PDF in the BHPS1 model is 40\% larger than in our fit.
Comparing the momentum fractions in Table~\ref{tab:momsr}, our fitted charm 
result with EMC data prefers a rather lower momentum fraction than was 
considered in Refs.~\cite{Hou:2015emq,Dulat:2015mca}. In fact it seems 
that the BHPS model, with normalization reduced by 40\% or so from
that used in BHPS1, might be in reasonable agreement with our fit at
large $x$. 
We also find that results 
contradict the claim from the authors of Ref.~\cite{Jimenez-Delgado:2014zga} (based on the JR
PDF fit framework)
that values of the charm momentum fraction of $C(Q)$ at the 0.5\% level are
excluded at the four-sigma level. Note, however, that none of these
models reproduce the features of our best-fit charm at small $x$, and
specifically the undershoot in comparison to the perturbative
behaviour discussed above.

Our general conclusion is thus that if the EMC data are reliable, then
charm is compatible with perturbative behaviour at small $x\lsim 0.1$,
where it vanishes at a scale which at NLO turns out to be $Q_0\sim1.6$~GeV,
while it has an intrinsic component at large $x$  which carries about
a percent of the proton momentum at low scale. Not including a fitted
charm component with $m_c=1.47$~GeV 
would thus bias the PDF determination  both at small
and large $x$, with the large $x$ bias localized at low scale and the
small $x$ bias also affecting high-scale physics. The small $x$ bias
would however mostly disappear if PDFs were provided with uncertainties
related to missing higher order corrections and the value of the charm
mass, or if the mass value was raised.

\begin{figure}[ht!]
\begin{center}
  \includegraphics[width=0.46\textwidth]{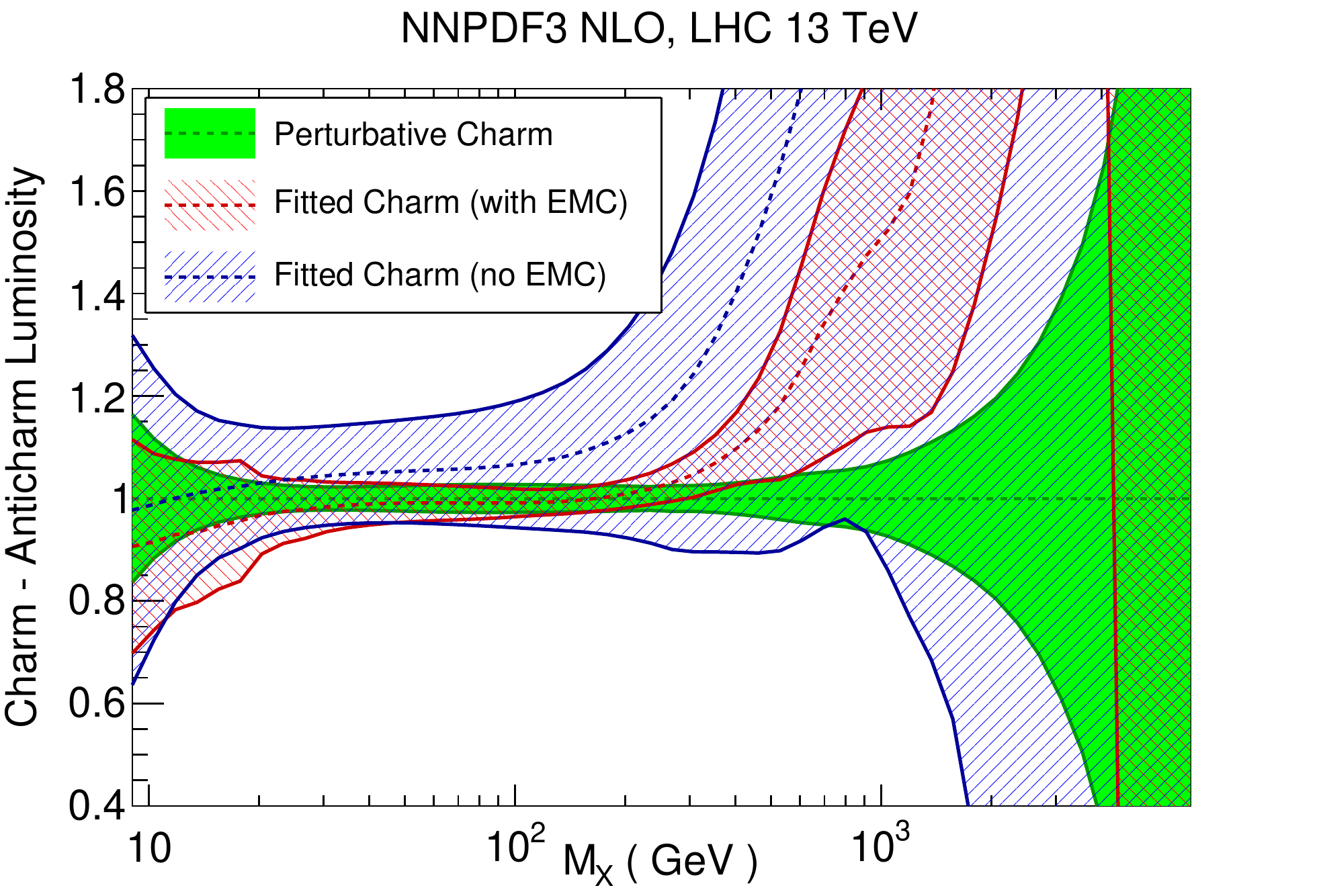}
  \includegraphics[width=0.46\textwidth]{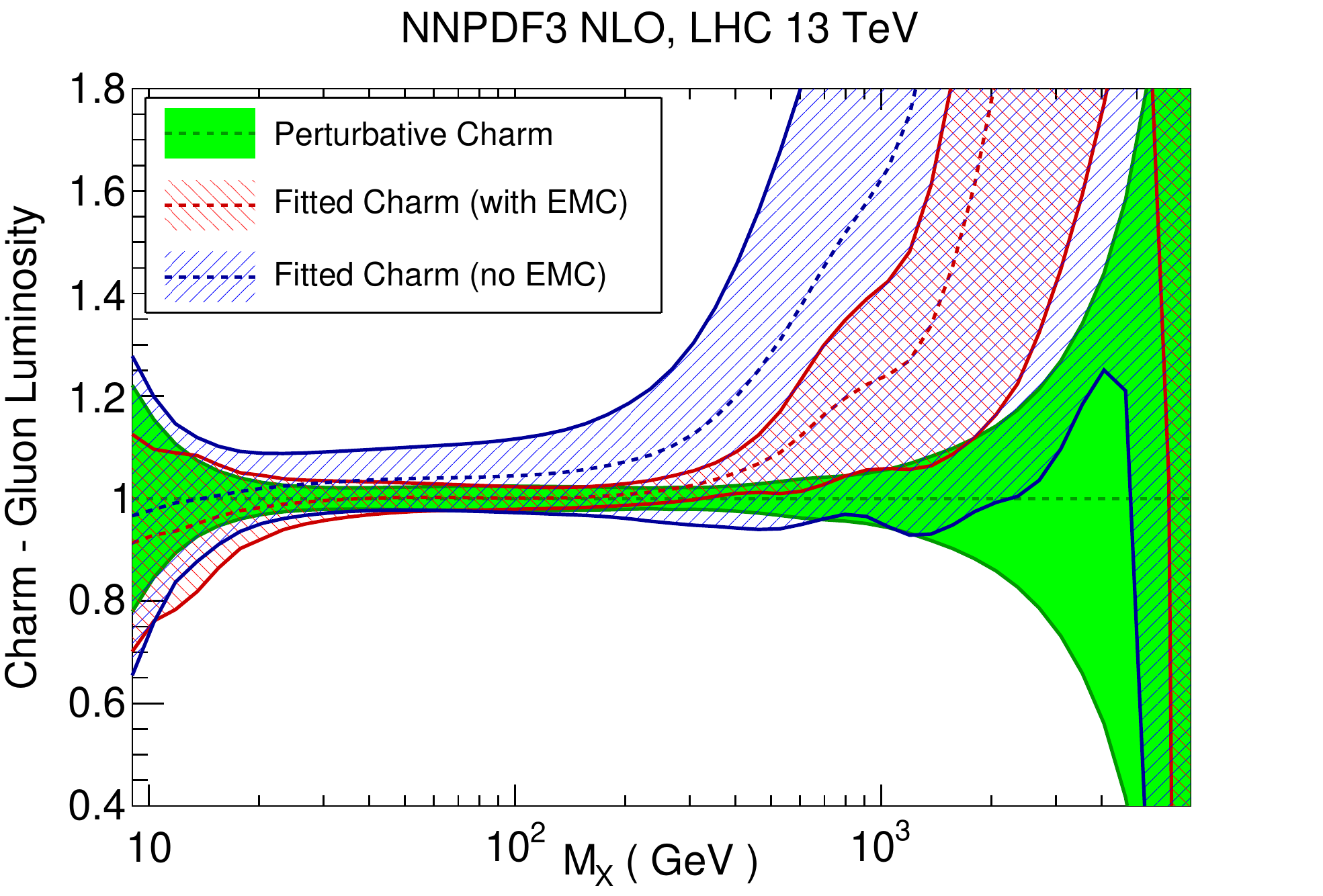}
 \end{center}
\vspace{-0.3cm}
\caption{\small \label{fig:lumis2}
  Parton luminosities at the LHC 13 TeV as a function of the invariant
  mass $M_X$ of the  final state, computed using the PDF sets  with
  perturbative charm, and with fitted charm with and without EMC data.
  The  charm-anticharm (left) and charm-gluon luminosities (right) are
  shown.
   }
\begin{center}
  \includegraphics[width=0.46\textwidth]{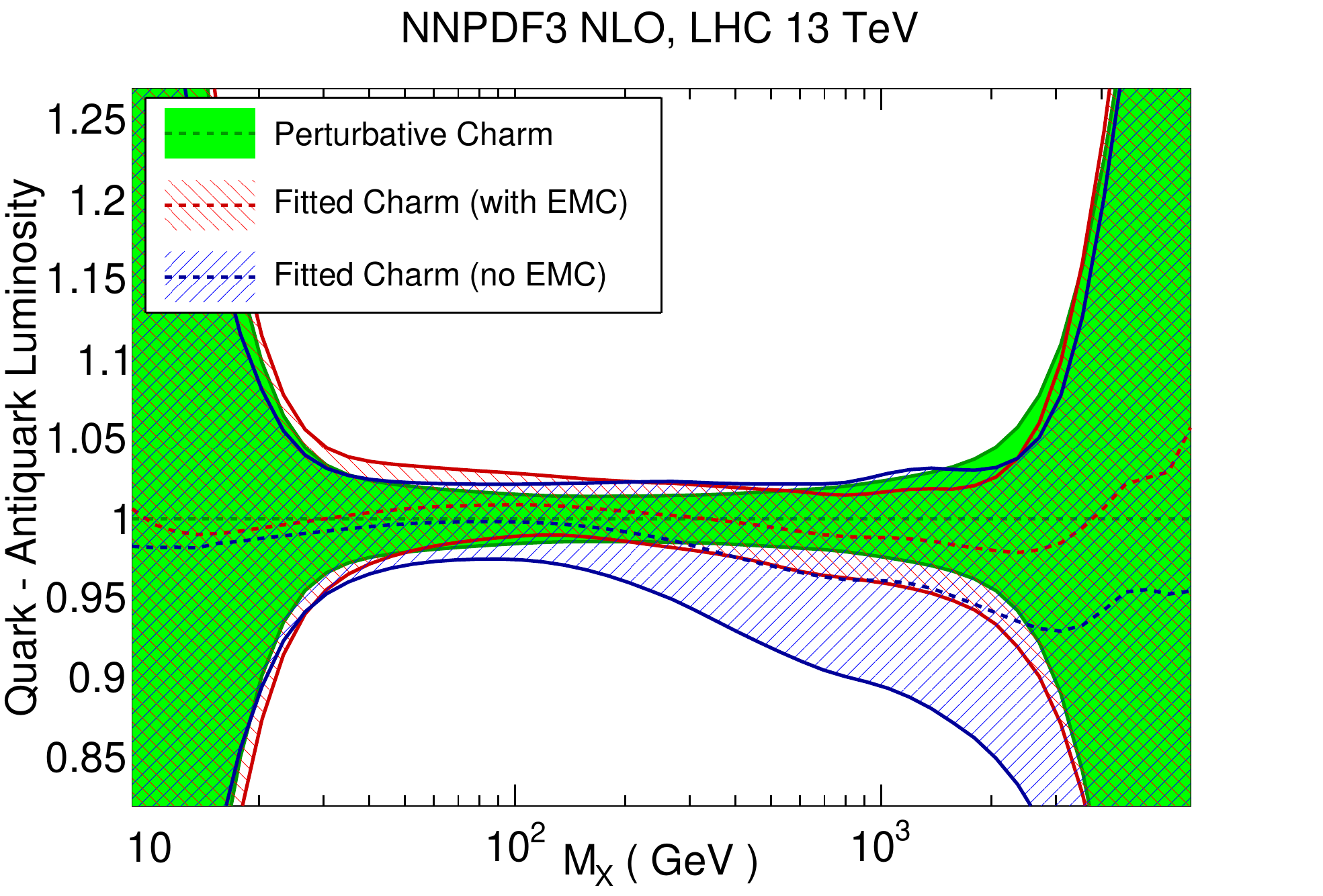}
  \includegraphics[width=0.46\textwidth]{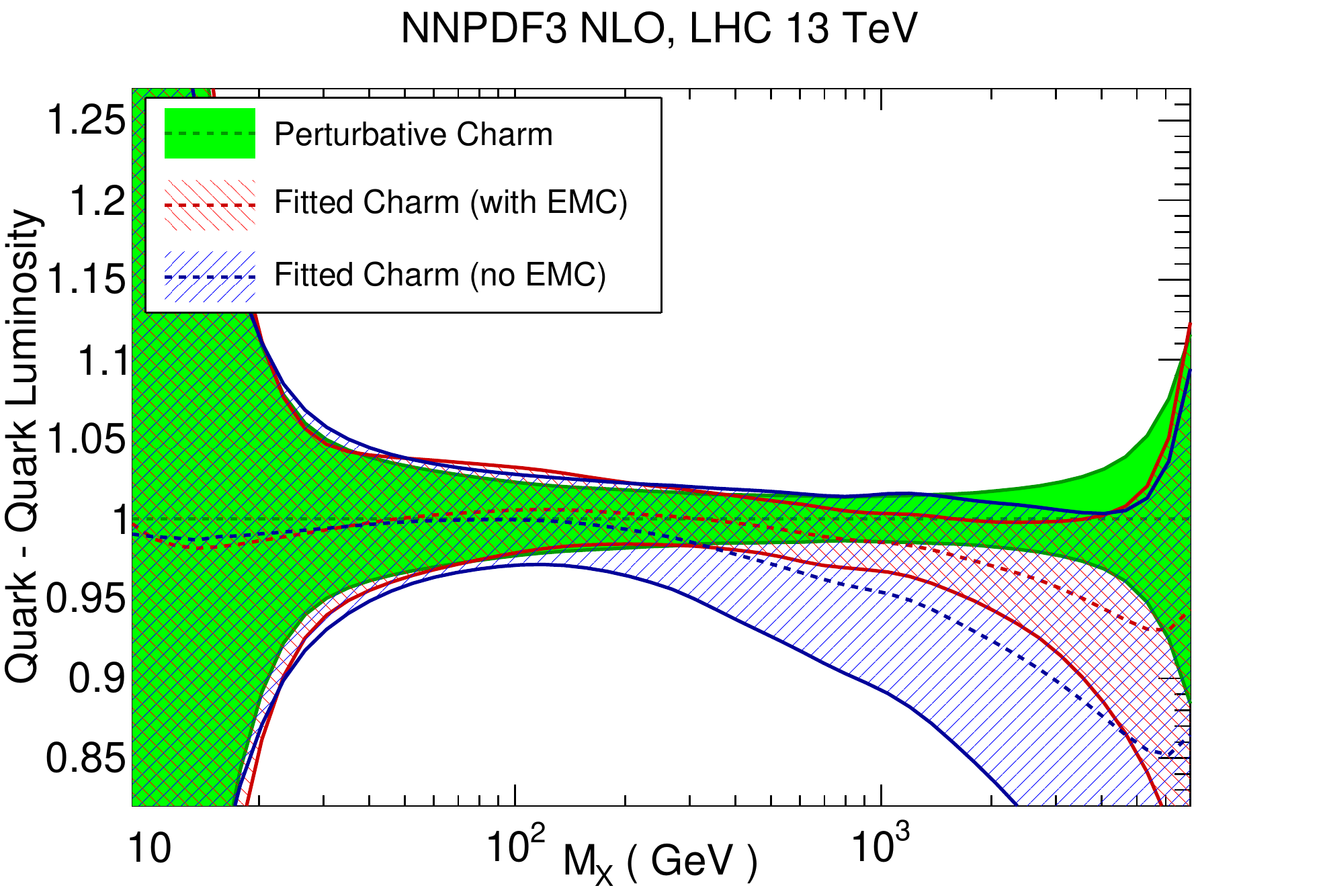}
  \includegraphics[width=0.46\textwidth]{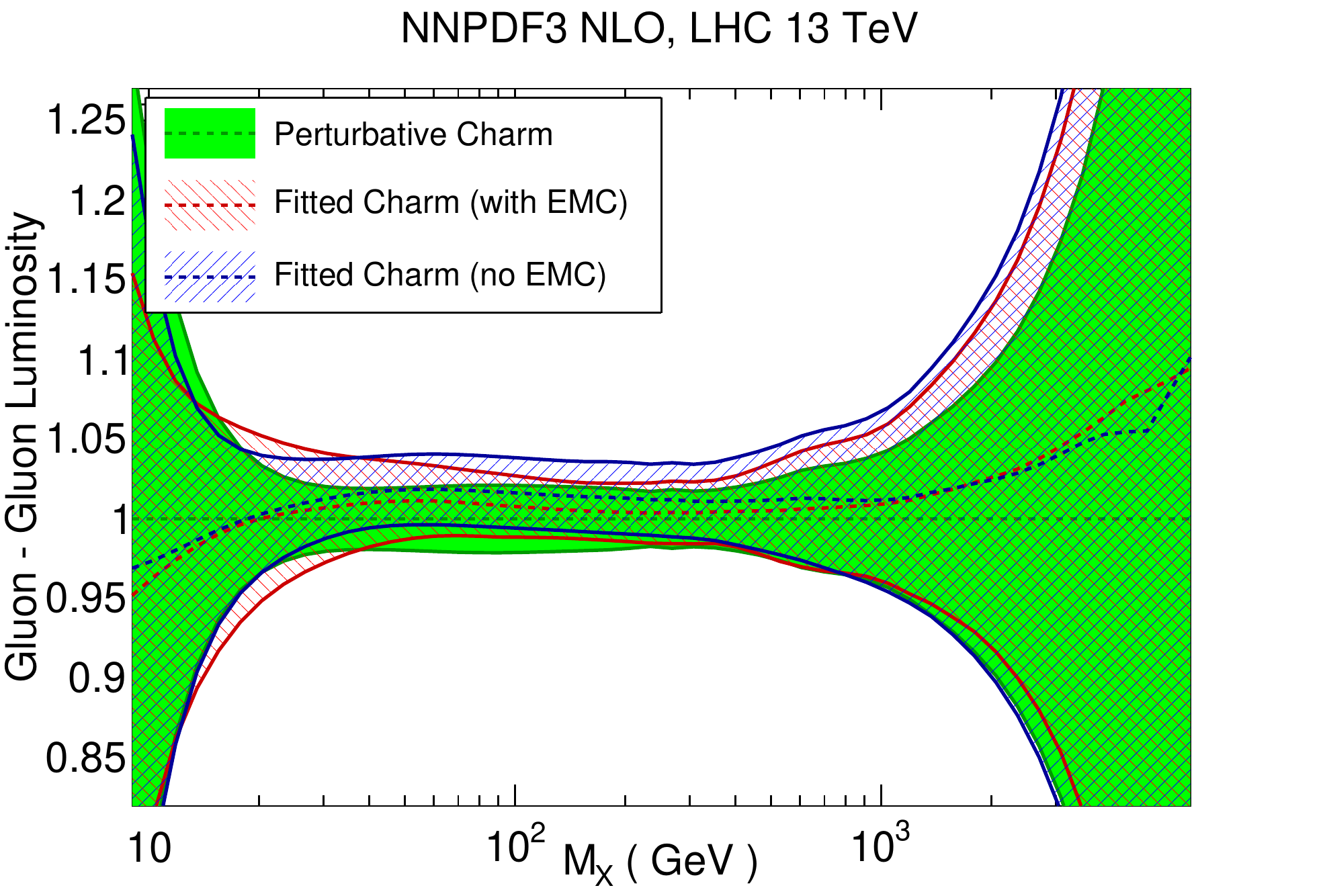}
  \includegraphics[width=0.46\textwidth]{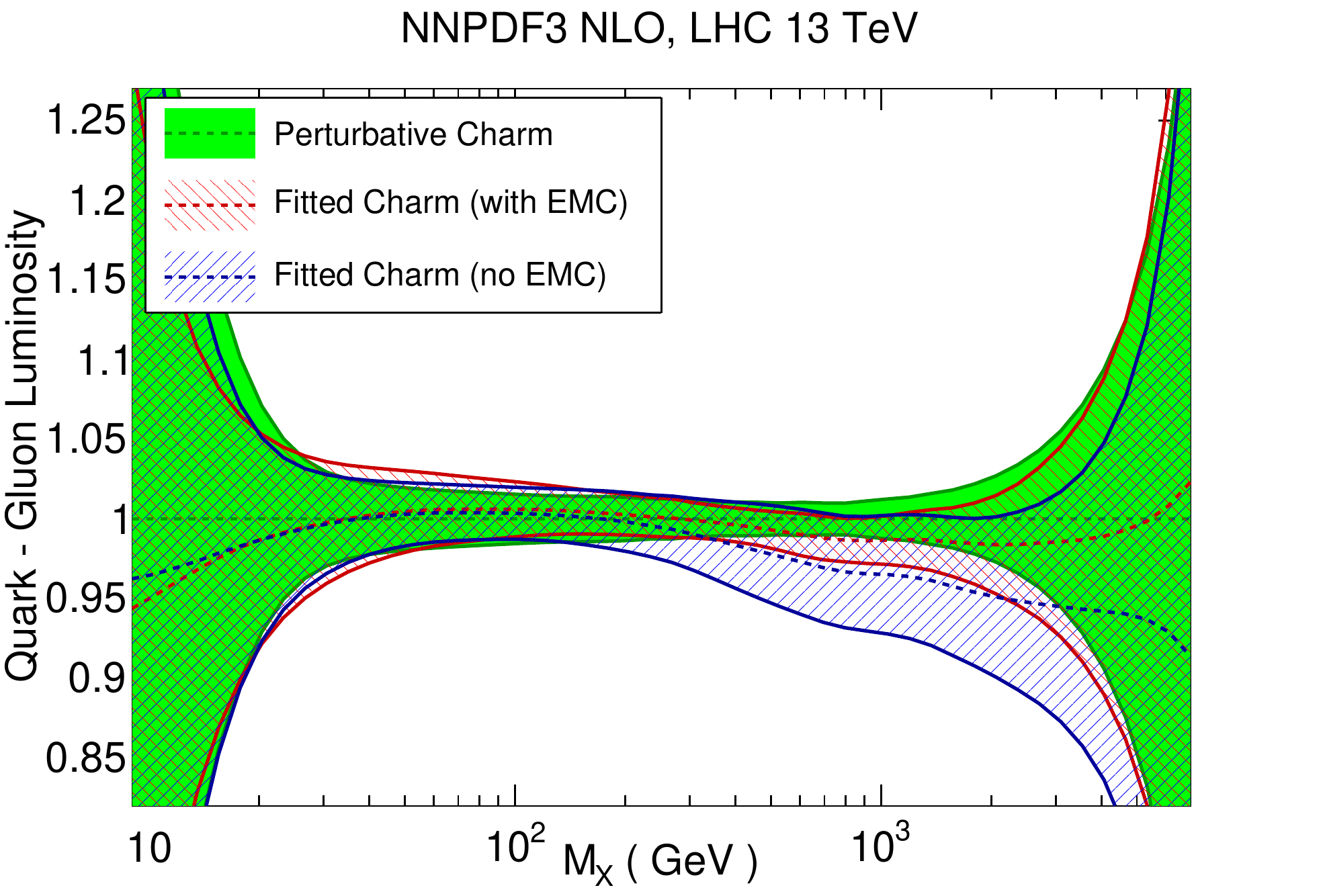}
\end{center}
\vspace{-0.3cm}
\caption{\small \label{fig:lumis1}
Same as Fig.~\ref{fig:lumis2}, but for quark-antiquark (top left),
quark-quark (top right),
  gluon-gluon (bottom left) and quark-gluon (bottom right) luminosities.
   }
\end{figure}

\section{LHC phenomenology}
\label{sec:pheno}

We now discuss the implications of fitting charm for LHC
phenomenology.
First, we compare parton luminosities computed with fitted or
perturbative charm, and specifically show at the level of luminosities
the improved stability upon variation of the charm mass
that was already discussed in Sect.~\ref{sec:hqmass} at the level of
PDFs.
We then turn to specific processes: first, we discuss the 
effect of fitting charm on standard candles, thereby showing 
that fitting charm is
advantageous for more robust uncertainty estimation. Then, we consider
representative LHC processes which are sensitive to charm and could be used
for a more accurate charm PDF determination:
charm quark pair production and $Z$ production
in association with charm quarks.

\begin{figure}[t!]
\begin{center}
  \includegraphics[width=0.46\textwidth]{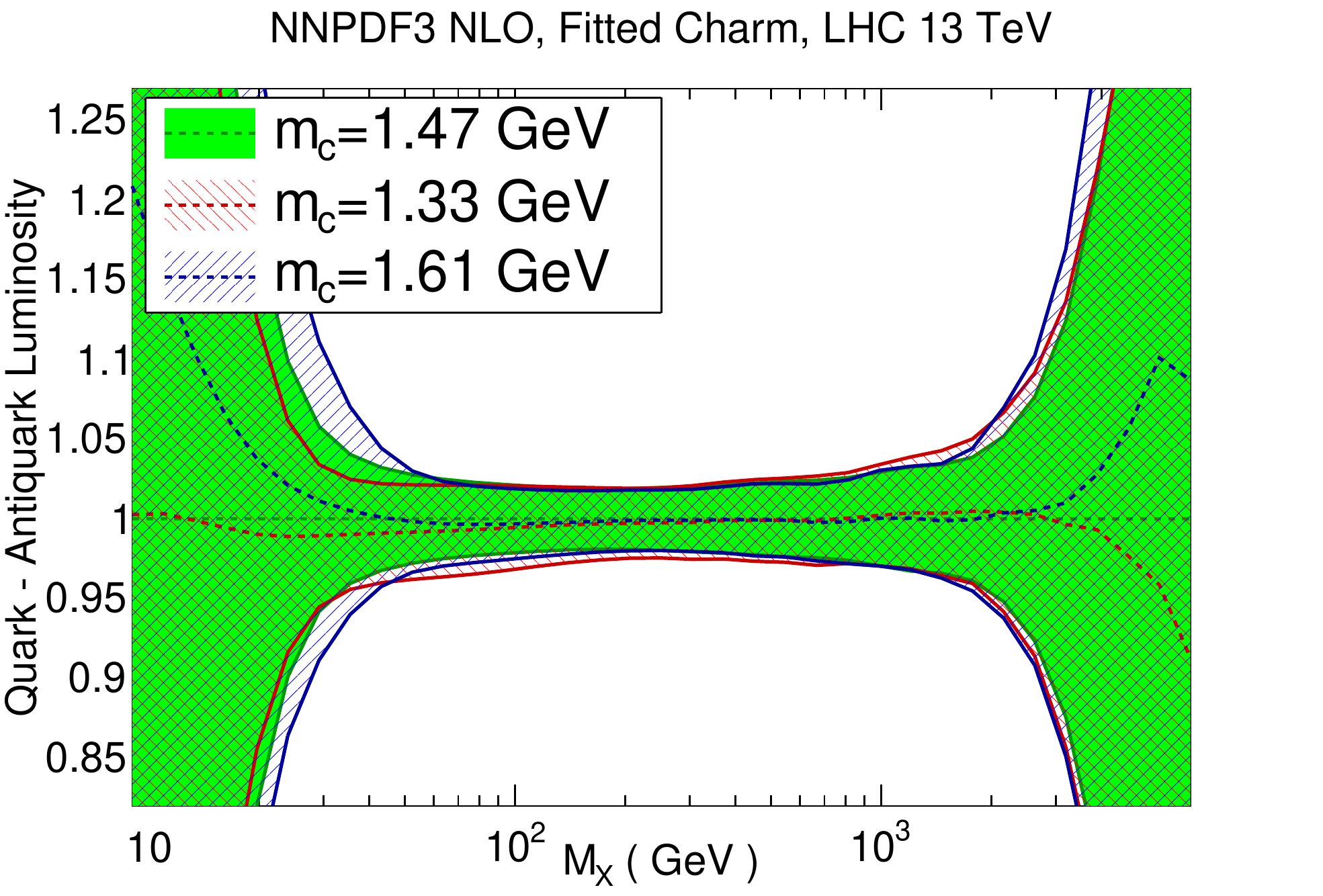}
  \includegraphics[width=0.46\textwidth]{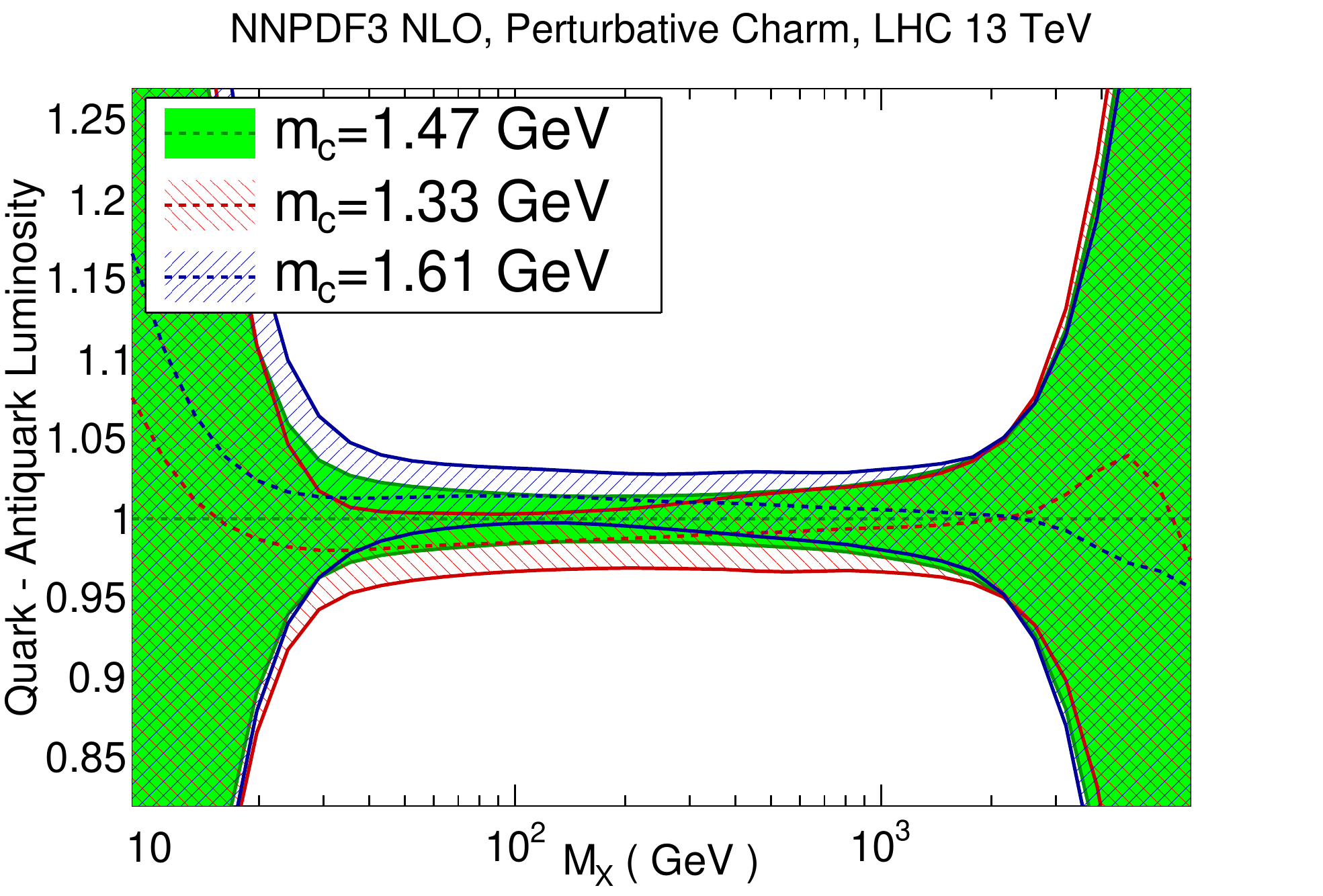}
  \includegraphics[width=0.46\textwidth]{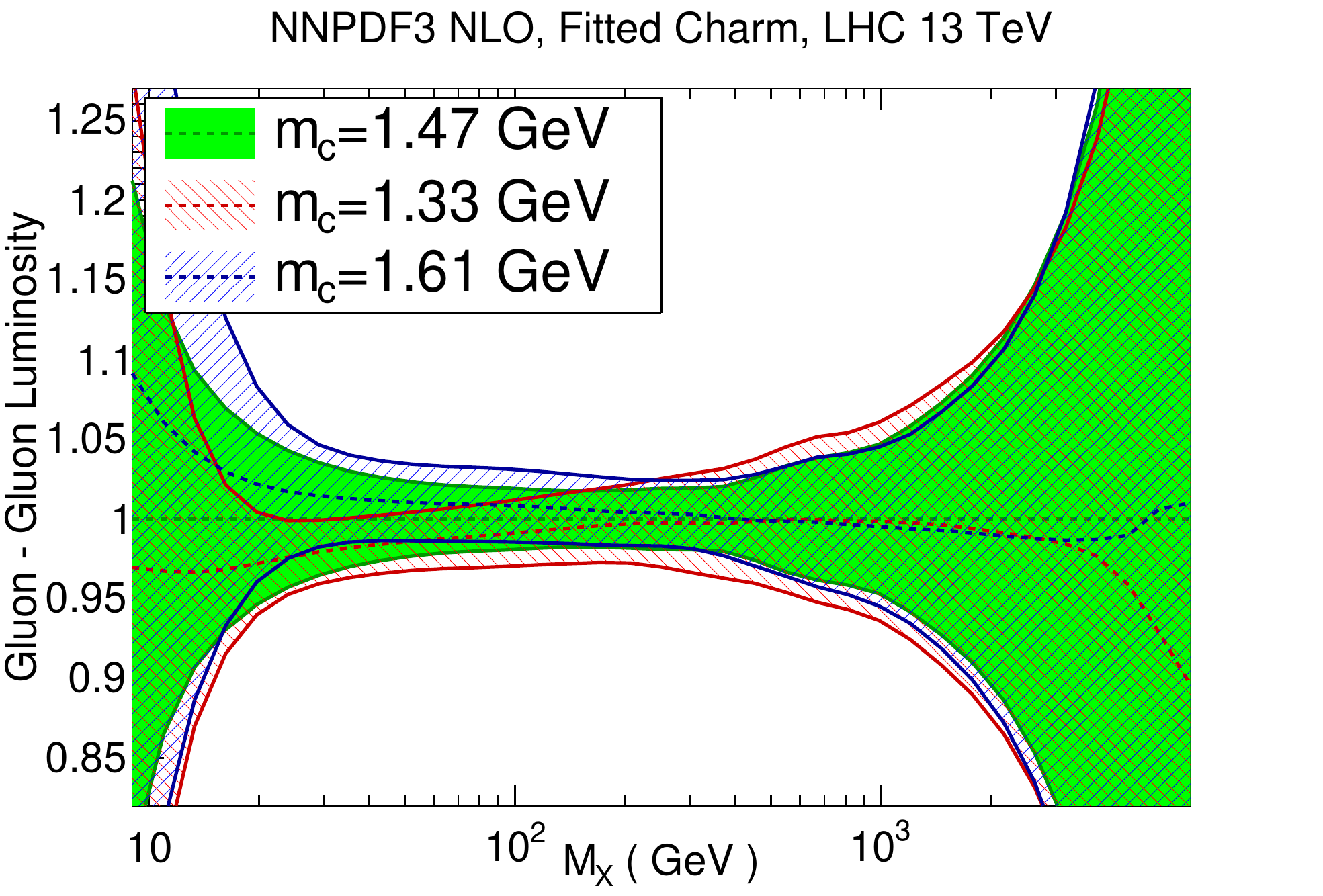}
  \includegraphics[width=0.46\textwidth]{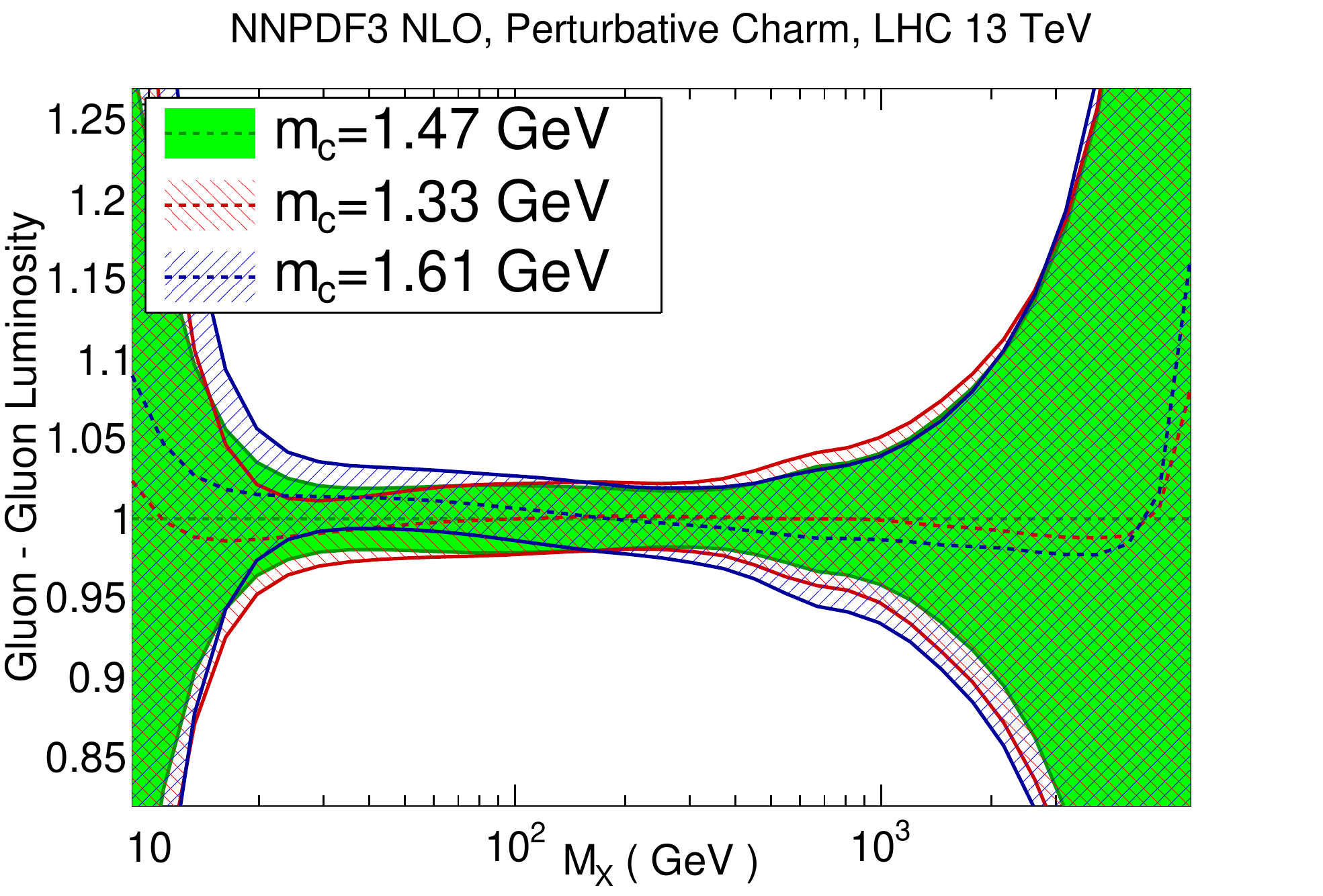}
\end{center}
\vspace{-0.6cm}
\caption{\small \label{fig:lumis1-mcstab}
  Same as Fig.~\ref{fig:lumis1}, but now comparing the quark-antiquark
  (top) and gluon-gluon (bottom) luminosities for different values of
  the pole charm mass, for fitted (left) and perturbative charm (right).
   }
\vspace{-0.2cm}
\begin{center}
  \includegraphics[width=0.46\textwidth]{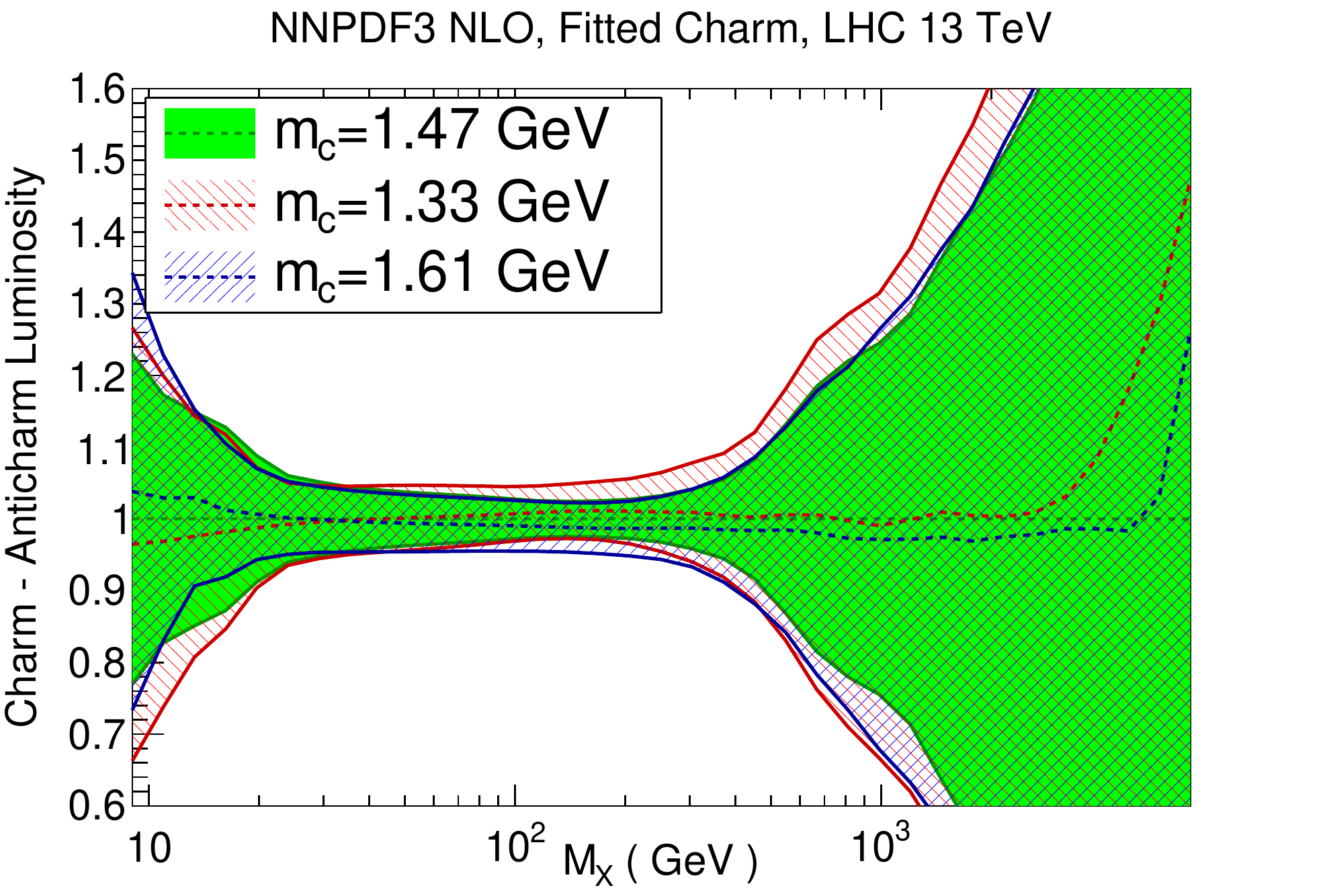}
  \includegraphics[width=0.46\textwidth]{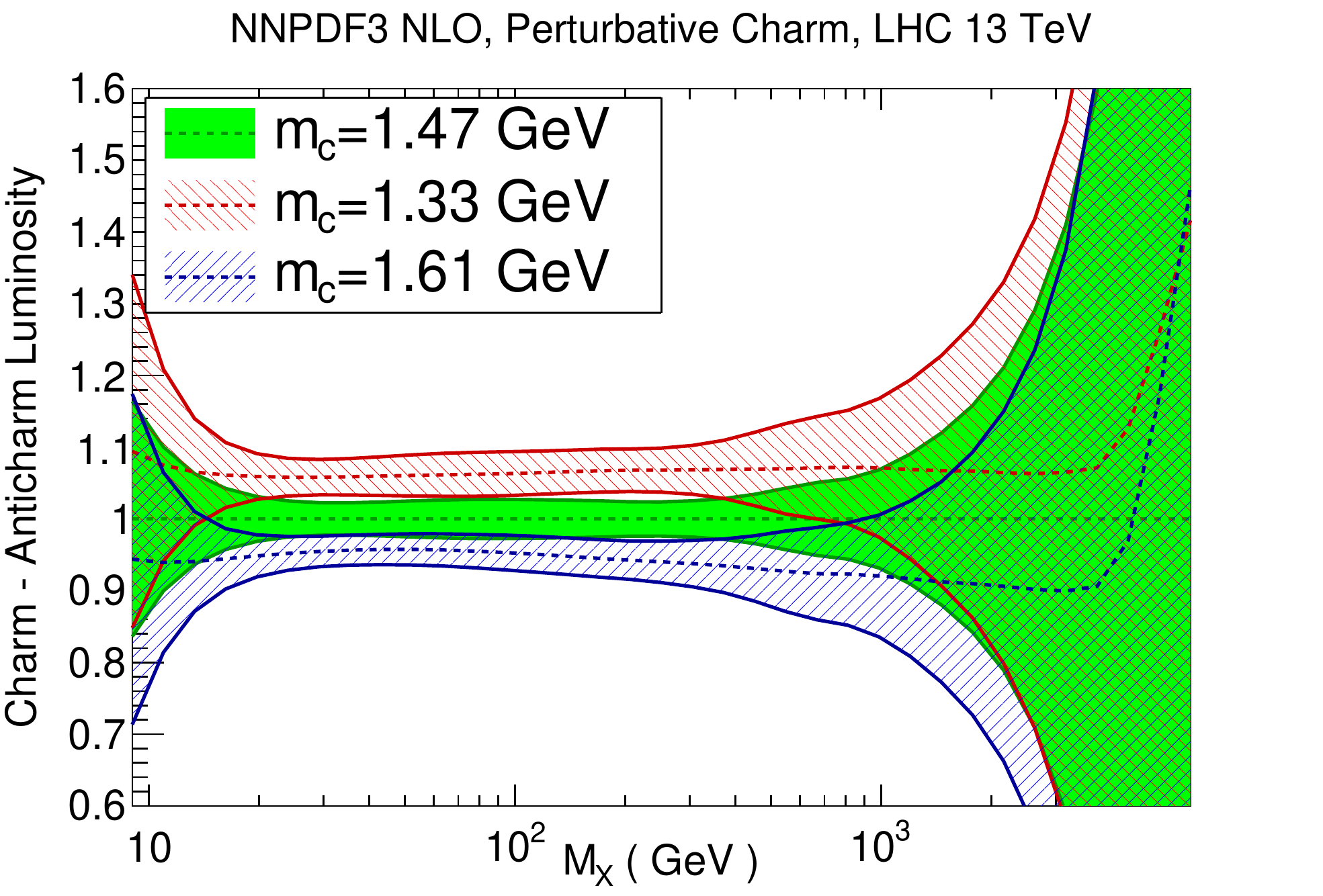}
  \includegraphics[width=0.46\textwidth]{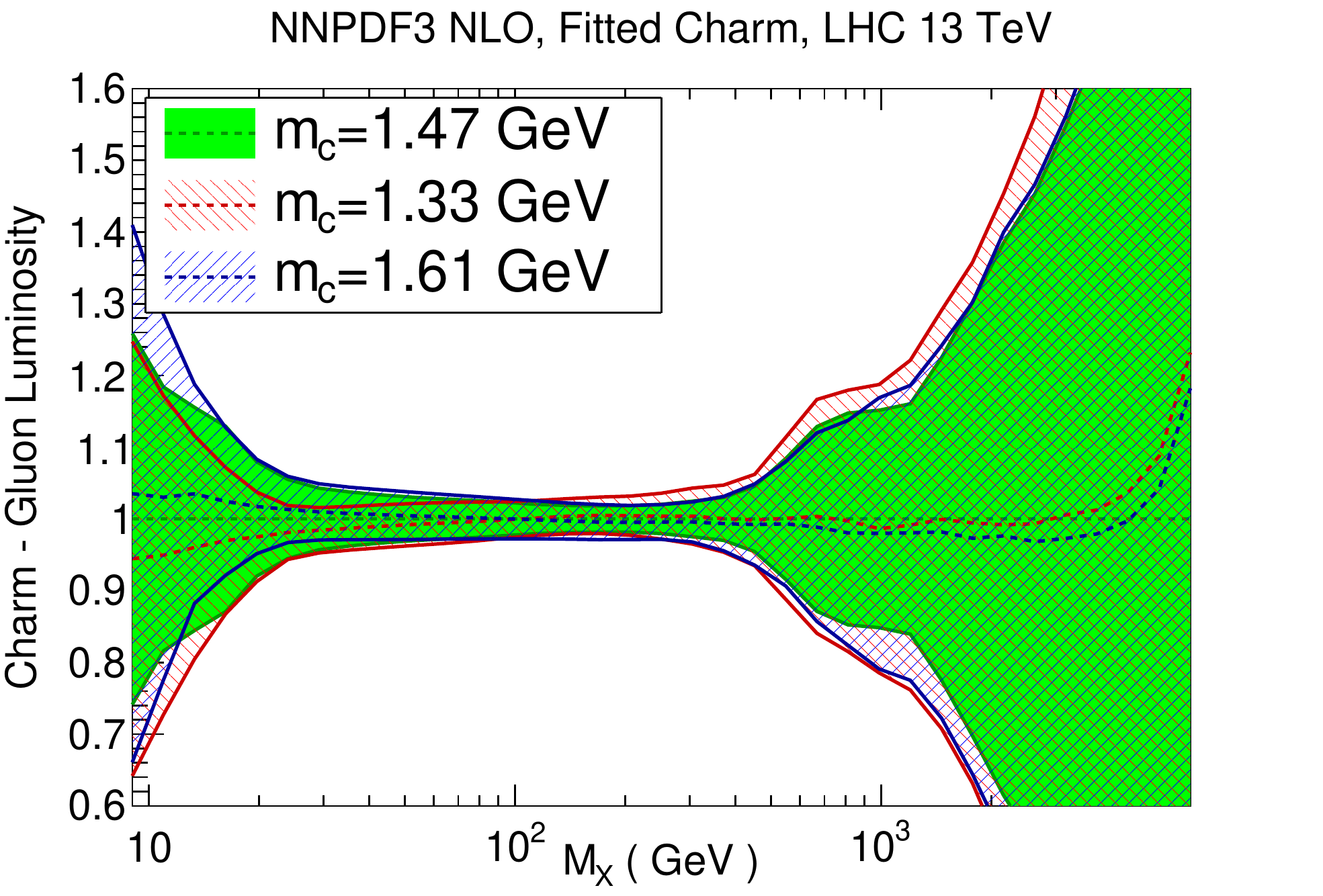}
  \includegraphics[width=0.46\textwidth]{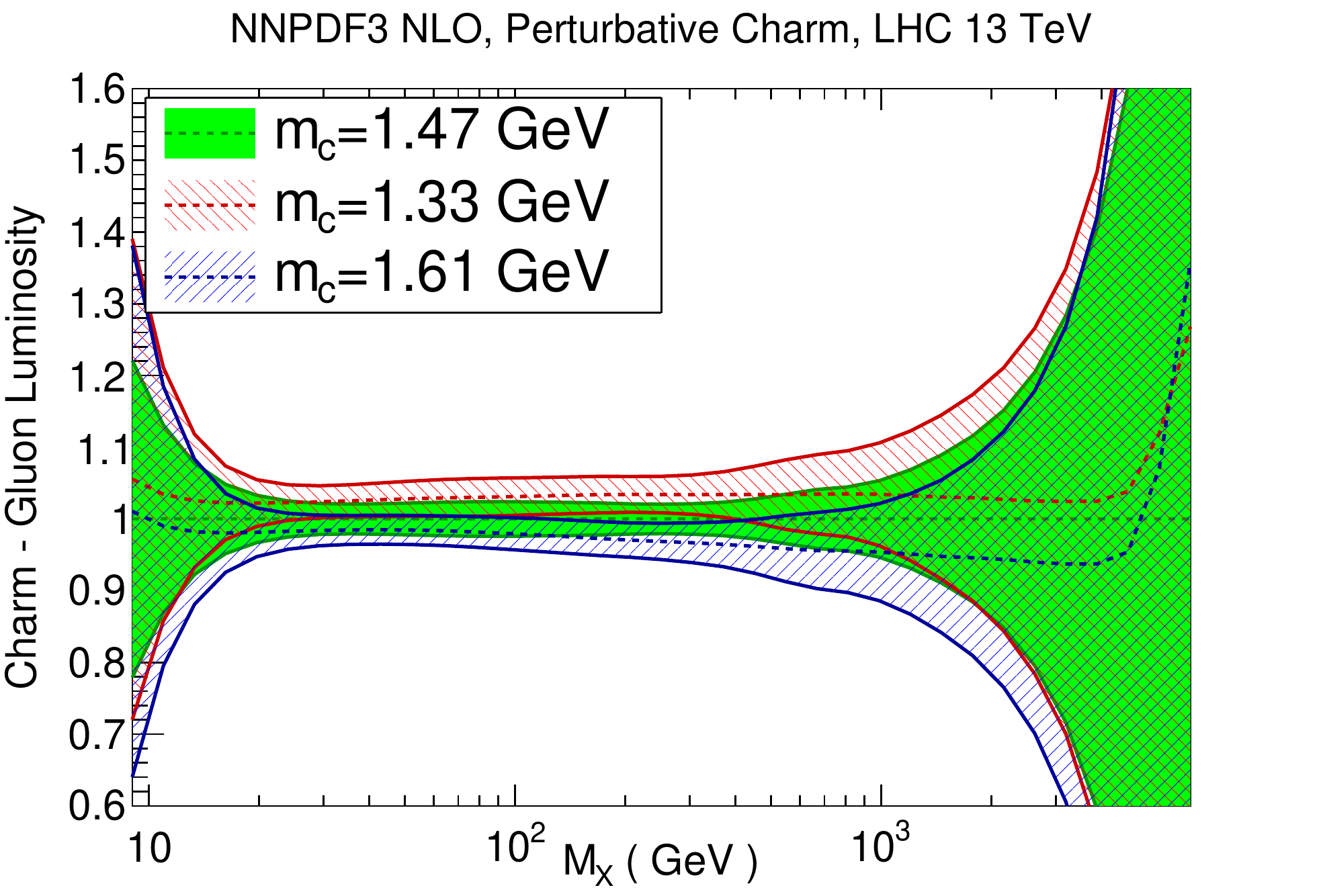}
 \end{center}
\vspace{-0.6cm}
\caption{\small \label{fig:lumis2-mcstab}
  Same as Fig.~\ref{fig:lumis1-mcstab} for charm-anticharm
  (top) and charm-gluon luminosities (bottom).
   }
\end{figure}

\subsection{Parton luminosities}
\label{sec:lumi}

In Fig.~\ref{fig:lumis2} we compare 
parton luminosities (defined as in Ref.~\cite{Campbell:2006wx}) involving charm
at the LHC 13 TeV, plotted as a function of the invariant
  mass $M_X$ of the final state, for the PDF sets with perturbative
  and fitted charm, with and without EMC data, discussed in
  Sect.~\ref{sec:results}. We show the charm-anticharm and charm-gluon
  luminosities, which are relevant for charm-dominated processes at the LHC, such
as $D$ meson production at large $p_T$ and rapidity,
where the $c\bar{c}$ process becomes important,
or $\gamma/Z+D$ production, which at the Born level is driven by the
$gc$ luminosity.
We find that when the EMC data are included, the uncertainty in the
luminosity with fitted charm is similar to that when charm is
perturbative for scales $M_X\sim100$~GeV, and larger than it by a
factor three or four for higher or lower scale, while if EMC data are
not included the uncertainty with fitted charm is substantially
larger for all scales. 
This suggests that the determination of the luminosities with
purely perturbative charm might be unreliable, with underestimated uncertainty 
and possibly a biased central value, particularly at high invariant masses.
In  Fig.~\ref{fig:lumis1} we show luminosities involving light quarks
and gluons. In this case, uncertainties are similar with fitted or
perturbative charm provided EMC data are included. 

The difference
between fitted and perturbative charm is particularly apparent 
in the dependence of 
luminosities on the value of the charm mass, which is shown in
Fig.~\ref{fig:lumis1-mcstab}  (for the light quark-antiquark and the
gluon-gluon luminosity). A marked increase in stability is seen in
the   $q\bar{q}$ luminosity for all $M_X$ when charm is fitted. This means that if charm is not fitted,
the choice of charm mass is a possible source of bias. 
The reduced dependence  on the value of $m_c$ 
becomes especially striking for luminosities involving charm: as shown
in Fig.~\ref{fig:lumis2-mcstab},
the spread in central values for the charm-anticharm luminosity as the
charm mass is varied  is about 15\% for perturbative  charm and about
2\% for fitted charm for all 20~GeV$<M_X<$1~TeV.
Similar conclusions hold for the $cg$ luminosity.

\subsection{LHC standard candles}
\label{sec:candles}

We now study the impact of the fitted charm PDFs for the calculation
of standard candles at the LHC.
We start with total cross-sections and then consider
some differential distributions, all at the LHC~13~TeV.

\subsubsection{Total cross-sections}

\begin{figure}[t!]
  \begin{center}
    \includegraphics[width=0.46\textwidth]{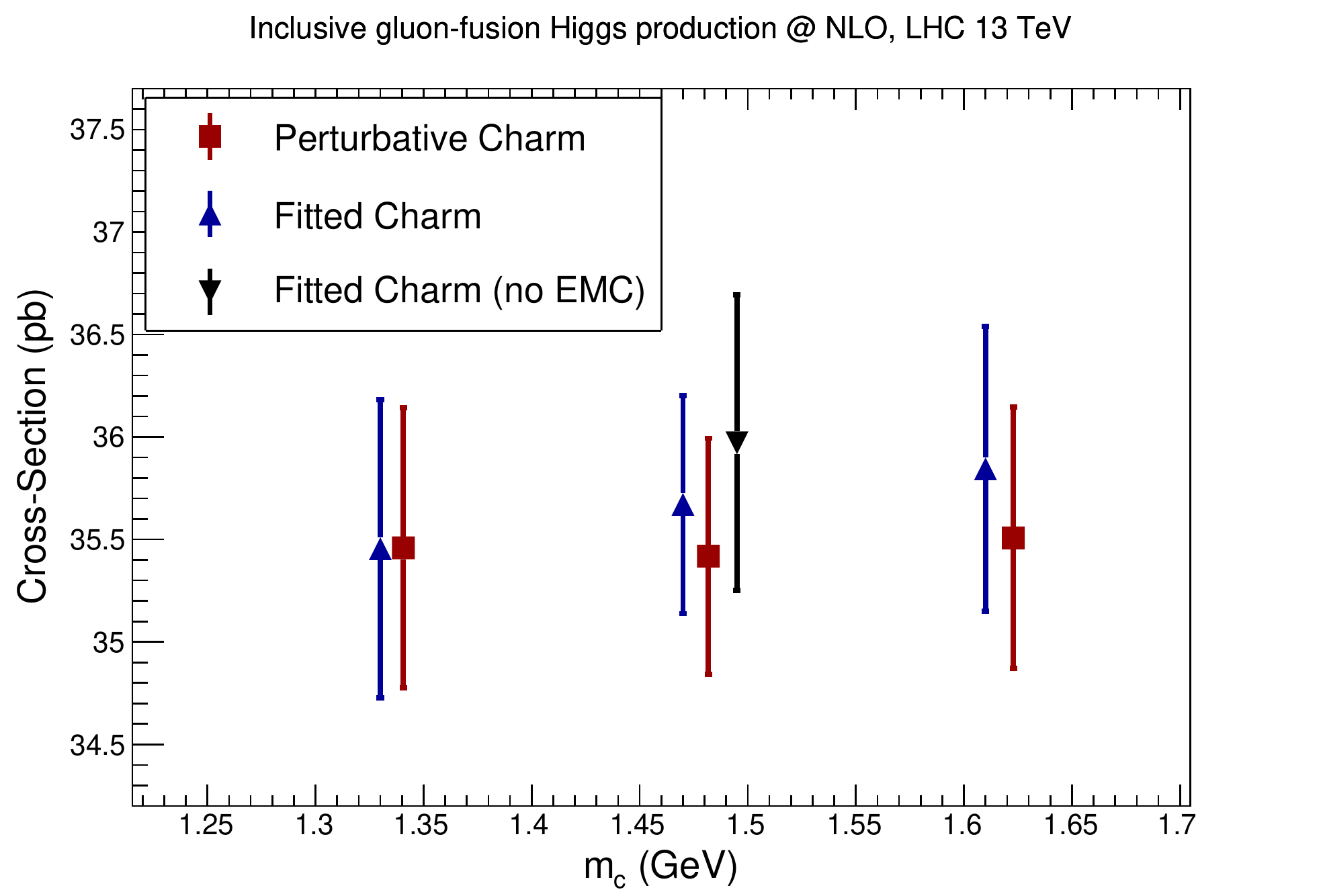}
    \includegraphics[width=0.46\textwidth]{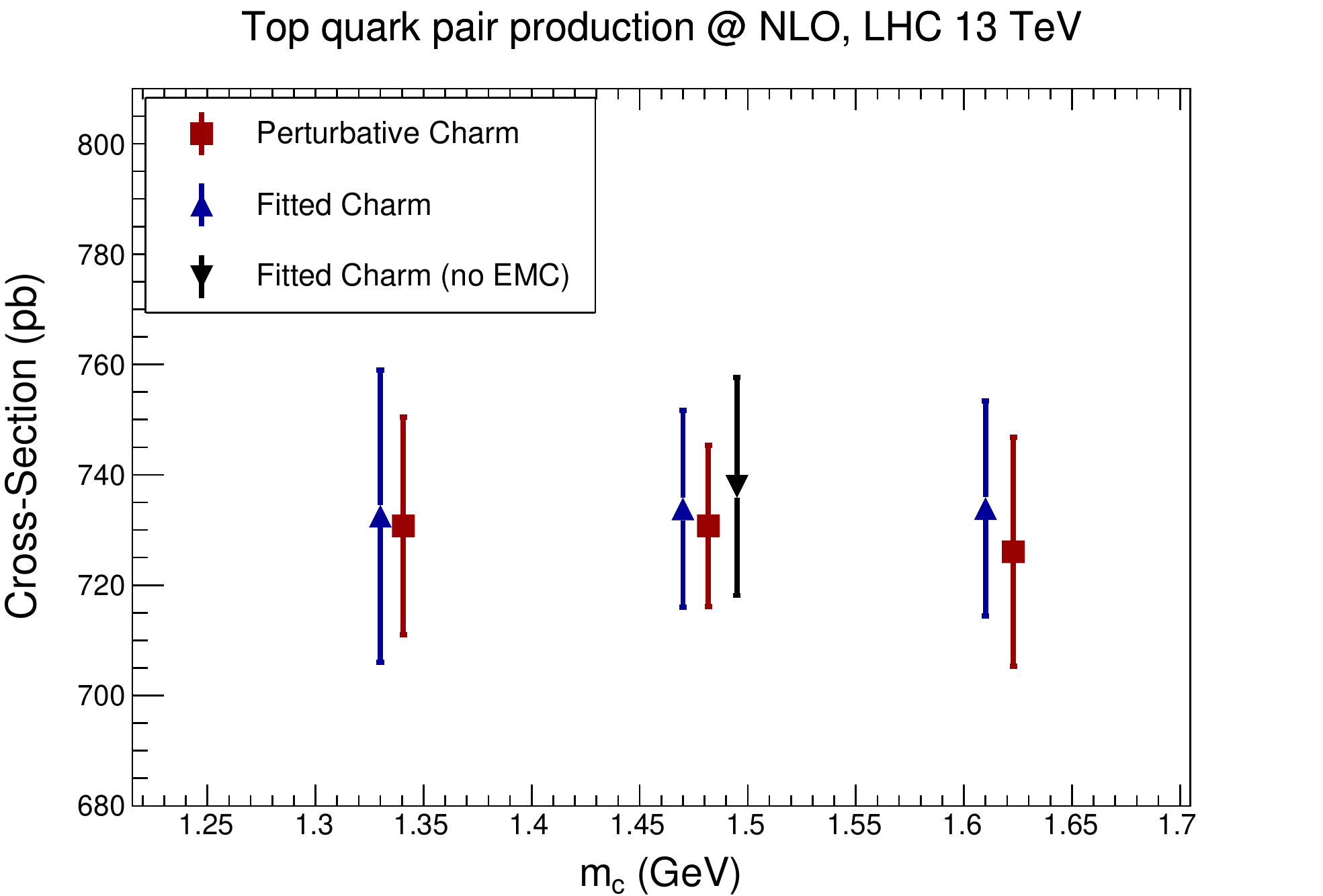}
\end{center}
\vspace{-0.3cm}
\caption{\small \label{fig:ggHiggs}
  The NLO cross-sections for Higgs production in gluon fusion (left)
  and inclusive top quark pair production (right) at the
   LHC~13~TeV
  with fitted or perturbative charm and
  $m_c^{\rm pole}= 1.33,
  1.47$ and 1.61 GeV.
  We also show the result with fitted charm and no EMC data for
   $m_c^{\rm pole}= 1.47$ GeV.
  The uncertainty shown is the PDF uncertainty
  only (not including i.e.\ $\alpha_s$ variations).
}
\begin{center}
  \includegraphics[width=0.46\textwidth]{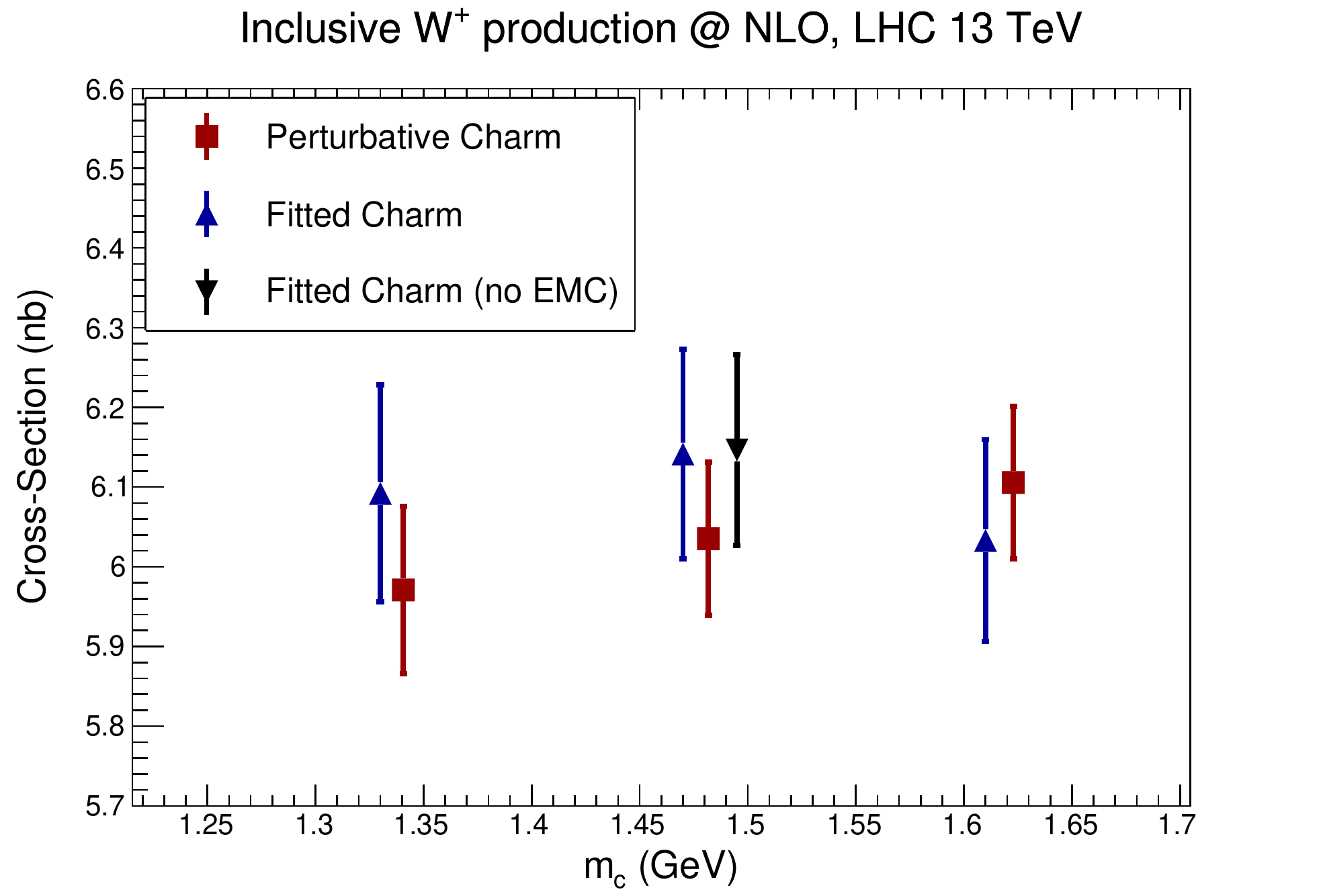}
  \includegraphics[width=0.46\textwidth]{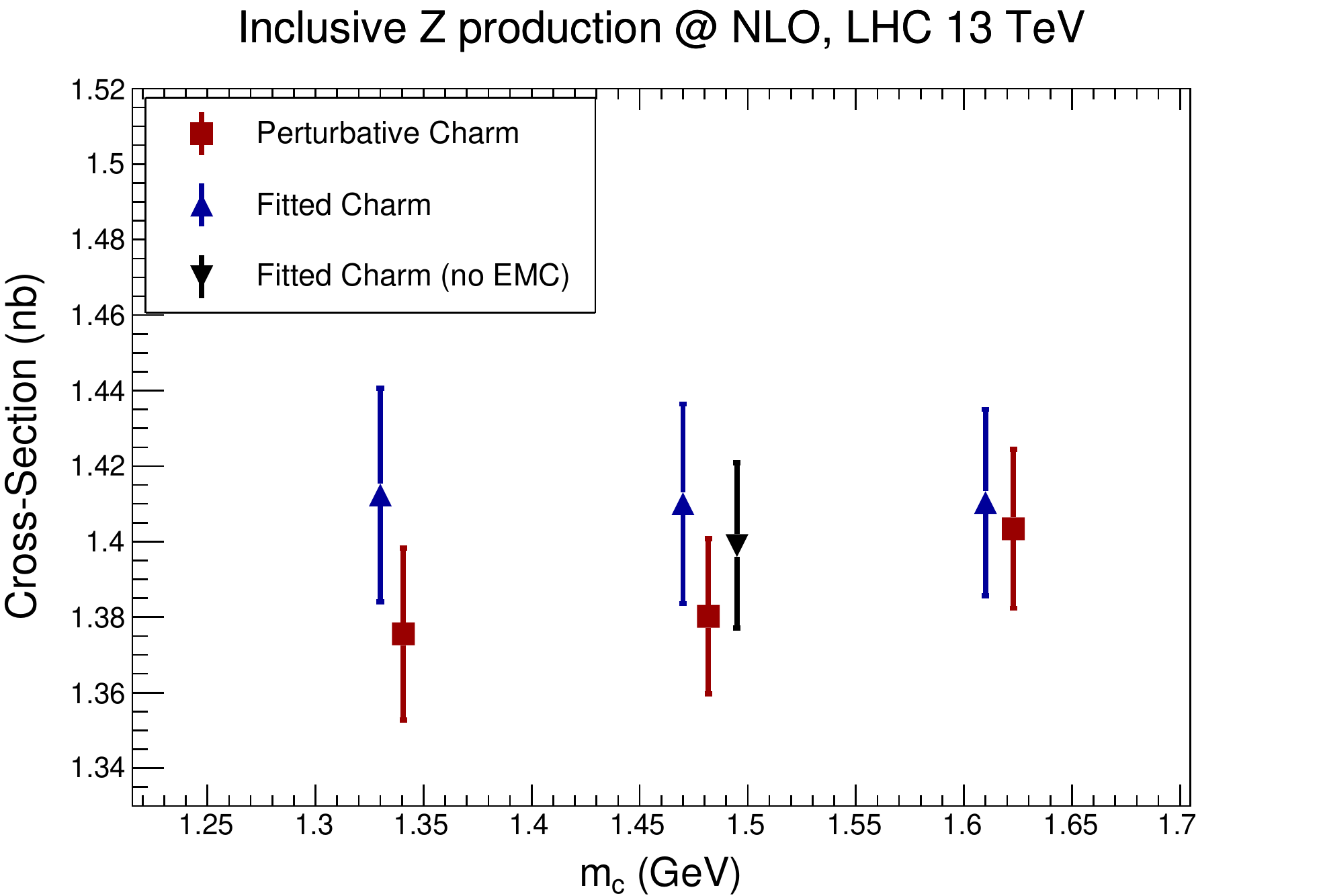}
\end{center}
\vspace{-0.5cm}
\caption{\small \label{fig:WZ} Same as Fig.~\ref{fig:ggHiggs} for
  the cross section for  the inclusive production of $W^+$ (left)
  and $Z$ (right) bosons at the LHC 13 TeV, including
  leptonic branching fractions and standard acceptance cuts.
   }
\end{figure}

We first consider Higgs and top production.
We have computed the total inclusive Higgs production cross-section in
the gluon fusion channel
using the {\tt ggHiggs}
code v3.2~\cite{ggHiggs} to NLO, including full dependence on the top,
bottom and charm masses, for $\mu_F=\mu_R=m_H/2$ and $m_H=125$~GeV.
We have also computed the inclusive top quark pair production cross-section at NLO
using {\tt top++} v2.0~\cite{Czakon:2011xx}.
Results  are collected in Table~\ref{table:mctable} and
represented in Fig.~\ref{fig:ggHiggs}. Note that the
uncertainty shown is the PDF uncertainty only (not including
$\alpha_s$ variation).
In both cases, the impact of fitting charm on
the cross-section is moderate, both for central values and
uncertainties, and while the cross-section is almost
independent of the charm mass for perturbative charm, it varies a
little
more when 
the charm is fitted. The overall uncertainty is thus a little 
larger with fitted charm, reflecting the slightly increased uncertainty 
in the gluon-gluon luminosity.

\begin{table}[h!]
  \centering
  \small
  \begin{tabular}{|c|c|c|c|c|}
    \hline
    Process  & Charm PDF   &   $m_c=1.33$ GeV  &  $m_c=1.47$ GeV  &   $m_c=1.61$ GeV  \\
    \hline
    \hline
     \multirow{3}{*}{$\sigma(gg\to h)$ [pb]}  & Fitted  & $ 35.5\pm 0.7 $ &
     $  35.7 \pm 0.5  $&   $ 35.8 \pm 0.7 $ \\
     & Fitted (no EMC) & - & $ 36.0 \pm 0.7 $   &  - \\
    & Perturbative  &  $ 35.5 \pm 0.7  $ &
     $ 35.4 \pm 0.6 $&  $ 35.5 \pm 0.6  $  \\
     \hline
        \multirow{3}{*}{$\sigma(t\bar{t})$ [pb]}  & Fitted  & $ 733\pm 26 $ &
     $ 734 \pm 18  $&   $ 734 \pm 20 $ \\
     & Fitted (no EMC) & - & $ 738 \pm 20$   &  - \\
    & Perturbative  &  $ 731 \pm 20 $ &
     $ 731 \pm 15  $&  $  726\pm 21  $  \\
     \hline
     \multirow{3}{*}{$\sigma(W^+\to l^+\nu)$ [nb]}
     & Fitted  & $6.09\pm 0.14$  &  $6.14\pm 0.13$ & $6.04\pm 0.13$ \\
          & Fitted (no EMC)  & -   &  $6.15\pm 0.12$ &  - \\
    & Perturbative  & $5.97\pm 0.10$ & $6.03\pm 0.10$ & $6.11\pm 0.10$  \\
    \hline
    \multirow{3}{*}{$\sigma(W^-\to l^-\nu)$ [nb]}
    & Fitted  & $4.42\pm 0.10$  &  $4.43\pm 0.09$& $4.40\pm 0.09$  \\
    & Fitted (no EMC) & -  &  $4.44\pm 0.08$& -  \\
    & Perturbative  & $4.38\pm 0.07$  &  $4.41\pm 0.07$  & $4.47\pm 0.07$ \\
    \hline
    \multirow{3}{*}{$\sigma(Z\to l^+l^-)$ [nb]}
    & Fitted     &   $1.412\pm 0.028$  & $1.410\pm 0.026$ & $1.410\pm 0.025$  \\
    & Fitted (no EMC)    &   -   & $1.400\pm 0.023$ & -  \\
    & Perturbative  & $1.376\pm 0.022$  & $1.380\pm 0.021$  &$1.5403\pm 0.021$  \\
    \hline
  \end{tabular}
  \caption{\small
Numerical values for the cross-sections represented in Figs.~\ref{fig:ggHiggs}-\ref{fig:WZ}. 
\label{table:mctable}
  }
\end{table}

Next, we have computed
the total cross-section for $W$ and $Z$ production at NLO at
the  LHC 13 TeV using {\tt MCFM}~\cite{Campbell:2004ch}. We include
the leptonic decays of the gauge bosons, and 
we impose standard acceptance requirements for the final-state
leptons, 
namely $p_T^l \ge 20$ GeV and $|\eta_l|\le 2.5$.
Results are presented in Fig.~\ref{fig:WZ} and collected in
 Table~\ref{table:mctable}. Here, while again results with
 perturbative or fitted charm are very similar,
an improvement in stability with
 respect to the choice of  $m_c$ when charm is fitted is clearly
 visible for $Z$ production.
 Also, we see that whether or not we include the EMC data makes very little 
 difference to these standard candles.

As a general conclusion, we find that the variation of total
cross-section for LHC standard candles as the charm mass is varied in
a very conservative range is a small fraction of the PDF uncertainty.
This conclusion is in agreement with
previous studies of the dependence of global fit results on the charm
mass (but with perturbative charm only)
presented in Refs.~\cite{Ball:2011mu,Martin:2010db,Gao:2013wwa,Harland-Lang:2015qea}.

\subsubsection{Differential distributions}

We now turn to differential distributions for
 Higgs production in gluon fusion, top-pair
production and $W,Z$ electroweak gauge-boson production at 13 TeV.
All  calculations have been performed at NLO using {\tt MadGraph5\_aMC@NLO}~\cite{Alwall:2014hca} interfaced
to {\tt aMCfast}~\cite{amcfast} and {\tt APPLgrid}~\cite{Carli:2010rw}.
The choice of binning, kinematical cuts and final-state decays in these processes
are the same as those used in the
{\tt SM-PDF} study~\cite{Carrazza:2016htc},
to which we refer for further information. In each case, we compare
results obtained with perturbative charm, and with fitted charm when
EMC data are included or not. All uncertainties shown are PDF
uncertainties only.
In addition, we also compare results for fitted charm (with and without EMC data)
obtained with different values of the charm mass, and the corresponding
results in case of fits with perturbative charm.

\begin{figure}[ht!]
\begin{center}
  \includegraphics[width=0.46\textwidth]{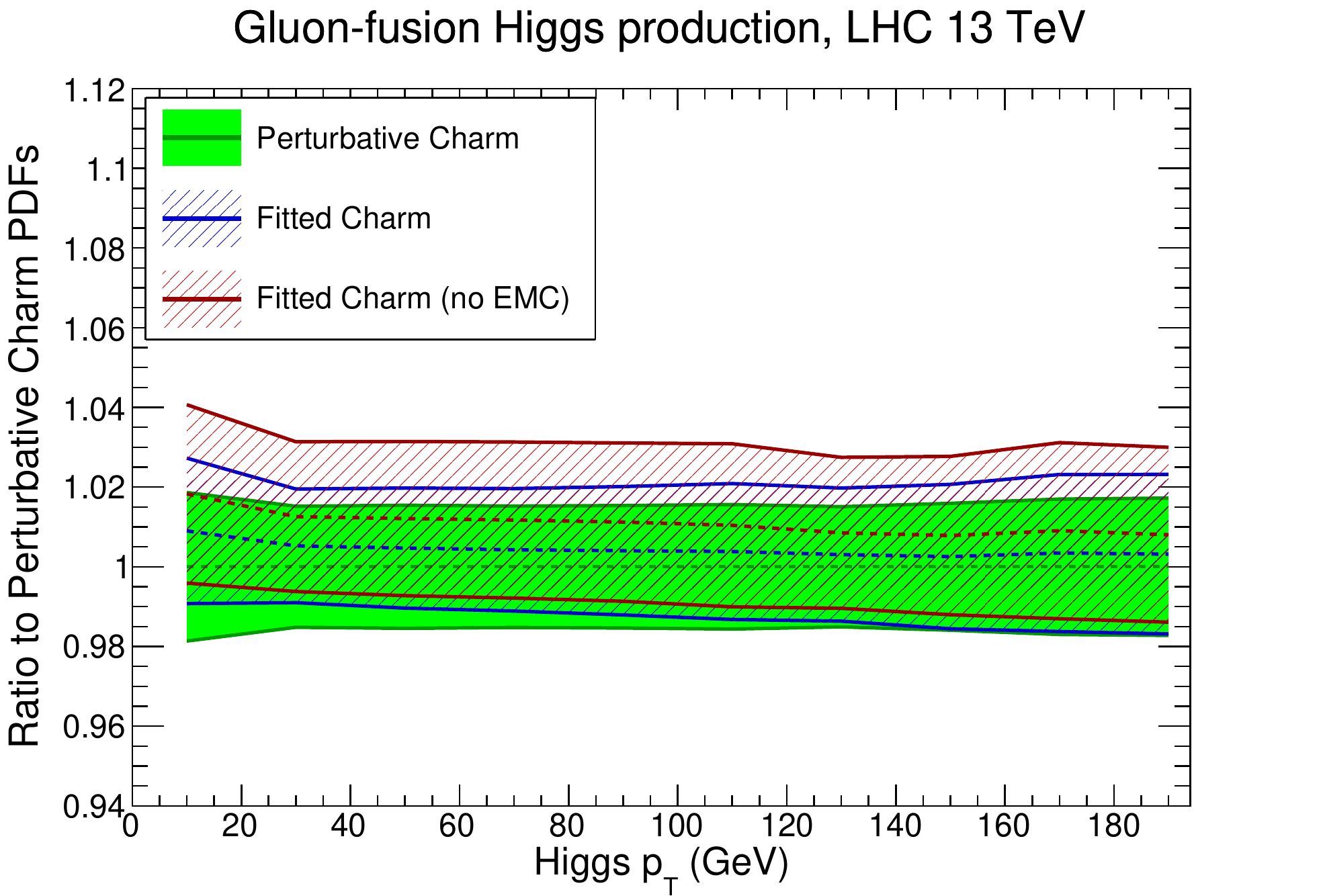}
  \includegraphics[width=0.46\textwidth]{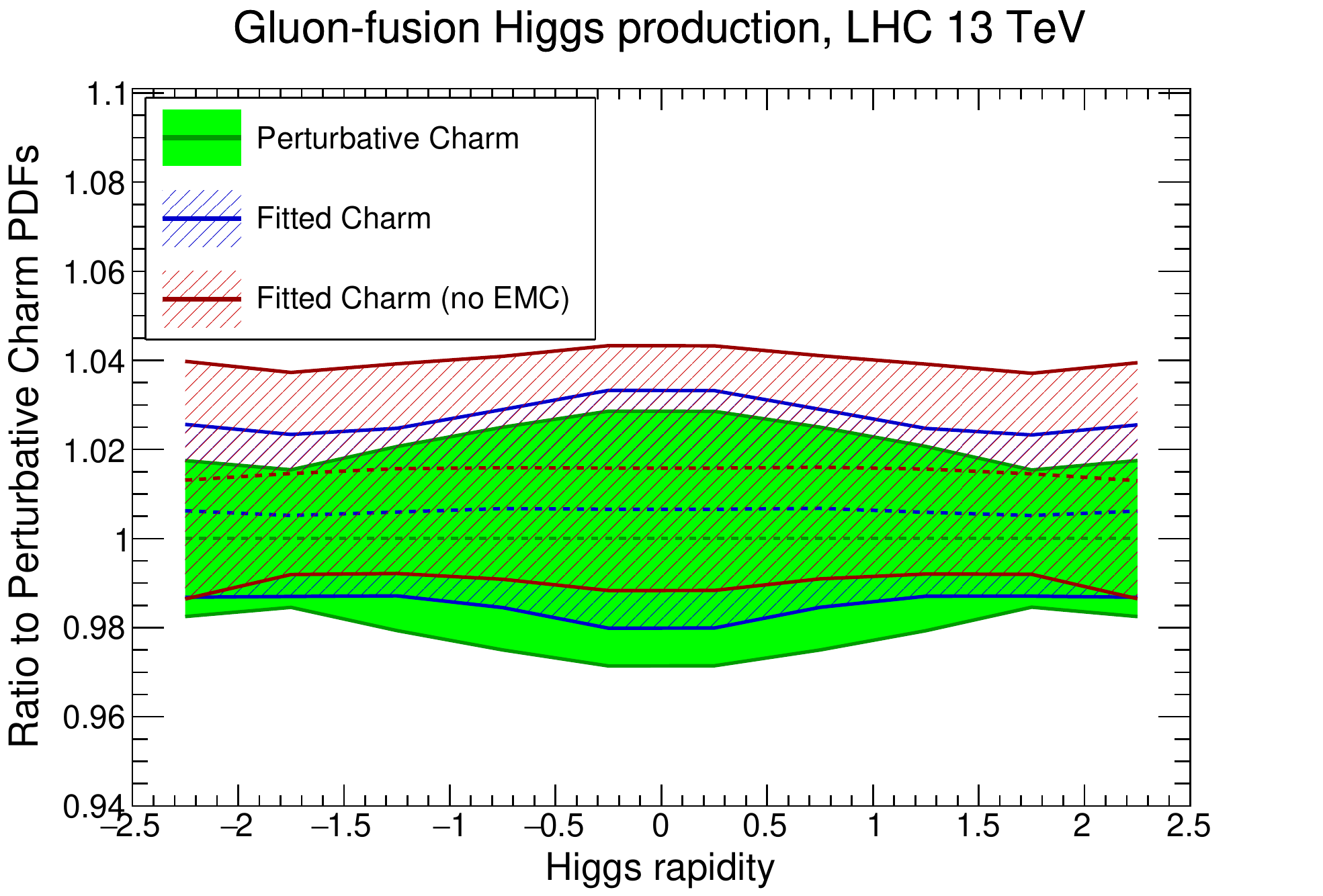}
  \includegraphics[width=0.46\textwidth]{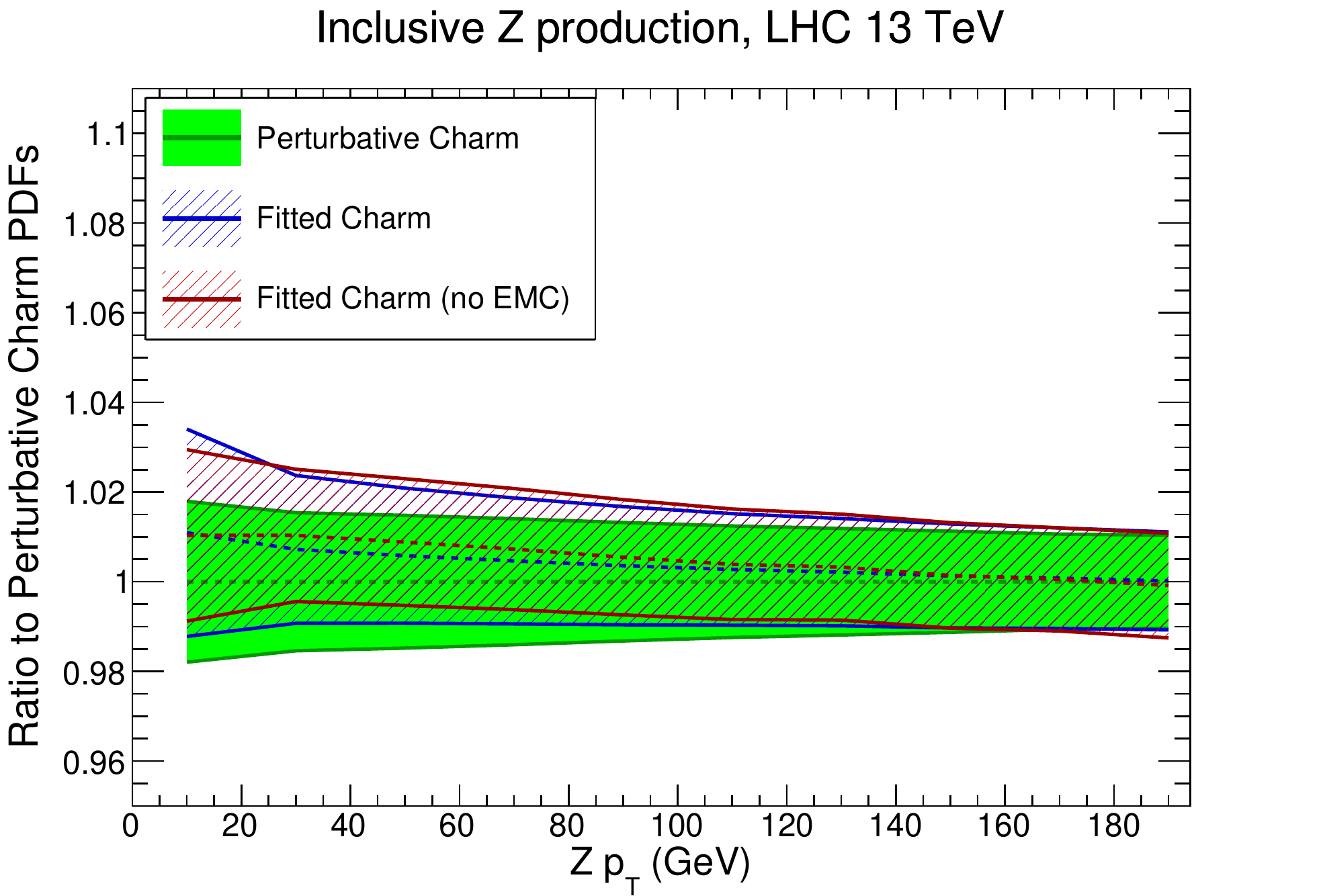}
  \includegraphics[width=0.46\textwidth]{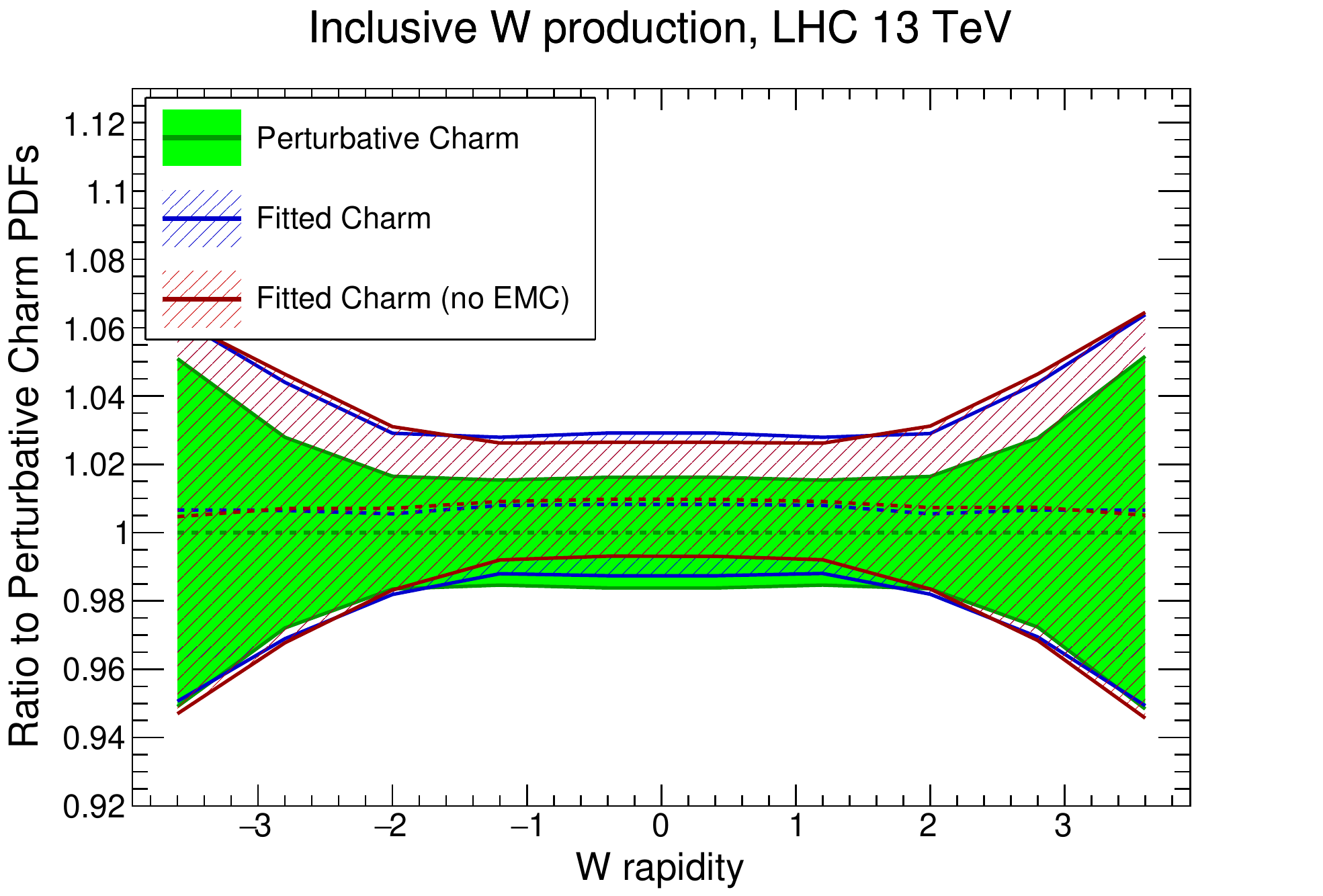}
  \includegraphics[width=0.46\textwidth]{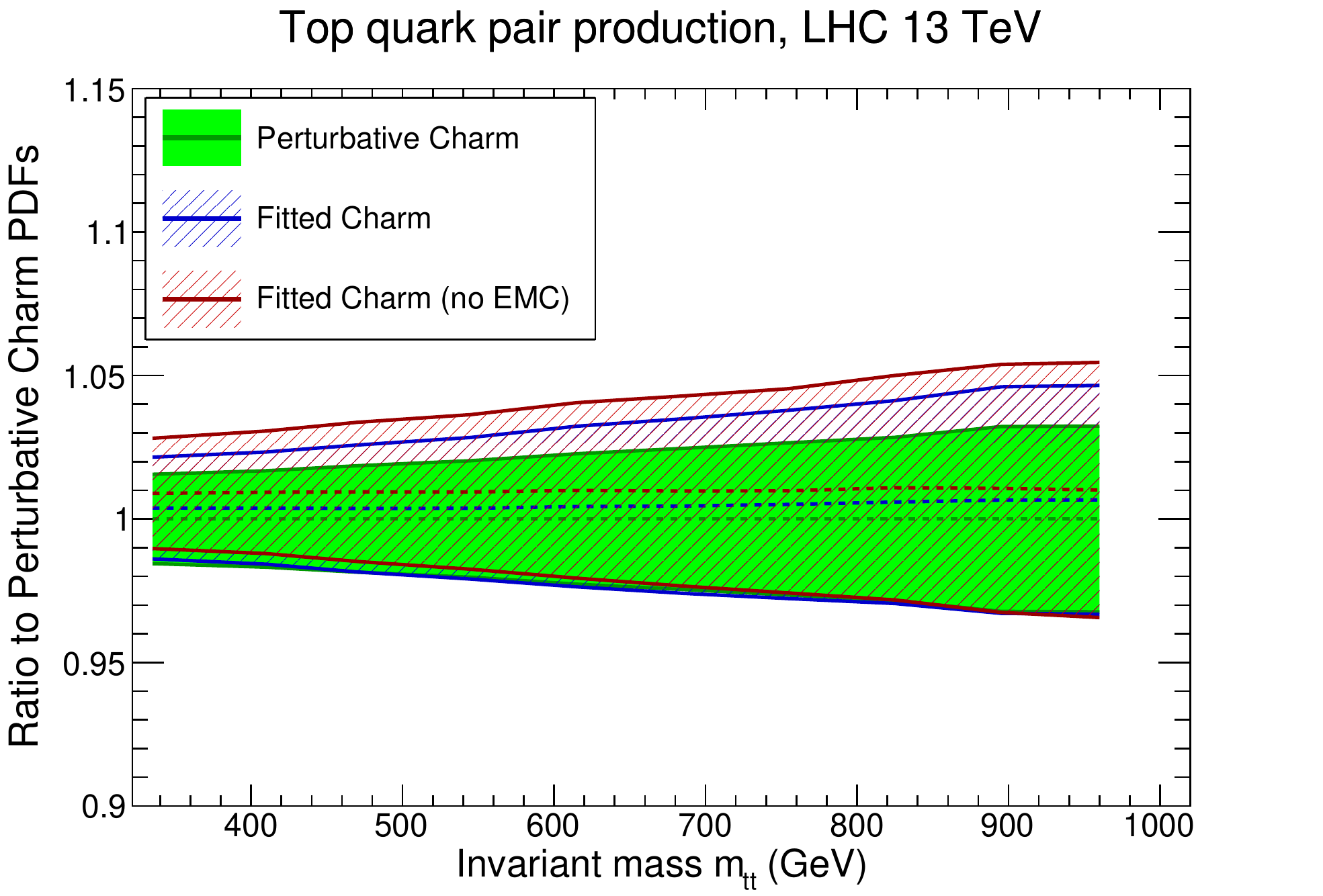}
  \includegraphics[width=0.46\textwidth]{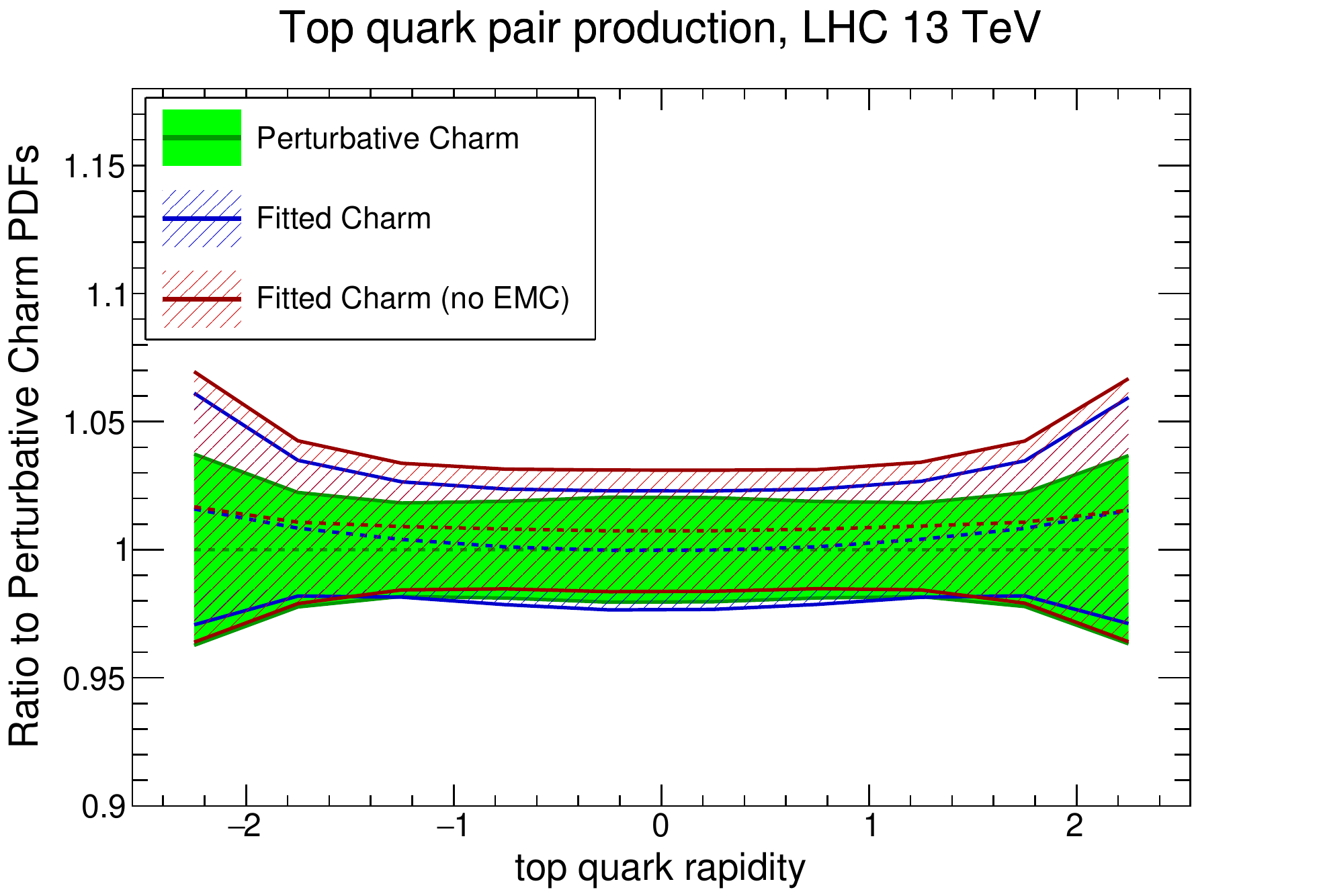}
\end{center}
\vspace{-0.5cm}
\caption{\small \label{fig:Higgs_13TeV} \label{fig:WZ_13TeV}
  Comparison of the results of the baseline fit with perturbative charm with the corresponding
  fitted charm PDFs, with and without the EMC data included for NLO differential
  distributions at 13 TeV.
  From top to bottom and from left to right
  we show the  Higgs transverse momentum and rapidity, the $p_T$
  of the $Z$ boson,
  the rapidity of the $W$ boson, and $m_{t\bar{t}}$ and $y_t$
  in $t\bar{t}$ production.
   }
\end{figure}
\begin{figure}[ht!]
\begin{center}
\includegraphics[width=0.46\textwidth]{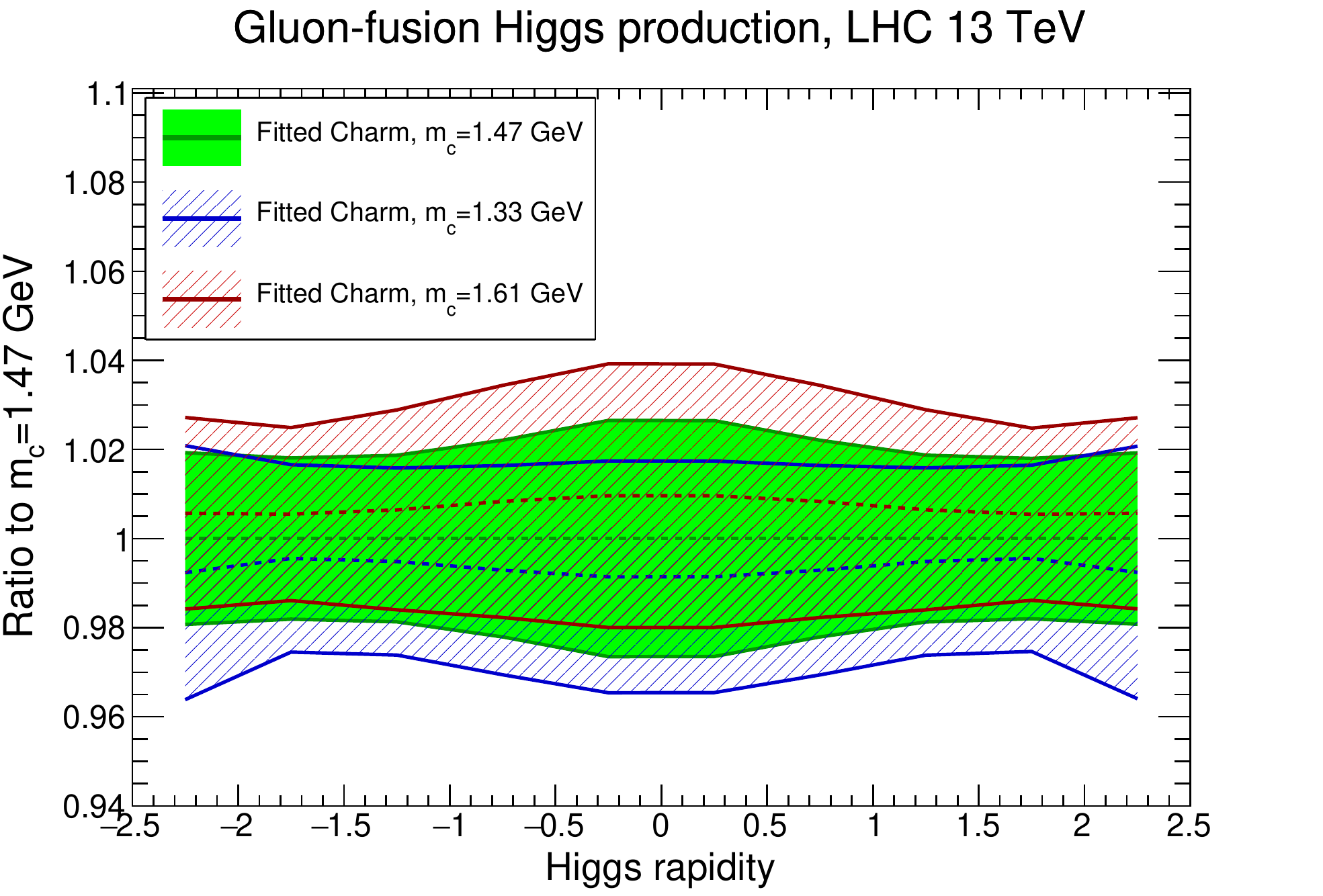}
\includegraphics[width=0.46\textwidth]{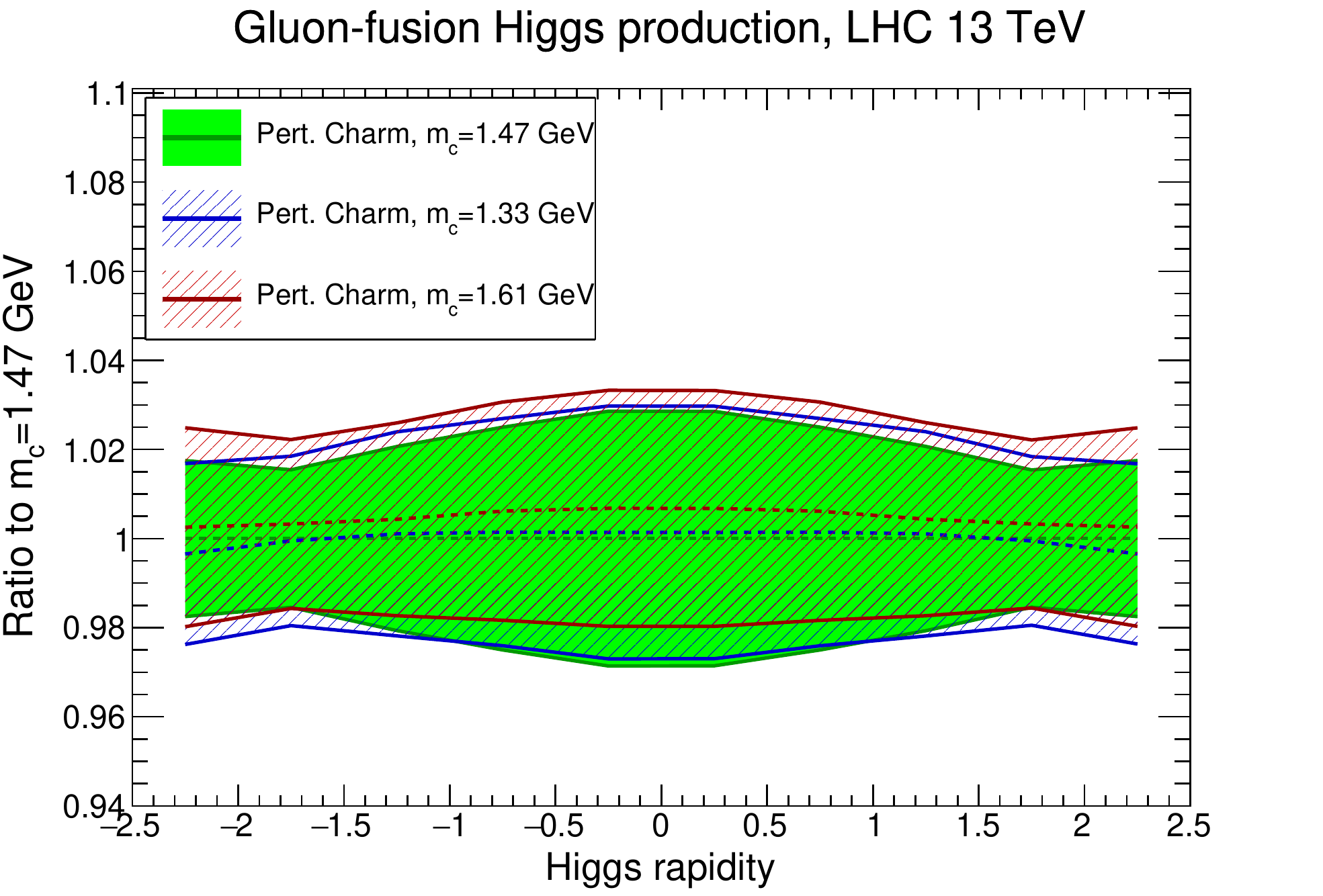}
\includegraphics[width=0.46\textwidth]{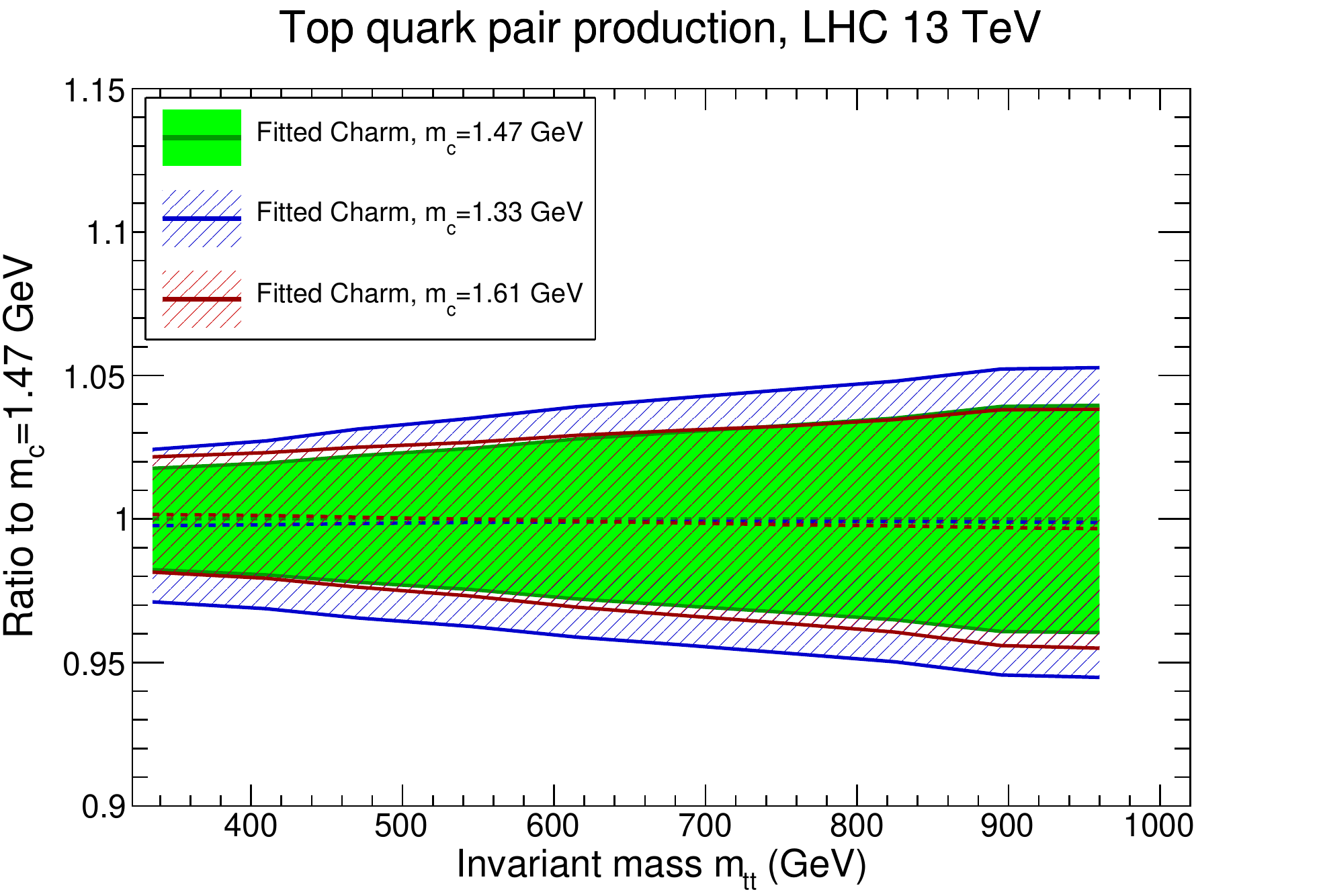}
\includegraphics[width=0.46\textwidth]{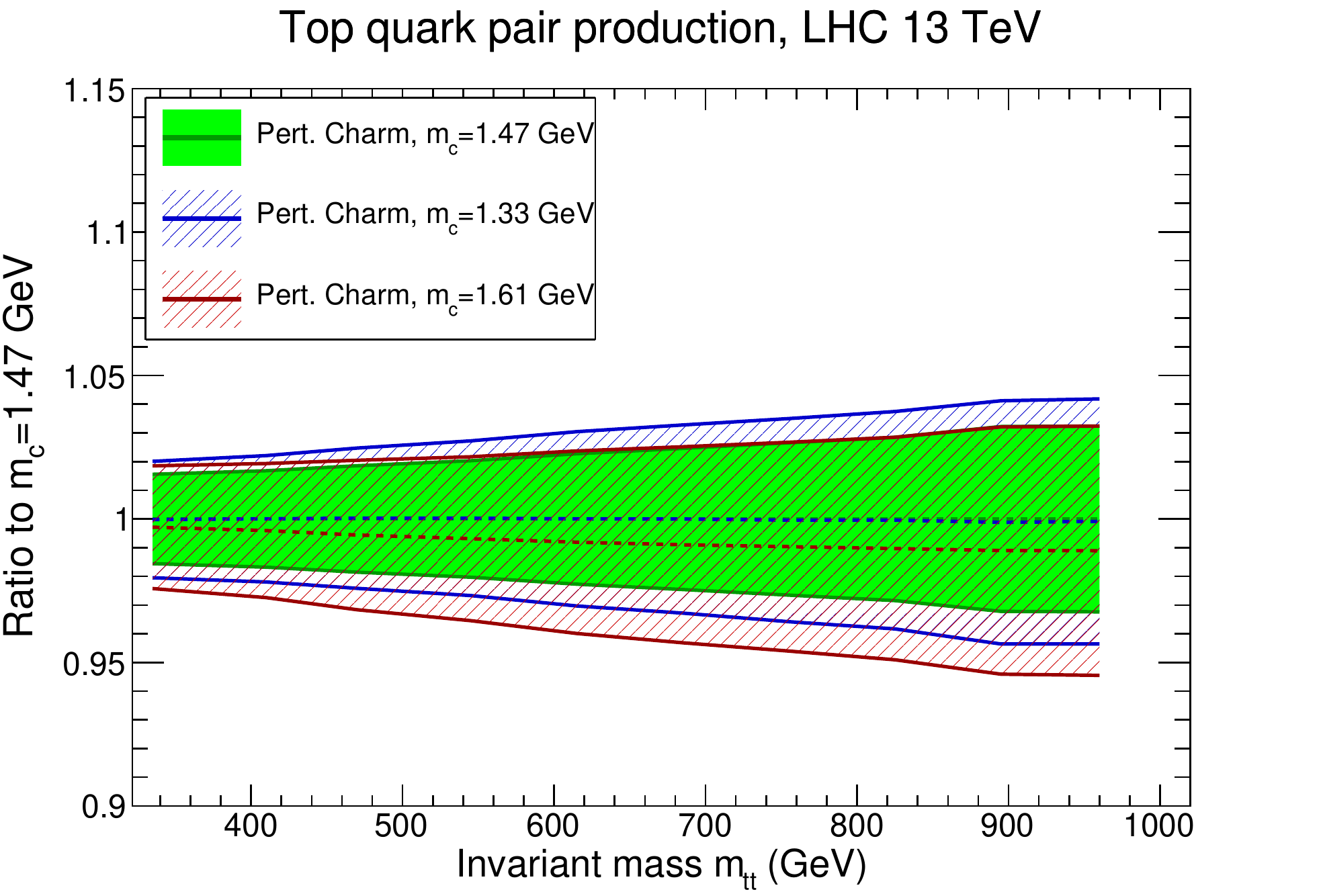}
\end{center}
\vspace{-0.6cm}
\caption{\small \label{fig:Higgs_13TeV_FCmc} 
  The Higgs rapidity distribution and the invariant mass distribution of
  top quark pairs in $t\bar{t}$ production, same as in 
Fig.~\ref{fig:Higgs_13TeV}, but
  now comparing different values of the charm mass with  fitted charm (left)
  and perturbative charm (right).
   }
\vspace{-0.1cm}
\begin{center}
\includegraphics[width=0.46\textwidth]{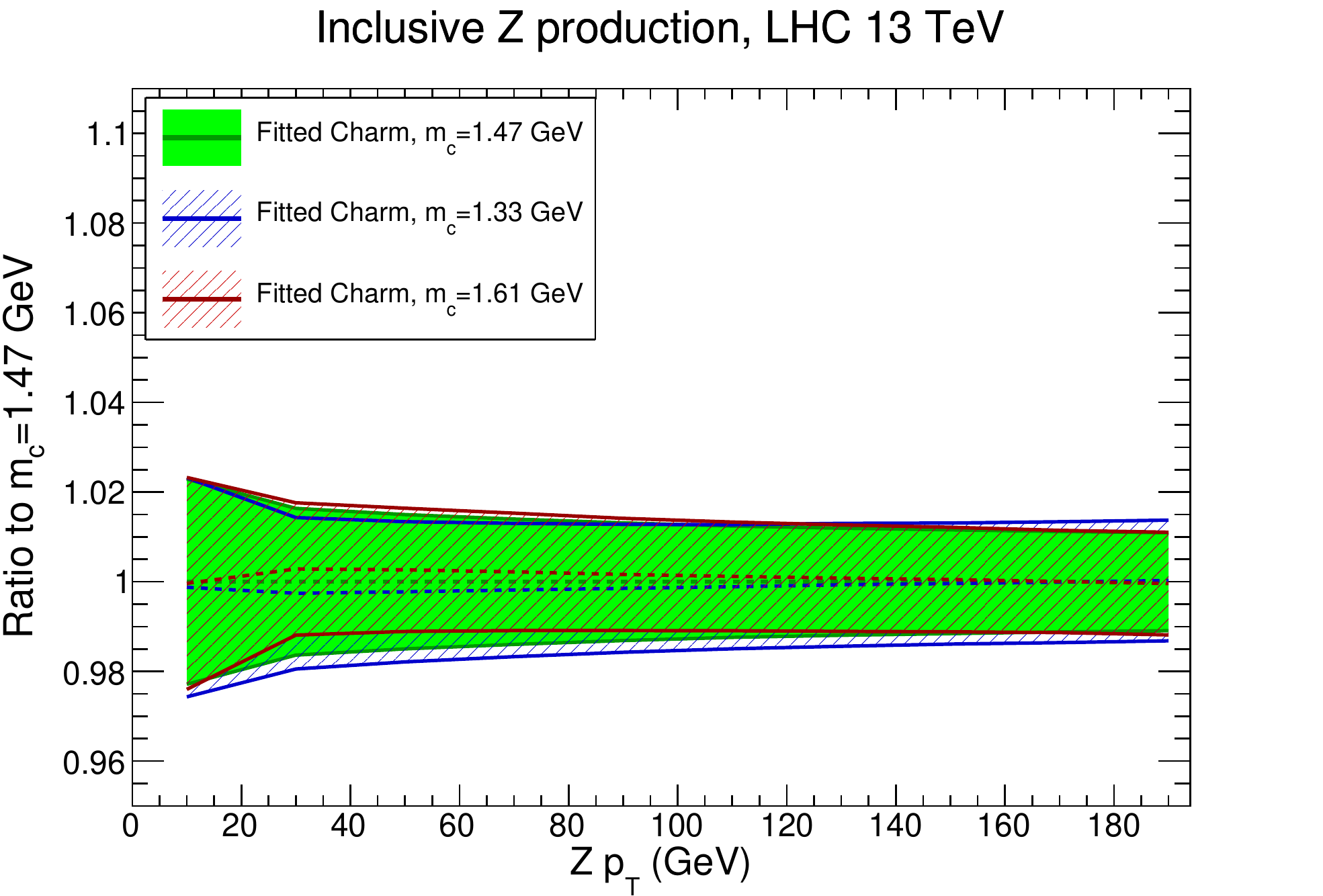}
\includegraphics[width=0.46\textwidth]{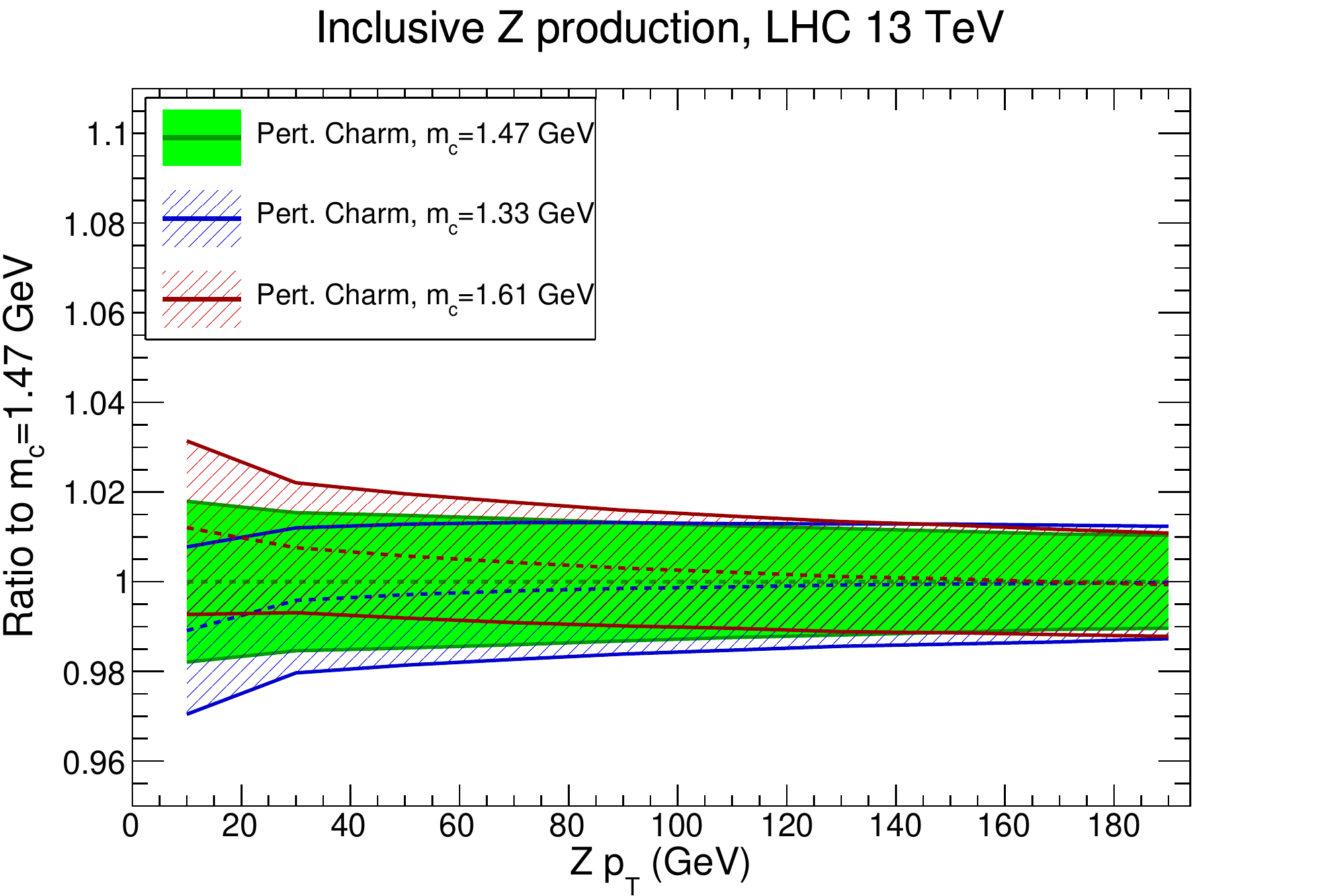}
\includegraphics[width=0.46\textwidth]{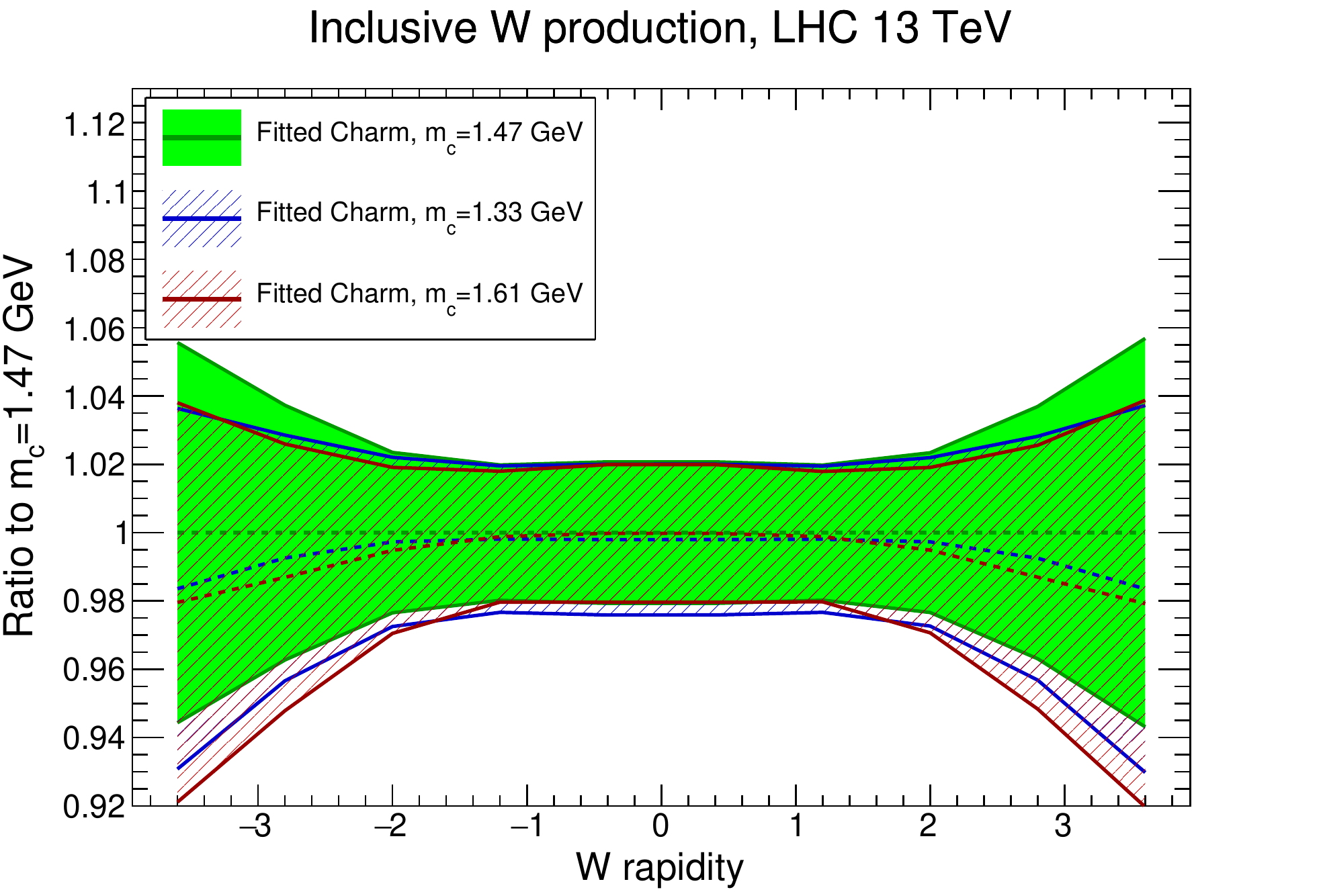}
\includegraphics[width=0.46\textwidth]{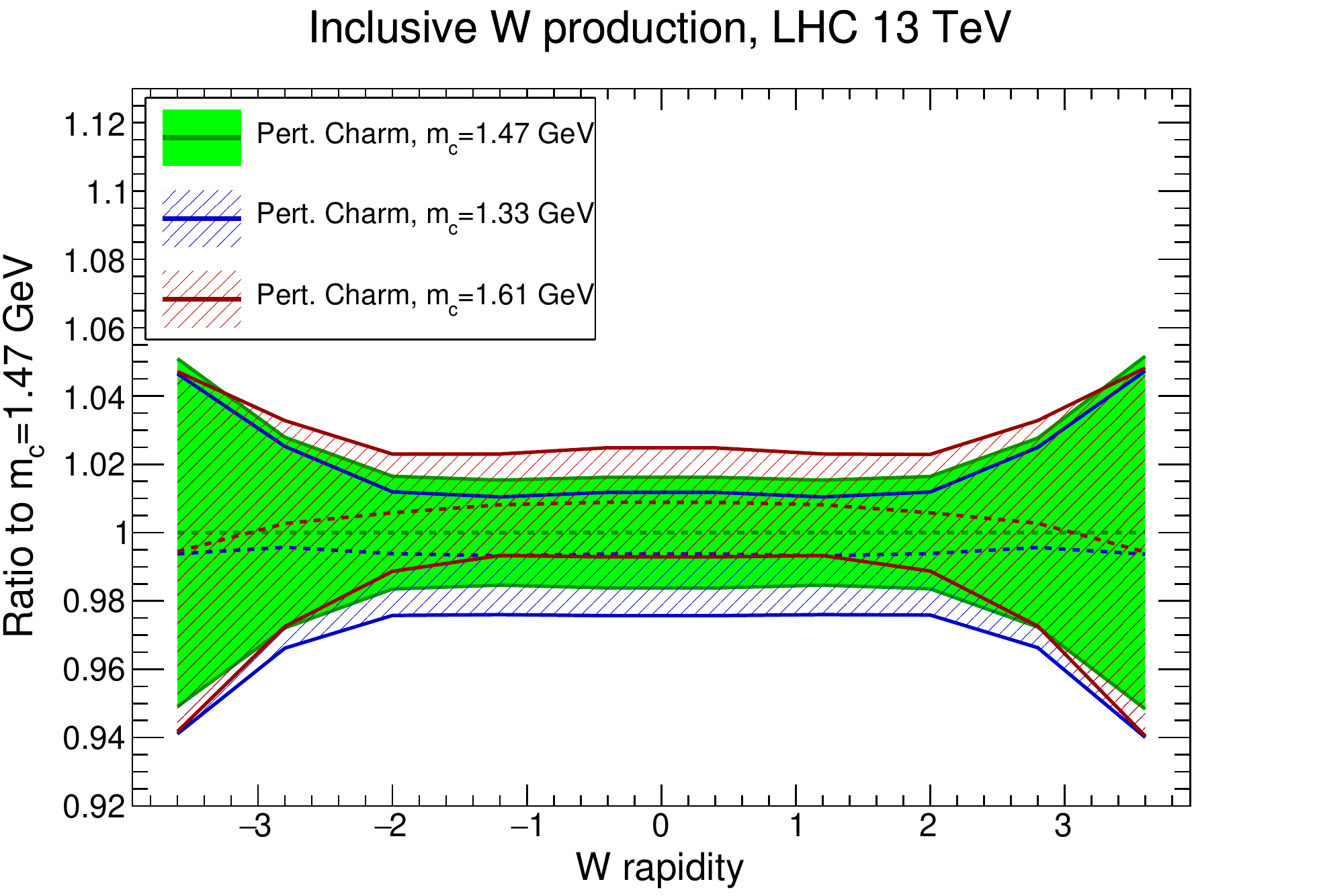}
\end{center}
\vspace{-0.6cm}
\caption{\small \label{fig:WZ_13TeV_FCmc}
  Same as Fig.~\ref{fig:Higgs_13TeV_FCmc} for the $p_T$ of the $Z$ boson
  and the rapidity of the $W$ boson.
   }
\end{figure}

%
In Fig.~\ref{fig:Higgs_13TeV} we show the
 the  Higgs transverse momentum and rapidity, the transverse momentum
  of the $Z$ boson,
  the rapidity of the $W$ boson, and the invariant mass and top quark
  rapidity
  in $t\bar{t}$ production.
In all cases, we observe considerable stability of central values when
moving from perturbative to fitted charm, with only a small increase
in uncertainty for fitted  charm, and no significant difference found
when EMC data are excluded.

Then in  Figs.~\ref{fig:Higgs_13TeV_FCmc} and~\ref{fig:WZ_13TeV_FCmc}
we show the comparison of the differential distributions of Fig.~\ref{fig:Higgs_13TeV}
upon variations of the charm quark mass, both for fitted and perturbative
charm PDFs.
For the gluon-initiated processes ($ggH$ and $t\bar{t}$) the results with fitted
and perturbative charm are quite similar: the main effect of fitted charm is to give a more conservative estimate of the overall uncertainty. 
For quark-induced processes ($W$ and $Z$) we see a marked improvement in the 
stability upon charm mass variations for fitted charm, particularly at low $p_T$ and at central rapidities: this is a direct reflection of the reduced sensitivity to charm mass variations in the medium $x$ region when charm is fitted.

We conclude that for LHC observables which do not depend directly on the
charm PDF, both at the inclusive and differential level, the impact of
fitting charm is moderate: for gluon dominated processes it provides a
more conservative error estimate, while for quark-induced processes it
offers a reduction in the (already quite weak) dependence on the 
value of $m_c$. 

\clearpage

\subsection{Probing charm at the LHC}

We now turn to LHC observables which do depend directly on the charm
PDF, and which could thus be used for its determination. 
Such observables include prompt photon production
in association with $D$
mesons~\cite{Stavreva:2009vi,Rostami:2015iva,Bednyakov:2013zta}, 
$Z$ boson production together with charm quarks~\cite{Boettcher:2015sqn,Bailas:2015jlc,Beauchemin:2014rya} and open $D$ meson production~\cite{Kniehl:2004fy,Bugaev:2009jw,Vogt:1994zf,Kniehl:2009ar,Kniehl:2012ti}, as well as more exotic processes such as double charmonium
production~\cite{Vogt:1995tf} and inclusive and diffractive
Higgs production~\cite{Brodsky:2006wb,Brodsky:2007yz}.
Here we  concentrate on two illustrative
cases, namely $Z$+charm and $c\bar{c}$ production.
We will specifically discuss the kinematic regions which are
sensitive to the charm PDF at large $x$, and which could therefore be
used to confirm our first evidence, discussed in
Sect.~\ref{sec:intrinsic}, for an `intrinsic' component of charm: as we
will see, these are  the regions of 
large $p_T$ or large rapidity.

\subsubsection{$Z$ production in association with charm quarks}

\begin{figure}[t]
\begin{center}
    \includegraphics[height=0.3\textwidth]{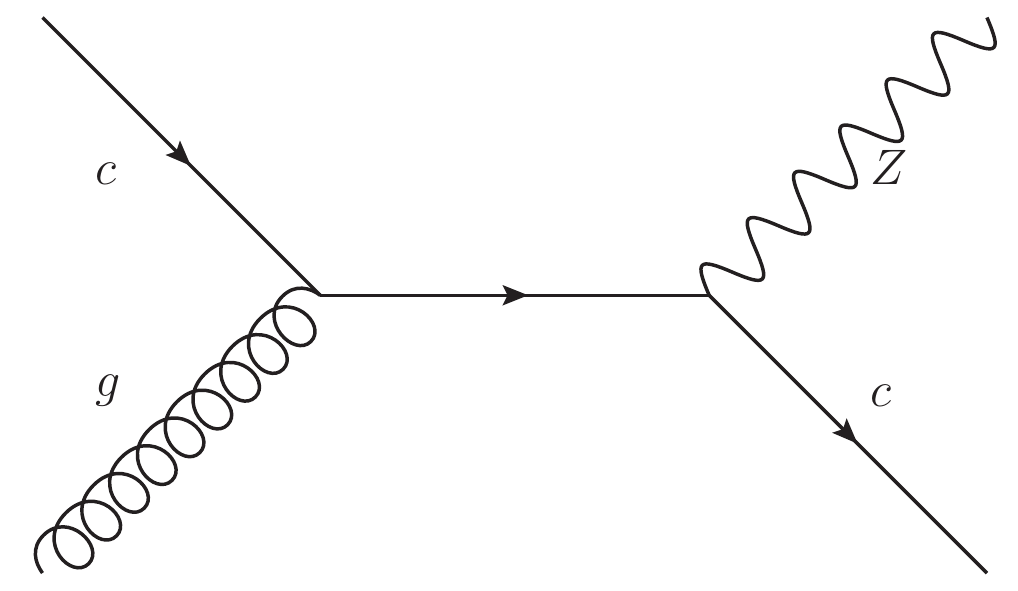}
    \hspace{1cm}
  \includegraphics[height=0.3\textwidth]{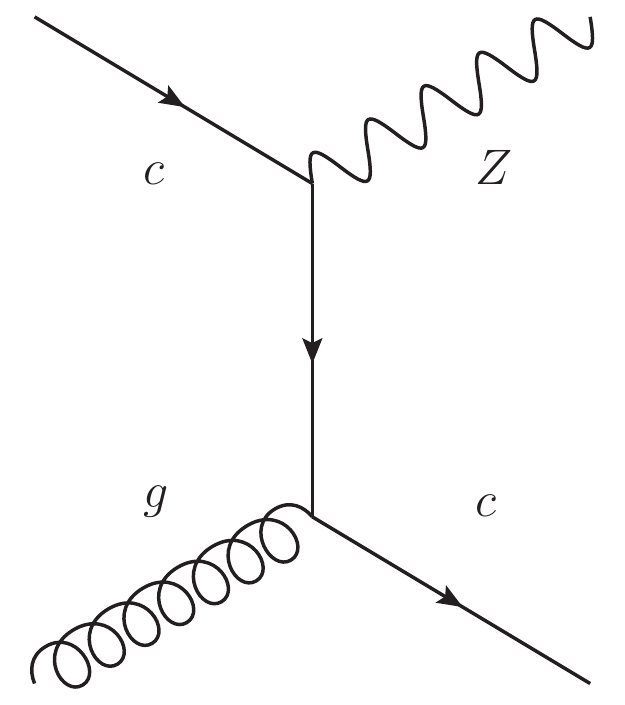}
\end{center}
\vspace{-0.3cm}
\caption{\small \label{fig:zcharm}
 Representative leading order Feynman diagrams for the production of a $Z$ boson
 in association with a charm quark at hadron colliders.
}
\end{figure}

\begin{figure}[t!]
\begin{center}
  \includegraphics[width=0.49\textwidth]{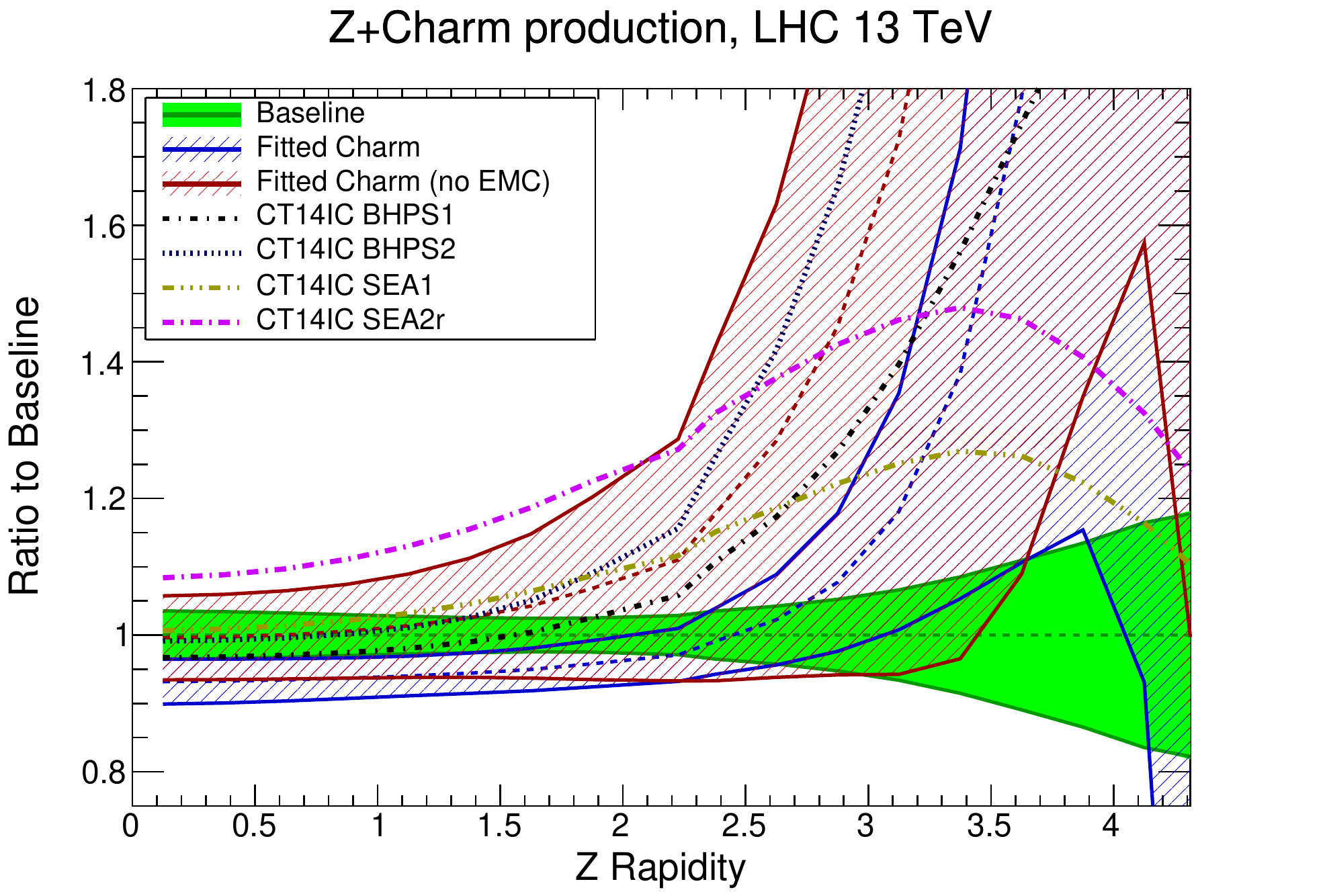}
  \includegraphics[width=0.49\textwidth]{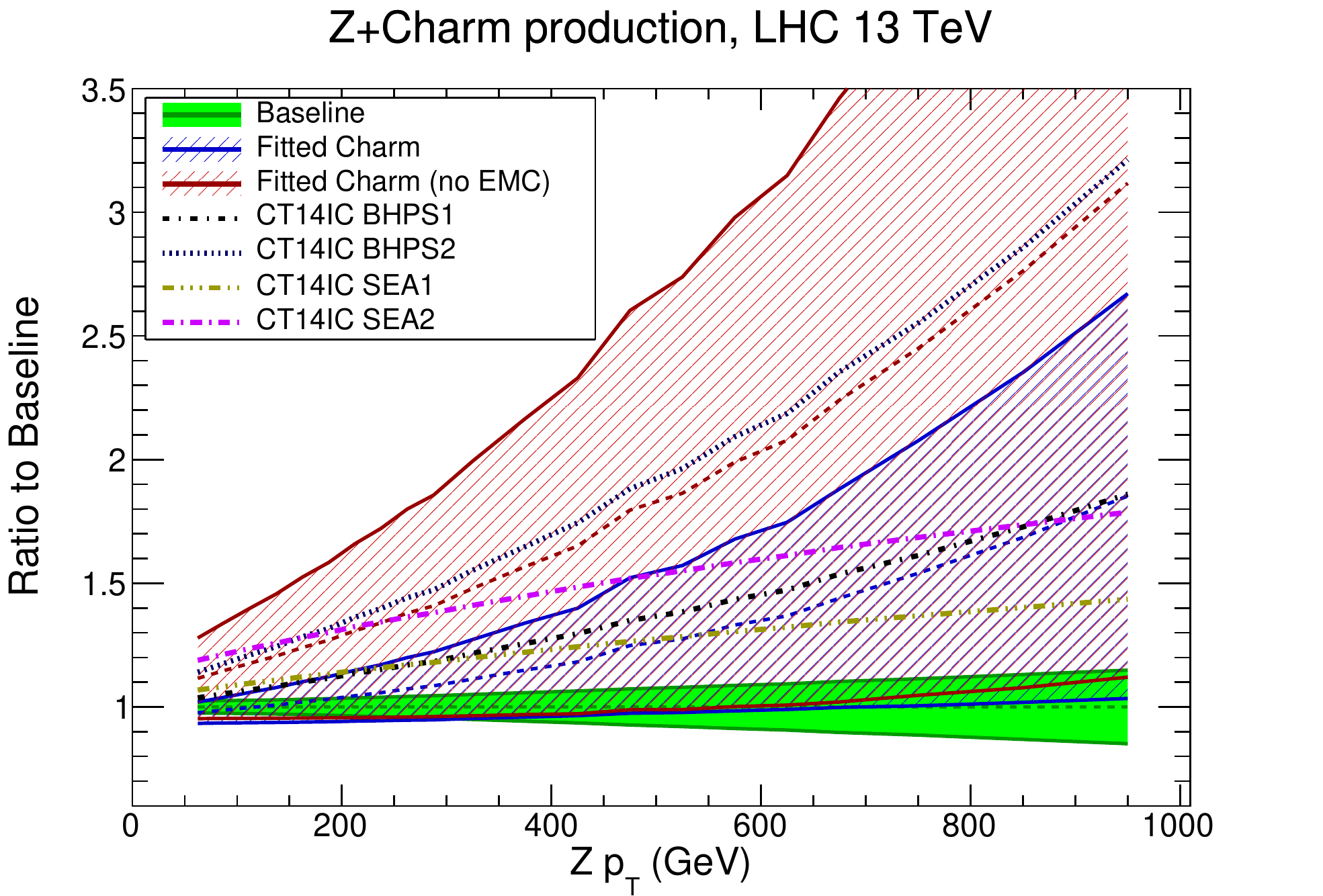}
\end{center}
\vspace{-0.3cm}
\caption{\small \label{fig:ZD_yZ_13TeV_nnpdf3IC}
  The $Z$ boson rapidity (left) and transverse momentum
  (right) distributions for  $Z$ production
  in association with charm at the LHC 13 TeV, computed using the
  NNPDF sets with perturbative or fitted charm, and the CT14 IC PDFs
  shown in Fig.~\ref{fig:ct14}. Results are shown as a ratio to the
  NNPDF perturbative charm set.
}
\end{figure}

In Fig.~\ref{fig:zcharm} we show representative
leading order Feynman diagrams for the production of a $Z$ boson
in association with a charm quark at hadron colliders, driven
by the $cg$ luminosity.
The calculation of this process at NLO has been performed with {\tt MCFM}
interfaced to {\tt APPLgrid}, and
cross-checked with  {\tt MadGraph5\_aMC@NLO}
interfaced to 
{\tt aMCfast}.
It is beyond the scope of this paper to perform a complete feasibility study
of this measurement, so we neglect
the hadronization of the charm quark into a $D$ meson,
which does not significantly affect 
the sensitivity of this process to the charm PDF.
In Fig.~\ref{fig:ZD_yZ_13TeV_nnpdf3IC}
we show the rapidity distribution 
and the transverse momentum of the $Z$ boson in $Z+c$ production
   at the LHC 13 TeV.
   We compare the results of perturbative or fitted charm PDFs, in the latter
   cases with and without the EMC data included.
  We also show predictions obtained using the four CT14NNLO  sets
  discussed in Sect.~\ref{sec:intrinsic}.

  %
  In the case of the $Z$ rapidity distribution percentage
 differences in central values are moderate at central rapidity
  but increase substantially in the forward region.
  In particular, in the LHCb acceptance region, $2.0 \le y_Z \le 4.5$,
  an enhancement of the cross-section by a factor two or more is possible in the case of
  fitted charm, compared to the baseline result with perturbative
  charm (for a
    recent study of charm PDF
  constraints in $Z+c$ production at LHCb see e.g.\
  Ref.~\cite{Boettcher:2015sqn}). 
  However in this region PDF uncertainties in the fitted charm case are large,
  and the three NNPDF sets shown
agree with each other at the one sigma level
  in the entire range of $y_Z$. This means that more accurate data for
  this observable could provide a useful constraint on the charm
  PDF. 

In the case of the transverse momentum  distribution of $Z$ bosons,
 the NNPDF  sets with  fitted charm and the CT14 sets based on the
 BPHS model
     exhibit a substantial enhancement of the cross-section at
large  $p_T^Z$ in comparison to the perturbative charm baseline. 
For the fitted charm NNPDF3 PDFs with EMC data, this
  enhancement could be as large as a factor two (at the one-sigma
  level) for $p_T^Z\simeq 700$ GeV.
Once again, however, results obtained with perturbative and 
fitted charm PDFs are consistent with each other within the large
uncertainties, so also in this case more accurate measurements could
provide a useful constraint.
  
Turning things around, an accurate measurement at high rapidity and
transverse momentum could rule out perturbative charm. Also, in the
central rapidity region, an accurate enough measurement could confirm
the undershoot in the fitted charm case which is seen is
Fig.~\ref{fig:ZD_yZ_13TeV_nnpdf3IC}, and though smaller in absolute
terms, it is as significant as the large rapidity excess on the scale
of present-day uncertainties. A full NNLO analysis will be required in
order to arrive at a definite conclusion, especially in view of the
fact that, as discussed in Sect.~\ref{sec:intrinsic}, the fitted
charm might be reabsorbing higher-order corrections.

\subsubsection{Charm quark pair production}

\begin{figure}[t!]
  \begin{center}
    \includegraphics[height=0.25\textwidth]{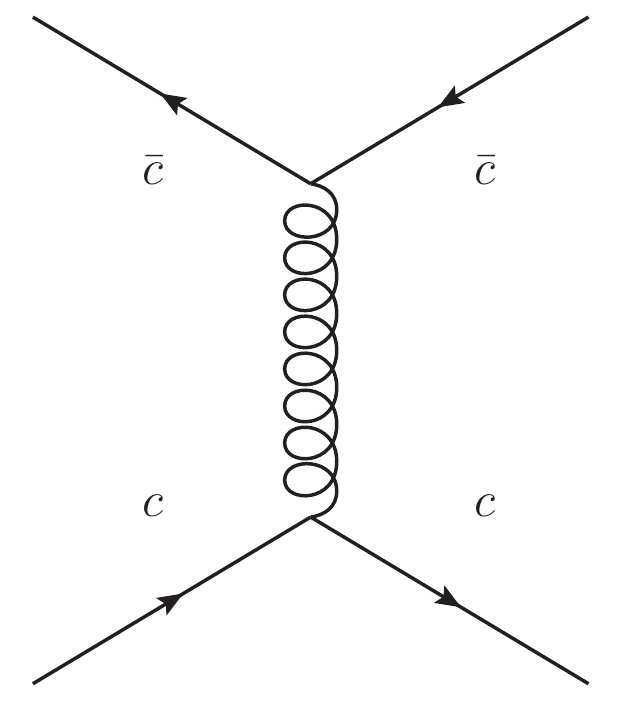}
    \hspace{1cm}
  \includegraphics[height=0.25\textwidth]{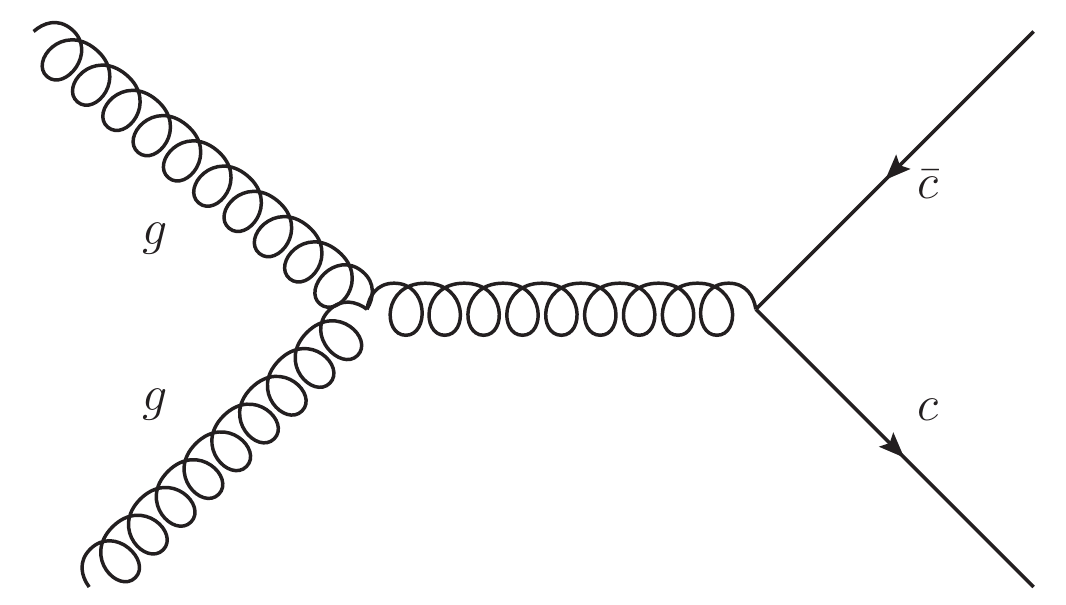}
\end{center}
\vspace{-0.3cm}
\caption{\small \label{fig:ccbarFeyn}
  Representative leading order Feynman diagrams for the production of a charm-anticharm
  pair at hadron colliders, initiated either by charm quarks (left)
  and by gluons (right).}
\end{figure}

At hadron colliders, heavy quark pair
production is driven by the $gg$ and $q\bar{q}$ luminosities.
The relative importance of the two channels  depends on the kinematics.
For instance, for the total inclusive cross-section in
top quark pair production~\cite{Czakon:2013tha},
the $gg$ process is dominant at the LHC 13 TeV (90\%),
while it is only 14\% at the Tevatron (where instead 86\% of the cross-section comes
from quark-initiated contributions).
In the case of charm quark pair production, at low transverse momentum $p_T^c$, the
cross-section is entirely dominated by gluon-initiated processes~\cite{Gauld:2015yia}.
However, in the case of fitted charm the $c\bar{c}$ channel can eventually
become dominant for high enough transverse momentum of the charm
quark $p_T^c$, or for high enough rapidity $y_c$: in these cases, 
large values of $x$
are probed, where the fall-off of the charm PDF is less steep than
that of the gluon, especially if charm has an intrinsic component.
 Representative leading-order Feynman diagrams for the production of a charm-anticharm
  pair at hadron colliders are shown in  Fig.~\ref{fig:ccbarFeyn}.

In the following, we use the {\tt FONLL} code~\cite{Cacciari:1998it} for the calculation
of the double-differential cross-section $d^2\sigma_{c\bar{c}}/dp_Tdy$ for the production
of a charm-anticharm pair at hadron colliders.
The {\tt FONLL} calculation combines a fixed-order  massive result,
accurate at small $p_T$, with a resummed next-to-leading log
prediction in which the charm mass is neglected.
As in the case of deep-inelastic scattering, the massive fixed-order
calculation should be modified in the presence of a fitted charm
component~\cite{Ball:2015dpa,Ball:2015tna}. This modification is not
included in the code Ref.~\cite{Cacciari:1998it}; 
here, however, we will only consider the large $p_T\gg m_c$ region,
where the FONLL computation coincides with the massless one and this
extra contribution is negligible.
Since our aim is only to illustrate how differences in the charm PDF
affect the charm pair production 
cross-section, we  do not include  final state effects
such as  hadronization of charm quarks into $D$ mesons and their
subsequent decay. 
%

\begin{figure}[t!]
\begin{center}
  \includegraphics[width=0.46\textwidth]{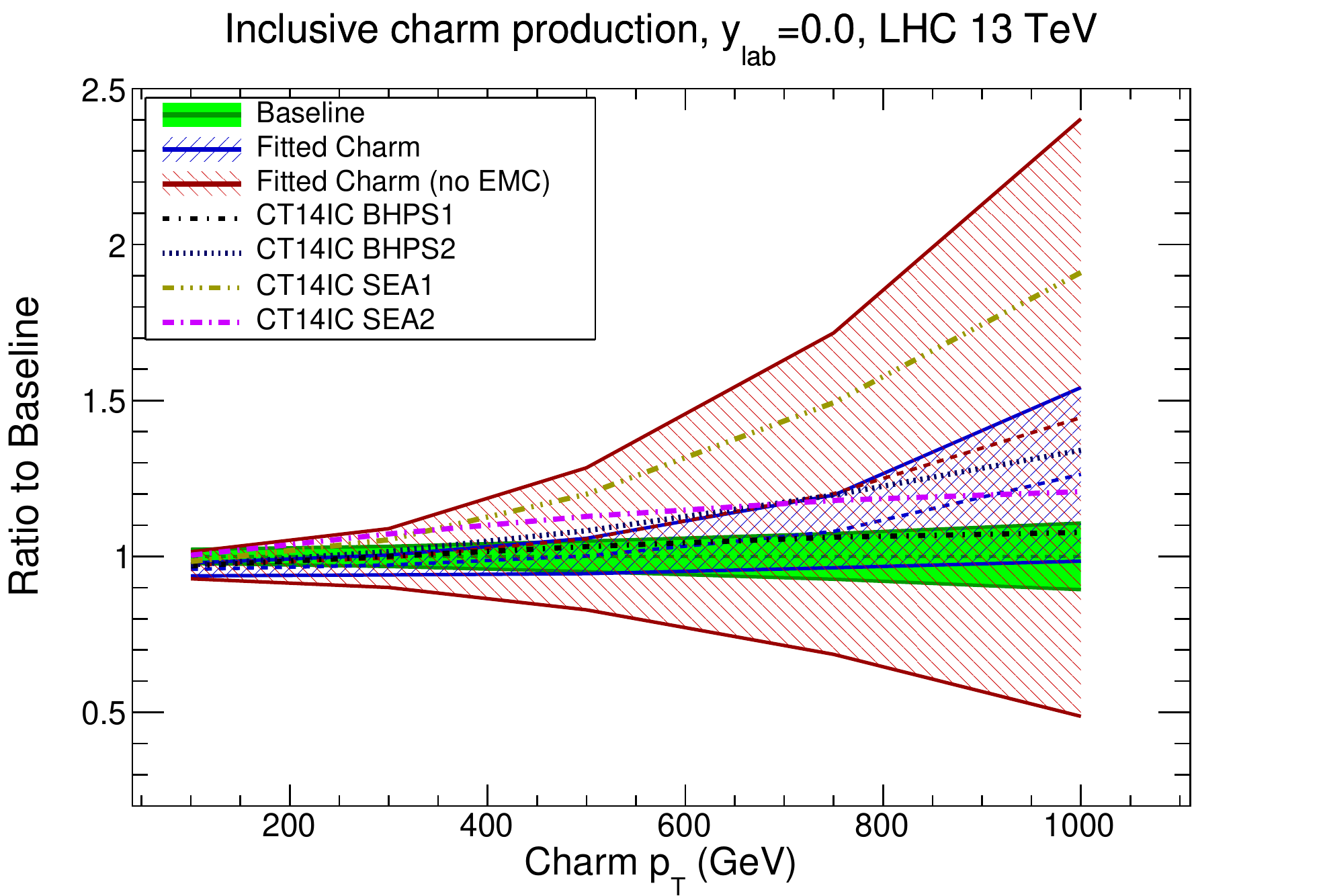}
  \includegraphics[width=0.46\textwidth]{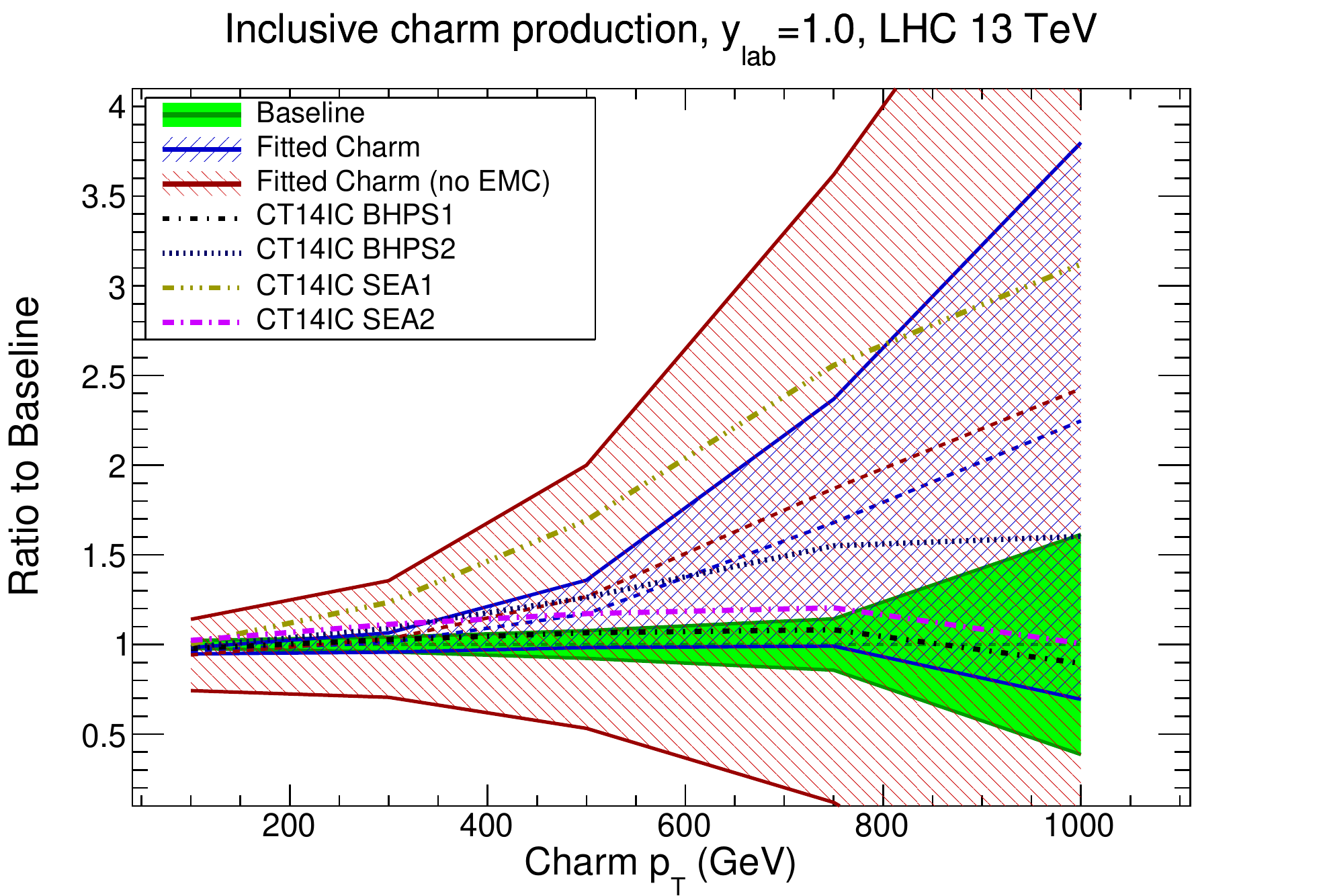}
  \includegraphics[width=0.46\textwidth]{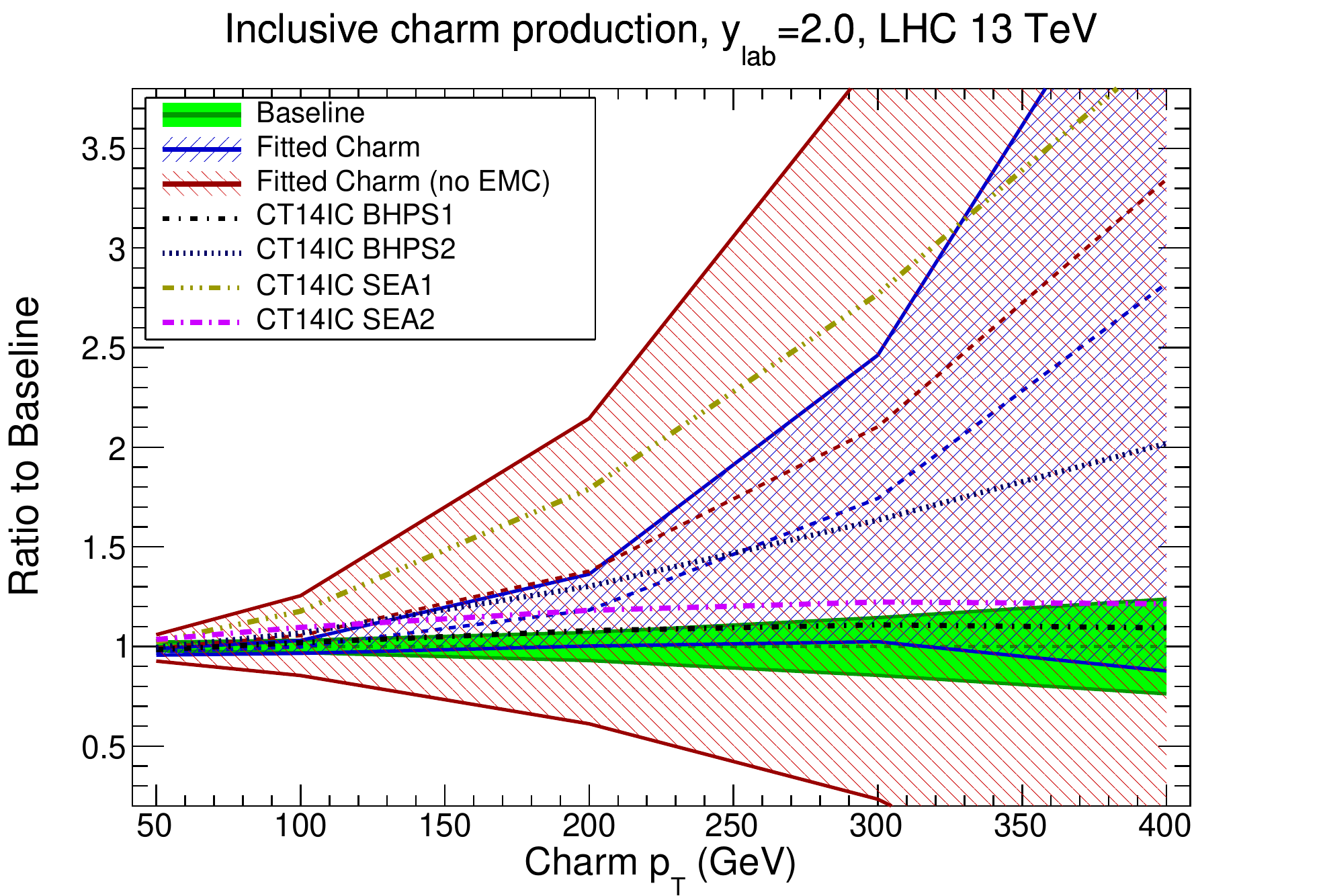}
  \includegraphics[width=0.46\textwidth]{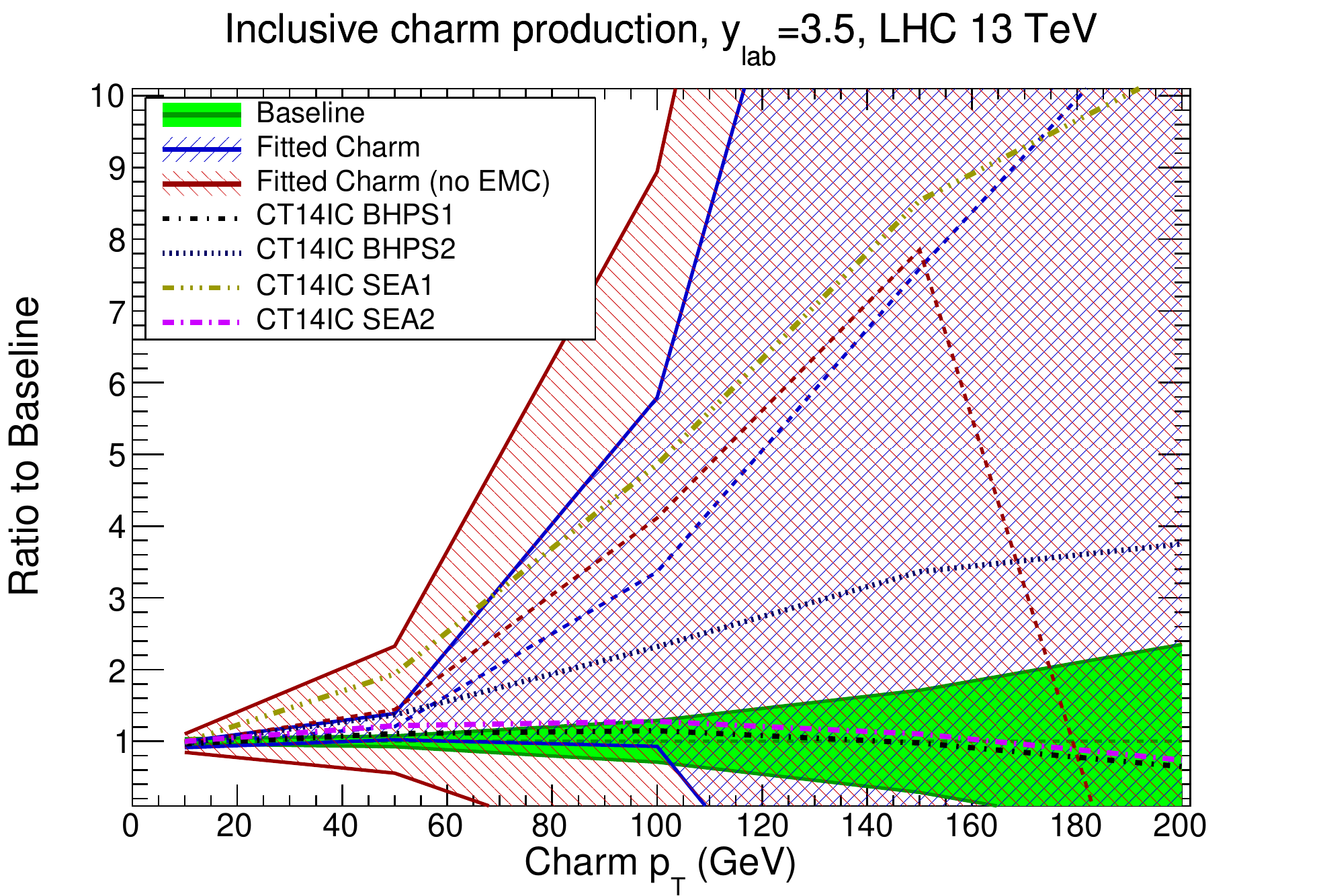}
\end{center}
\vspace{-0.3cm}
\caption{\small \label{fig:ccbar_13TeV}
  The double-differential cross-section $d^2\sigma_{c\bar{c}}/dp_Tdy$ for
  charm-anticharm pair production at the LHC 13 TeV, as a function of the
  charm quark $p_T$ for different values of its rapidity $y$.
  From top to bottom and from left to right, we show the results for $y_c=0,1.0,2.0$ and
  3.5.
 We compare results obtained using the
  NNPDF sets with perturbative or fitted charm, and the CT14 IC PDFs
  shown in Fig.~\ref{fig:ct14}. Results are shown as a ratio to the
  NNPDF perturbative charm set.
   }
\end{figure}

In Fig.~\ref{fig:ccbar_13TeV} we show the
double-differential cross-section $d^2\sigma_{c\bar{c}}/dp_Tdy$ for
charm-anticharm pair production at the LHC 13 TeV, as a function of the
  charm quark transverse momentum for different values of its rapidity $y_c$.
The impact of different charm PDFs
becomes more important at large $p_T^c$ and for large $y_c$.
For instance, for $y_c=3.5$, intrinsic charm can enhance the cross-section
for charm production 
by up to one order of magnitude for $p_T^c=200$ GeV.

While $D$ meson production in the forward region has been measured by LHCb,
available data only cover the kinematic region up to $p_T^D=8$ GeV at 7 TeV~\cite{Aaij:2013mga}
and up 15 GeV at 13 TeV~\cite{Aaij:2015bpa}, where the differences between the
fitted and perturbative charm predictions
are small.
Future LHCb $D$ meson production data with higher integrated luminosity and a higher reach in $p_T^D$ could be used to constrain the charm content of
the proton.
Similarly, $D$ meson production in the central region $|\eta_D|\lsim
2$, but higher  $p_T^D$ values than currently available could provide
valuable constraints. Note that likewise current  ATLAS $D$ meson
measurements at 7 TeV~\cite{Aad:2015zix} extend only 
up to $p_T^D=100$ GeV, so data at 13 TeV with increased luminosity would
also be required here.

\section{Delivery and outlook}

\label{sec:delivery}

We have presented a first model-independent determination
of the charm content of the proton in the NNPDF framework.
Our results suggest that, if the EMC data are taken at face value,  
the charm PDF is compatible with perturbative behaviour for
$x\lsim0.1$, in that it vanishes for all $x$ in this region  around
$Q_0\approx 1.6$~GeV, while  it has an
`intrinsic' large $x$ component which peaks for $x\sim0.5$, and carries
$0.7\pm 0.3$\% of the nucleon momentum at the 68\% CL at a low scale
$Q=1.65$~GeV.
The perturbative component of our fitted charm is quite stable upon
variation of the charm mass, and thus  lies significantly
below perturbatively generated charm if the central PDF value
$m_c=1.47$~GeV is adopted. This could possibly be due to
missing higher order corrections, which are expected to be of
comparable size. This suggests that PDF sets (including NNPDF3.0), in
which charm is perturbatively generated but no theoretical
uncertainties are provided, may be significantly underestimating the
uncertainty on the charm PDF at small $x$, and missing its intrinsic
component at large $x$. These results hold even if the uncertainty on
the EMC charm data is considerably inflated, and in fact at the level of
central values they still hold even with the EMC data excluded
altogether, though in that case they lose statistical significance.

Perhaps more interestingly, our results show that the widely held
opinion (see e.g.~Ref.~\cite{Jimenez-Delgado:2014zga} and
Refs. therein) that the EMC data cannot be included in  a global fit
because they are in tension with other datasets,
i.e. they cannot be adequately fit at leading-twist taking both data
and theory at face value, is untenable. Indeed, we show that if we take
the published EMC $F_2^c$ data and simply include them  in an NLO fit
based on the
FONLL-B scheme with a fitted charm PDF we can fit them perfectly, with a
$\chi^2$ per data point equal to $\chi^2/N_{\rm dat}=1.09$. In other
words, regardless of their reliability, the EMC data provide us with
an interesting test-case scenario which demonstrates that the perturbative
treatment of charm in current PDF fits may fail to satisfy the
accuracy standards that are required in order to match the  high
precision  that current PDF uncertainties suggest.

When  charm is fitted on the same footing as the
other light PDFs, we find a small but non-negligible
general improvement in global
fit quality, and a very significant improvement in the description of
large $x$ charm structure function data from EMC, which cannot be
fitted otherwise. The dependence of the charm PDF on the
value of the charm mass is significantly reduced, and 
there is also a more modest reduction in the charm mass dependence 
of light quark PDFs.
We also find that while with fitted charm  
overall uncertainties on gluon-induced LHC cross-sections are 
a little more conservative, the charm mass dependence of quark-induced 
processes can be reduced at central rapidity and low $p_T$.
This suggests that the fitted charm PDF will lead to more
reliable phenomenology at the LHC, eliminating a possible source of
bias from assumptions about the origin of charm and the value 
of the charm mass.

An immediate consequence of our results is  that existing determinations
of the charm quark mass from deep-inelastic structure
functions~\cite{Gao:2013wwa,Abramowicz:1900rp,Alekhin:2012un,
  Behnke:2015qja,Alekhin:2012vu,xFitter::2016pbr} might be affected by underestimated theory
uncertainties due to the assumption that charm is generated
perturbatively.
With this motivation, we plan to perform in the near future a direct determination
of the charm mass in the global NNPDF analysis both with fitted and
with perturbative charm, using the same approach as for the determination
of the strong coupling
constant~\cite{Lionetti:2011pw,Ball:2011us}.

Inclusion of a
fitted charm PDF is planned for future general-purpose global PDF sets
from the NNPDF collaboration. Further measurements which might constrain 
fitted charm, in particular 
$Z+c$ and $c\bar{c}$ production at high $p_T$ and high rapidity, are expected 
at LHC Run 2.  We expect the accuracy of the charm determination to
improve substantially in the near future, and the issue of the
reliability of the EMC data to be finally settled by these measurements.

\bigskip
\bigskip
\begin{center}
\rule{5cm}{.1pt}
\end{center}
\bigskip
\bigskip

The NLO PDFs presented here  are available in the
{\tt LHAPDF6} format~\cite{Buckley:2014ana} from the NNPDF
{\tt HepForge} webpage:
\begin{center}
\url{https://nnpdf.hepforge.org/html/nnpdf3ic/nnpdf3ic.html}
\end{center}
In particular, we make available the following PDF sets:

\begin{itemize}

  \item PDF sets with fitted charm, for three different values
  of the pole charm mass:
\begin{flushleft}
  \tt NNPDF3\_IC\_nlo\_as\_0118\_mcpole\_1330\\
  NNPDF3\_IC\_nlo\_as\_0118\_mcpole\_1470\\
  NNPDF3\_IC\_nlo\_as\_0118\_mcpole\_1610
\end{flushleft}

\item PDF sets with identical theory settings as those above, 
  with the only differences being that the charm PDF is perturbatively
  generated and that the EMC data are excluded, for the same three values
  of the charm mass:
\begin{flushleft}
  \tt NNPDF3\_nIC\_nlo\_as\_0118\_mcpole\_1330\\
  NNPDF3\_nIC\_nlo\_as\_0118\_mcpole\_1470\\
  NNPDF3\_nIC\_nlo\_as\_0118\_mcpole\_1610
  \end{flushleft}

\item A PDF set with fitted charm and the central value of the charm
  quark pole mass $m_c^{\rm pole}=1.47$ GeV
  without the EMC charm data included:
  \begin{flushleft}
  \tt  NNPDF3\_IC\_nlo\_as\_0118\_mcpole\_1470\_noEMC\\
  \end{flushleft}

\item PDF sets with fitted charm, for three different values
  of the running $\overline{{\rm MS}}$ charm mass:
\begin{flushleft}
  \tt NNPDF3\_IC\_nlo\_as\_0118\_mc\_1150\\
  NNPDF3\_IC\_nlo\_as\_0118\_mc\_1275\\
  NNPDF3\_IC\_nlo\_as\_0118\_mc\_1400
\end{flushleft}

\item  PDF sets with identical theory settings as those above, 
  with the only differences being that the charm PDF is perturbatively
  generated and that the EMC data are excluded, for the same three values
  of the charm mass:
\begin{flushleft}
  \tt NNPDF3\_nIC\_nlo\_as\_0118\_mc\_1150\\
  NNPDF3\_nIC\_nlo\_as\_0118\_mc\_1275\\
  NNPDF3\_nIC\_nlo\_as\_0118\_mc\_1400
  \end{flushleft}

\item A PDF set with fitted charm and central theory settings but
  without the EMC charm data included:
  \begin{flushleft}
  \tt NNPDF3\_IC\_nlo\_as\_0118\_mc\_1275\_noEMC
  \end{flushleft}

\end{itemize}

These PDF sets are not meant to be used for general-purpose
applications, for which the NNPDF3.0 PDF sets are still recommended,
but rather for  studies related to the charm content of the proton.
Specifically,  
fitted charm PDFs should always be compared with the corresponding
baseline fits presented in this publication, in order to have a consistent comparison
 of two PDF sets with identical theory and methodological settings and only
 differing in the treatment of charm and the inclusion or not of the EMC
 charm data.

\noindent
\subsection*{Acknowledgements}

We thank
J.~Feltesse and A.~Cooper-Sarkar for comments on the EMC data,
P.~Nadolsky and R.~Thorne for several stimulating comments on many aspects of our
results, and especially  M.~Arneodo for detailed information on
the EMC charm structure function data.

R.~D.~B.\ is funded by an STFC Consolidated Grant
ST/J000329/1;
S.~C.\ is supported by the HICCUP ERC Consolidator grant (614577);
S.~F.\ is  supported by the Executive Research Agency (REA) of the European
Commission under the Grant Agreement PITN-GA-2012-316704 (HiggsTools);
J.~R.\ is supported by an STFC Rutherford Fellowship and
Grant ST/K005227/1 and ST/M003787/1;
V.~B., M.~B., N.~H.,  J.~R.\  and L.~R.\ are
supported by an European Research Council Starting Grant "PDF4BSM".
%


\bibliography{nnpdf31IC.bbl}

\end{document}